\documentclass{aa}  
\usepackage{natbib}

\usepackage{graphicx}
\usepackage{txfonts}
\usepackage{amssymb}
\begin{document}

   \title{Structure of Herbig AeBe disks at the milliarcsecond scale}   \subtitle{A statistical survey in the H band using PIONIER-VLTI}
   
%======================================================================================
   
\author{
        B.~Lazareff          \inst{\ref{ipag}}
        \fnmsep\thanks{email: \texttt{bernard.lazareff@univ-grenoble-alpes.fr}}
\and    J.-P.~Berger         \inst{\ref{eso},\ref{ipag}}
        \fnmsep\thanks{email: \texttt{jean-philippe.berger@univ-grenoble-alpes.fr}}
\and    J.~Kluska            \inst{\ref{exeter}}
\and    J.-B.~Le Bouquin     \inst{\ref{ipag}}
\and    M.~Benisty           \inst{\ref{ipag}}
\and    F.~Malbet            \inst{\ref{ipag}}
\and    C.~Koen              \inst{\ref{uwc}}
\and    C.~Pinte             \inst{\ref{umifca}}
\and    W.-F.~Thi            \inst{\ref{mpe}}
\and    O.~Absil             \inst{\ref{ul}}\fnmsep\thanks{F.R.S.-FNRS Research Associate}
\and    F.~Baron             \inst{\ref{georgia}}
\and    A.~Delboulbé         \inst{\ref{ipag}}
\and    G.~Duvert            \inst{\ref{ipag}}
\and    A.~Isella            \inst{\ref{rice}}
\and    L.~Jocou             \inst{\ref{ipag}}
\and    A.~Juhasz            \inst{\ref{instast}}
\and    S.~Kraus             \inst{\ref{exeter}}
\and    R.~Lachaume          \inst{\ref{pontifica},\ref{mpia}}
\and    F.~Ménard            \inst{\ref{ipag}}
\and    R.~Millan-Gabet      \inst{\ref{ipac},\ref{nasaexo}}
\and    J.~D.~Monnier        \inst{\ref{umich}}
\and	T.~Moulin            \inst{\ref{ipag}}
\and    K.~Perraut           \inst{\ref{ipag}}
\and	S.~Rochat            \inst{\ref{ipag}}
\and    F.~Soulez            \inst{\ref{cral},\ref{epfl}}
\and    M.~Tallon            \inst{\ref{cral}}
\and    E.~Thiébaut          \inst{\ref{cral}}
\and    W.~Traub             \inst{\ref{jpl}}
\and    G.~Zins              \inst{\ref{esochile}}
}

\institute{
Univ. Grenoble Alpes, IPAG, F-38000 Grenoble, France
CNRS, IPAG, F-38000 Grenoble, France 
\label{ipag}
\and
ESO
Karl-Schwarzschild-Strasse 2
D-85748 Garching bei München, Germany
\label{eso}
\and
University of Exeter
Department of Physics and Astronomy
Stocker Road, 
Exeter, Devon EX4 4QL
UK      \label{exeter}
\and
Department of Statistics, 
University of the Western Cape, 
Private Bag X17, 
7535 Bellville, 
South Africa
\label{uwc}
\and
STAR, Université de Liège, 19c allée du Six Août, 4000 Liège, Belgium
\label{ul}
\and
UMI-FCA, 
CNRS / INSU France, 
and Departamento de Astronomía,
Universidad de Chile, 
Casilla 36-D Santiago, 
Chile
\label{umifca}
\and
Max-Planck-Institut für Extraterrestrische Physik, 
85748 Garching, Germany
\label{mpe}
\and
Center for High Angular Resolution Astronomy, 
Georgia State University, PO Box 3969, Atlanta, GA 30302, USA
\label{georgia}
\and
Institute of Astronomy, University of Cambridge, 
Madingley Road, Cambridge CB3 0HA, UK
\label{instast}
\and
Department of Physics and Astronomy, Rice University, 
6100 Main Street, Houston, TX 77005, USA
\label{rice}
\and
Centro de Astroingeniería, Instituto de Astrofísica, 
Facultad de Física, Pontificia Universidad Católica de Chile, 
Casilla 306, Santiago 22, Chile
\label{pontifica}
\and
Max-Planck-Institut für Astronomie, Königstuhl 17, 
D-69117 Heidelberg, Germany
\label{mpia}
\and
Department of Astronomy, University of Michigan, 
1085 S. University Ave, 311 West Hall, 
Ann Arbor, MI 48109, USA
\label{umich}
\and
Infrared Processing and Analysis Center, California
Institute of Technology, Pasadena, CA 91125, USA
\label{ipac}
\and
NASA Exoplanet Science Institute,
California Institute of Technology, 
770 S. Wilson Ave., Pasadena, CA 91125, USA
\label{nasaexo}
\and
CRAL, Observatoire de Lyon, CNRS, Univ. Lyon 1, 
École Normale Supérieure de Lyon, 69364 Lyon, France
\label{cral}
\and
Biomedical Imaging Group, 
Ecole polytechnique fédérale de Lausanne, Switzerland
\label{epfl}
\and
Jet Propulsion Laboratory
M/S 321-100
4800 Oak Grove Drive 
Pasadena, CA 91109, USA
\label{jpl}
\and
ESO Vitacura,
Alonso de Córdova 3107, 
Vitacura, Casilla 19001, 
Santiago, Chile 
\label{esochile}
}
   
% ===========================================================================================   

\date{Received 15 July 2016 / accepted 04 November 2016}

\abstract
{It is now generally accepted that the near-infrared excess of Herbig AeBe stars originates in the dust of a circumstellar disk.  }
{The aims of this article are to infer the radial and vertical structure  of these disks at scales of order  $1\,\mathrm{au}$, and the properties of the dust grains.}
% BL
{The program objects (51 in total) were observed with the H-band (1.6$\mu$m) PIONIER/VLTI interferometer. The largest baselines allowed us to resolve (at least partially) structures of a few tenths of an au at typical distances of a few hundred parsecs. Dedicated UBVRIJHK photometric measurements were also obtained. Spectral and 2D geometrical parameters are extracted via fits of a few simple models: ellipsoids and broadened rings with azimuthal modulation. Model bias is mitigated by parallel fits of physical disk models. Sample statistics were evaluated against similar statistics for the physical disk  models to infer properties of the sample objects as a group.}
{We find that dust at the inner rim of the disk has a sublimation temperature $T_\mathrm{sub}\approx 1800\mathrm{K}$. A ring morphology is confirmed for approximately half the resolved objects; these rings are wide $\delta\,r/r\geq 0.5$.  A wide ring favors a rim  that,  on the star-facing side, looks more like a knife edge than a doughnut. The data are also compatible with a the combination of a narrow ring and an inner disk of unspecified nature inside the dust sublimation radius.  The disk inner part has a thickness $z/r\approx 0.2$,  flaring to $z/r\approx 0.5$ in the outer part. 
We confirm the known luminosity-radius relation; a simple physical model is consistent with both the mean luminosity-radius relation and the ring relative width; however, a significant spread around the mean relation is present. In some of the objects we find a halo   component, fully resolved at the shortest interferometer spacing, that is related to the HAeBe class.}
{}

%http://www.aanda.org/component/content/article?id=170

%       \thanks{Based on observations at ESO with large programme 190.C-0963}
        
   \maketitle

\section{Introduction}

Herbig Ae Be stars are intermediate-mass ($2 - 8 M_{\odot}$) pre-main
sequence stars. They are surrounded by accretion disks that are
believed to be the birth places of planetary systems.  The remarkable diversity of discovered exo-planetary
systems has renewed the strong interest in understanding the structure
of such disks at the astronomical unit (au) scale. The sublimation front at the inner rim  contributes to define the physical conditions  in the dusty planet-forming region. 

\citet{1992ApJ...397..613H,1992ApJ...393..278L} have recognized that
the spectral energy distributions (SED) of Herbig Ae Be stars display
near-infrared excesses in the $2$--$7\, \mu\mathrm{m}$ region that are too
strong to be reproduced with the flaring hydrostatic disk models
that are successfully applied to T Tauri stars \citep[see
e.g.,][]{1987ApJ...323..714K,1997ApJ...490..368C}. 
The initial proposition by \citet{1992ApJ...397..613H} that
accretion heating could provide the near infra-red (NIR)
 excess flux can be ruled out on
the arguments that, firstly,~such accretion rates are not measured and secondly,~they
are incompatible with the presence of an optically thin inner cavity
\citep{1993ApJ...407..219H}. 

The first survey of Herbig Ae/Be stars using near-infrared optical
interferometry by \citet{2001ApJ...546..358M} concluded that the
near-infrared emission size was indeed incompatible with the standard
flared disk models. In order to explain both interferometric and
photometric measurements \citet{2001A&A...371..186N} have proposed
that the rim vertical structure had to be
taken into account in the global stellar energy reprocessing
balance. Such a rim is frontally illuminated and therefore naturally
hotter. It should emit at the dust sublimation temperature and puff up
\citep{2001ApJ...560..957D}. Aperture-masking observations of a
massive young star by \citet{2001Natur.409.1012T} revealed a doughnut
shape NIR emission and offered a first confirmation of this
idea. \citet{2002ApJ...579..694M,2005ApJ...624..832M} assembled
the available interferometric data to confirm that the measured sizes 
were well correlated with the central stars'
luminosities and consistent with the  dust sublimation radius. 
This trend was observed over more than four decades of
stellar luminosity. The advent of spectrally-resolved interferometry
has further revealed the presence of an additional hot component emitting from within the sublimation radius in a significant sample of objects \citep{2007A&A...464...43M,2007A&A...464...55T, 2008ApJ...676..490K, 2008A&A...489.1157K, 2008A&A...483L..13I,2008ApJ...677L..51T,2009ApJ...692..309E,2014MNRAS.443.1916E}.

Several teams have proposed alternative explanations for the  $2$-$7\,\mu$m excess: an extended
(spherical) envelope \citep{1993ApJ...407..219H, 2006ApJ...636..348V},
disk wind \citep{2007ApJ...658..462V, 2012ApJ...758..100B}, magnetically lifted
 grains  \citep{2012ApJ...745...60K}, and
supported disk atmosphere \citep{2014ApJ...780...42T}.

The availability of spatially resolved observations of the inner rims
has prompted research on the processes that might structure them.
\citet{2005A&A...438..899I} have shown that the
dependence of dust sublimation temperature on gas density causes the sublimation
front to be curved. \citet{2007ApJ...661..374T} pointed out that
the dependence on grain size of cooling efficiency and
dust settling results in a broader NIR emission region. \citet{2009A&A...506.1199K} and \citet{2016arXiv160404601F}, taking into account the pressure dependence of the sublimation temperature, but not grain settling, find rim profiles similar to  \citet{2007ApJ...661..374T}, that is, a wedge pointing to the star.  
\citet{2014A&A...566A.117V} addressed the issue of the thickness of the rim;  considering a comprehensive list of processes such as viscous
heating, accretion luminosity, turbulent diffusion, dust settling, gas
pressure, he concluded that the resulting inner rim thickness was too small to explain the infrared excess. The reader is referred to
\citet{2010ARA&A..48..205D} or \cite{2015Ap&SS.357...97K} for  reviews of the structure of the inner circumstellar disks. 

The goal of  this article is to provide a
statistical view of the milli-arcsecond morphology of the
near-infrared emission around Herbig AeBe stars. For that purpose, we
took advantage of the improved performance offered by the visitor
instrument PIONIER (sensitivity, precision and efficiency) at the Very
large Telescope Interferometer (VLTI). We established a Large Program
(090C-0963) that had the following questions in mind:
\begin{enumerate}
\item Can we constrain the radial and vertical structure of the inner disk?
\item Can we constrain the rim dust temperature and composition?
\item Can we constrain the presence of additional emitting structures
  (e.g., hot inner emission, envelopes)
\item Can we reveal deviations from axisymmetry in the emission that could be linked to radiative transfer effects?
\end{enumerate}

This article is organized as follows. Section \ref{observations} introduces the sample and the immediate data reduction, that produces calibrated data. In  Section \ref{dataprocess}, we submit the observed data to simple model fits, in a systematic and uniform way, condensing  hundreds of visibilities and phase closures into a dozen or fewer parameters for each object. In Section \ref{discussion} we use the processed data to infer some of the physical properties of the sample objects, striving to avoid  possible biases. In the conclusion, we wrap up our main findings, place them into perspective with current models, and consider the outlook for follow-up work.

\section{Observations}

\label{observations}

\subsection{Instruments}
The interferometric observations were made using the four 1.8m auxiliary telescopes (AT) of the VLTI (see \citet{2014SPIE.9146E..0JM})  at ESO's Paranal observatory, and the PIONIER instrument \citep{2011A&A...535A..67L}, a four-telescope recombiner operating in H-band ($1.55$–$1.80\,\mu$m). The three AT configurations available at the epoch of the observations provided horizontal baselines in the range $11$m-$140$m. The photometric observations were obtained with two telescopes at the South African Astronomical Observatory (SAAO) Sutherland Observatory: 0.50m for UBVRI bands, and 0.75m for JHK bands. 

\subsection{Sample}
\label{sample}

Our sample is based on the lists of \citet{1998A&A...331..211M}, \citet{1994A&AS..104..315T}, and \citet{2003AJ....126.2971V}. One should note that a)~the definition of HAeBe objects does not rest on quantitative criteria; b)~our list of targets does not meet any completeness criterion. Our sample is limited at $m_\mathrm{H}\sol 8$, where the fainter objects are intended to test the sensitivity limit of PIONIER/VLTI. The program objects are listed in Table~\ref{table_objectlist}. 

\subsubsection*{Spectral types.}  
\begin{figure}
 \includegraphics[width=85mm]{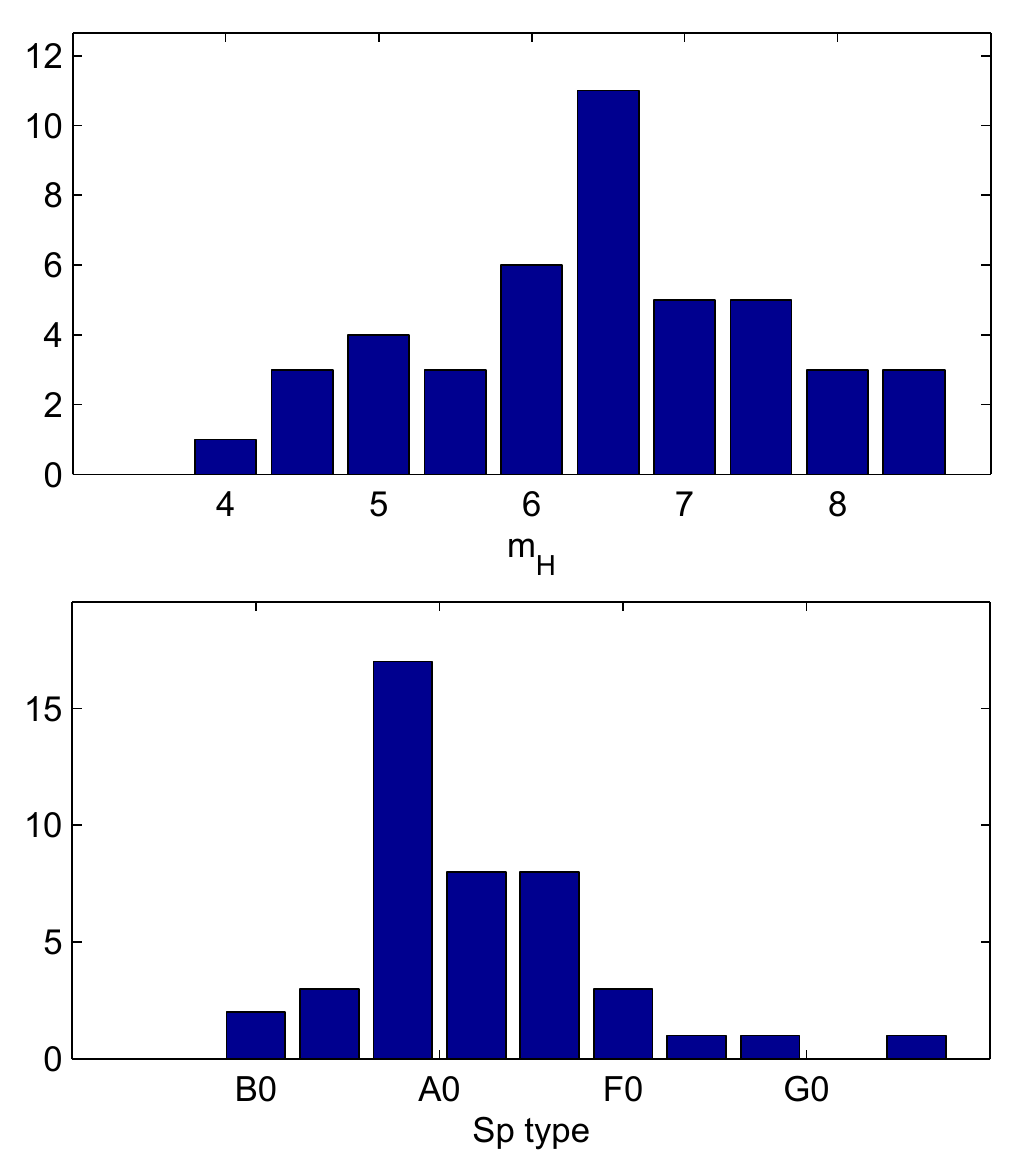}
\caption{Statistics of  H magnitudes (top) of program objects  and their spectral types (bottom).}
\label{hmagspstats}
\end{figure}

They were obtained from the SIMBAD database and checked against the Catalog of Stellar Spectral Classifications (Skiff, 2009-2016)\footnote{http://vizier.u-strasbg.fr/viz-bin/Cat?B/mk}; in case of discrepancy, we used a recent measurement from the compilation of Skiff, and we quote the original reference. When available, we adopted the determination by \citet{2015MNRAS.453..976F}, that also includes determinations of bolometric luminosity (the values of $T_\mathrm{eff}$ were converted to spectral types using the tables of \citet{2013ApJS..208....9P}). Statistics of H magnitude and spectral types are shown Fig.~\ref{hmagspstats}. 

\subsubsection*{Distances.} Parallactic distances were obtained from the Gaia \citep{GaiaInPress} and Hipparcos \citep{1997A&A...323L..49P} catalogs, and, failing that, from detailed spectroscopic studies; see notes of 
Table~\ref{table_objectlist}. 

\subsubsection*{Binarity.} For T Ori see \citet{1994ASPC...62...55S}; for HD~53367 see \citet{2006A&A...452..551P};  for HD~104237 see \citet{2013MNRAS.431.3485C}; for AK~Sco see \citet{1989A&A...219..142A}; 
concerning VV~Ser, \citet{2004ApJ...613.1049E} mention that their data do not rule out a binary model; binarity is not supported by our (more extensive) data. HD~36917 is often quoted as SB2, referring to \citet{1976PASP...88..712L}, who actually mention a composite spectrum, which is not confirmed by \citet{2013MNRAS.429.1001A}. HD~145718 is listed by Simbad as an Algol binary. However, following \citet{2006A&A...457..581G} and references therein, this remains unconfirmed. 

\subsubsection*{Coordinates of HD~56895.} We provide the following information for the record, and to avoid others a similar mistake. We list in Table~\ref{HD56895} the coordinates for HD~56895 or HD~56895B from various sources. It appears that the object listed by \citet{1998A&A...331..211M} as HD~56895B  has the coordinates (within a couple of arcsec) and the V magnitude of HD~56895, while its NIR magnitudes are those of a nearby (13\arcsec) object with much redder colors. Our interferometric observations were made at the position of HD~56895, not what was intended. As a consequence, this object is not included in the data analysis. 

\begin{table*}[p]
\caption{ 
List of objects in the present survey. 
}
\label{table_objectlist}
\centering
\begin{tabular}{llllcrrcclll}
\hline \hline
\multicolumn{1}{|c|}{\rule{0pt}{3ex}Object} & \multicolumn{1}{|c|}{RA (2000)}   & \multicolumn{1}{|c|}{Dec (2000)}   & 
\multicolumn{2}{|c|}{Spectral type}& \multicolumn{3}{|c|}{Distance} & 
\multicolumn{2}{|c|}{HAeBe Status}& \multicolumn{1}{|c|}{nHQ} & \multicolumn{1}{|c|}{Bin}    \\
\multicolumn{1}{|c|}{}            & \multicolumn{1}{|c|}{}    & \multicolumn{1}{|c|}{}     &
\multicolumn{1}{|c}{Value}        & \multicolumn{1}{c|}{Ref.} & 
\multicolumn{1}{|c}{Value} & \multicolumn{1}{c}{$\pm$} & \multicolumn{1}{c|}{Ref.} &
\multicolumn{1}{|c}{Ref.}   & \multicolumn{1}{c|}{Group} &
\multicolumn{1}{|c|}{} & \multicolumn{1}{|c|}{} \\
\multicolumn{1}{|c|}{(1)}            & \multicolumn{1}{|c|}{(2)}    & \multicolumn{1}{|c|}{(3)}     &
\multicolumn{1}{|c}{(4)}        & \multicolumn{1}{c|}{(5)} & 
\multicolumn{1}{|c}{(6)} & \multicolumn{1}{c}{(7)} & \multicolumn{1}{c|}{(8)} &
\multicolumn{1}{|c}{(9)}   & \multicolumn{1}{c|}{(10)} &
\multicolumn{1}{|c|}{(11)} & \multicolumn{1}{|c|}{(12)} \\
\hline 
\noalign{\vskip 1mm}                              
HD 17081    & 02 44 07.35 & -13 51 31.3 & B7   &       & 135 &   16 &  Hi & M98 & Ia  & Ur    &      \\
AB AUR      & 04 55 45.85 & +30 33 04.3 & A0   &       & 153 &   10 &  Ga & M98 &     &       &      \\
HD 31648    & 04 58 46.27 & +29 50 37.0 & A5   &       & 142 &    7 &  Ga & M98 & IIa & Sc    &      \\
HD 293782   & 05 04 29.99 & -03 47 14.3 & A3   &  F15  & 346 &   34 &  Ga & T94 &     & Bb    &      \\
HD 34282    & 05 16 00.48 & -09 48 35.4 & A0   &  F15  & 325 &   31 &  Ga & M98 & Ib  & Bb    &      \\
CO ORI      & 05 27 38.34 & +11 25 39.0 & F7   &       & 435 &   93 &  Ga & (u) &     & Bb,Sc &      \\
HD 35929    & 05 27 42.79 & -08 19 38.4 & F1   &  F15  & 385 &   38 &  Ga & M98 & IIa & Ur    &      \\
MWC 758     & 05 30 27.53 & +25 19 57.1 & A5   &  S06  & 151 &    9 &  Ga & M98 & Ia  & Sc,Ur &      \\
HD 244604   & 05 31 57.25 & +11 17 41.5 & A2   &  F15  & 433 &   64 &  Ga & M98 & IIa & Bb    &      \\
HD 36917    & 05 34 46.98 & -05 34 14.6 & B9   &       & 495 &   64 &  Ga & M98 & IIb & Ur    &      \\
T ORI       & 05 35 50.44 & -05 28 34.9 & A2   &  F15  & 750 &  141 & F15 & T94 &     &       & E    \\
HD 37258    & 05 36 59.25 & -06 09 16.4 & A0   &  F15  & 424 &   23 & F15 & M98 & IIa & Bb    &      \\
HD 37357    & 05 37 47.08 & -06 42 30.3 & A0   &  F15  & 344 &   10 & F15 & M98 & IIa & Bb    & 2FP  \\
V 1247 ORI  & 05 38 05.25 & -01 15 21.7 & A7   &  F15  & 319 &   27 &  Ga & V03 &     & Bb    &      \\
HD 37411    & 05 38 14.51 & -05 25 13.3 & A0   &  F15  & 358 &  108 & F15 & M98 & IIb & Bb    &      \\
HD 37806    & 05 41 02.29 & -02 43 00.7 & A1   &  G98  & 420 &   49 &  Ga & M98 & IIa &       &      \\
HD 38087    & 05 43 00.57 & -02 18 45.4 & B3   &       & 199 &   75 &  Hi & M98 &     &       & VHA  \\
HD 39014    & 05 44 46.38 & -65 44 07.9 & A7   &       &  44 &    1 &  Hi & M98 &     & Ur    &      \\
HD 250550   & 06 01 59.99 & +16 30 56.7 & B9   &  F15  & 637 &  154 &  Ga & T94 & Ia  & Bb    &      \\
HD 45677    & 06 28 17.42 & -13 03 11.1 & B2   &       & 610 &  123 &  Ga & T94 &     &       &      \\
HD 259431   & 06 33 05.19 & +10 19 20.0 & B6   &       & 610 &  115 &  Ga & T94 & Ia  &       &      \\
MWC 158     & 06 51 33.40 & -06 57 59.5 & B8   &  G98  & 340 &   47 &  Ga & M98 & IIa &       &      \\
HD 53367    & 07 04 25.53 & -10 27 15.7 & B0.5 &  F15  & 247 &   82 &  Hi & T94 &     &       & SB1  \\
HD 56895    & 07 18 31.79 & -11 11 33.9 & F0   &       & 167 &    8 &  Ga & M98 &     & Ur    &      \\
HD 58647    & 07 25 56.10 & -14 10 43.6 & B9   &       & 292 &   38 &  Ga & M98 & IIa &       &      \\
HD 85567    & 09 50 28.54 & -60 58 03.0 & B8   &  F15  & 907 &  164 & F15 & M98 & IIa &       &      \\
HD 95881    & 11 01 57.62 & -71 30 48.4 & B9.5 &  F15  & 935 &  227 &  Ga & M98 & IIa &       &      \\
HD 97048    & 11 08 03.32 & -77 39 17.5 & B9.5 &  F15  & 179 &    8 &  Ga & T94 & Ib  &       &      \\
HD 98922    & 11 22 31.67 & -53 22 11.5 & B9.5 &  F15  & 649 &  156 &  Ga & M98 & IIa &       &      \\
HD 100453   & 11 33 05.58 & -54 19 28.5 & F0   &  F15  & 103 &    3 &  Ga & M98 & Ib  &       &      \\
HD 100546   & 11 33 25.44 & -70 11 41.2 & A0   &  F15  & 109 &    4 &  Ga & M98 & Ia  &       &      \\
HD 104237   & 12 00 05.08 & -78 11 34.6 & A6   &  F15  & 104 &    3 &  Ga & M98 & IIa &       & SB2  \\
HD 139614   & 15 40 46.38 & -42 29 53.5 & A7   &  F15  & 131 &    5 &  Ga & M98 & Ia  &       &      \\
HD 141569   & 15 49 57.75 & -03 55 16.3 & A0   &  F15  & 111 &    5 &  Ga & M98 & IIb & Sc,Ur &      \\
HD 142666   & 15 56 40.02 & -22 01 40.0 & A8   &  F15  & 150 &    5 &  Ga & T94 & IIa &       &      \\
HD 142527   & 15 56 41.89 & -42 19 23.3 & F5   &  F15  & 156 &    6 &  Ga & M98 & Ia  &       &      \\
HD 143006   & 15 58 36.91 & -22 57 15.3 & G8   &       & 166 &   10 &  Ga & M98 &     &       &      \\
HD 144432   & 16 06 57.95 & -27 43 09.8 & A8   &  F15  & 253 &   95 &  Hi & M98 &     &       &      \\
HD 144668   & 16 08 34.29 & -39 06 18.3 & A3   &  F15  & 208 &   38 &  Hi & M98 & IIa &       &      \\
HD 145718   & 16 13 11.59 & -22 29 06.7 & A6   &  F15  & 149 &    6 &  Ga & V03 &     &       &      \\
HD 149914   & 16 38 28.65 & -18 13 13.7 & B9.5 &       & 149 &   13 &  Ga & M98 &     & Ur    &      \\
HD 150193   & 16 40 17.92 & -23 53 45.2 & A2   &  F15  & 145 &    6 &  Ga & M98 & IIa &       &      \\
AK SCO      & 16 54 44.85 & -36 53 18.6 & F7   &  F15  & 143 &    5 &  Ga & T94 &     &       & SB2  \\
HD 158643   & 17 31 24.95 & -23 57 45.5 & A0   &       & 131 &   15 &  Hi & M98 &     &       &      \\
HD 163296   & 17 56 21.29 & -21 57 21.9 & A1   &  F15  & 117 &    4 &  Ga & M98 & IIa &       &      \\
HD 169142   & 18 24 29.78 & -29 46 49.4 & F1   &  M15  & 145 &   43 & V05 & M98 & Ib  &       &      \\
MWC 297     & 18 27 39.53 & -03 49 52.1 & B1.5 &  F15  & 170 &   34 & F15 & T94 &     &       &      \\
VV SER      & 18 28 47.87 & +00 08 39.8 & B6   &H04;M09& 330 &      & A04 & T94 & IIa &       &      \\
R CRA       & 19 01 53.66 & -36 57 07.9 & B5   &       & 130 &      & A04 & T94 &     &       &      \\
HD 179218   & 19 11 11.25 & +15 47 15.6 & A0   &       & 293 &   31 &  Ga & M98 & Ia  & Or    &      \\
HD 190073   & 20 03 02.51 & +05 44 16.7 & A2   &       & 855 &  351 &  Ga & M98 & IIa &       &      \\
 \hline
\end{tabular}
\tablefoot{
Columns: (5) spectral type from Simbad except as noted; (9) reference establishing that object belongs to HAeBe type; (10) group following \citet{2010ApJ...721..431J}; (11) cause of exclusion from HQ sub-sample \emph{Ur} unresolved (i.e. all $V^2\geq 0.8$); \emph{Sc}, only one baseline configuration; \emph{Bb}, only in broadband mode; \emph{Or}, over-resolved; (12) evidence for binarity: E, eclipsing; 2FP, double fringe packet; VHA, rapid variation of $V^2$ versus HA; SB1, SB2, usual meaning. 
References: F15 \citet{2015MNRAS.453..976F}; S06 \citet{2006A&A...458..173S}; G98 \citet{1998AJ....116.2530G}; 
H04 \citet{2004AJ....127.1682H}; M09 \citet{2009A&A...495..901M}; Hi (Hipparcos) \citet{1997A&A...323L..49P}; 
Ga (Gaia) \citet{GaiaInPress}; V05 \citet{2005A&A...437..189V}; A04 \citet{2004A&A...426..151A}; 
M98 \citet{1998A&A...331..211M}; T94 \citet{1994A&AS..104..315T}; V03 \citet{2003AJ....126.2971V};
(u) uncertain, see for example, \citet{2012ApJ...744..121Y}, \citet{2004AJ....128.1294C}, \citet{2002A&A...384.1038E}.
}
\end{table*}

\begin{table*}[ht]
\caption{Coordinates for HD~56895~(B) from various literature sources.}
\label{HD56895}
\centering
\begin{tabular}{llllllll}
\hline
Catalog & Name          & RA (2000)             & Dec (2000)    
        & V             & J             & H             & K             \\
\hline
Simbad  & HD 56895      & 07 18 31.792  & -11 11 33.92  & 8.42 \\
Simbad  & MC79 2-5      & 07 18 30.979  & -11 11 40.73  & 10.5 \\
2MASS   & HD 56895      & 07 18 31.799  & -11 11 33.90  &       & 7.632 & 7.526   & 7.427 \\
2MASS   & 07183097-1111407              
                                        & 07 18 30.979  & -11 11 40.73  &         & 6.465 & 5.049 & 4.171 \\
Malfait+1998    & HD 56895B 
                                        & 07 18 31.82   & -11 11 34.8   & 8.40    & 6.12  & 4.56  & 3.56  \\
This work               & HD 56895              & 07 18 31              & -11 11 33               \\
\hline
\end{tabular}
\tablefoot{
\centering
The coordinates from \citet{1998A&A...331..211M}, originally 1950.0, have been precessed to 2000.0.
}
\end{table*}

\subsection{Interferometric data}

The interferometric data were acquired with the PIONIER instrument  at the VLTI, during a total of 30 nights in six observing runs between December 2012 and June 2013, under Large Program 190.C-0963.  The log of observations is given in Appendix~\ref{logobs}.

We used all the available configurations offered by the ATs: A1--B2--C1--D0, D0--G1--H0--I1, A1--G1--J3--K0. Most of the data were dispersed over three spectral channels across the H band, providing a spectral resolving power of $\mathcal{R}\approx 15$, while the faintest objects were observed in broad H-band ($\mathcal{R}\approx 5$).

Data were reduced and calibrated with the \texttt{pndrs} package described in \citet{2011A&A...535A..67L}. Each observation block (OB) provides five consecutive files within a few minutes. Each file contains six visibilities squared  $V^2$ and four closure phases 
$\phi_\mathrm{cp}$ dispersed over the three spectral channels. Whenever possible, the five files were averaged together to reduce the final amount of data to be analyzed and to increase the signal-to-noise ratio. The statistical uncertainties typically range from $0.5\degr$ to $10\degr$ for the closure phases and from 2.5\% to 20\% for the  visibilities squared, depending on target brightness and atmospheric conditions.

Each observation sequence of one of our targets was immediately preceded and followed by the observation of a calibration star in order to master the instrumental and atmospheric response. \citet{2012SPIE.8445E..0IL} have shown that such calibration star(s) should be chosen close to the science object both in terms of position (within a few degrees) and magnitude (within $\pm1.5$ mag). We used the tool \texttt{SearchCal}\footnote{http://www.jmmc.fr/searchcal} in its FAINT mode \citep{2011A&A...535A..53B} to identify at least one suitable calibration star within a radius of 3 deg of each object of our sample.

Most of our objects are grouped into clusters in the sky. Consequently, the instrumental response could be cross-checked between various calibration stars. This allowed us to discover a few previously unknown binaries among the calibration stars. These have been reported to the bad calibrator list\footnote{http://apps.jmmc.fr/badcal}. We estimated the typical calibration accuracy to be $1.5\degr$ for the phase closures and 5\% for the  visibilities squared. The typical accuracy of the effective wavelength of the PIONIER channels is 2\%. Finally, the on-the-sky orientation of PIONIER has been checked several times and is consistent with the definition of \citet{2005PASP..117.1255P}.

\subsection{Photometry}

We made dedicated photometric measurements of most objects in our study, in the UBVRIJHK bands, at the  SAAO Sutherland 0.5~m and and 0.75~m telescopes (now decommissioned);   the time parallax relative to the interferometric observations is smaller (at most two years) than for  photometry available in the literature. Whenever SAAO measurements could not be made, previous measurements from the literature were used with weighting following their error estimates; see Tables~\ref{photcats} and~\ref{table_photomeas}.  The NOMAD catalog is not a primary source, and was used only for B and V magnitudes of MWC~297 and VV~Ser from the unpublished USNO YB6 catalog. The Wise measurements are used only for plotting on SED graphs. 

\begin{table*}[htbp]
\caption{Photometric measurements used to determine the SED of our program objects.}
\label{photcats}
\centering
\begin{tabular}{llll}
\hline\hline
Reference       & Bands & Key & Epoch  \\
\hline
\text{\citet{2003AJ....125..984M} (USNO)}       & B2 R2 I               & 1 & 1974 - 2000  \\
\text{\citet{2004AAS...205.4815Z} (NOMAD)}      & B V R                 & 2 & unknown              \\
\text{\citet{2000A&A...355L..27H} (TYCHO)}      & B V                   & 3 & 1989 - 1993  \\
\text{\citet{2003yCat.2246....0C} (2MASS)}      & J H K                 & 4 & 1998 - 2001  \\
\text{\citet{2012yCat.2311....0C} (WISE)}       & 3.3, 4.6, 12, 22 $\mu$m       & &2010    \\
\text{\citet{2001A&A...379..564O} (EXPORT)}     & U B V R I             & 5 & 1998 - 1999  \\
\text{\citet{1998A&A...331..211M}               }       & U B V J H K L M& 6 & 1989 - 1992  \\
\text{\citet{2012A&A...543A..59M}}                      & J H K L M     & 7 & 1992 - ?             \\
\text{\citet{2008ApJ...689..513T}}                      & U B V R I J H K       & 8 & 2004-2006    \\
SAAO 0.5m and 0.75m telescopes                          & U B V R I J H K       & & 2014   \\
\hline
\end{tabular}
\tablefoot{
SAAO measurements are used by default. WISE measurements are used only for plots but not for the SED fits. The Key field is a pointer from Table~\ref{table_photomeas}.
}
\end{table*}

\begin{table*}
\caption{Photometry data for the program objects.}
\label{table_photomeas}
\centering
\begin{tabular}{lrrrrrrrrlc}
\hline \hline
    Object  &     U &      B &      V &      R &      I &      J &      H &      K & Key & Var \\
\hline
     HD~17081 &       &  4.12 &  4.23 &  4.30 &  4.42 &  4.47 &  4.52 &  4.55 & 1, 3  & \\
       AB~AUR &  7.18 &  7.17 &  7.05 &  7.00 &  6.80 &  5.96 &  5.17 &  4.30 & 1, 3, 4, 8 & \\
     HD~31648 &  7.93 &  7.89 &  7.70 &  7.58 &  7.45 &  6.81 &  6.23 &  5.41 & 1, 3, 4, 5, 6, 7, 8 & \\
    HD~293782 & 10.30 & 10.04 &  9.76 &  9.62 &  9.45 &  8.86 &  8.00 &  7.12 & & * \\
     HD~34282 & 10.22 & 10.07 &  9.88 &  9.79 &  9.65 &  9.15 &  8.48 &  7.80 & & \\
       CO~ORI & 13.41 & 12.92 & 11.76 & 11.05 & 10.31 &  8.70 &  7.52 &  6.61 & & * \\
     HD~35929 &  8.69 &  8.54 &  8.11 &  7.85 &  7.59 &  7.29 &  6.90 &  6.58 & & \\
      MWC~758 &  8.74 &  8.61 &  8.30 &  8.13 &  7.92 &  7.31 &  6.49 &  5.70 & & \\
    HD~244604 &  9.77 &  9.64 &  9.43 &  9.32 &  9.18 &  8.69 &  7.98 &  7.23 & & \\
     HD~36917 &  8.22 &  8.17 &  8.00 &  7.86 &  7.66 &  7.33 &  7.10 &  6.66 & & \\
        T~ORI & 10.98 & 10.96 & 10.49 & 10.13 & 10.49 &  8.70 &  7.41 &  6.34 & & * \\
     HD~37258 &  9.78 &  9.72 &  9.59 &  9.51 &  9.40 &  8.94 &  8.37 &  7.68 & 4, 6 & \\
     HD~37357 &  9.07 &  9.02 &  8.89 &  8.83 &  8.74 &  8.37 &  7.92 &  7.33 & 4, 6 & \\
   V~1247~ORI & 10.31 & 10.22 &  9.88 &  9.68 &  9.44 &  8.88 &  8.20 &  7.41 & 4 & \\
     HD~37411 & 10.11 & 10.00 &  9.83 &  9.72 &  9.56 &  9.05 &  8.36 &  7.57 & 4, 6 & \\
     HD~37806 &  7.83 &  7.98 &  7.92 &  7.84 &  7.73 &  7.33 &  6.48 &  5.58 & & \\
     HD~38087 &  8.04 &  8.45 &  8.30 &  8.15 &  7.96 &  7.71 &  7.43 &  7.38 & & \\
     HD~39014 &  4.68 &  4.55 &  4.32 &  4.19 &  4.05 &  3.90 &  3.77 &  3.72 & & \\
    HD~250550 &  9.55 &  9.75 &  9.60 &  9.46 &  9.26 &  8.47 &  7.53 &  6.63 & 4 & \\
     HD~45677 &  6.81 &  7.43 &  7.39 &  7.21 &  7.11 &  6.87 &  6.09 &  4.48 & & * \\
    HD~259431 &  8.57 &  9.05 &  8.73 &  8.39 &  8.07 &  7.42 &  6.71 &  5.71 & 4, 6, 8 & \\
      MWC~158 &  6.34 &  6.66 &  6.64 &  6.56 &  6.48 &  6.01 &  5.22 &  4.32 & & \\
     HD~53367 &  6.86 &  7.47 &  7.00 &  6.65 &  6.23 &  5.81 &  5.54 &  5.31 & & * \\
     HD~56895 &  8.82 &  8.78 &  8.38 &  8.12 &  7.81 &  6.36 &  4.93 &  3.91 & & \\
     HD~58647 &  6.74 &  6.86 &  6.80 &  6.76 &  6.68 &  6.50 &  6.16 &  5.46 & & \\
     HD~85567 &  8.17 &  8.69 &  8.54 &  8.33 &  8.09 &  7.60 &  6.69 &  5.75 & & \\
     HD~95881 &  8.44 &  8.40 &  8.25 &  8.12 &  7.96 &  7.32 &  6.45 &  5.51 & & * \\
     HD~97048 &  9.20 &  8.94 &  8.55 &  8.30 &  8.02 &  7.42 &  6.85 &  6.10 & & \\
     HD~98922 &  6.78 &  6.80 &  6.75 &  6.69 &  6.61 &  6.19 &  5.40 &  4.45 & & \\
    HD~100453 &  8.12 &  8.09 &  7.79 &  7.62 &  7.44 &  7.11 &  6.55 &  5.78 & & \\
    HD~100546 &  6.59 &  6.68 &  6.67 &  6.65 &  6.66 &  6.51 &  6.18 &  5.69 & & \\
    HD~104237 &  7.02 &  6.92 &  6.67 &  6.51 &  6.32 &  5.81 &  5.17 &  4.56 & & \\
    HD~139614 &  8.55 &  8.52 &  8.27 &  8.12 &  7.97 &  7.68 &  7.22 &  6.62 & & \\
    HD~141569 &  7.25 &  7.22 &  7.10 &  7.04 &  6.94 &  6.78 &  6.62 &  6.55 & & \\
    HD~142666 &  9.39 &  9.17 &  8.67 &  8.34 &  8.00 &  7.49 &  6.68 &  5.92 & & \\
    HD~142527 &  9.25 &  9.01 &  8.29 &  7.85 &  7.40 &  6.58 &  5.78 &  5.09 & & \\
    HD~143006 & 11.21 & 10.94 & 10.12 &  9.63 &  9.17 &  8.33 &  7.52 &  6.86 & & \\
    HD~144432 &  8.65 &  8.54 &  8.18 &  7.95 &  7.69 &  7.16 &  6.53 &  5.94 & & \\
    HD~144668 &  7.76 &  7.53 &  7.16 &  6.90 &  6.63 &  5.82 &  5.12 &  4.37 & & \\
    HD~145718 &  9.53 &  9.24 &  8.78 &  8.51 &  8.21 &  7.79 &  7.31 &  6.75 & & \\
    HD~149914 &  7.12 &  7.02 &  6.74 &  6.54 &  6.30 &  5.89 &  5.75 &  5.70 & & \\
    HD~150193 &  9.67 &  9.36 &  8.81 &  8.43 &  7.97 &  6.93 &  6.06 &  5.29 & & \\
       AK~SCO &       &  9.81 &  9.17 &  8.72 &  8.38 &  7.67 &  7.03 &  6.38 & 1, 3 & \\
    HD~158643 &  4.79 &  4.82 &  4.80 &  4.77 &  4.72 &  4.63 &  4.56 &  4.28 & & \\
    HD~163296 &  6.93 &  6.93 &  6.84 &  6.77 &  6.69 &  6.24 &  5.46 &  4.61 & & \\
    HD~169142 &  8.38 &  8.42 &  8.13 &  7.95 &  7.74 &  7.42 &  7.07 &  6.86 & & \\
      MWC~297 & 15.53 & 14.53 & 12.23 & 10.39 &  8.97 &  6.26 &  4.46 &  3.06 & & \\
       VV~SER & 13.71 & 13.33 & 12.33 & 11.66 & 10.92 &  8.73 &  7.38 &  6.27 & & * \\
        R~CRA & 13.96 & 13.69 & 12.77 & 11.93 & 10.83 &  7.34 &  5.22 &  3.56 & & \\
    HD~179218 &  7.58 &  7.50 &  7.39 &  7.34 &  7.28 &  7.03 &  6.60 &  5.93 & & \\
    HD~190073 &  7.91 &  7.95 &  7.83 &  7.75 &  7.62 &  7.25 &  6.60 &  5.77 & & \\
 \hline
\end{tabular}
\tablefoot{
Unless otherwise mentioned, the data listed were acquired at the SAAO Sutherland observatory 0.5m and 0.75m telescopes in the scope of the present work. When such measurements are not available, we resort to published measurements as indicated by the \emph{Key} field, that is a pointer to  Table~\ref{photcats}. The \emph{Var} column signals with an asterisk objects with evidence for photometric variability.
}
\end{table*}

\section{Data processing}

\label{dataprocess}

In this section, we present steps of intermediate data processing, from reduced and calibrated data (magnitudes, visibilities, phase closures) to meaningful derived parameters. The first subsection below deals with model fits of photometric data, while the remaining subsections deal with model fitting of interferometric data. 

\subsection{Model fits of the spectral energy distribution}
\label{analysphotom}

For each object, we performed a least squares fit of the observed fluxes (actually, the logarithms) in the eight Johnson-Cousins \citep{1998A&A...333..231B} bands U, B,... K, according to a simple model of a stellar photosphere plus dust emission (modeled as a single temperature blackbody). The maximum of $\lambda F_{\lambda}$, for $T=1500\cdots 1800\mathrm{K}$ is at 
$\lambda\approx 2.4\cdots 2.0\mu\mathrm{m}$ in K band; the other, shorter $\lambda$ bands used in the SED fit discriminate against cooler dust. 

\begin{equation} \begin{split}
F_{\mathrm{mod}}(\nu_k) &= \big[
F_{\mathrm{s} V} \: 10^{0.4(m_{V}-m_{k})_\mathrm{s}} +\\
& F_{\mathrm{d} H} \: B(\nu_k,T_\mathrm{dp})
/ B(\nu_H,T_\mathrm{dp}) \big] \, 
10^{-0.4 A_V\, r_k} 
\end{split} ,\end{equation}
where
\begin{itemize}
\item $k = 1 \cdots 8$;
\item $(m_{V}-m_{k})_\mathrm{s}$ are the intrinsic stellar colors derived from the published (SIMBAD) spectral type and an interpolation in the tables of \citet{2013ApJS..208....9P};
\item $r_k$ is the extinction ratio $A_k / A_V$ from the extinction law of \citet{1989ApJ...345..245C} with $R=3$.
\end{itemize}
The fit parameters are 
\begin{itemize}
\item $F_{\mathrm{s} V}$, the absorption-free stellar flux in V band
\item $F_{\mathrm{d} H}$, the absorption-free dust flux in H band
\item $A_V$
\item $T_\mathrm{dp}$, the blackbody temperature of the \textbf{d}ust component as inferred from \textbf{p}hotometry
\end{itemize}
Three of these ($F_{\mathrm{s} V}$, $F_{\mathrm{d} H}$, $T_\mathrm{dp}$) will be used in the Discussion section; also, $f_\mathrm{d}$, the fraction of the observed flux at $1.63\,\mu$m contributed by the dust component, will be used in the interpretation of interferometric data, see Section~\ref{degeneracy} below. 

%A derived parameter of interest is $f_\mathrm{d}$, the fraction of the observed flux at $1.63\,\mu$m contributed by the dust component, which will be compared with a related quantity derived from interferometric data. The procedure and its result are independent of the distance of the object and of the star's luminosity class (except for a possible small dependence of \emph{colors} on luminosity class).

What value of the extinction ratio $R_V$ should we adopt? \citet{2004AJ....127.1682H}, in a study of 75 HAeBe objects, provide solid evidence for a value $R_V=5$ fitting the data better than the standard value $R_V=3.1$. This conclusion rests predominantly on objects with   a high visual extinction $1 \lessapprox A_V \lessapprox 6$. On the other hand, \citet{2009A&A...495..901M} conclude that $R_V=3.1$ results in a better fit to the SED of VV~Ser ($A_V\approx 3$) and adopt this value for other objects with lower extinction; \citet{2015MNRAS.453..976F} take a similar approach; see also \citet{2006A&A...456.1045B}, where among those 60 objects having $A_V\leq 1.0$, the preferred value of $R_V$ is 3.1 in 50 cases.  Our sample has an $A_V$ distribution with a  median value 0.36; accordingly, we decided to perform our analysis assuming $R_V = 3$. 

\begin{figure}
\includegraphics[width=85mm]{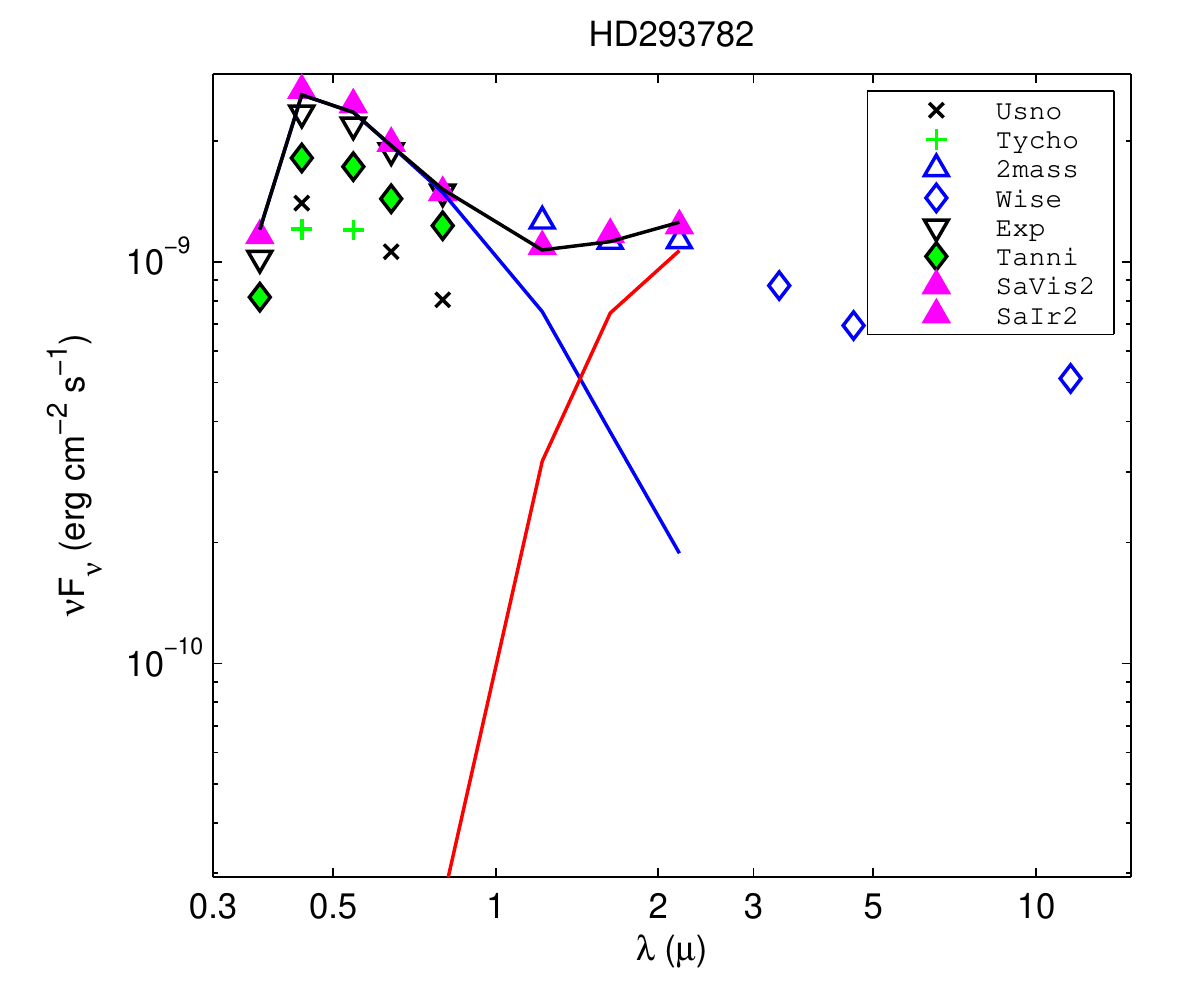}
\caption{Example of a fit to photometric data. Only the bands UBVRIJHK are used in the fit; the Wise data are for reference only. The three lines connect the points of the best-fit model for the star, dust, and total fluxes. The data from the recent SAAO-Sutherland photometry have been given precedence over older data. Also, this object shows strong evidence of variability.}
\label{ExampleSedFit}
\end{figure}

The numerical method used for fitting the SED is described in Sect. \ref{parametricgeneral} below, and the 1-$\sigma$ error bars are derived from the marginal distributions of the Markov chain. The fit results are displayed in Table~\ref{tablePhotomFit}. A sample fit is shown in Fig.~\ref{ExampleSedFit}. 

\subsection{Interferometric data: at first glance}
\label{qualitative}

\begin{figure}
\includegraphics[width=85mm]{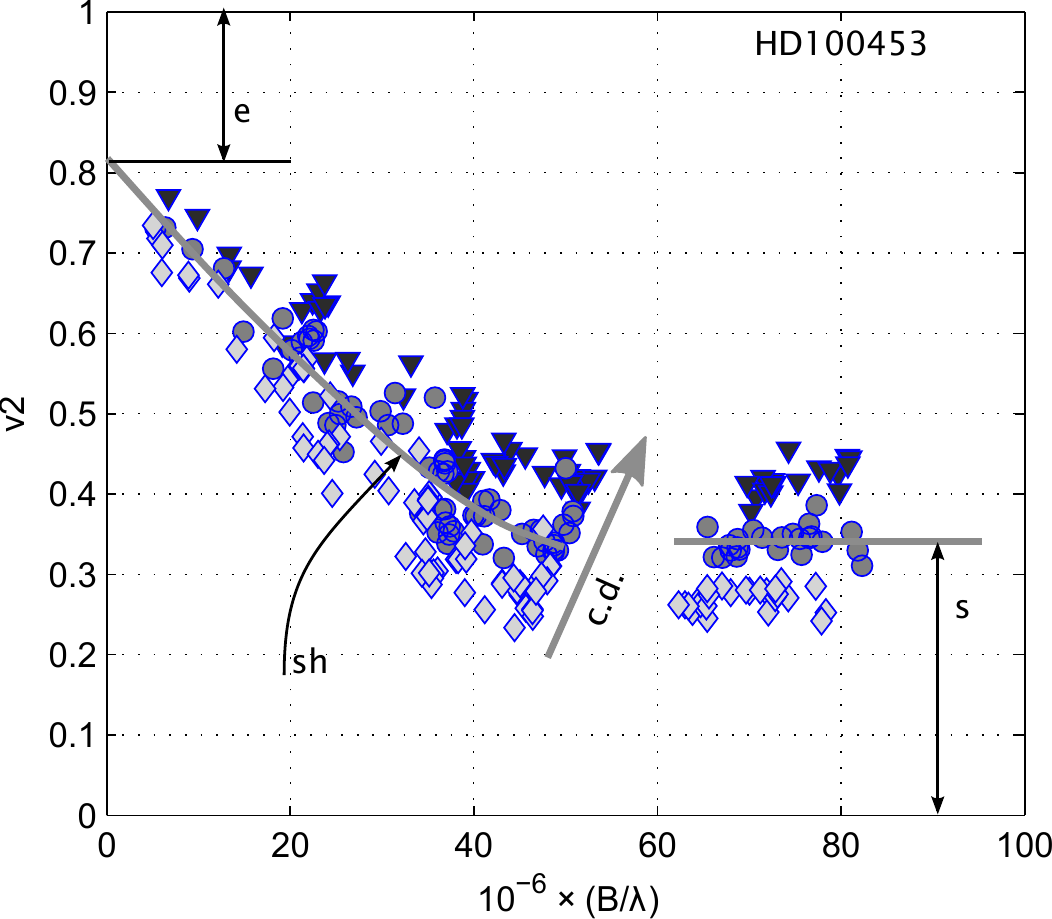}
\caption{Visibility squared versus $B/\lambda$. Data points from the three wavelength channels of PIONIER are identified by the shape and shading of the symbols; diamonds: 1.77$\mu$m; circles: 1.68$\mu$m; triangles: 1.59$\mu$m. Legend for annotations as follows: \textsf{sh}: shape of the visibility decrease; \textsf{c.d.}: chromatic dispersion; \textsf{s}: asymptotic visibility; \textsf{e}: deficit of visibility extrapolated to zero baseline.
\label{VisDiagPedagogical}
}
\end{figure}

Before embarking into numerical model fitting, we should have a qualitative feel for the relationship between measurement results and object properties. We start with a plot of visibilities for the object HD100453, see Fig.~\ref{VisDiagPedagogical}. Following common use, the plot is for visibility squared $V^2$ --a shorthand for 
$\left| V \right|^2$--, but we will discuss visibilities $V$. Assuming that the star itself is unresolved at all observed baselines, and that we observe at a single  wavelength $\lambda_0 = 1.68\,\mu$m, the combined complex visibility of the star and circumstellar matter is: 

\begin{equation} 
\label{visibilityEqSimple}
\begin{split}
V(u,v) &= f_\mathrm{s} \: V_\mathrm{s}(u,v) 
+ f_\mathrm{c} \: V_\mathrm{c}(u,v)  \\
&= (1-f_\mathrm{c}) + f_\mathrm{c} \: V_\mathrm{c}(u,v) 
\end{split} 
,\end{equation}
where $V_\mathrm{c}(u,v)$ is the visibility that the circumstellar component alone would have, $f_\mathrm{c}$ the fraction of the total flux (at $\lambda_0$, inside the diffraction-limited field of view of one VLTI telescope) contributed by that component; likewise for $f_\mathrm{s}$ and $V_\mathrm{s}$ , with generally $V_\mathrm{s}=1$.

To first order, and ignoring azimuthal dependency, the variation of the combined object visibility $V$ as a function of the radius in the $(u,v)$ plane $b=\sqrt{u^2+v^2}$, reflects the variation of the visibility of the circumstellar component $V_\mathrm{c}(u,v)$. The part of the diagram where $V^2$ decreases with increasing $b$ contains information on the characteristic size and even the shape of the circumstellar component, see legend $\mathsf{sh}$ in Fig.~\ref{VisDiagPedagogical}.  At large enough baselines, the circumstellar component is fully resolved, the only coherent flux is from the star, diluted by the circumstellar flux: $V=(1-f_\mathrm{c})=\sqrt{\mathsf{s}}$, see Fig.~\ref{VisDiagPedagogical}. Conversely, if one extrapolates the curve of $V^2$ to zero baseline, the intercept may be at a value 
$V_0=\sqrt{1-\mathsf{e}}$ less than one. Actually, the visibility must return to unity at zero baseline, but it does so at a baseline significantly smaller than those sampled by PIONIER/VLTI, i.e. at an angular scale larger than those analyzed by our observations. We will call this fully resolved component the ``halo'' without otherwise presuming a specific morphology; see also \citet[Sec.~3.4]{2003A&A...400..185L}. 

The visibility curve is used as an a priori criterion to define objects with high quality data ``HQ''): a necessary condition being that the squared visibility drop to 80\% or less of its zero-baseline intercept. This, together with binarity and other exclusion clauses, appears in the footnotes to Table~\ref{table_objectlist}. 

Finally, we note that the three wavelength channels appear to be spread diagonally in the $b$-$V^2$ diagram. When observing with a geometrical baseline $B$, the three spectral channels are sampling different values of  $b\equiv\sqrt{u^2+v^2}=B/\lambda$. From that cause alone, the various points would populate a unique curve following Eq.~\ref{visibilityEqSimple}. 
To proceed, we introduce in the visibility equation: (a)~the dependence on wavelength; (b)~the explicit hypothesis that the circumstellar component's visibility can be factored into a scalar wavelength term and a wavelength-independent spatial part; and (c) we assume the star to be fully unresolved $V_\mathrm{s}(u,v)=1$:
\begin{equation} 
\label{visibilityEqInterm}
\begin{split}
V(u,v,\lambda) &= f_\mathrm{s}(\lambda) 
+ f_\mathrm{c}(\lambda) \: V_\mathrm{c}(u,v)  \\
&= (1-f_\mathrm{c}(\lambda)) + f_\mathrm{c}(\lambda) \: V_\mathrm{c}(u,v) 
\end{split} 
.\end{equation}
In the H band the stellar flux rises at shorter wavelengths, the dust flux decreases, therefore 
$f_\mathrm{c}$ decreases, and the combined visibility $V$ increases. We refer to this as chromatic dispersion (noted $\mathsf{c.d.}$ in Fig.~\ref{VisDiagPedagogical}). We note that an anisotropic  brightness distribution can also cause a spread in the 
$b$-$V^2$ diagram, which is different from chromatic dispersion because it manifests itself also in monochromatic data. 

When the observations reach clearly into the regime where the circumstellar emission is fully resolved, as in the example we have chosen, it is possible to infer from the values of \textsf{s} in each channel the respective values of 
$f_\mathrm{c}$. Returning to the other observed values of $V^2$, using Eq.~\ref{visibilityEqSimple}, and under the assumption that $V$ is real positive, it is possible to isolate the visibility of the circumstellar component $V_\mathrm{c}(u,v)$. Because these conditions are rarely all met, we  cannot generally analyze a program source on the back of an envelope, and we must resort to  parametric modeling. 

\subsection{Parametric model fitting: general methods}
\label{parametricgeneral}

We use parametric models to  extract morphological and statistical information from our  interferometric dataset. Compared with image reconstruction, this has drawbacks, mainly that a functional form is assumed for the source spatial structure. Only a few of our objects have a dataset  rich enough to allow for image reconstruction to be attempted; and image reconstruction itself is not totally free from assumptions - regularization, for example. On the other hand, parametric fitting has the advantage that within a family of models, formal error bars can be assigned to the inferred parameter values. Also, in principle, the adequacy of respective models can be gauged from the 
$\chi^2$ that they achieve. 

Building upon the observations of Sect.~\ref{qualitative}, we construct a  parametric model with the following components:
(1)~the star, assumed unresolved at all observed baselines;
(2)~circumstellar emission with a model-specific and wavelength-independent spatial structure;
(3)~an extended halo component, defined by its spatial scale: fully resolved at all observed baselines ($\ell\gtrsim 20\,\mathrm{mas}$), but inside the PIONIER field of view, defined by its single-mode coupling to the Auxiliary telescopes ($\ell\lesssim 200\,\mathrm{mas}$); in the context of this work, this presumes nothing concerning the shape of that component. 

All fitting was performed in Fourier space; sky images were derived only for illustration purposes. Within the limits of the H band, analyzed by PIONIER in three spectral bands (1.59, 1.68, and 1.76$\mu$m), the spectral dependence of either the star or the circumstellar component is modeled as a power law, defined by the respective spectral indices $k_\mathrm{s}$ and $k_\mathrm{c}$, 
where $k=\mathrm{d}\,\log\,F_{\nu} / \mathrm{d}\,\log\,{\nu}$. 
Since at large distances from the star, the H band is in the far Wien region of the grains' thermal radiation, we assume that the halo component arises from diffusion of starlight (see e.g., \citet{2008ApJ...673L..63P})  and has the same spectral index. We verified that making the alternate assumption (same spectral index as the thermal radiation from the near circumstellar component) has a minimal impact on the fit results.  Returning to Eq.~\ref{visibilityEqSimple}, but now considering a range of wavelengths and including the halo component as discussed above, the visibility can be expressed as:
 
\begin{equation} 
\label{visibilityEqComplete}
V(u,v,\lambda) = \frac{
f_\mathrm{s} \, (\lambda_0/\lambda)^{k_\mathrm{s}} 
+ V_\mathrm{c}(u,v) \, f_\mathrm{c} \, (\lambda_0/\lambda)^{k_\mathrm{c}} 
}{
(f_\mathrm{s}  + f_\mathrm{h})\, (\lambda_0/\lambda)^{k_\mathrm{s}} 
+ f_\mathrm{c}\, (\lambda_0/\lambda)^{k_\mathrm{c}} 
}
,\end{equation}
where   $\lambda_0$ is a reference wavelength (conveniently that of the central PIONIER channel, 1.68$\mu$m) at which the flux fractions of the three components are defined: $f_\mathrm{s}$, $f_\mathrm{c}$, $f_\mathrm{h}$ for star, circumstellar, and halo respectively ($f_\mathrm{s}+f_\mathrm{c}+f_\mathrm{h}=1$), and, in contrast with Eq.~\ref{visibilityEqInterm}, the wavelength dependence is explicitly parameterized.  The visibility of the circumstellar component $V_\mathrm{c}(u,v)$, is specific to each model, as concerns both the functional form (chosen) and the numerical values of the parameters (the result of the fitting procedure). Several functional forms are introduced below in sections  \ref{ellipsoids}, \ref{ring},  and \ref{powerlaw}. 

The value of $k_\mathrm{s}$ is found assuming that  the object's central star radiates in H band as a blackbody at $T_\mathrm{eff}$. $T_\mathrm{eff}$ itself is derived from the spectral type, using Table~5 of \citet{2013ApJS..208....9P}. 

%The fitting procedure minimizes:
%
%\begin{equation} 
%\label{chi2def}
%\chi^2 = \sum_{i_\mathrm{v2}} \left(\frac{v2_\mathrm{mod}-v2_\mathrm{obs}}{\sigma_\mathrm{v2}}\right)^2
%+ \sum_{i_\mathrm{cp}} \left(\frac{cp_\mathrm{mod}-cp_\mathrm{obs}}{\sigma_\mathrm{cp}}\right)^2
%\end{equation}
%
%where the sum runs over all observed data ($uv$ points combined with spectral resolution elements), $v2_\mathrm{mod}$ and $cp_\mathrm{mod}$ are the visibility squared and closure phase computed for a candidate model, a function of the parameter set $(p_1 \cdots p_n)$. 

The 1-$\sigma$ error estimates for the observed quantities, visibility squared and closure phase, 
are computed by the data reduction software, and  are derived from the internal scatter during one exposure (typically on 1~min timescale). 

The minimization of $\chi^2$ proceeds in three steps. First the shuffled complex evolution (SCE) algorithm \citep{SCE} performs a global search for a minimum of $\chi^2$.  Next, the breakdown of the value of $\chi^2$ is examined, and the contributions from $V^2$ and $\phi_\mathrm{cp}$ are allocated weighting factors, such that (a)~they are balanced; (b)~the reduced $\chi^2$ is close to unity. The SCE search  is then restarted. Finally the Monte-Carlo Markov chain (MCMC) algorithm \citep{MCMC}, still with rescaled $\chi^2$, is used to (a)~derive 1-$\sigma$ error bars for each fit parameter from the marginal distributions of the Markov chain; (b) improve slightly the best-fit $\chi^2$. As in \citet{2009AJ....138.1667B}, the rescaling of $\chi^2$ is motivated by some error bars on observables being likely underestimated, with a risk of underestimating the error bars on the fitted parameters. Moreover, some of the errors on the observables might be correlated (per-night calibration errors). An over-estimated $\chi^2$ carries the risk of the minimum search algorithm being caught in a local minimum. 
The unmodified $\chi^2$ value is listed in the results as an indication of how the model captures the complexity of the data.

%We cannot rule out (quasi-)static  errors (even if significant progress has been made since PIONIER was commissioned). In particular, we suspect that there are  errors correlated between the three spectral channels. So, the error bars on the observed quantities might be underestimated; as a consequence, the 
%$\chi^2$ for a candidate model might be overestimated, and the error bars on the model parameters \textit{underestimated}. In order to mitigate (imperfectly) the latter risk, a correction factor $\frac{1}{1.5}$ is applied to expression \ref{chi2def}. 
%Because some errors might be correlated across spectral bands and/or nearby points in the ($u,v$) plane, leading to an underestimation of the error estimations on the fit parameters, we apply a (somewhat arbitrary) $\frac{2}{3}$ factor to the $\chi^2$ supplied to the MCMC algorithm. 

\subsection{Size-flux degeneracy}
\label{degeneracy}

\begin{figure}
 \includegraphics[width=85mm]{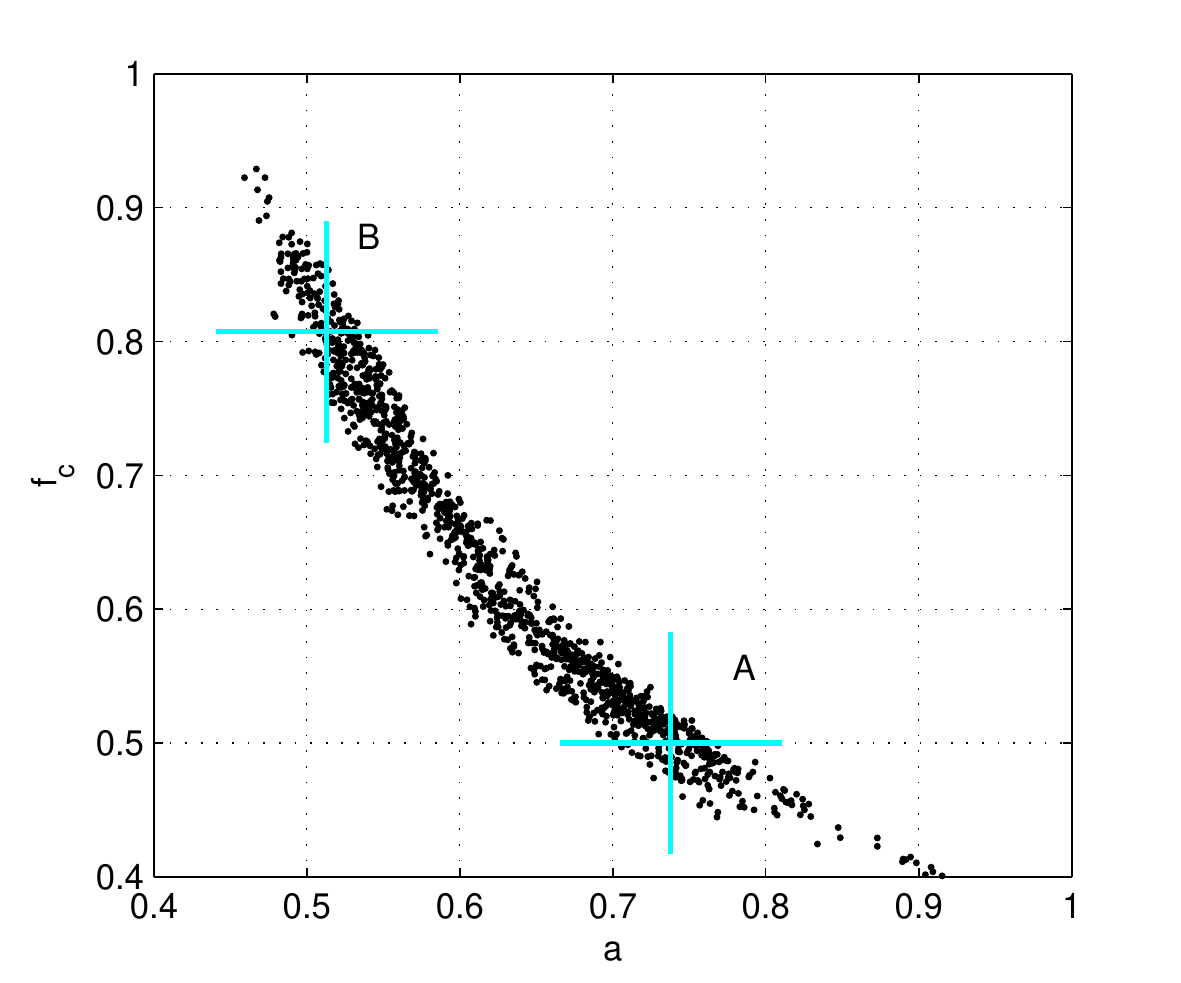}
 \includegraphics[width=85mm]{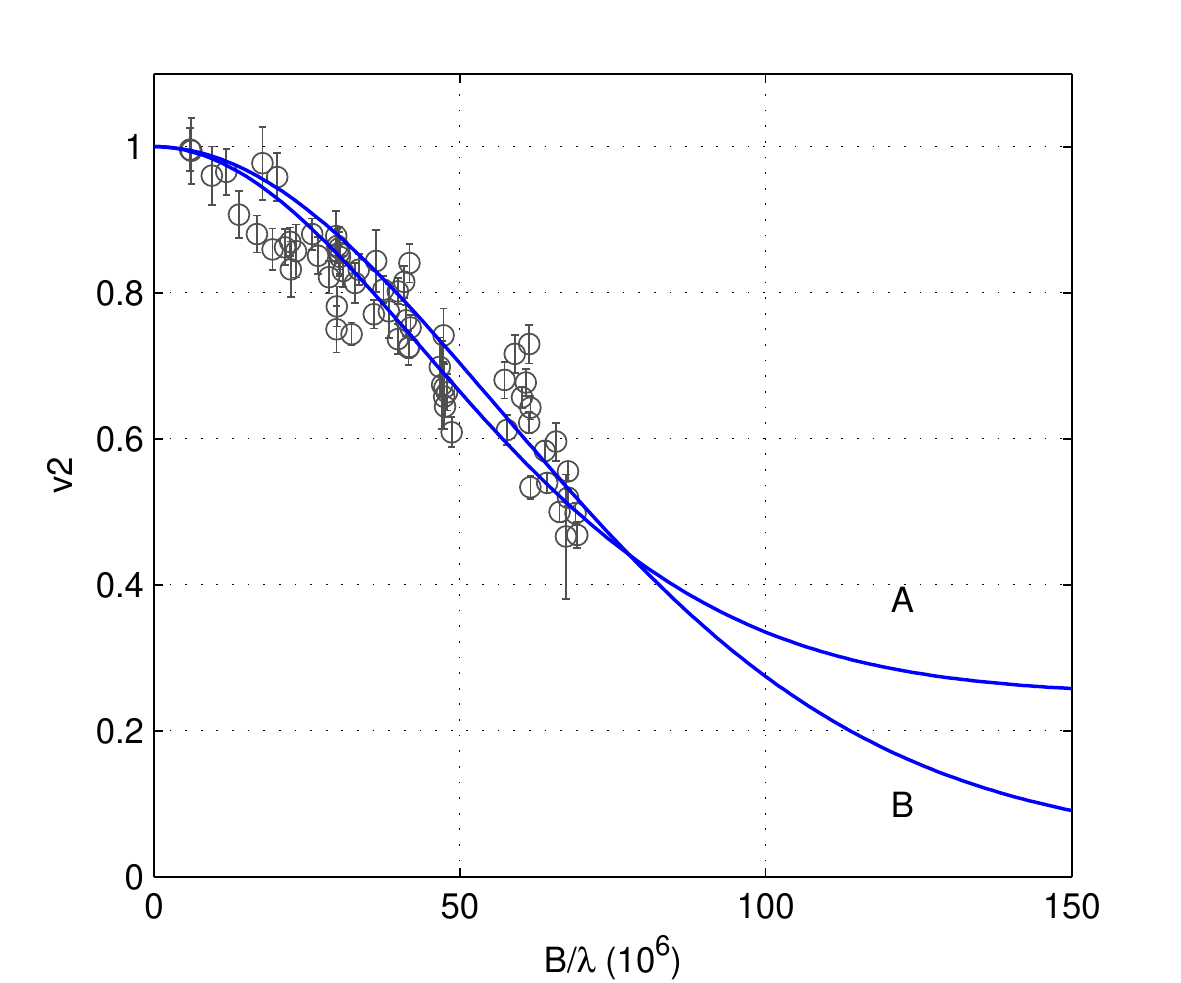}\caption{Illustrating, in the case of HD85567, the degeneracy that arises when fitting a simple model (star+circumstellar) to interferometric visibility data. Top panel: radius $a$ and flux fraction $f_\mathrm{c}$ (in the MCMC chain) of the circumstellar component, crosses labeled \textit{A} and \textit{B} mark the 
10\% and 90\% points in the respective marginal distributions.  Bottom panel: observed data for $V^2$ (middle spectral channel only), and visibility profiles for (isotropic) Gaussians with the parameters at points \textit{A} and \textit{B} respectively. }
\label{HD85567DegenExample}
\end{figure}

Our initial fits of simple models to the interferometric observations (we skip for the moment  some details of the model, to be found in Sec.~\ref{ellipsoids} below) revealed, as one might expect, that some parameter pairs are correlated. Apart from the obvious correlation arising from: $f_\mathrm{s}+f_\mathrm{c}+f_\mathrm{h}=1$, the clearest correlation was generally found to be between the characteristic size and the fractional flux of the circumstellar component. This is illustrated in 
Fig.~\ref{HD85567DegenExample}. The top panel shows the scatter plot (in the MCMC chain) for the variables $a$, the half-flux radius of the  circumstellar component, and 
$f_\mathrm{c}$, its fractional flux. Points \textit{A} and \textit{B} at 10\% and 90\% of the respective marginal distributions are also shown; note that the respective values differ by a factor of more than 1.5.  In the bottom panel, one can see that the visibilities of two Gaussian models for parameter values at \textit{A} and \textit{B} are quite similar within the range of observed $B/\lambda$, and could be discriminated only by observations at longer baselines. 

\begin{figure}
\includegraphics[width=85mm]{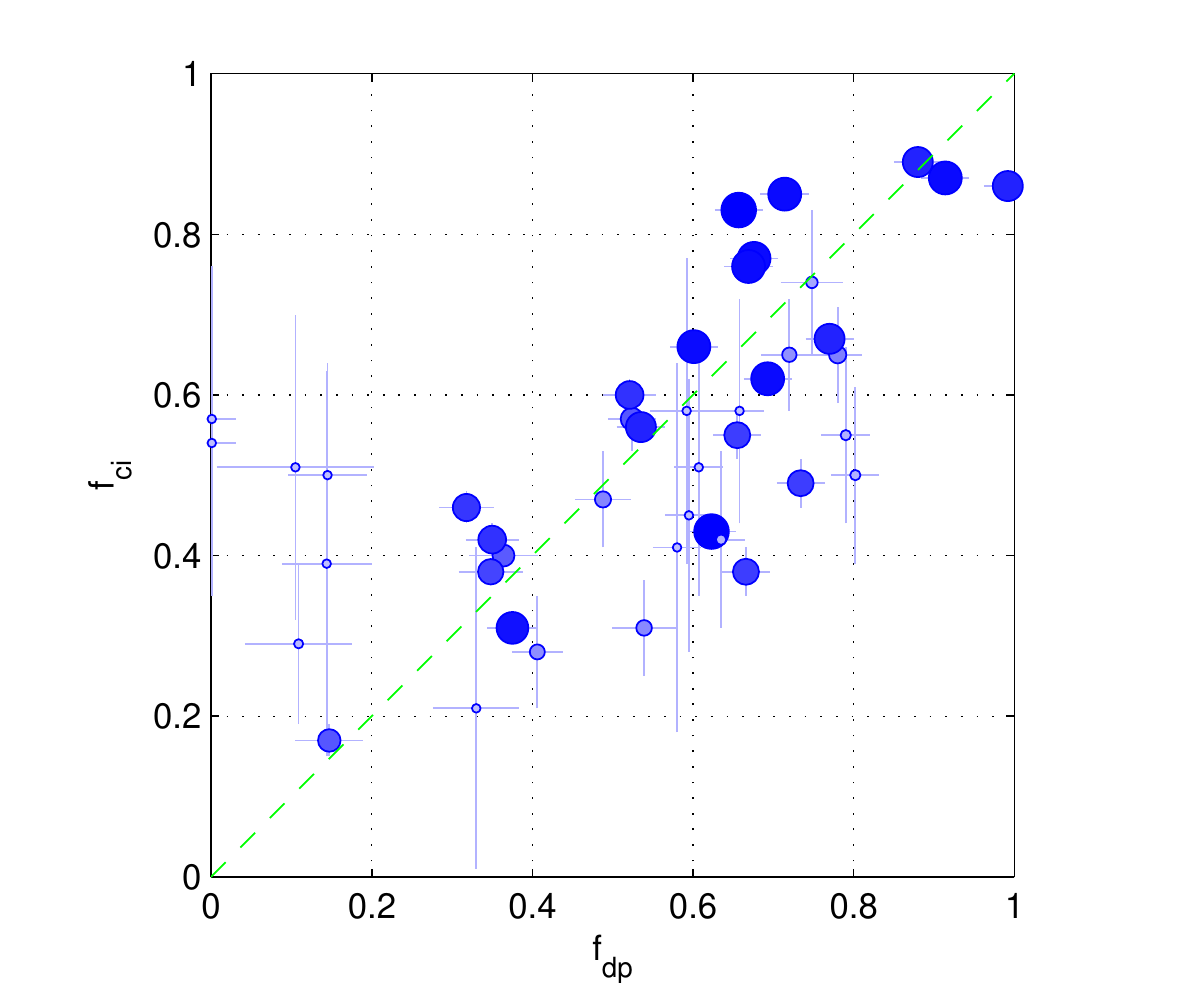}
\caption{Dust flux fraction 
$f_\mathrm{d}$ (H-band) from photometric fits, versus circumstellar flux fraction $f_\mathrm{c}$  from ellipsoid fits to the interferometric data. The size and color of each symbol reflect the combined accuracy of the  $f_\mathrm{d}$  and  $f_\mathrm{c}$ determinations. The error bars are 1-$\sigma$}
\label{fdfc}
\end{figure}

We see that a degeneracy between the size and the flux fraction of the circumstellar component plagues the model fitting of a partially resolved object, as already noted by \citet{2003A&A...400..795L}. This begs the question: are the dust component (identified in the spectral domain)  and the circumstellar component (identified by its spatial properties), related? The correlation in Fig.~\ref{fdfc} shows that this is indeed the case. A similar plot of $f_\mathrm{c}+f_\mathrm{h}$ versus $f_\mathrm{c}$ (not shown) results in a correlation that is neither better nor worse. 

\begin{figure}
\includegraphics[width=85mm]{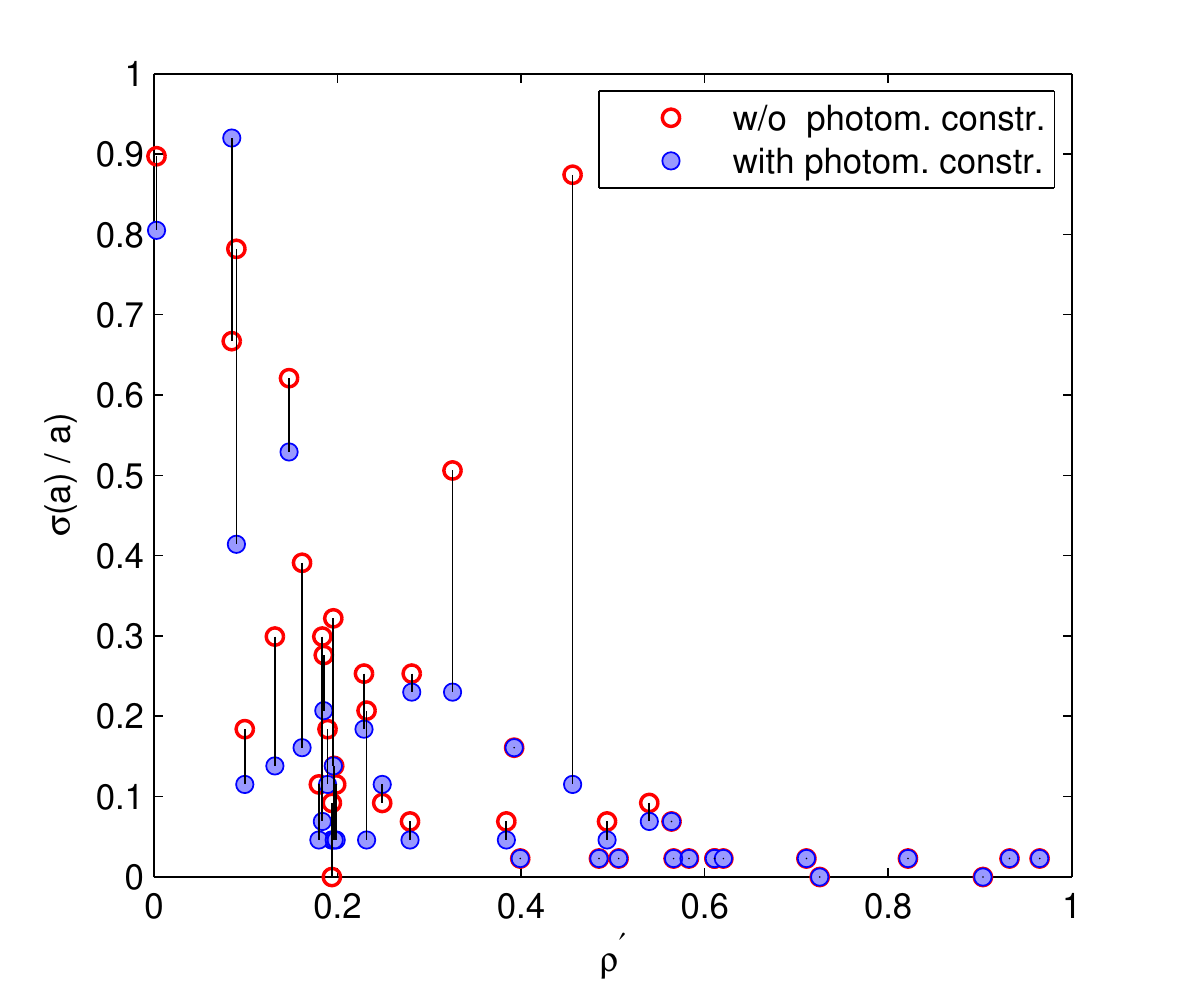}
\caption{Relative error estimate for the radius versus the dimensionless resolution $\rho^\prime = (B/\lambda)_{max} \times a$, without and with  the photometric constraint in the interferometric fit.  The improvement is most marked for the smaller, marginally resolved objects. 
}
\label{photimprove}
\end{figure}

Seeing that $f_\mathrm{c}$ deviates the most from $f_\mathrm{d}$ (reminder: the photometric flux fraction of the dust component at $1.68\mu\mathrm{m}$, defined in \ref{analysphotom} above) for objects where the former is poorly constrained by interferometric data,  we will assume, as a working hypothesis, that $f_\mathrm{d}$ and $f_\mathrm{c}$ measure the same (circumstellar) component of the objects. Accordingly, in order to account for this additional constraint, we introduce in the expression of $\chi^2$ that drives the fitting of interferometric data an extra term $\left((f_\mathrm{c}-f_\mathrm{d})/\sigma(f_\mathrm{d})\right)^2$, where the value of $f_\mathrm{d}$ is  from the photometric fit. Figure~\ref{photimprove} shows the improvement in the size standard error when the photometric constrain is imposed in the fit of interferometric data.   Such a photometric constraint is applied in all subsequent model fits. 

\subsection{Ellipsoids}
\label{ellipsoids}

The first and simplest model that we use for the circumstellar emission has a brightness distribution continuously decreasing  outwards, with elliptic isophotes.  
%In the $(u,v)$ space let $(u_r,v_r)$ be the rotated frame with the $u$ coordinate aligned with the ellipse (of some isophote) major axis. Then define the anamorphic coordinates: $u_c=u_r$ and $v_c=(b/a)u_r$. The visibility is now isotropic, a function $\mathcal{V}(q)$ of $q=\sqrt{u_c^2+v_c^2}$. 

The Gaussian radial brightness distribution is widely used in modeling. But it is peculiar in having a strong asymptotic decay, faster than any power law or plain exponential, which might not be a good description of reality. So we also use a distribution with an opposite behavior, having an $r^{-3}$ asymptotic decrease (the smallest integer exponent with converging flux integral). That is the pseudo-Lorentzian (Lor) distribution as shown in Table~\ref{hankelpairs},  which has a simple analytic Hankel (radial Fourier) transform. 
% See also Fig.~\ref{hankelpairsplots}. 

Starting from a face-on model with circular symmetry, an inclined model with arbitrary axial ratio 
$\cos i$ and position angle $\theta$ can be derived by simple transforms in the $(u,v)$ coordinates. 

\begin{table}
\caption{Hankel transform pairs for the Gaussian and pseudo-Lorentzian radial distributions.}
\label{hankelpairs}
\centering
\begin{tabular}{lcc}
\hline\hline
                & $\mathcal{F}(r)$      & $\mathcal{V}(q)$      \\
\hline\noalign{\smallskip}
Gauss   & $\displaystyle \frac{\ln (2)}{\pi a^2}\, \exp \left( -\left( \frac{r}{a}\right)^2 \ln (2) \right)$
                & $\displaystyle \exp \left( - \frac{(\pi\,a\,q)^2}{\ln (2)} \right)$        \\
                \noalign{\smallskip}
Lor     & $\displaystyle \frac{a}{2\,\pi\,\sqrt{3}}\,\left(\frac{a^2}{3} + r^2 \right)^{-3/2}$   
                & $\displaystyle \exp\left(-\frac{2\,\pi\,a}{\sqrt{3}}\,q \right)$        \\
\noalign{\smallskip}
\hline
\end{tabular}
\tablefoot{
Radial scaling is such that $a$ is the half-flux radius. The distributions are scaled to unit flux, i.e. $\mathcal{V}(0)=1$.
}
\end{table}

%\begin{figure}
% \includegraphics[width=85mm]{./graphs/HankelPairs_crop.pdf}
%\caption{Plots of the two radial brightness profiles used in models. Left: sky plane; right: $(u,v)$ plane. Continuous line: Gauss profile; dashed line: Lor profile. Note the different behaviors of 
%$\mathcal{V}(q)$ near the origin. Despite appearances, half of the flux of the Lor profile lies at $r>a$. }
%\label{hankelpairsplots}
%\end{figure}

In exploratory tests, it was found that some objects ``favor'' (achieving a lower $\chi^2$) the Gauss profile, while other objects favor the Lor distribution. Following that, we introduced $f_\mathrm{Lor}$ as one of the fit parameters,  allowing intermediate radial distributions, such that the circumstellar component visibility in Eq~\ref{visibilityEqComplete} is

\begin{equation}
\label{hybridkernel}
V_\mathrm{c} = (1-f_\mathrm{Lor}) \, \mathcal{V}_\mathrm{Gauss} + f_\mathrm{Lor} \, \mathcal{V}_\mathrm{Lor}
.\end{equation}

We proceed with the fits of the program objects. Table~\ref{parfitell} gives the list of fit parameters; the values are listed in Table~\ref{tableellFit}. 

\begin{table}
\caption{
Parameters for fits of ellipsoidal brightness distribution. 
}
\label{parfitell}
\centering
\begin{tabular}{ll}
\hline
$k_\mathrm{c}$          & spectral index of circumstellar component             \\
$f_{\mathrm{c}}$        & fractional flux of circumstellar component    \\
$f_{\mathrm{h}}$        & fractional flux of halo component             \\
$f_\mathrm{Lor}$        & weighting for radial profile          \\
$l_a$                           & (log) half flux semimajor axis                \\      
$\cos i$                        & axis ratio of isophotes                               \\
$\theta$                        & position angle of major axis, East from North           \\
\hline
\end{tabular}
\tablefoot{
$l_a$ is shorthand for  $log_{10}(a/1mas)$. In practice, the redundant parameter $f_\mathrm{s}$, fractional flux of the star, is also included, together with a constraint 
$f_\mathrm{s}+f_\mathrm{c}+f_\mathrm{h}=1$
}
\end{table}

\subsection{Ring}
\label{ring}

This type of model is intended to model the emission from a bright rim,  for example, the inner rim of a sublimation-bounded dust disk. We start with a ``wireframe'' distribution 
\begin{equation}
\mathcal{F}_0(\mathbf{r}) = \textstyle{\frac{1}{2 \pi}} \, \delta(r-a_\mathrm{r})
,\end{equation}
normalized to unit flux. Its complex visibility is 
\begin{equation}
\mathcal{V}_0(\mathbf{q}) = J_0(2\pi\, q\, a_\mathrm{r})
.\end{equation}
Next we introduce azimuthal modulation: 
\begin{equation}
\mathcal{F}(\mathbf{r}) = \mathcal{F}_0(\mathbf{r}) \times 
\left(1+\sum_{j=1}^m (c_j \cos j \phi + s_j \sin j \phi)\right)
\label{modulatedring}
,\end{equation}
where $\phi$  is the polar angle, $m$ is the order of azimuthal modulation, and the sum is absent if $m=0$. 
The underlying motivation for such a modulation comes from models such as those of \citet{2001ApJ...560..957D}, \citet{2005A&A...438..899I}, where the inner rim of a disk is viewed inclined: the rim contour appears, to first order, as an ellipse, with the far side being more luminous from the vantage point of the observer. Leaving aside the radial structure, the angular modulation in such a model involves only sine terms of odd order. Because we want our analysis to remain model-agnostic as much as possible, we do not include such a restriction in Eq.~\ref{modulatedring}. 

Transforming the modulation amplitude to polar representation: 
$c_j + i\, s_j = \rho_j \exp(i\,\theta_j)$ and likewise for the coordinates in the $(u,v)$ plane: 
$q_u + i\, q_v = q \exp(i\,\psi)$, the complex visibility for the modulated skeleton is:
\begin{equation}
\mathcal{V}(\mathbf{q}) = J_0(2\pi \, q \, a_\mathrm{r})
+\sum_{j=1}^m (-i)^j \, \rho_j \cos\big(j\,(\psi-\theta_j)\big) \, J_j(2\pi\, q \, a_\mathrm{r})
\label{modtrans}
.\end{equation}

Just as for ellipsoids, (sect.~\ref{ellipsoids}) the wireframe distribution is squeezed by a factor $\cos i$ along the direction of the minor axis, and rotated by the position angle. The above-mentioned azimuthal modulation, Eq.~\ref{modulatedring}, is defined with the angle origin along the major axis. 

Because real ring-like images (if they exist) have a finite width, the wireframe image is convolved by an ellipsoidal kernel, with a semi-major axis $a_\mathrm{k}$, the same axial ratio $\cos i$ as the wireframe, and a radial distribution that is a hybrid between a Gaussian and a pseudo-Lorentzian, as described above in Eq.~\ref{hybridkernel} and Table~\ref{hankelpairs}. 

\begin{table}
\caption{Parameterizations of broadened elliptical ring}
\label{paramradiuswidth}
\begin{tabular}{lll}
\hline
\noalign{\vskip 1mm} 
Geometrical             & $a_\mathrm{r}$                & Ring angular radius $^{(a)}$\\
                                        & $a_\mathrm{k}$                & Kernel angular radius \\
\noalign{\vskip 1mm} 
\hline
\noalign{\vskip 1mm} 
Physical                        & $a=\left(a_\mathrm{r}^2+a_\mathrm{k}^2\right)^{1/2}$  & Half light radius $^{(b)}$\\
                                        & $w=a_\mathrm{k} / a$  & Kernel / half-light \\
\noalign{\vskip 1mm}
\hline
\noalign{\vskip 1mm} 
Fit                             & $l_a = \log_{10}(a/1\,\mathrm{mas})$                  & $-1 \le l_a \le +1.5$   \\
                                & $l_\mathrm{kr} = \log_{10}(a_\mathrm{k}/a_\mathrm{r})$        & $-1 \le l_{kr} \le +1$  \\
\noalign{\vskip 1mm}
\hline
\multicolumn{3}{p{8cm}}{\footnotesize $(a)$ A single notation is used for radii, whether angular (radians, milliarcseconds) or linear (AU);  the context resolves ambiguities. }\\
\multicolumn{3}{p{8cm}}{\footnotesize $(b)$ Actually, an approximation to the half light radius valid to 5\% or better for our radial profiles.}\\\end{tabular}
\end{table}

Irrespective of the particular parameterization, this model can describe both thin rings ($a_\mathrm{k} \ll a_\mathrm{r}$) and, asymptotically ($a_\mathrm{k} \gg a_\mathrm{r}$) ellipsoids,  since the convolution kernel is identical to the ellipsoid model.  

The geometrical (see Table~\ref{paramradiuswidth}) parameter pair  ($a_\mathrm{r}$, $a_\mathrm{k}$) becomes degenerate in the limit  of ellipsoid-like cases where only $a_\mathrm{k}$ is relevant. The physical pair is more meaningful, especially when the object is marginally resolved, and the parameter $w$ spans a finite range $0 \leq w \leq 1$. It is not, however fully suitable for the MCMC search because (by default) the method assumes a uniform prior over the allowed search interval: most of the $0 \leq w \leq 1$ interval corresponds to a distinct ring structure, leading to a bias towards detection of such a structure.  The ``fit'' parameters in Table~\ref{paramradiuswidth} cover a wide range, with a scale-free prior. 
The fit parameters number one more (excluding azimuthal modulation) than for the ellipsoid model; they are listed in Table~\ref{parfitring}. 

\begin{figure}
\includegraphics[width=85mm]{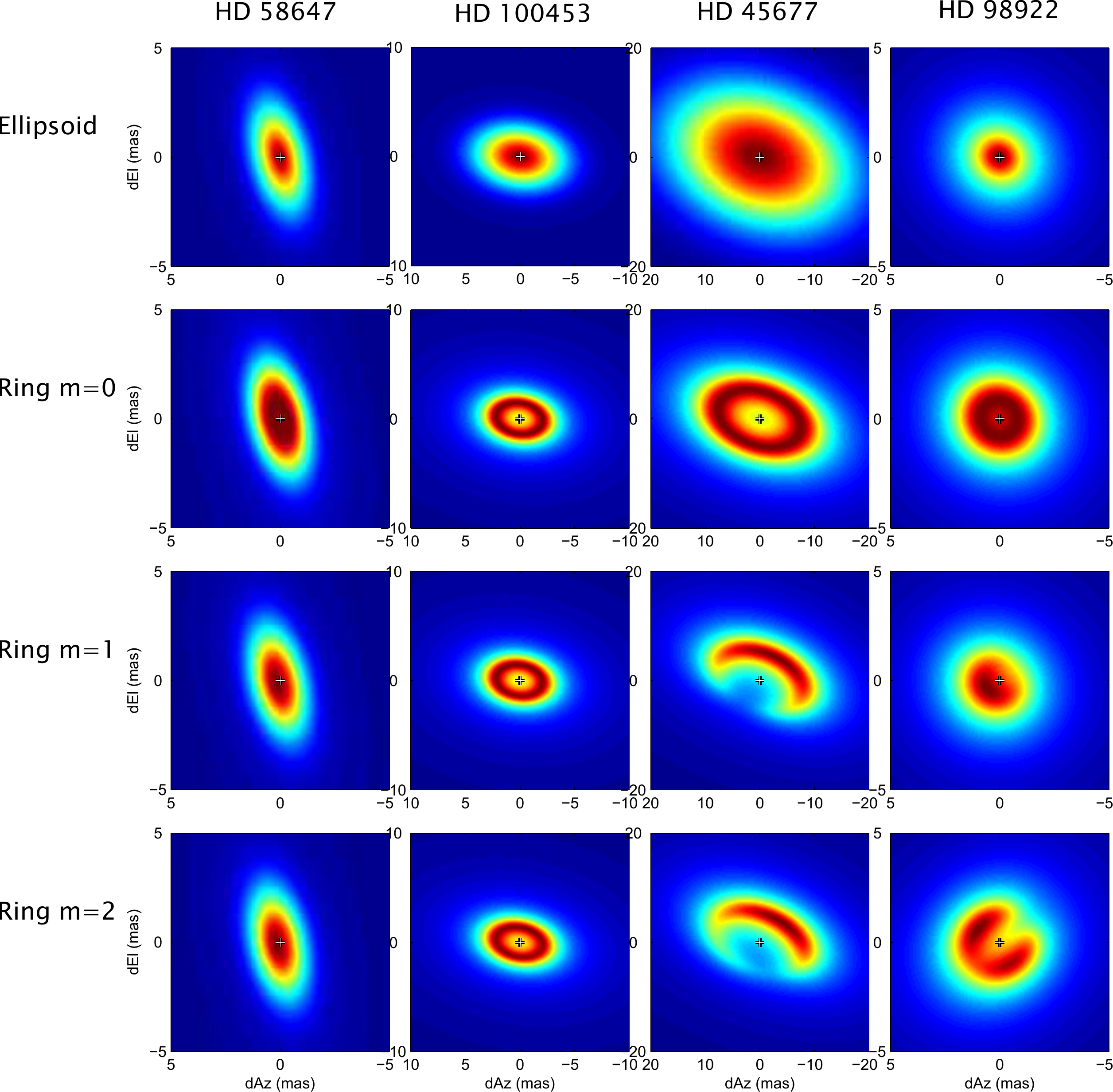}
\caption{Examples of model fits, showing four objects and fit results at four levels of complexity of the fit model. The symbol $m$ is the order of azimuthal modulation; see Eq.~\ref{modulatedring}. 
}
\label{fitexamples}
\end{figure}

Figure \ref{fitexamples} shows examples of fits for some program objects, and shows how a given object - observed at a given resolution - requires a certain degree of complexity, but not more. 

\begin{table}
\caption{
Parameters for Ring fits. Redundant parameter $f_\mathrm{s}$ is also included, see Table~\ref{parfitell}
}
\label{parfitring}
\centering
\begin{tabular}{ll}
\hline
$k_\mathrm{c}$          & spectral index of circumstellar component             \\
$f_{\mathrm{c}}$        & fractional flux of circumstellar component    \\
$f_{\mathrm{h}}$        & fractional flux of halo component             \\
$f_\mathrm{Lor}$        & weighting for radial profile          \\
$l_a$                           & (log) half light semi major axis      \\
$l_\mathrm{kr}$                 & (log) ratio (kernel radius) / (ring radius)   \\
$\cos i$                        & axis ratio of isophotes                               \\
$\theta$                        & position angle of major axis, East from North           \\
$c_j$, $s_j$            & cosine and sine amplitudes for  mode $j$ \\
\hline
\end{tabular}
\tablefoot{
Redundant parameter $f_\mathrm{s}$ is also included, see Table~\ref{parfitell}.
}
\end{table}

\subsection{Power law with inner boundary}
\label{powerlaw}

In some, if not all, current models for inner disk rims, one expects the brightness distribution to have a sharper cutoff inwards than outwards. Accordingly, we tested models with a radial distribution:
\begin{equation}
\mathcal{F}_0(r) = 
\begin{cases}
0 & \text{if } r < a_\mathrm{r} \\
\frac{a_\mathrm{r}}{2\pi} r^{-3} & \text{if } r\geq a_\mathrm{r}
\end{cases}
,\end{equation}
and similar models with an exponent different from $-3$, or an exponential decay, with or without convolution by a kernel (see \ref{ellipsoids} above). None of these showed a significant advantage, as measured by the $\chi^2$, over the already listed model types. 

\section{Analysis and discussion}
\label{discussion}

In this section we use the results of parametric fits performed in the previous section (Data processing) to address the properties of the objects themselves. Because the observations provide (at best) a 2D (interferometry) or 1D (photometry) projection of the objects, and because the parametric models are arbitrary, we attempt to reduce these shortcomings by comparisons with physical models. 

We issue up front two disclaimers: firstly, our goal  is not to  fine-tune a physical model and declare that it represents reality; at best we can hope to support statements on some large-scale properties of the real objects. Secondly, 
 we discuss our sample globally, and derive generic properties, which makes the best use of our limited statistics, but does not imply that real objects come from a single mold. 
Depending on the issue being discussed, the sample may be either the 44 non-binary objects ``noBin'' or the 27 so-called high-quality objects (HQ) : in short, those non-binary objects that are suitably resolved, with $V^2$ dropping below 0.8 at the long baselines. 

\subsection{Dust temperatures}
\label{dust_temp}

\subsubsection{Temperatures from photometry}
\label{dust_temp_photom}

\begin{figure}
\includegraphics[width=85mm]{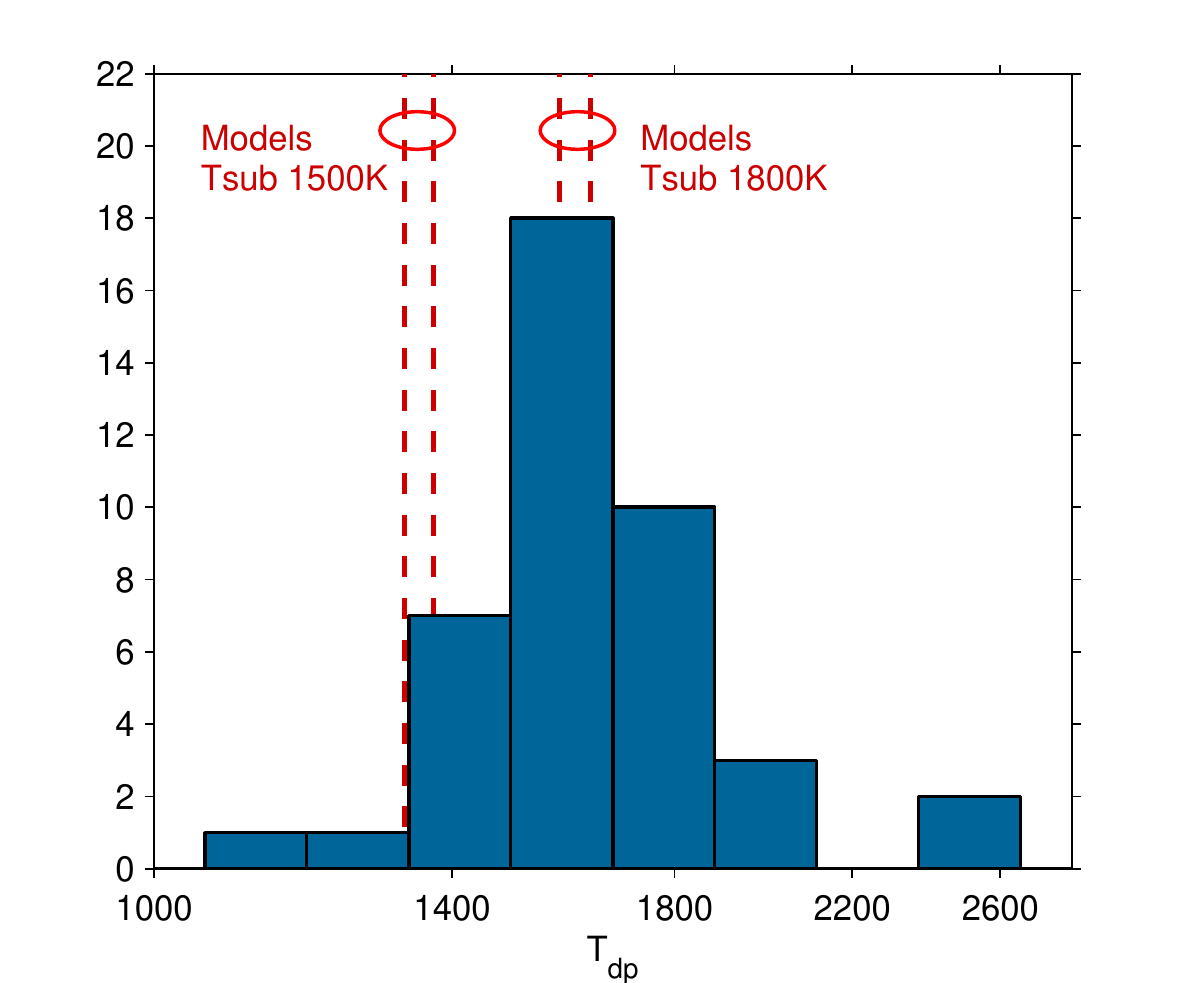}
\caption{Histogram of $T_\mathrm{dp}$ (log scale) derived from fits to photometric data, restricted to non-binary objects having a fractional contribution of the dust to the H-band flux $f_\mathrm{d} \ge 0.1$. Dashed lines: values of $T_\mathrm{dp}$ obtained when fitting, with same procedure, models with sublimation temperature $1500\:\mathrm{K}$  or $1800\:\mathrm{K}$; see text and table~\ref{mcfostModTab}}
\label{histTdP}
\end{figure}

Figure~\ref{histTdP} shows the histogram of fitted dust temperatures $T_\mathrm{dp}$, subject to $f_\mathrm{d}>0.1$ to exclude objects where a fit of dust properties is not meaningful. 
The histogram of $T_\mathrm{dp}$ shows a few possible outliers above 2200~K or below 1200~K; to avoid any excessive influence of such outliers, we make use of the median statistic. The median value of $T_\mathrm{dp}$ is $\simeq 1650\:\mathrm{K}$ 
($\log_{10}(T_\mathrm{dp})=3.22$).  The median of the error estimates on $\log_{10}(T_\mathrm{dp})$ is: 
$0.026$, consistent with an estimate of the sample dispersion as: 
$\operatorname{median}\big[|\log_{10}(T_\mathrm{dp})-\operatorname{median}(\log_{10}(T_\mathrm{dp}))|\big] = 0.030$, and with the dispersion being mostly from measurement uncertainties. 

\begin{figure}
\includegraphics[width=85mm]{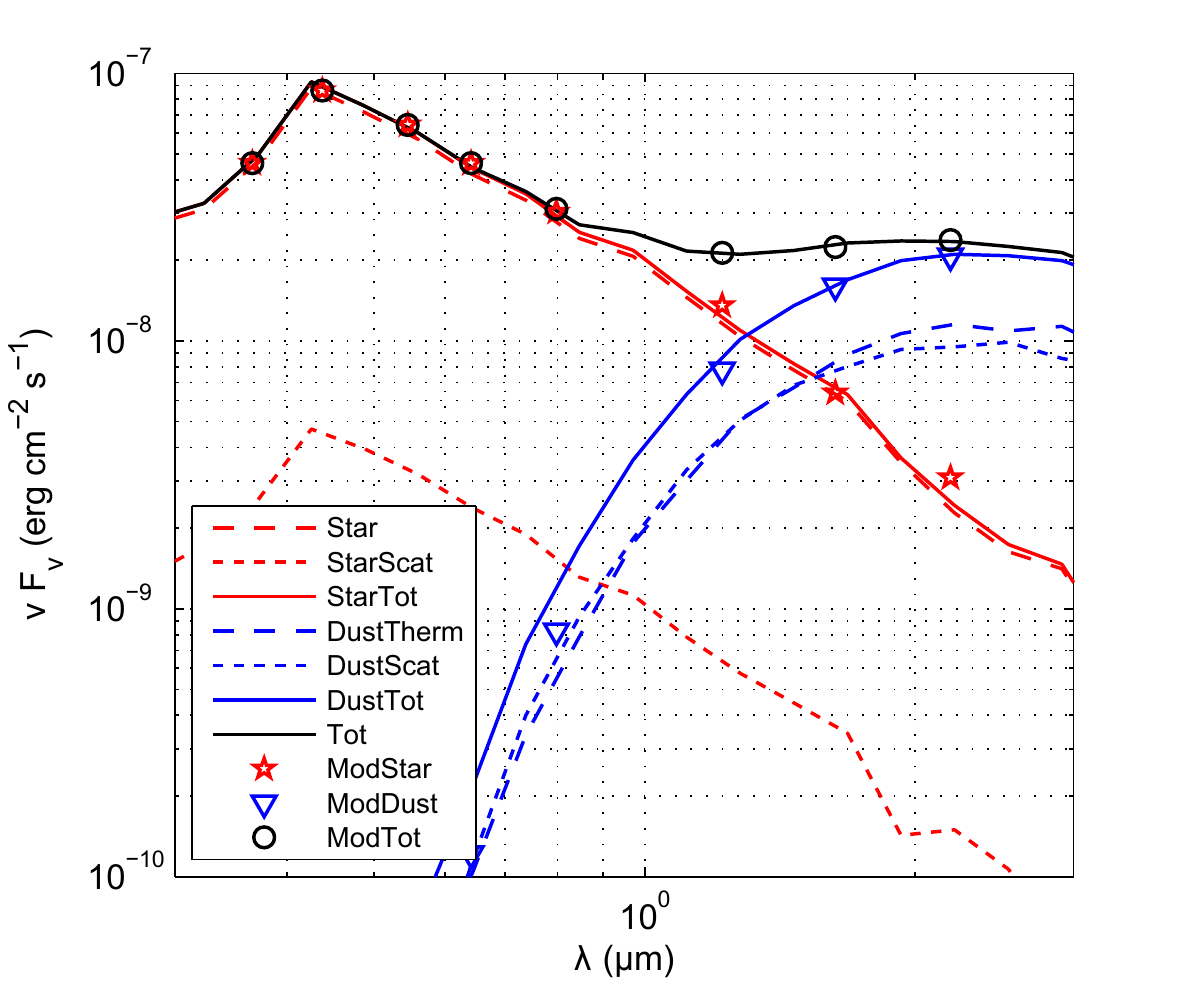}
\caption{Sample fit of a model SED (A1c) obtained with the MCFOST program. The lines show various contributions to the SED of the MCFOST model: direct thermal radiation (dash); diffusion (dash-dot); total (continuous) for either the stellar or the dust contributions. The symbols show the equivalent flux values in the eight photometric bands (U$\cdots$K) for the fitted star+dust model. A model with a single blackbody for the dust appears to provide an acceptable fit to the near-infrared spectrum, even though the emitting dust spans a range of temperatures.}
\label{sedModelFit}
\end{figure}

The median value for $T_\mathrm{dp}$  appears to be in the range of dust sublimation temperatures, in agreement with the current ideas regarding disks around HAeBe stars; see \citet{2010ARA&A..48..205D} and references therein,  or the disk-wind model of \citet{2012ApJ...758..100B}. However, in any such model, one expects dust to exist over a range of temperatures below the sublimation temperature; furthermore, our fitting procedure might introduce some bias. To address these two concerns, we  use the MCFOST radiative transfer code  \citep{2006A&A...459..797P} to generate the SED from a few simple models of HAeBe objects  combining two spectral types (for the central star) with two types of grains. The inner rim radius of each model is adjusted  so that the maximum grain temperature would be (approximately) the sublimation temperature $T_\mathrm{sub}$ for the respective grain types: 1500~K for silicates, and 1800~K for carbon.  These models are used only to discuss dust temperatures, and are distinct from those defined in Sect.~\ref{simu_obs}. The values of 
$\nu F_\nu$ in the eight Johnson bands (U-K) are used as input to the same fitting procedure that is used for deriving $T_\mathrm{dp}$ from the observations. A sample fit is shown in Fig.~\ref{sedModelFit}.  
The results are summarized in Table~\ref{mcfostModTab}; the four values of $T_\mathrm{d,fit}$ are plotted on Fig.~\ref{histTdP}; the models with carbon grains appear to agree better with the observed values of $T_\mathrm{dp}$; in fact, the difference in $\log_{10} \, (T_\mathrm{dp})$ between carbon and silicate models, 0.0765, is 2.7 times larger than the (median-based) sample dispersion. 
We conclude that: 
(1)~the best-fit blackbody temperature for the NIR emission is close to 1650~K;
(2)~based on simple models, that value (1650~K) is consistent with 
$T_\mathrm{sub} = 1800\:\mathrm{K}$ (carbon?);
(3)~the data rule out that the grains responsible for the NIR bump have $T_\mathrm{sub} = 1500\:\mathrm{K}$ (silicates?).

Our finding that the sublimation-bounded inner rim of Herbig object disk is populated predominantly by 
%Carbon grains is not in contradiction with models for the  composition and evolution of interstellar grains} \citep{2004ApJS..152..211Z,2013A&A...558A..62J 
grains with a high sublimation temperature can be explained by the survival of the fittest in a hostile environment. A similar conclusion has been reached by  \citet{2014A&A...567A..51C} in their study of HD~135344B. 

\begin{table}
\caption{Properties of simple models and the respective fits to the dust temperature. 
}
\label{mcfostModTab}
\begin{tabular}{lcccccc}
\hline 
\hline
Model   & $T_\mathrm{eff}$      & $L_\mathrm{b}$                & $R_\mathrm{in}$       & Grains  & $T_\mathrm{d,max}$    & $T_\mathrm{d,fit}$ \\
                & (K)           & $L_\sun$      & AU            &                       & K                               & K                             \\
\hline
\noalign{\vskip 2pt}    
A1s             & 9250          & 36            & 0.50          & silicate      & 1520                    & 1370  \\
A1c             &                       &                       & 0.36          & carbon  & 1880                  & 1580  \\
\hline
\noalign{\vskip 2pt}
B2s             & 20600         & 2280          & 4.5           & silicate      & 1460                    & 1330  \\
B2c             &                       &                       & 2.9           & carbon  & 1800                  & 1637  \\
\hline
\end{tabular}
\tablefoot{
Common properties as follows. Viewing angle $\cos i = 0.55 \quad (i\approx 56\,\degr)$; distance 122~pc;  $R_\mathrm{out}=125$~AU; scale height 0.07~AU at $r=1$~AU; flaring exponent 1.10; dust mass $0.65\,10^{-3} M_\sun$; 
grain size: $a_\mathrm{min}=0.1\,\mu$m, $a_\mathrm{max}=1000\,\mu$m, exponent 3.5.
}
\end{table}

\subsubsection{Temperatures from interferometric data}

When spectrally dispersed interferometric data has been acquired, the model fit to the interferometric data  provides an estimate of the slope, across the H band, of the radiation intensity from the (partly) resolved component. More precisely, the fit provides the differential slope between the extended and the unresolved (stellar) components; the latter being known from the star's spectral type. 

Define for any component the spectral index (across the H band) as:
\begin{equation}
k = \frac{\mathrm{d}\,\log F_\nu}{\mathrm{d}\,\log \nu},
\end{equation}
that can be converted easily to and from blackbody temperatures. The spectral slope of the circumstellar component, determined from interferometric data, $k_\mathrm{c}$, can be converted to an equivalent temperature $T_\mathrm{ci}$.  

\begin{figure}
\includegraphics[width=85mm]{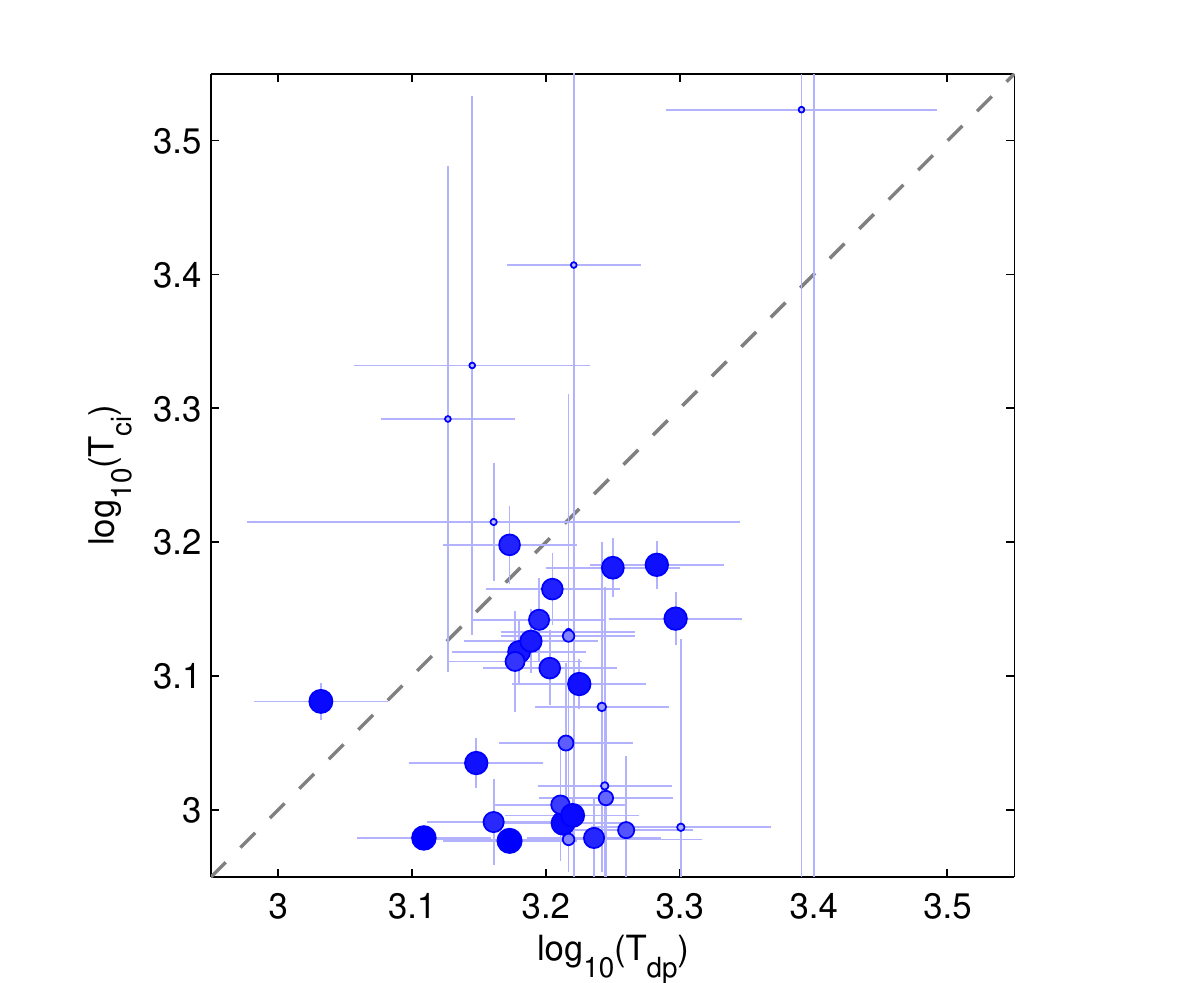}
\caption{
Dust temperatures derived from fits to, respectively, the photometric data $T_\mathrm{dp}$ and the interferometric data $T_\mathrm{ci}$. Values from the Ellipsoid fits to the subset of 44 non-binary objects, with some restrictions (see text). The size of the symbols reflects the combined accuracy for both variables. 
\label{TdpTdi}
}
\end{figure}

In Fig.~\ref{TdpTdi} we present a scatter diagram of $T_\mathrm{ci}$ versus 
$T_\mathrm{dp}$. The data are from the Ellipsoid fits to the 44 non-binary program objects, with extra restrictions: (i) dust flux fraction (H-band photometry) $f_d\geq 0.1$; (ii) circumstellar flux fraction (H-band interferometry) $f_c\geq 0.1$; (iii) object HD~56895 excluded (see \ref{sample} above); (iv) object must be observed in dispersed mode. A total of 35 objects meet these conditions. 

First, the ranges of $T_\mathrm{ci}$ and $T_\mathrm{dp}$ are similar and not far from expected grain sublimation temperatures. Second, there is no clear correlation between these two variables. Third, the values of $T_\mathrm{ci}$ are predominantly lower than $T_\mathrm{dp}$ for the same object. We remind the reader that  $T_\mathrm{dp}$ is derived by decomposing the light from a Herbig object in the spectral domain, while 
$T_\mathrm{ci}$ relies on a decomposition in the spatial domain, where the circumstellar component is assumed to be dominated by radiation from dust. The identity between the respective stellar and non-stellar parts in each of these decompositions is, however,  an assumption. 

Following our caveat above on the respective definitions of $T_\mathrm{ci}$ and $T_\mathrm{dp}$, one possible cause for the observed discrepancy would be a non-photospheric and unresolved component of the object. For instance the inner gaseous disk that should be present inside the dust sublimation radius, at least if the star is still accreting. The determination of $T_\mathrm{ci}$ relies on the differential slope in the H band of the (partly) resolved component relative to the unresolved component. If a non-photospheric emission is present with a flux rising (towards short wavelengths) faster than the photospheric flux, then assigning the photospheric value to the slope of the unresolved component is an under-estimate, which propagates to an under-estimate of  the dust temperature. The existing models for such an inner gaseous disk, for example, \citet{2004ApJ...617..406M} (their Fig.~11), or \citet{2010ARA&A..48..205D} (their Fig.~13) based on the opacities of \citet{2009MNRAS.398..985H} show that at least potentially, its SED can upset the slope of the photospheric emission.

\subsection{Overcoming some limitations of parametric fits via substitution weighing}
\label{simu_obs}

In the previous section, parametric fits of the observed data were used to obtain quantitative information in preparation for the discussion and statistical analysis of the present section. The values of fitted parameters can be subject to biases: either identifiable like the difference between the radius of a Ring model (mean radius) and the  rim radius (innermost), or, worse, unforeseen and undetected. 

To overcome as much as possible these issues, we use an approach analogous to substitution weighing (that allows accurate measurements using a balance with unbalanced or even unequal arms). We  generate a small number of physical disk models, using the MCFOST code. 

Each model is then  ``observed'' at various inclinations  and distances.  Here, to observe means to generate visibilities squared and closure phases for a set of wavelengths and baselines typical of a program object (HD~144668). A realistic noise is added, based on the estimates derived from extensive analysis of PIONIER data; for details see Appendix~\ref{McfostModelDetails}. These data are fed to the same fitting procedure as the real data. The comparison between the fit parameters from, respectively, real and model objects, helps to interpret the fit results in a way that eliminates possible systematic biases from the fitting procedure. 

Our models are loosely inspired by two published models: \citet{2005A&A...438..899I}, hereafter IN05, and \citet{2007ApJ...661..374T}, hereafter THM07; we take full responsibility for the shortcomings of our imitations;  our goal is mainly to sample two different morphologies. 
Each of IN05 and THM07 models derives a consistent solution for radiative transfer, thermal equilibrium, hydrostatic equilibrium, and dust sublimation. Furthermore, THM07 considers two grain sizes and size-dependent settling, as considered in detail by \citet{2004A&A...421.1075D}.
The smaller grains, having a lower cooling efficiency, reach the sublimation temperature at larger distances than the larger grains. 

\begin{figure*}
\sidecaption
\includegraphics[width=12cm]{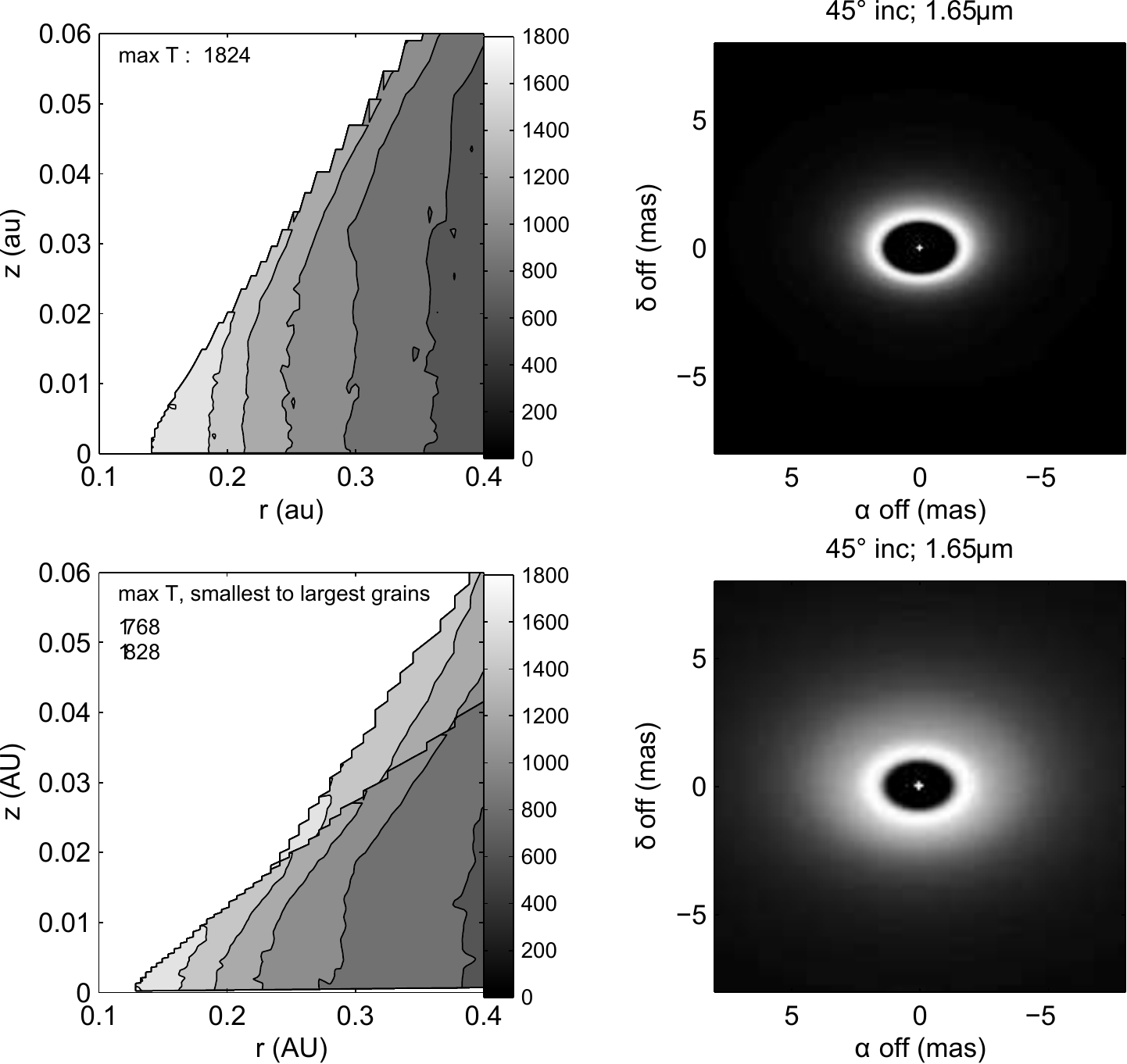}
\caption{Graphical summary of models IN (above) and THM (below). Left: temperature profiles; in the case of the THM model, the temperature distributions for large grains is overlaid onto that for small grains. Right: sky images at $\lambda = 1.65{\mu}m$, $i=45\degr$; a mild enhancement of low brightness regions has been realized by a square root scaling. The angular scale is for the object located at a distance of 100 parsecs.}
\label{inthimages}
\end{figure*}

Our IN05-like model, IN hereinafter, has a single grain radius: 1.2~$\mu$m  while our THM07-like model, THM hereinafter, has two grain radii: 0.35~$\mu$m and 1.2~$\mu$m, resulting in a broader radial distribution of H-band emission. In contrast with the IN05 and THM07 originals, our models solve only for  thermal equilibrium of dust and radiative transfer, given an arbitrary dust distribution with azimuthal symmetry.  See Fig.~\ref{inthimages}.

Variants of each of the base models were produced for three types of central stars: A8V, A0V, and B2V. Spatial scales $\ell$ for each model were adjusted 
($\ell\propto L_\mathrm{bol}^{1/2}$ to first order) to bring the maximum grain temperature close to 1800~K, while densities were adjusted $\rho \propto \ell^{-1}$  to preserve opacities at analogous points. Anticipating the observed value of the fraction $f_\mathrm{rep}$ of stellar flux intercepted by the dusty disk (section~\ref{reproc}), we have tuned the relative thickness $z/r$ of our models to match $f_\mathrm{rep}\approx 0.2$ for the A8 and A0 variants, while we targeted $f_\mathrm{rep}\approx 0.05$ for the B2 variants, based on Fig.~\ref{histrepfrac} (right panel). More details are provided in Appendix~\ref{McfostModelDetails}. 

Additionally, to test the prevalence of false positives (detection of ring structure) in our fitting procedure,  we generated  the ELL
series of images, that are ellipsoids, having the same values for the stellar flux, the circumstellar flux, and the half-light radius as the THM models; in other words, the ``same'', but no ring structure. 
These images are generated from a prescription for their light profile, not from a radiative transfer calculation, and should not be confused with the Ellipsoid fit model introduced in \ref{ellipsoids}.

\subsection{Is there evidence for ring structure ?}
\label{rim_exist}

\begin{figure}
\includegraphics[width=85mm]{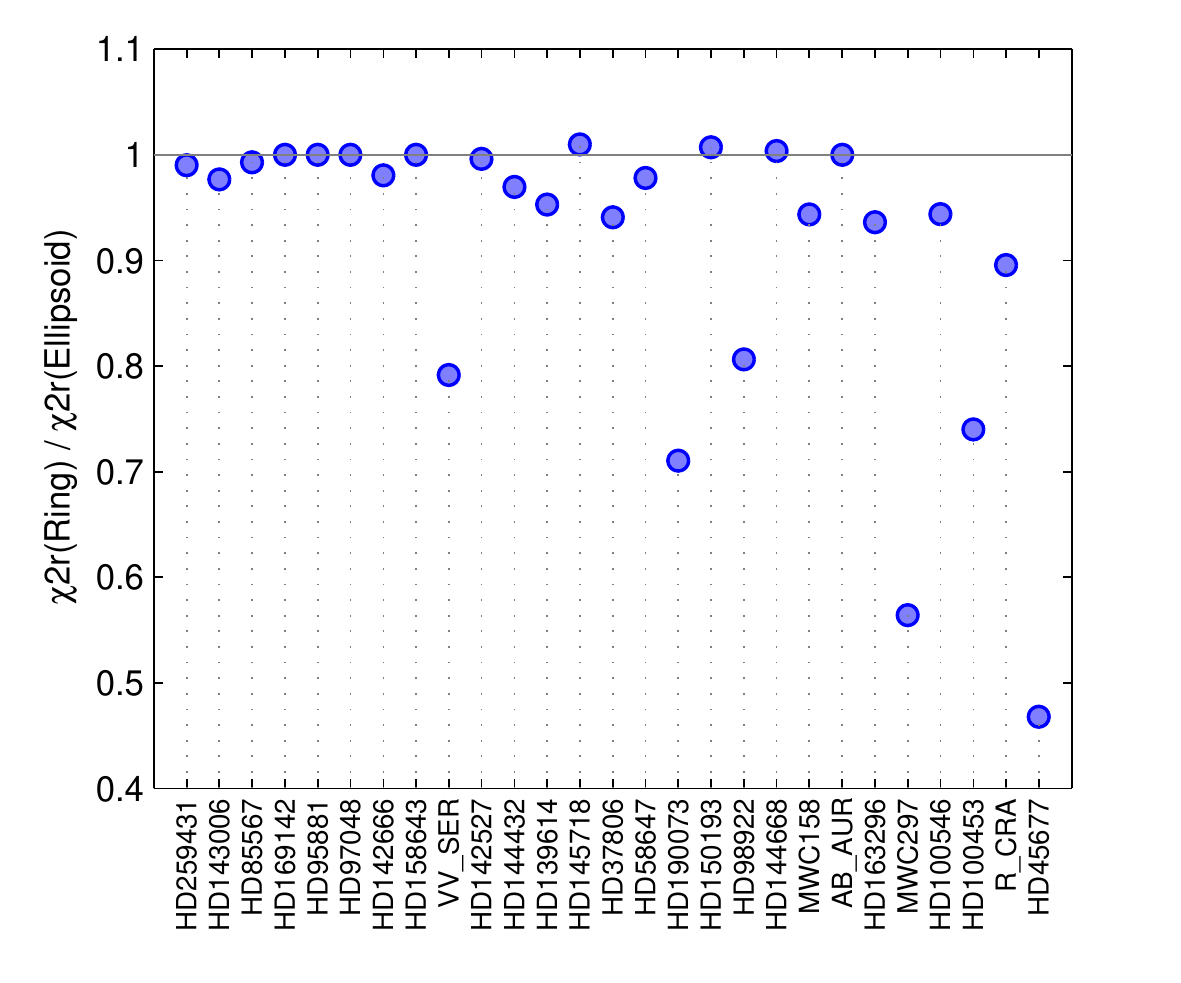}
\caption{
Showing, for the  HQ subsample, the improvement (or not) of $\chi^2_r$ between an Ellipsoid  fit and a Ring fit. The objects have been sorted by increasing size. 
\label{chi2ringnoring}
}
\end{figure}

One possible way to answer that question is to compare the  
$\chi^2_r$ values for an Ellipsoid fit and a Ring fit in its simplest form, with no azimuthal modulation. Figure~\ref{chi2ringnoring} shows the results, over the HQ sub-sample, sorted by size. One can see  that:
firstly, six objects express a preference for the ring structure:  VV~Ser, HD~190073, HD~98922, MWC~297, HD~100453, HD~45677;
secondly, a large angular size favors, but does not guarantee that the ring structure results in a lower  $\chi^2_r$.
A natural (but not unique) interpretation of these results is that (i) a significant fraction of our program objects have a ring structure; (ii) the detection of such a structure is difficult or impossible in marginally resolved objects. 

\begin{figure*}
\includegraphics[width=85mm]{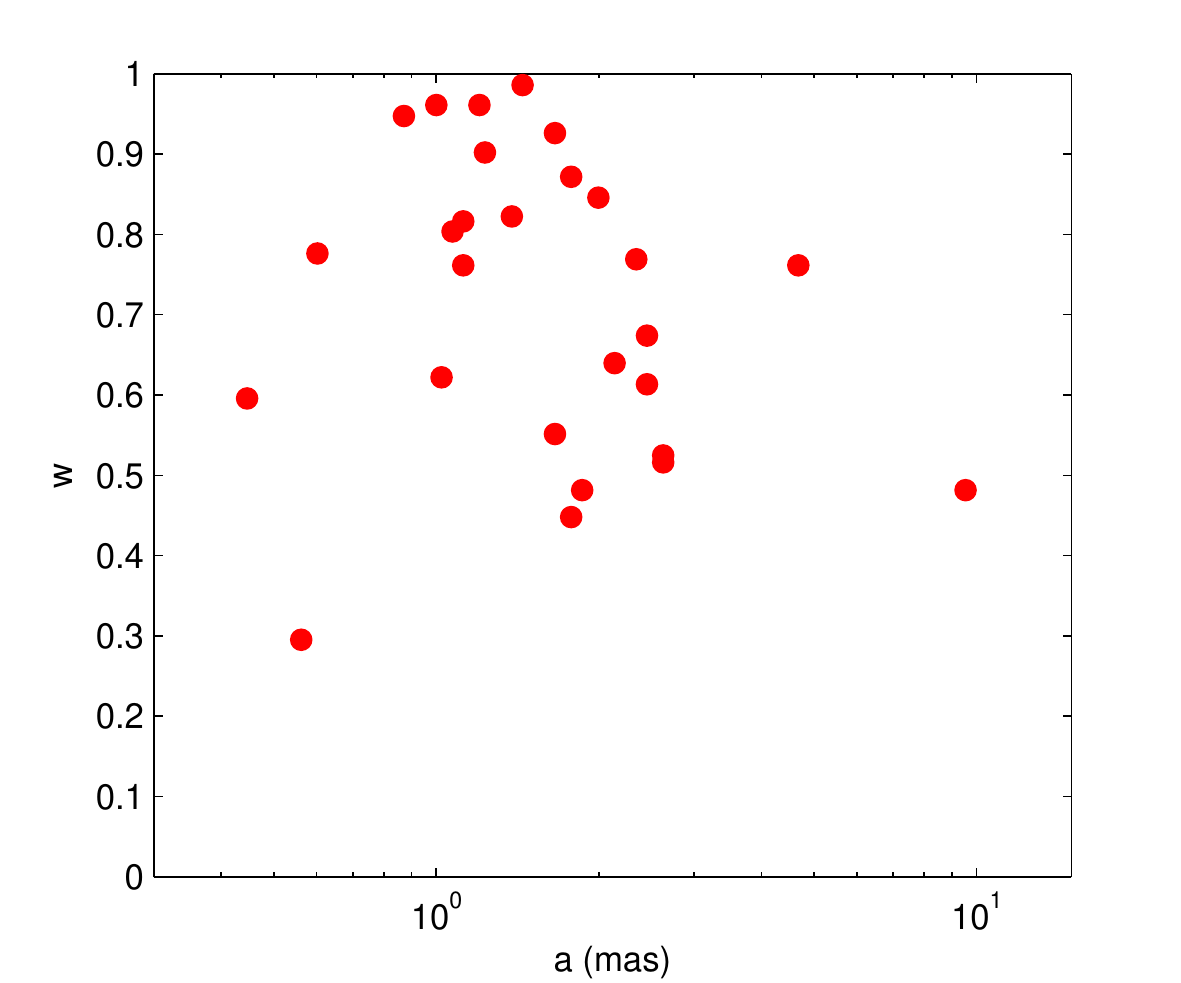}
\hfill
\includegraphics[width=85mm]{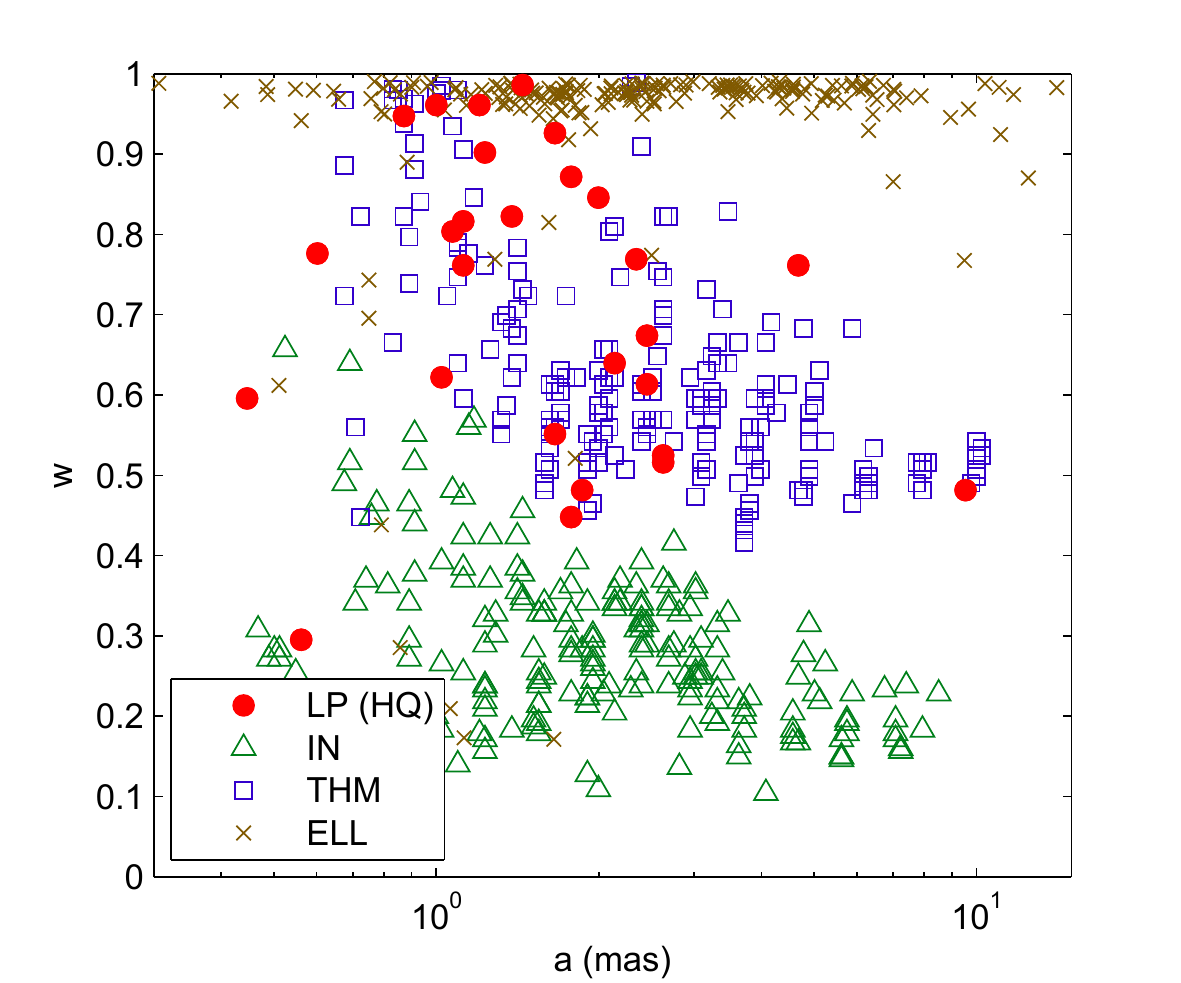}
\caption{Left panel: parametric fit results for the 27 HQ objects. Normalized ring width $w$ versus half flux radius $a$. See text. Right panel: with fits of MCFOST models  IN,  THM  and ELL added. }
\label{wrLp}
\end{figure*}

The same question - whether our data show evidence for ring structures in the program objects--- can be addressed in a different way. Remember that the Ring model fits comprise the Ellipsoid  fits as a subset, since as 
$\log_{10}(a_\mathrm{k}/a_\mathrm{r}) \rightarrow \infty$, $w \rightarrow 1$, and the ring structure degenerates into an ellipsoid; see Tables~\ref{paramradiuswidth} and~\ref{parfitring}. So, instead of examining the $\chi^2_r$ of a ring fit, we examine the value of $w$, and, since we suspect that the ability to detect ring structures is hampered by the limited angular resolution, we plot $w$  versus $a$;  see Fig.~\ref{wrLp} (left panel). The following results  emerge:
(1)~a negative correlation between $w$ and $a$, i.e. the detection of a ring structure is less likely for small radii, as expected; 
(2)~for a significant fraction of the objects, a ring structure is detected by the fit, but...
(3)~the rings seem to be wider than expected models with a rounded rim, as proposed by \citet{2005A&A...438..899I} and pictured in a number of publications.  
At this point, we must consider the following possibilities:
(a)~our parametric fit is biased, the fitted ring is wider than reality;
(b)~actual rings are wide ($w$ of order 0.5);
(3)~there are no rings in the H-band images of  HAeBe objects circumstellar matter, and the  fit algorithm was biased towards finding them.

Figure~\ref{wrLp} (right panel) shows, alongside the HQ objects, the fit results for the IN models at various inclinations and distances. Although we cannot compare one-to-one the fitted ring width with the real width (if only because of the lack of a unique definition for the latter), this plot makes it clear that the putative rings of the LP objects are wider than those of the IN model. 

Also on Figure~\ref{wrLp}, we have plotted the results of fitting THM models.  This shows that at least one simple model of dust distribution and radiative transfer  can result in a ($a$, $w$) diagram, if not identical, at least much closer to the observed one than the IN model. 

We remind the reader that the IN model, inspired by \citet{2005A&A...438..899I}, has a rim with a rounded shape, with a (relatively) large radius of curvature, and a single grain size, while the THM model, inspired by \citet{2007ApJ...661..374T}, has a shape closer to a knife-edge and two grain sizes; its sloping surface under oblique illumination leads to a more radially extended emission.  

Finally, performing fits of plain ellipsoid images ELL (defined in Sect.~\ref{simu_obs}) we find only a small rate of false ring detection: 90\% of the $w$ values are $>0.925$, and 80\% are $>0.975$; while $w<0.7$ occurs only for small angular radii 
$a<1\,\mathrm{mas}$. 

We conclude that, using the substitution weighing approach, we have eliminated possibilities (1) and (3) as listed above, leaving only (2): our interferometric data support the presence, in a significant fraction (at least 50\%) of the HQ program objects, of a ring-like structure. Moreover, these ring structures are significantly wider than what results from a simple rounded rim as our IN model, and a fortiori for a flat vertical wall. 

The THM model, however, fits the observed ($a$, $w$) diagram rather well. Some doubt concerning a sharp inner rim had already been voiced by \citet{2008ApJ...677L..51T}; these authors interpreted the lack of a strong bounce in the visibility curve as supporting the presence of emission inside the sublimation rim. It is noteworthy that, besides THM07, \citet{2009A&A...506.1199K} and \citet{2016arXiv160404601F} also perform a self-consistent modeling of the rim including grain sublimation, and that most of their models result in a wedge-shaped (as opposed to rounded) rim. The photospheric contours in their Figs.~6 and 7 (bottom rows) bear a striking similarity with  those of our THM model (Fig.~\ref{inthm_maps}, bottom row, middle panel), considering that our thm model (not physically self-consistent)  was adjusted specifically to match the large value of the  observed width of the H-band emission region. 
\citet{2010ARA&A..48..205D} (their Fig.~10 and Eq.~13) also explicitly derive a physical model with a large radial extension of the sublimation front. 

%\begin{figure}
%\includegraphics[width=85mm]{./graphs/Width_vs_Radius_Lp_ell.pdf}
%\caption{Similar to Fig.~\ref{wrLp}, but with fits of  the ELL (purely geometrical ellipsoids) added.}
%\label{wrLpell}
%\end{figure}

\subsection{Do we see azimuthal modulation?}
\label{az_mod}

To  answer this question, we have performed, over the HQ sub-sample, three types of model fits, beginning with the Ring model with no azimuthal modulation; see Table~\ref{tableRingAzFits}. 

\begin{table}[h]
\caption{Fits performed with or without azimuthal modulation }
\label{tableRingAzFits}
\centering
\begin{tabular}{lll}
\hline \hline
Id                      &       Number                  &        $\chi^2$ \\
                        &       of az. modes    &       driven by       \\
\hline
\noalign{\vskip 2pt}    
Ring            &       $m = 0$                 &       $V^2$   \\
RingM1          &       $m = 1$                 &       $V^2 + \phi_\mathrm{cp}$      \\
RingM2          &       $m = 2$                 &       $V^2 + \phi_\mathrm{cp}$      \\
\hline
\end{tabular}
\end{table}
Because we keep a record of the partial $\chi^2$ contributions, we are able a posteriori to choose to include  the closure phase ($\phi_\mathrm{cp}$) contribution  in the final value of $\chi^2$ for the $m = 0$ model, allowing a fair comparison with the other two models. 
% Why do we include azimuthal modes up to $m=2$? An $m=1$ modulation, being odd with respect to center symmetry, affects only the imaginary part in the $uv$ space, and therefore: (a)~does not impact visibilities to first order; (b)~affects the closure phases. Conversely, an m=2 modulation (alone) affects visibilities but not closure phases, to first order.

\begin{figure}
\includegraphics[width=85mm]{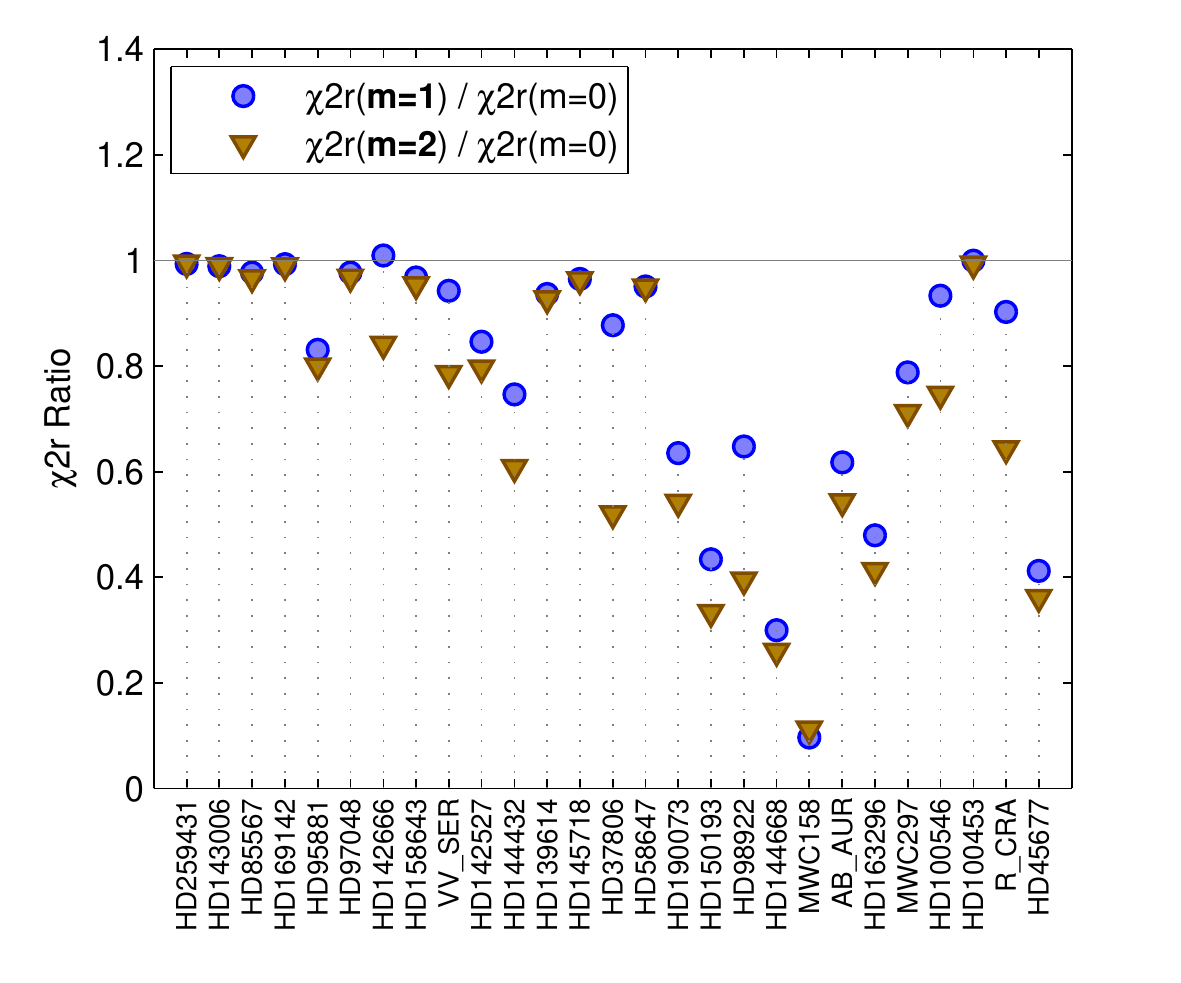}
\caption{
Improvement  of $\chi^2$  (or lack of-) when increasing the number of azimuthal modes in the fit model, the $\chi^2$ containing both the terms from $V^2$ and $\phi_\mathrm{cp}$. We note that, in contrast with Fig.~\ref{chi2ringnoring}, the $\chi^2$ for the un-modulated ring model includes the closure residuals, to be comparable to the other two fit models, with azimuthal modulation. As for  
Fig.~\ref{chi2ringnoring}, the objects are sorted in order of increasing size. 
\label{chi2_v2t3_m012}
}
\end{figure}

The values of the three models listed in Table~\ref{tableRingAzFits} are compared in Fig.~\ref{chi2_v2t3_m012}, with the un-modulated model serving as a reference. As in Fig.~\ref{chi2ringnoring}, the objects are ordered by increasing angular size (value of $a$). One can see that most, but not all, of the better-resolved objects show an improvement of 
$\chi^2$ when modulation is introduced; $m=2$ modulation seems to provide a smaller incremental improvement  than $m=1$. 

A deviation from centrosymmetry (odd $m$ modulation) is expected when a rim is viewed at non-zero inclination. But a spatial offset between the circumstellar emission and the star has the same effect. So, we propose a slightly more stringent test in the next subsection. 

\subsection{Is the azimuthal modulation aligned as we expect?}

In any model featuring an inner disk rim heated by the central star, one expects, at intermediate inclinations, the far side of the rim to be more luminous than the near side. Such an asymmetry involves, in Eqs.~\ref{modulatedring} and \ref{modtrans}, only sine terms of odd order; other terms are included because the model must be unbiased. 

\begin{figure*}
\includegraphics[width=170mm]{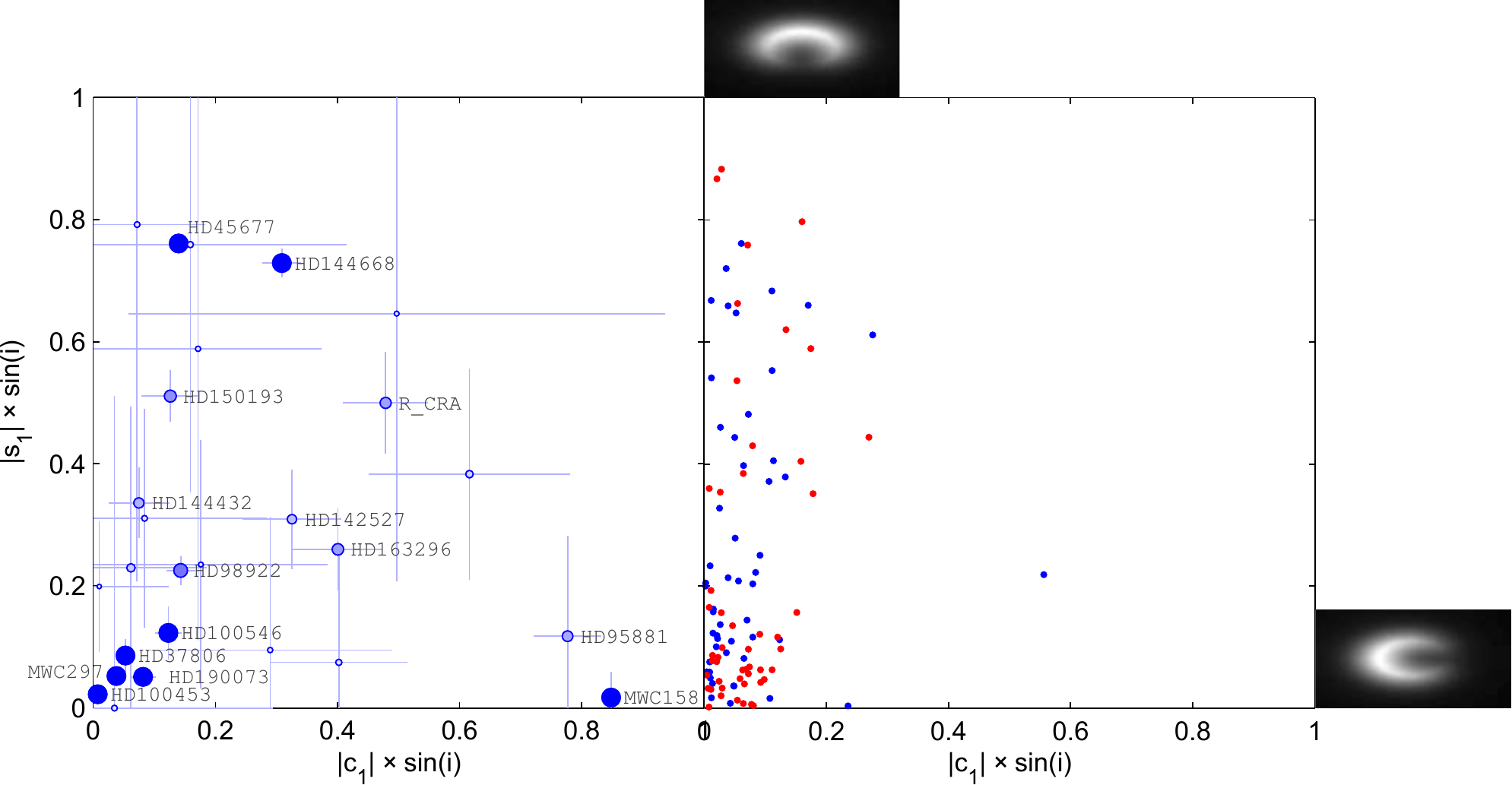}
\caption{
Left:  Cartesian plot of the azimuthal modulation parameters for the HQ objects.  The size of the symbols varies with the quality (small error bars) of the data; objects above a set quality level are labeled with their name. Right: similar diagram for models $in\_A0$ (blue) and $thm\_A0$ (red); a cutoff $\cos i\geq 0.45$ has been applied, similar to what is observed in actual observations. As a visual reminder, two thumbnail images of models  are shown along the corresponding axes. 
}
\label{modalignLP}
\end{figure*}

We test the presence of such a far-side dominance using a plot based on the modulation coefficients $c_1$ and $s_1$. We remind the reader that the modulation is defined in the rotated frame that is aligned with the apparent shape of the rim, and note that the position angle of that ellipse is (nearly) indeterminate for (near) zero inclination (pole-on). Accordingly, we represent each object at the coordinates 
$(x=|c_1| \,\sin i ,\: \, y=|s_1| \,\sin i)$. See Fig.~\ref{modalignLP}. The asymmetry between the near and far sides of a rim image is expected to result in points aligned along the vertical axis. As a reference, the right panel shows the result of fits to $in\_A0$ and $thm\_A0$ models. The scatter diagram on the left panel of Fig.~\ref{modalignLP} show a majority of points closer to the vertical than to the horizontal axis, similar to, but less marked than the model diagram on the right panel. 

One notable exception is MWC~158 (at graph coordinates $(0.85,0)$); this object has been observed with PIONIER before and during the Herbig Large Program, over a time span of $\approx$2~years, and has been shown \citep{2016A&A...591A..82K} to have  a variable structure on 
$\approx 2\:\mathrm{mas}$ scale, possibly corresponding to the orbital motion of a bright condensation around the central star. 

Following this  special examination of MWC~158, which does not fit into the expected pattern for modulation coefficients, we must examine with equal attention the objects that do conform to the expected pattern. 
Twoe objects dominate the alignment along the vertical axis in 
Fig.~\ref{modalignLP}:  HD~45677 and HD~144668. The values of $w$ for these two objects are:
%\begin{tabbing} 
%\hspace{3cm} \= \kill
%HD~45677  \> 0.47      \\
%VV~Ser    \> 0.64      \\
%HD~144668 \> 0.84      \\
%\end{tabbing}
\begin{center}
\begin{tabular}{lc}
HD~45677  & 0.48        \\
HD~144668 & 0.85        \\
\end{tabular}
\end{center}

Figure~\ref{wrLp} (test for false positives with ELL images) shows that a $w$ value below 0.95 corresponds to a ring structure with a high degree of confidence. Therefore, the visual alignment pattern in  
Fig.~\ref{modalignLP} does arise from modulated rings, not modulated ellipsoids (in which case it would be essentially an offset). 

\subsection{Closure phase statistics}
\label{closurestats}

Closure phases provide  a model-independent way to measure the degree of asymmetry  of the objects. We generate plots of closure phase 
$\phi_\mathrm{cp}$ versus resolution $\rho'$, as follows:
(1)~we measure the resolution by the largest of the three projected baselines of the triangle that defines each closure phase measurement; this is an arbitrary way to subsume a triangle by a scalar; 
(2)~the resulting scatter plot is converted to an ordinary plot of closure phase versus resolution by computing percentiles in a sliding resolution window (width 0.2 in $\log_{10}$ representation);
(3)~the effective resolution must take into account the angular radius of the object, and, to first order, its anisotropy.
Accordingly, for an object having  half-light semi-major angular diameter 
$a$ at position angle $\theta$, with an axis ratio $\cos i$, and observed over a projected baseline with coordinates $(u,v) = (b_x,b_y)/\lambda$,  we define the dimensionless resolution $\sigma'$ along that baseline:
\begin{equation}
\begin{aligned}
\alpha  &=& \pi/2 - \theta \\
u'              &=& u \, \cos \alpha + v \, \sin \alpha  \\
v'              &=& \cos i \; ( - u \, \sin \alpha + v \, \cos \alpha ) \\
\sigma' &=& a \; (u'^2+v'^2)^{(1/2)} 
\end{aligned}
,\end{equation}
and for a closure phase measured over a triangle, the value of $\sigma'$ for the largest side:
\begin{equation}
\rho' = max(\sigma'_k) \quad k = 1 \cdots 3
.\end{equation}

\begin{figure*}
\includegraphics[width=85mm]{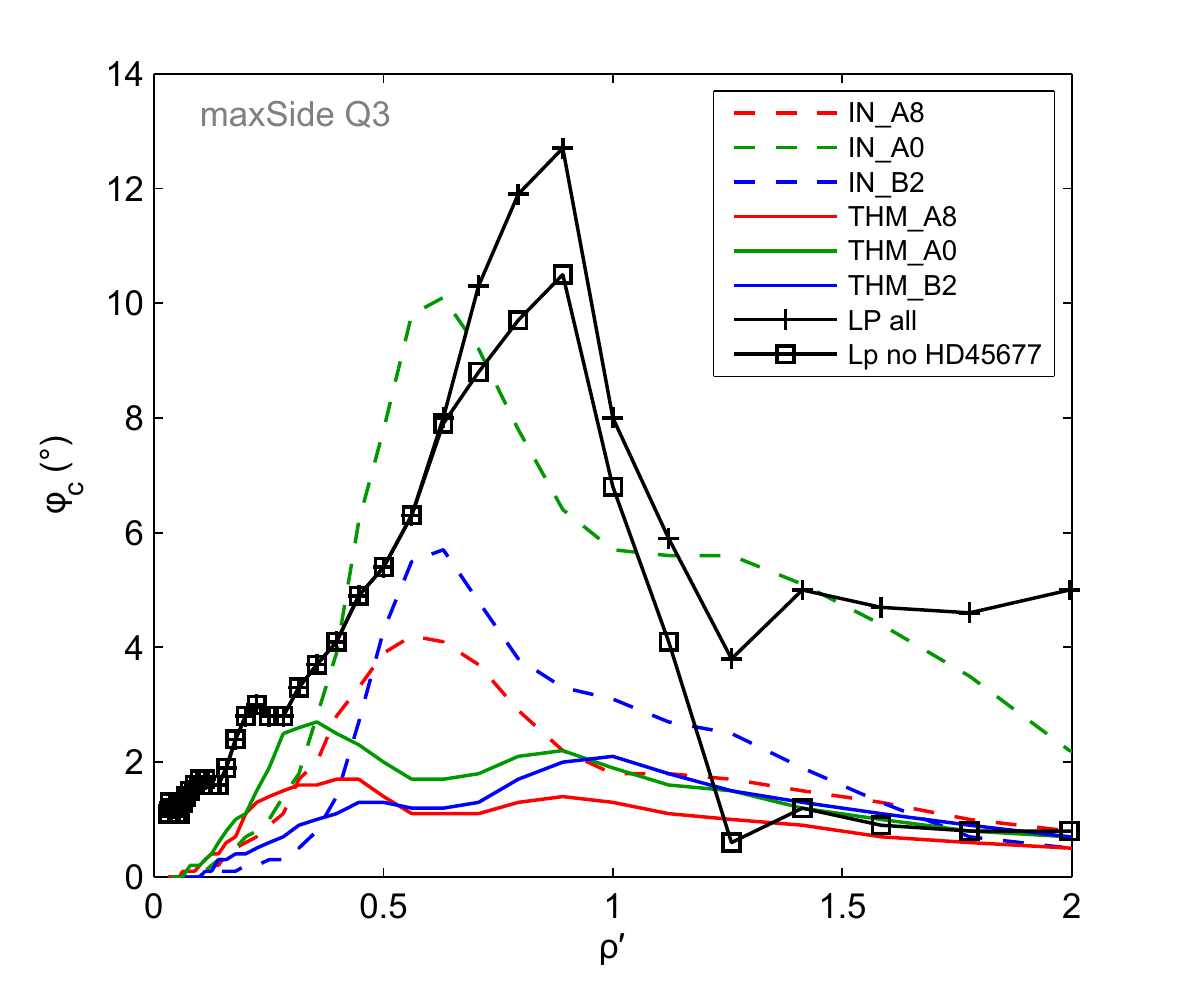}
\includegraphics[width=85mm]{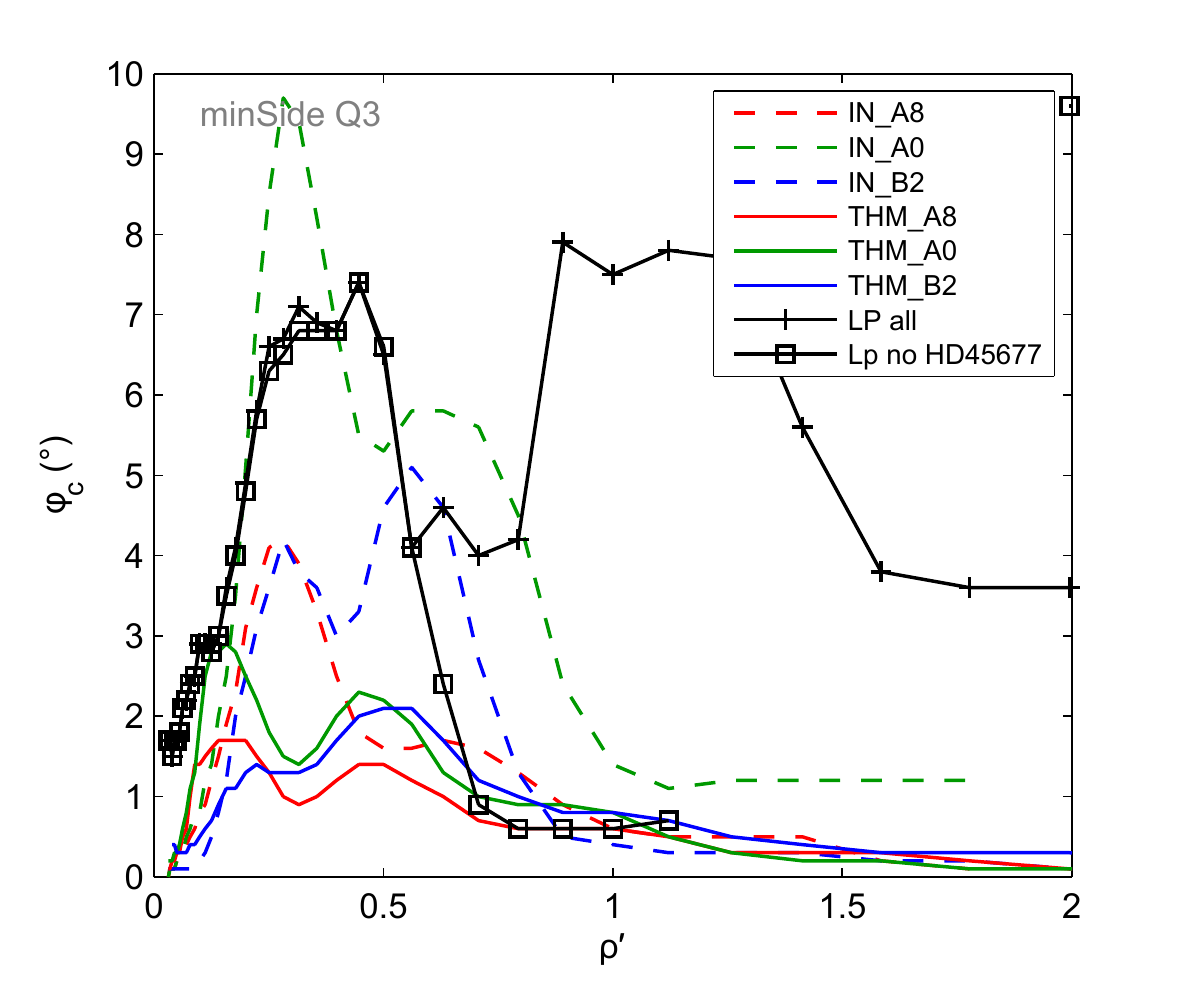}
\caption{
Closure phase versus dimensionless resolution. Left side: resolution defined by largest side of triangle; right side: resolution measured by smallest side of triangle. Black, crosses: LP (non-binary) objects; black, squares:  excluding HD~45677 (see text). Similar plots are also shown for the IN and THM MCFOST models in the three variants according to the central star. 
}
\label{closuresresolution}
\end{figure*}

Figure~\ref{closuresresolution} shows (black lines) plots of the sliding third quartile (see above) of $\phi_\mathrm{cp}$ versus $\rho'$ either for the totality of the 44 non-binary objects, or excluding HD~45677. 
\citet{2006ApJ...647..444M} used similar plots, but based, at each resolution value, on the maximum  closure phase, which may have a low probablity of being measured in actual observations. We show two variants, with the measure of dimensionless resolution based either on the largest or the smallest side of each closure triangle. 

The  plot  that includes HD~45677 shows higher values of 
$\phi_\mathrm{cp}$ above $\rho'\approx 1$. 
HD~45677 is the object with the second largest angular radius ($\approx 10\:\mathrm{mas}$) in our sample, almost four times larger than the next largest object. Image reconstruction \citep{2014IAUS..299..117K} and parametric models up to azimuthal order $m=5$ agree to show substructure in the sublimation rim of HD~45677; we believe that the closure phases contributed by HD~45677 data above $\rho'\approx 1$ belong to such substructure rather than to the ring asymmetry. Finally, the object with the largest angular radius,
HD~179218 ($\approx 25\:\mathrm{mas}$) does now show a clear closure signal (see Fig.~\ref{summplotNonHQ}). 

Closure phase peaks around $\rho'\approx 0.5$, as already noted by \citet{2006ApJ...647..444M}, and the peak value for our data set is 
$13\degr$ or $7.5\degr$ respectively for the two versions of the graph. Aiming to relate these values with properties of the sublimation rim, we also show on Fig.~\ref{closuresresolution} similar plots, for the  MCFOST models, IN and THM, already used to calibrate our parametric fits. Borrowing from the results of 
Sect.~\ref{inclination}, we have applied to these simulated observations the same $\cos i > 0.45$ cutoff as found for actual data.  One can see that none of the six model variants matches the observed 
$\phi_\mathrm{cp}$ versus $\rho'$ plot. Possibly $in\_A0$ comes closest, but the IN model has been discarded, on the basis of the ring width, in favor of the THM model. On the other hand, the THM model fails to reproduce the observed closure phases. One can also note that the value of 
$\rho'$ where $\phi_\mathrm{cp}$ peaks is not uniform, and can vary between 
$\approx 0.3$ and $\approx 0.7$. Clearly the closure-resolution diagram is a promising sieve to reject or validate models. But, in keeping with our disclaimer in \ref{simu_obs}, we do not attempt, in this paper, to tune a model to fit all our observables.  

\subsection{Disk thickness}
\label{thickness}
We attempt to gain information on the disk thickness ($z/r$) via two approaches: the statistics of inclinations and of reprocessed flux fraction.  

\subsubsection{Inclination statistics}
\label{inclination}

\begin{figure}
\includegraphics[width=85mm]{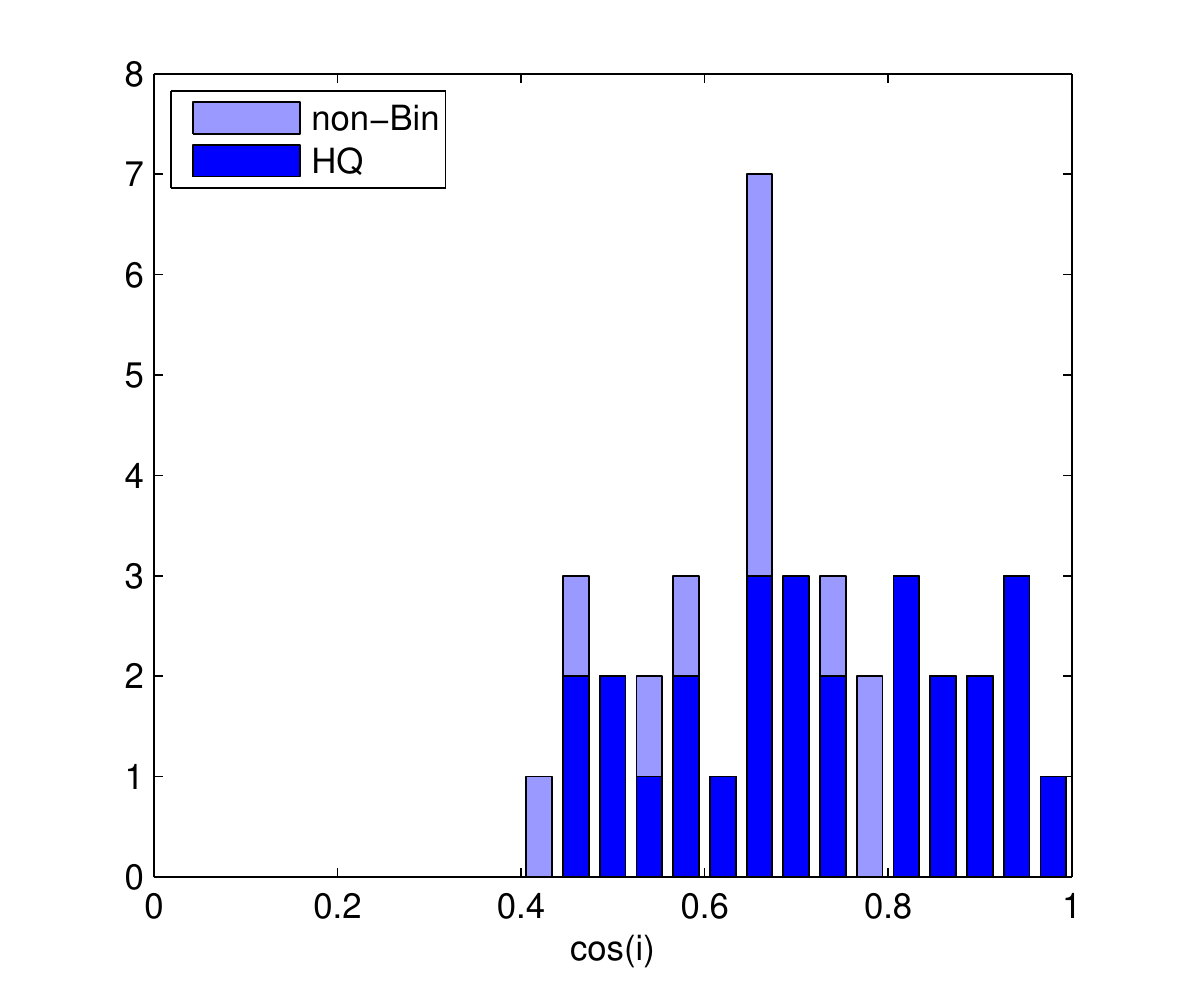}
\caption{
Histogram of fit values for $\cos i$.  The histogram for the 27 objects of the HQ sub-sample, fitted with the Ring model is superimposed over the histogram for the 44 non-binary objects, fitted with the Ellipsoid model. Parameter $\cos i$ is in fact robust: its value is not sensitive to the particular fit model used. 
\label{histcosi}
}
\end{figure}

Figure~\ref{histcosi} shows the histogram of $\cos i$ fit values for either  all non-binary objects, or the  HQ subsample.   A flat histogram over the full $(0,1)$ range would be expected for a population of thin rings with random orientations. Apart from the small-number fluctuations between the two fit models, we note that: a)~there is no significant pileup around $\cos i=1$, as would result from a contamination by a population of intrinsically spherical objects; b)~there is a lower cutoff around $\cos i \approx 0.45$. We interpret this cutoff as due to self-obscuration, i.e. objects do exist at higher inclination, of course, but are absent from the sample because either the central star or the H-band dust emission is obscured. Let 
$\eta = {\cos i}_\mathrm{cutoff}$, 
then the relative thickness at which obscuration sets in is given by:
$\zeta = z/r = \eta / \sqrt{1-\eta^2}$ 
and $\zeta(\eta=0.45) = 0.50$. 
%We cannot determine from the $\cos i$ statistics alone whether that $z/r$ value is representative of the vicinity of the rim, or whether it arises in a more distant part of a flaring disk. 
We find no significant correlation of the fitted $\cos i$ with the spectral type. 

%We have used the THM model to test the presence of a possible bias of the fit procedure towards larger values of $\cos i$. In the case of the THM models, a small positive bias exists, which is less than 0.02, not significant in the preceding discussion. Remember, though, that the THM model is just one among many that are compatible with the ring width statistics.  

Fitting the THM models reveals a small ($<0.02$) positive bias in the values of $\cos i$,  which does not affect our interpretation. 

The observed cutoff may be a consequence of flaring and obscuration, by the outer disk of either the NIR radiation of the inner disk or the central star itself. We do not attempt to mimic this situation with our MCFOST models because our models extend radially only far enough (typically $10\times$ rim radius) to model the NIR emission up to K band. If indeed flaring is responsible for the observed cutoff, we can simply state (without the need for numerical modeling) that the radial direction having an extinction $A_\mathrm{V}\approx$~a~few has a slope (with respect to the equatorial plane) of order $z/r\approx 0.50$. 

\subsubsection{Reprocessing of stellar flux}
\label{reproc}

\begin{figure}
\includegraphics[width=85mm]{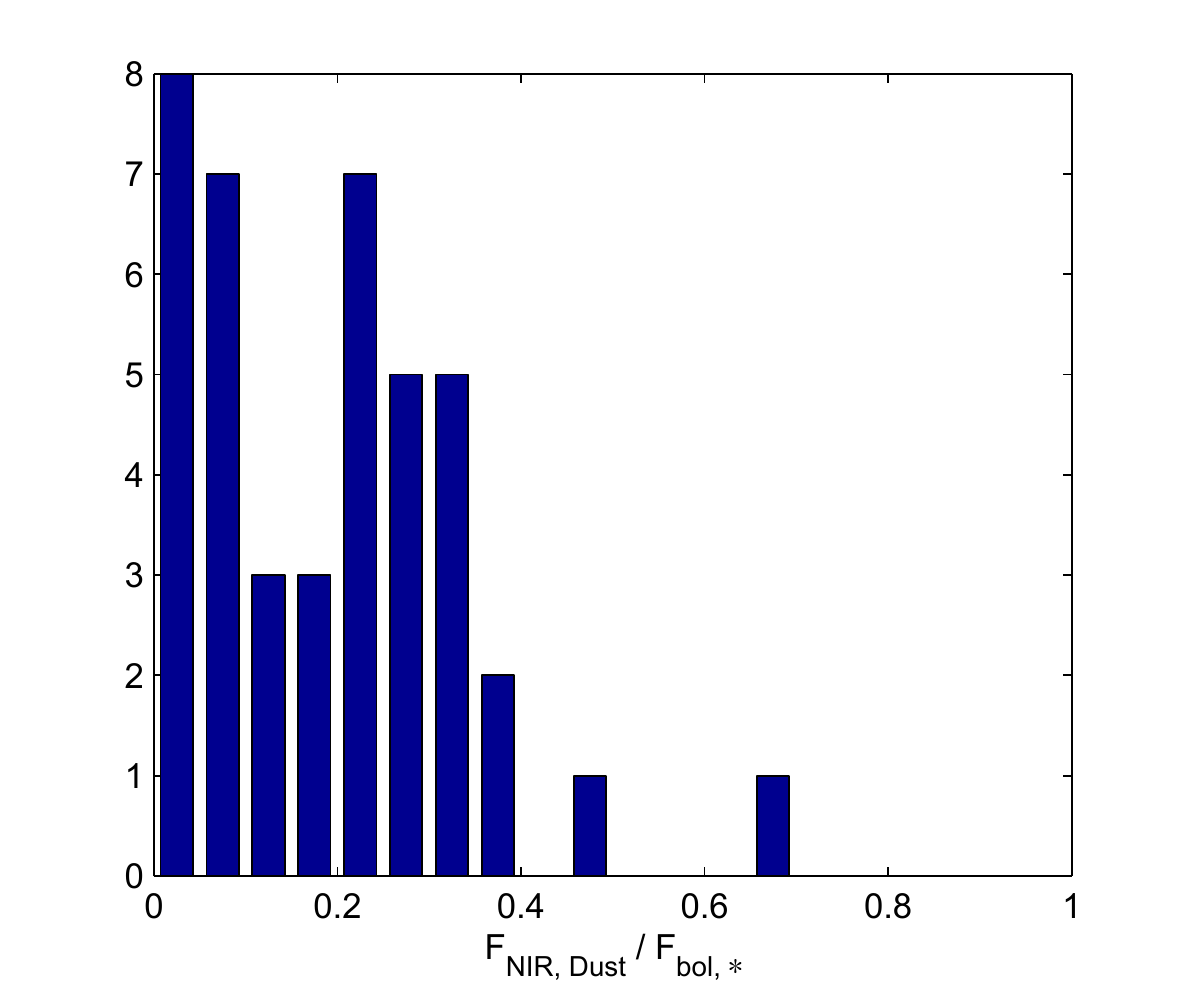}
\caption{Histogram of the fraction $f_\mathrm{rep}$ of stellar bolometric luminosity reprocessed to near infrared for non-binary objects. 
\label{histrepfrac}
}
\end{figure}

Another approach to the disk relative thickness is to consider the fraction of the stellar luminosity reprocessed to near-infrared. More specifically, we use the fits of the SED as discussed in sect.~\ref{analysphotom}, to derive: a)~the de-reddened stellar bolometric flux $F_{\mathrm{bol}\star}$; b)~the flux of the blackbody component $F_{\mathrm{NIR}}$. Remember that the data used in the fit do not extend longward of K band, so only hot dust is considered. 
We plot the histogram of the reprocessed energy fraction $f_\mathrm{rep} = F_{\mathrm{NIR}} / F_{\mathrm{bol}\star}$, for  the set of non-binary objects; the result is shown on Fig.~\ref{histrepfrac}. One can see that the histogram has a cutoff around 0.5.  One object, CO~Ori, is outside the limits of the histogram, at 
$f_\mathrm{rep}=1.64$; that object has a significant extinction $A_\mathrm{V}\approx 1.2$ and a spectral type G5, atypical for the HAeBe class.

%\begin{table}
%\caption{List of three outlier objects excluded fom the histogram on the left panel of Fig.~\ref{histrepfrac}, and \emph{additional} object is outside the bounds of the scatter plot in the right panel.}
%\label{repfracoutliers}
%\begin{tabular}{lccc}
%\hline
%Object         & $f_\mathrm{rep}$      & Sp.Type       & Sample        \\
%\hline
%\noalign{\vskip 2pt}    
%CO ORI         & 1.64                          & G5            & \\
%HD56895                & 2.45                          & F0            & \\
%R CRA          & 4.83                          & B5            & \\
%\hline
%\noalign{\vskip 2pt}    
%HD143006       & 0.46                          & G7            & HQ     \\
%\hline
%\end{tabular}
%\end{table}

The value of $f_\mathrm{rep}$ would have a straightforward geometrical interpretation in the case of a disk rim fully opaque to incident stellar radiation, but optically thin for re-emitted dust radiation; in that case, its value is simply the solid angle, as seen from the star, intercepted by the rim. Because of (a)~the self-obscuration of the disk in the NIR, and (b)~some  dependence of the emergent radiation  on the inclination relative to the local normal direction, one can expect the reprocessed fraction $f_\mathrm{rep}$ to depend on the inclination.  

We can, however, compare the median value of the observed sample with the similar metric for a model disk to derive a typical value. The median values of $f_\mathrm{rep}$ are:  for the 44 LP non-binary objects: 0.18; for the $in\_A0$ MCFOST models: 0.19; for the $thm\_A0$ models: 0.18.  The median for MCFOST models is restricted to inclinations for which the extinction of the central star $A_{V}<3$. Remember (section~\ref{simu_obs}) that the thickness of the IN and THM models was adjusted to achieve such an agreement. The scatter of  
$f_\mathrm{rep}$ seen in Fig.~\ref{histrepfrac} probably reflects the diversity of the objects included in our sample.

There is, in principle, a flaw in this procedure. In the previous subsection we argued that the flaring of the outer disk restricts the observed HAeBe disks  to the inclination range 
$1 \geq \cos i \geq 0.45$, while we are now calibrating the observed statistics of  $f_\mathrm{rep}$ against models without any outer flaring, that remain observable over a wider range of inclinations, typically $1 \geq \cos i \geq 0.2$. However, the model with the strongest dependence of $f_\mathrm{rep}$ versus inclination (a vertical-wall rim) is known, even before the present work, to be incompatible with the visibility curve. %, and even the less extreme IN model is excluded on the same basis; see Fig.~\ref{wrLp}. 
The favored THM model has only a weak dependence: 
$\Delta\,f_\mathrm{rep} / \Delta\,\cos i \approx 0.1$. 
%Models such as THM that have a radially extended emission also have a comparatively weaker dependence of $f_\mathrm{rep}$ versus inclination; see Fig.\ref{frepcosi}. 
This means that selection effects on 
$\cos i$ do not affect significantly our determination of the thickness 
$z/r$ of the inner disk. 

Another question is: why should the geometrical interpretation of the reprocessed fraction be correct? After all, the albedo of the grains in our models is close to 0.5 over the range of stellar photon wavelengths. But the diffusion is strongly forward-peaked; in MCFOST, with the Mie treatment, for $a =1.2\,\mu\mathrm{m}$  grains, and at 450~nm, the phase function drops to half the peak value within  $\approx 5\,\degr$ of the forward direction; so, diffusion will not hinder absorption. The situation is of course different for $1.65\,\mu\mathrm{m}$ stellar photons diffused off  grains (typically $\approx ~0.1\,\mu\mathrm{m}$) in the outer parts of the disk's surface. 

Summarizing the results of the previous and present section, we find typical values $z/r\approx 0.2$ for the inner disk, and $z/r\approx 0.45$ for the outer disk are needed to match the statistics of $f_\mathrm{rep}$ and 
$\cos i$ . The value $z/r\approx 0.2$ is known from previous work 
(see e.g., \citet{2006ApJ...636..348V}, \citet{2012A&A...539A...9M} and references therein) to be larger by a factor of approximately two than predicted by hydrostatic models. See discussion in \citet{2014A&A...566A.117V}. One possibility is that magnetic support increases the disk thickness over the hydrostatic value, as proposed by \citet{2014ApJ...780...42T}. 

\subsection{Luminosity-radius relation}
\label{lum_rad}

\begin{figure}
\centering
\includegraphics[width=85mm]{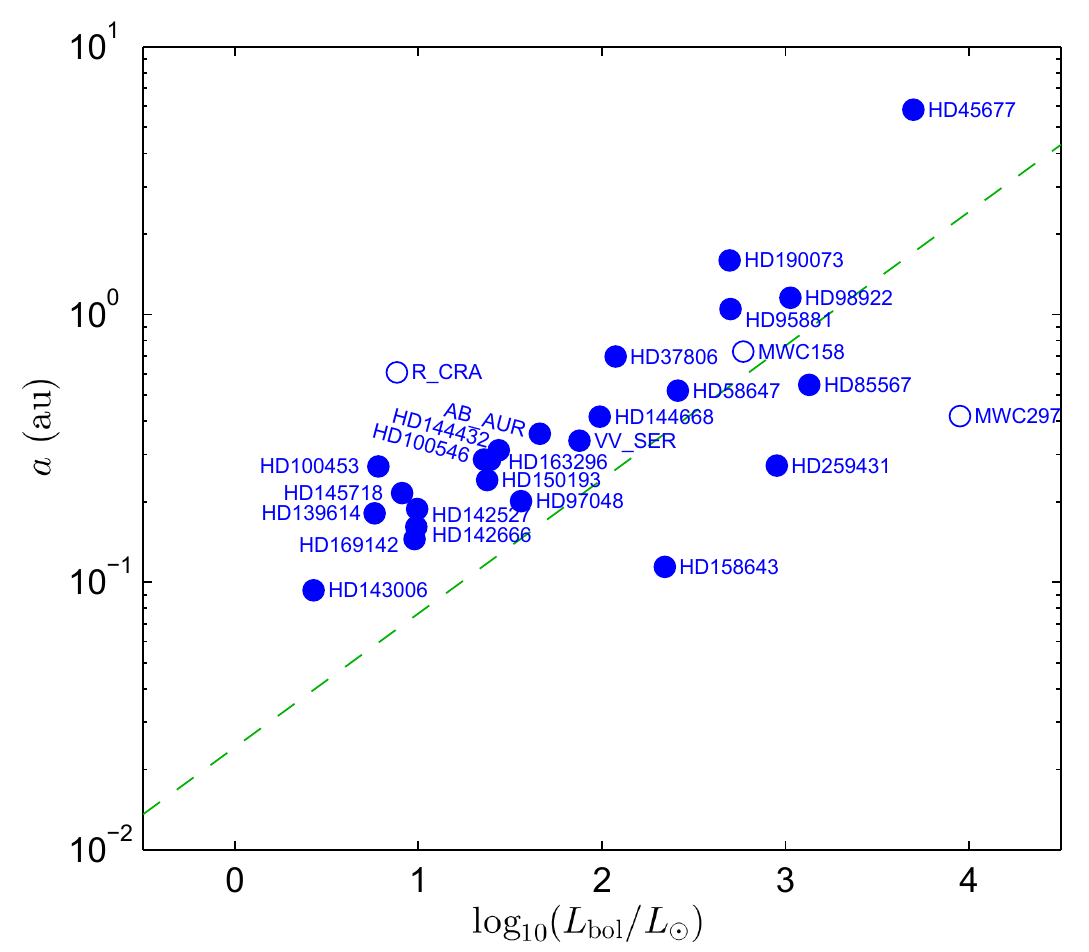}
\caption{Luminosity-radius plot for HQ objects. 
Open symbols denote three objects for which either the dust emission is dominant in H~band ($f_\mathrm{d}>0.90$) or the extinction is large ($A_V > 5$).
The dashed line shows, as a reference, the sublimation radius $a_\mathrm{sub,1800}$ for sublimation temperature $T_\mathrm{sub} = 1800\:\mathrm{K}$, cooling efficiency $\epsilon=1$, and backwarming factor
$Q_\mathrm{bw}=1$. }
\label{lumradius}
\end{figure}

The disks of Herbig AeBe stars are thought to have an inner radius bounded by dust sublimation. This can be tested quantitatively via the luminosity-radius relation \citep{2002ApJ...579..694M, 2005ApJ...624..832M}. We present in Fig.~\ref{lumradius} the radius-luminosity diagram for the 27 objects of the HQ sub-sample. It is prepared as follows: 
(a)~our SED fit provides a de-reddened $m_V$; (b)~using Simbad spectral types (Table~\ref{table_objectlist}) and  bolometric corrections from \citet{2013ApJS..208....9P}, we derive bolometric apparent magnitudes $m_b$; (c)~using parallaxes from  Hipparcos \citep{1997A&A...323L..49P} we convert $m_b$ to absolute luminosity $L_b$ and angular to linear radius. 
Some of our HQ central stars have benefited from detailed spectral studies: 15 of them by \citet{2015MNRAS.453..976F}, and one  by \citet{2009A&A...495..901M}; see Table~\ref{table_objectlist}. For these objects the value of $L_b$ derived as outlined above and the distance were superseded by the (presumably more reliable) values from these two references. 
%We plot data for the  HQ sub-sample, with the following additional conditions: (1)~$f_c >0.2$; (2)~spectral type A9 or earlier; (3)~luminosity class V or IV. 
Note that an error on the distance of an object moves the corresponding point in the diagram parallel to the theoretical $\log L - \log a$ line; so, accurate distances are not critical to establish a possible correlation. 

On the same plot, we have drawn the (or rather a) theoretical luminosity-radius relation.  Such a relation is defined by  \citet{2001Natur.409.1012T};  \citet{2002ApJ...579..694M} mention and neglect  backwarming, i.e. treat the equilibrium temperature of an isolated dust grain, while \citet{2009A&A...506.1199K} explicitly consider backwarming in the range from 
$C_\mathrm{bw}=1$ to $C_\mathrm{bw}=4$, the latter value being implicitly used by \citet{2001ApJ...560..957D}. The rim radius can be written as:
\begin{equation} 
\label{rsub}
\begin{aligned}
R_{rim} &= \frac{1}{2} \left(C_\mathrm{bw}/\epsilon\right)^{1/2} 
\left( T_{\ast}/T_\mathrm{sub}\right)^2 \, R_{\ast} \\
                &= \frac{1}{2}\left(C_\mathrm{bw}/\epsilon\right)^{1/2} 
\left( L_\mathrm{bol} / 4 \pi \sigma T_\mathrm{sub}^4\right)^{1/2}
\end{aligned} 
,\end{equation}
where $\epsilon=Q_\mathrm{abs}(T_\mathrm{sub}) / Q_\mathrm{abs}(T_{\ast})$, the cooling efficiency, is the ratio of Planck-averaged absorption cross-section for the considered grain species. 

The line drawn on Fig.~\ref{lumradius} expresses Eq.~\ref{rsub} with 
$T_\mathrm{sub}=1800\:\mathrm{K}$, $\epsilon=1$ and $C_\mathrm{bw}=1$. The sublimation temperature is fixed 
(see Sect.~\ref{dust_temp_photom} above).  The values  assigned to the other two are, however, arbitrary, and the relation as represented is best called a reference, rather than theoretical, relation. 

Several factors can cause the actual objects to deviate from that reference:
(1)~$\epsilon$ is expected to be $\leq 1$, reaching unity for large 
($a_\mathrm{grain} > \lambda_\mathrm{NIR}/(2\pi)$) grains. \citet{2001Natur.409.1012T} argue that the largest grains (largest $\epsilon$) are the last survivors, but one needs to assume that grain growth indeed proceeds to 
$\approx 0.5\mu\mathrm{m}$ size; 
(2)~the value adopted for the reference line, $C_\mathrm{bw}=1$ is the smallest possible;
(3)~the observed radius, with limited spatial resolution, is not the inner rim radius, but an effective  radius for the H-band emitting region, that will in all cases be larger than the inner rim radius. 

\begin{figure}
\centering
\includegraphics[width=80mm]{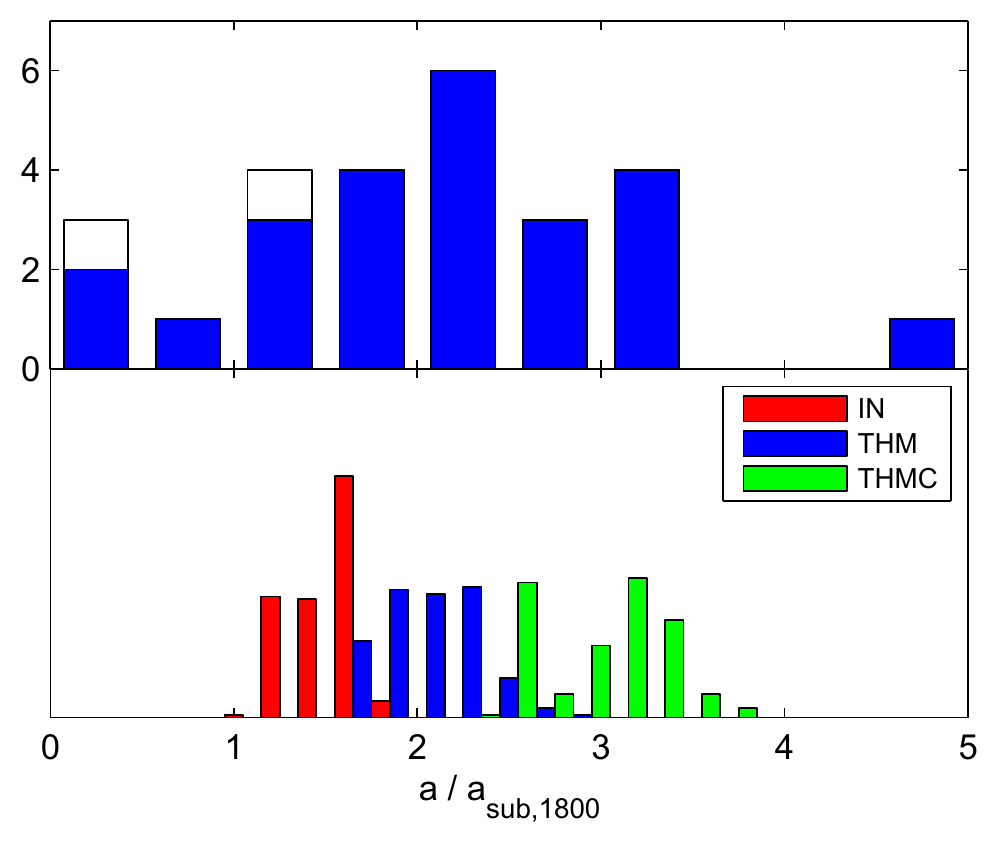}
\caption{
Top panel: Histogram of the ratio of the fitted half-light radius $a$ to $a_\mathrm{sub,1800}$, the rim radius for a sublimation temperature of 1800~K and no backwarming. The open blocks in the bars have the same meaning as in Fig.~\ref{lumradius}. One outlier, R~CrA is beyond the plot limits with $r_\mathrm{norm}\approx 9$. Bottom panel: histograms produced from simulated observations and parametric fits to three types of MCFOST models, each for three  central star types: A8, A0, and B2. 
\label{histradratio}
}
\end{figure}

All three of the listed factors tend to make the measured radius larger than the reference radius (at a given luminosity). To investigate this quantitatively, we display on Fig.~\ref{histradratio} the histogram of the ratio 
$r_\mathrm{norm}=a_\mathrm{fit}/a_\mathrm{sub,1800}$ of these two radii.The median value  is 
$\operatorname{median}(r_\mathrm{norm})=2.0$. Note that the sublimation temperature of the grain material is not used as an adjustment variable. 

Still using our procedure of substitution weighting, we process, with the same fitting procedure, images from three types of MCFOST models:
(1)~the IN model achieves values of $r_\mathrm{norm}$ in the range 1--2, mostly due to backwarming; 
(2)~the THM model achieves larger values of $r_\mathrm{norm}$ than the IN model, due to the larger radial extent of the H-band emitting region; it does not, however, seem to match the observed values; 
(3)~a third model, THMC, is similar to THM, but with smaller grain sizes and correspondingly smaller $\epsilon$ values:  $0.147\,\mu$m 
($\epsilon=0.14$) and $0.235\,\mu$m ($\epsilon=0.28$); the range of 
$r_\mathrm{norm}$ values is correspondingly larger. 

Taken together, the three models IN, THM, and THMC can account for a spread of $r_\mathrm{norm}$ over the range 1--4. HD~100453 ($r_\mathrm{norm} = 4.5$) exceeds that range by a small amount, while the five objects with 
$r_\mathrm{norm} < 1$ that can be identified on Fig.~\ref{lumradius} are more problematic.  The objects with the four lowest values of $r_\mathrm{norm}$, ranging between 0.18 and 0.59, all have dust temperatures (see \ref{analysphotom})  below 1760~K, so that anomalous refractory grains cannot be invoked. Anticipating the results displayed in Fig.~\ref{radiusringdisk}, we also see that an H-band emission component located inside the sublimation rim cannot plausibly explain a fit radius smaller than the reference radius. 

\subsection{Extended component}

Our data show the presence of a fully resolved flux in a significant fraction of our sample. This is quantified by the $f_\mathrm{h}$ fitted parameter, which is a fraction of the total H-band flux in the instrument's field of view.  This can be attributed to a spatially extended emission\footnote{larger than the PIONIER interferometric field of view (FOV), but smaller than its photometric FOV}; see for example,  \citet{2010PASJ...62..347F}, their Table 5.

Based on ISO spectra \citet{2001A&A...365..476M} defined two types of Herbig Ae Be stars: group I are stars for which the mid-infrared continuum spectrum can be modeled with one power law and one additional black-body to reproduce the rising mid-IR continuum; group II objects only require the power law. In subsequent mid-IR spectroscopy studies \cite{2005A&A...437..189V} and \cite{2010ApJ...721..431J} noticed that both groups could be sorted apart in a diagram involving the near-infrared (J, H, K, L and M) to far infrared flux ratio (IRAS 12, 25 and 60 $\mu\rm{m}$) and the IRAS color m$_{12}$- m$_{60}$\footnote{group I sources are defined as sources with
  $L_{\rm{NIR}}/L_{\rm{IR}} \leq (\rm{m}_{12} - \rm{m}_{60}) + 1.5 $}. These
parameters have been considered as indicators of respectively the ratio between the flux intercepted by the inner rim and the outer disk and the degree of
flaring.

\begin{figure}
\includegraphics[width=85mm]{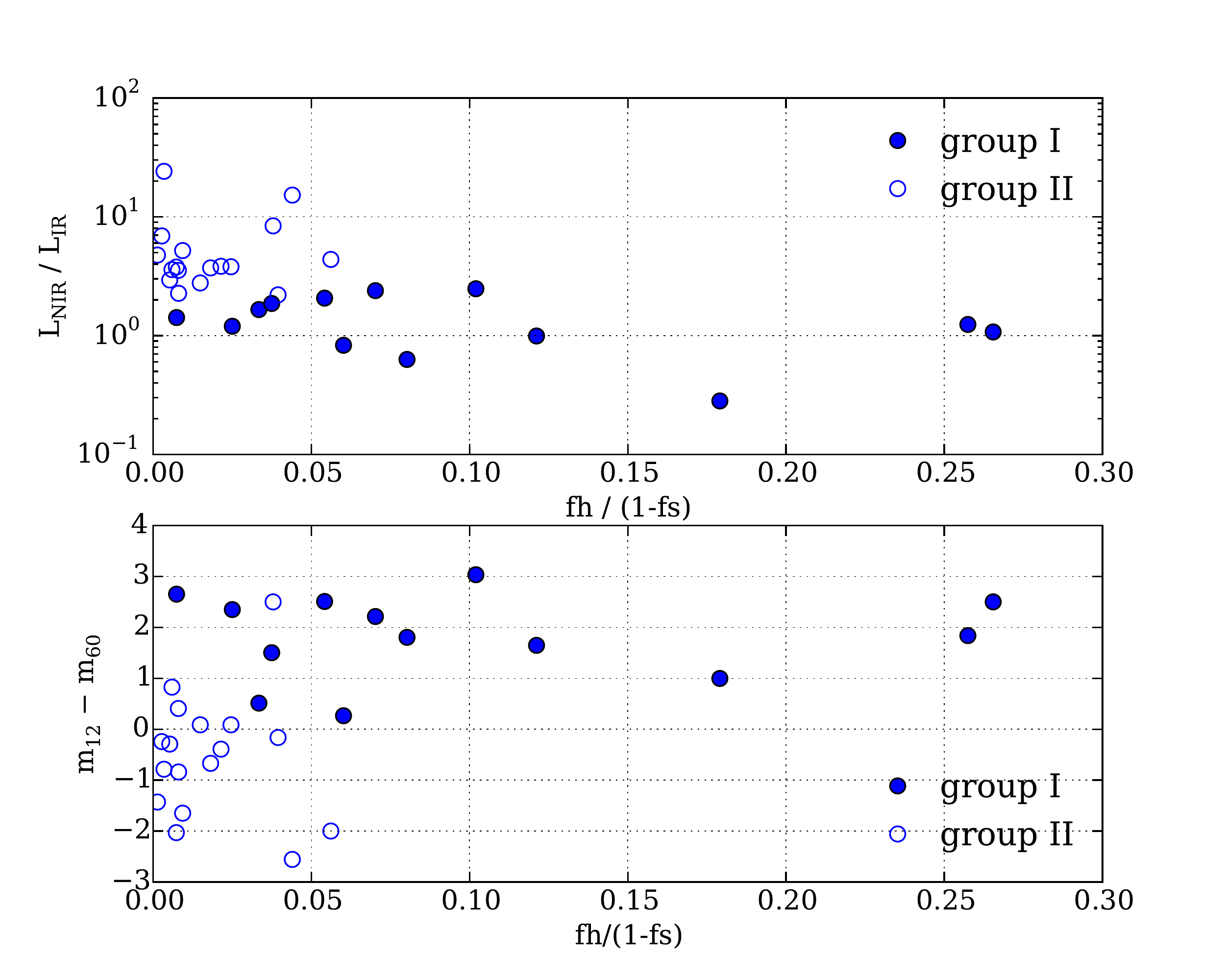}
\caption{
Fractional flux $f_\mathrm{h}$ of the halo component of objects versus their infrared color. 
}
\label{haloherbigclass}
\end{figure}

We explored whether we could find a correlation between such indicators
and our own estimator of extended flux (see Fig.~\ref{haloherbigclass}). The latter was constructed as the ratio between the extended flux fraction $f_\mathrm{h}$ and the global excess fraction $1-f_{s}$. We do find a clear indication that objects with a  higher ratio of near infrared to infrared flux  display small fully resolved flux while higher mid infrared color indices are correlated with higher extended flux.
This fits well in the scenario where indeed group I and group II are objects with significantly different large scale near to far infrared emission. 

We remind the reader of the discussions \citep{2013A&A...555A..64M,2015A&A...581A.107M} concerning the exact nature of the group I/II classification and the attempts to introduce notions of disk evolution (e.g., dust settling) and planet formation (e.g., gap clearing). It has even been proposed that the near-infrared  fully resolved flux might be caused partly by scattering off the inner rim of an outer disk in a transitional disk containing a large gap \citep{2010A&A...511A..75B,2011A&A...531A...1T}. We note that the $f_\mathrm{h}$ parameter provides a limited spatial information and do not wish to speculate more on its exact origin. In particular we recall that single-mode observations have a very restricted field of view, for PIONIER a Gaussian of roughly 200~mas field of view, which introduces a bias in the gap detection capability.  But, nevertheless, it would certainly be interesting to correlate it with other planet formation evidence.

\subsection{Inner gaseous disks}

The possibility of an inner gaseous disk located inside the dust sublimation rim of Herbig objects has drawn some attention in recent years, from several different viewpoints. Firstly, as logically unavoidable if the dusty disk must feed accretion at or near the stellar surface \citep{2004ApJ...617..406M}.  Second, a number of authors have directly detected gas inside the dust sublimation radius of various Herbig objects via spectrally resolved interferometry in the Br$\gamma$ line, for example: \citet{2008A&A...489.1157K,2009ApJ...692..309E,2015A&A...582A..44C,2016MNRAS.457.2236K,2015MNRAS.453.2126M,2014MNRAS.445.3723I}. Third, continuum (or low spectral resolution) observations, for example, \citet{2007ApJ...657..347E,2008A&A...483L..13I,2008ApJ...676..490K,2008ApJ...677L..51T,2008ApJ...689..513T,2010A&A...511A..74B},  infer the presence of an additional component inside the dust sublimation radius, that can be uniform, ring-like, or compact; note that the authors of the last paper argue in favor of an inner refractory dust disk.  Fourth, some authors have investigated the opacities and/or the emitted spectrum from a gaseous disk as might be found inside the dust sublimation radius; see \citet{2014A&A...568A..91M,2015MNRAS.454.2107V}, with the former authors concluding that the thermal equilibrium problem may be multi-valued. 

In view of the diversity of spatial distributions that have been proposed for the inner disk continuum emission, and that cannot at present be tied to a physical model, we could not submit the hypothesis of dust rim plus inner gas disk to the same systematic tests as we did in the previous subsections for the dust-rim-only hypothesis. The issue is not the presence of gas inside the dust sublimation radius, for which the Br$\gamma$ observations provide direct evidence. But, line opacities being typically several orders of magnitude larger than continuum ones, this does not prove that an inner disk should be a significant contributor to the continuum image and visibilities. 

Nevertheless, we did a simple test of a model comprising:
\begin{itemize}
\item a ring similar to the Ring model used above, except that an elastic constraint enforces $l_\mathrm{kr}=-0.3\pm0.15$, corresponding to a relative width $w'\approx 0.45^{+0.13}_{-0.11}$, intervals corresponding to $\Delta\bigl(\chi^2\bigr)=1$. Such rims are narrower than any of the fit results in Fig.\ref{wrLp}. 
\item a disk with a structure like the Ellipsoid model used above, and again with an elastic constraint $\log_{10}(a_\mathrm{disk}/a_\mathrm{ring})=0\pm0.1$, both $a$ values being half-light radii, and the interval given, again, for $\Delta\bigl(\chi^2\bigr)=1$. 
\end{itemize} 

\begin{figure}
\includegraphics[width=85mm]{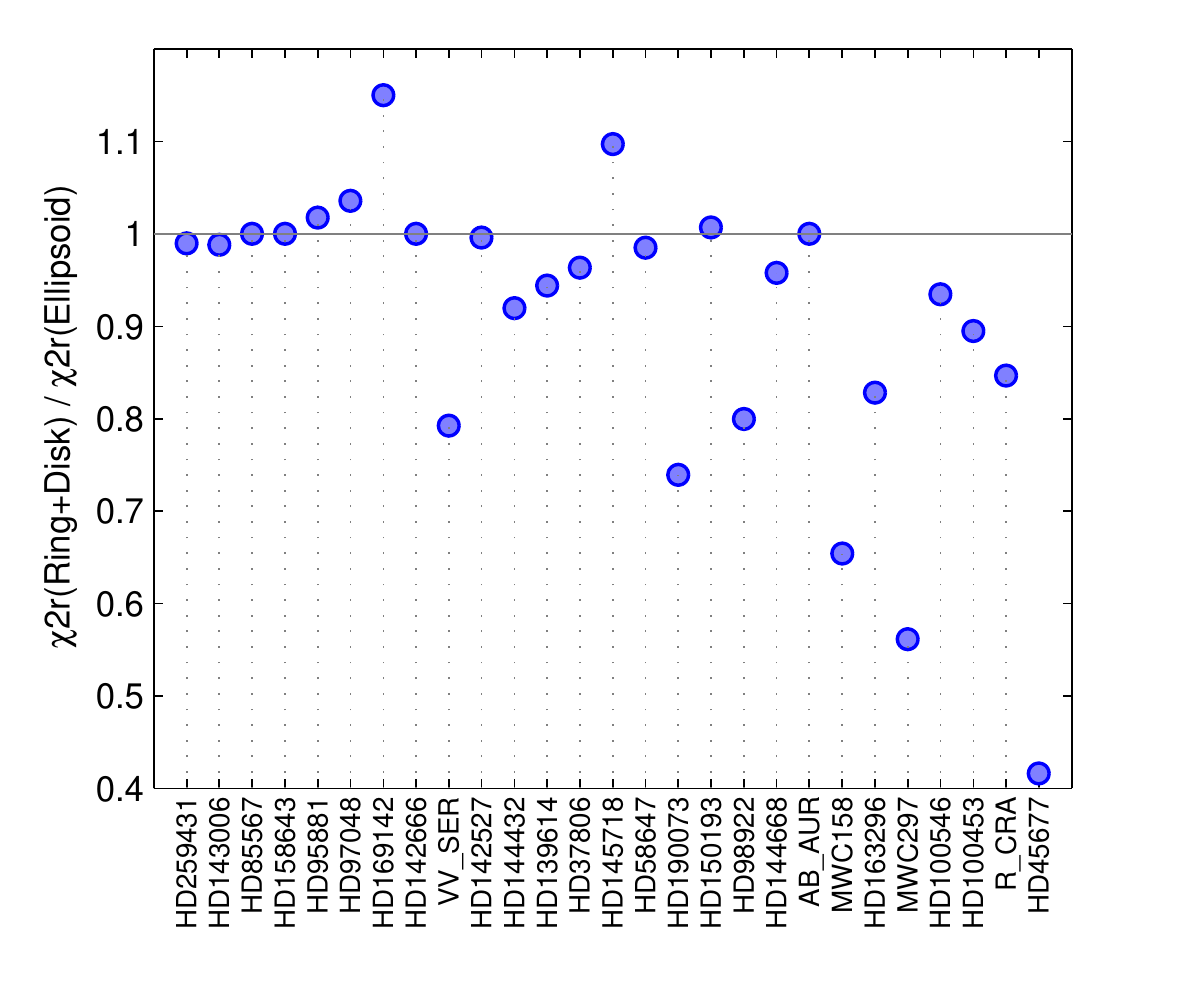} 
\caption{Value of $\chi^2$ achieved for the ring + inner disk models, normalized by the value for the Ellipsoid model; similar to Fig.\ref{chi2ringnoring}. The  $\chi^2$ achieved by the ring + inner disk model is, in some cases, worse than for the Ellipsoid model; this is because in contrast with the results shown in Fig.~\ref{chi2ringnoring}, the more complex model does not comprise the simpler one as a subset.}
\label{chi2ringdisk}
\end{figure}

\begin{figure}
\includegraphics[width=85mm]{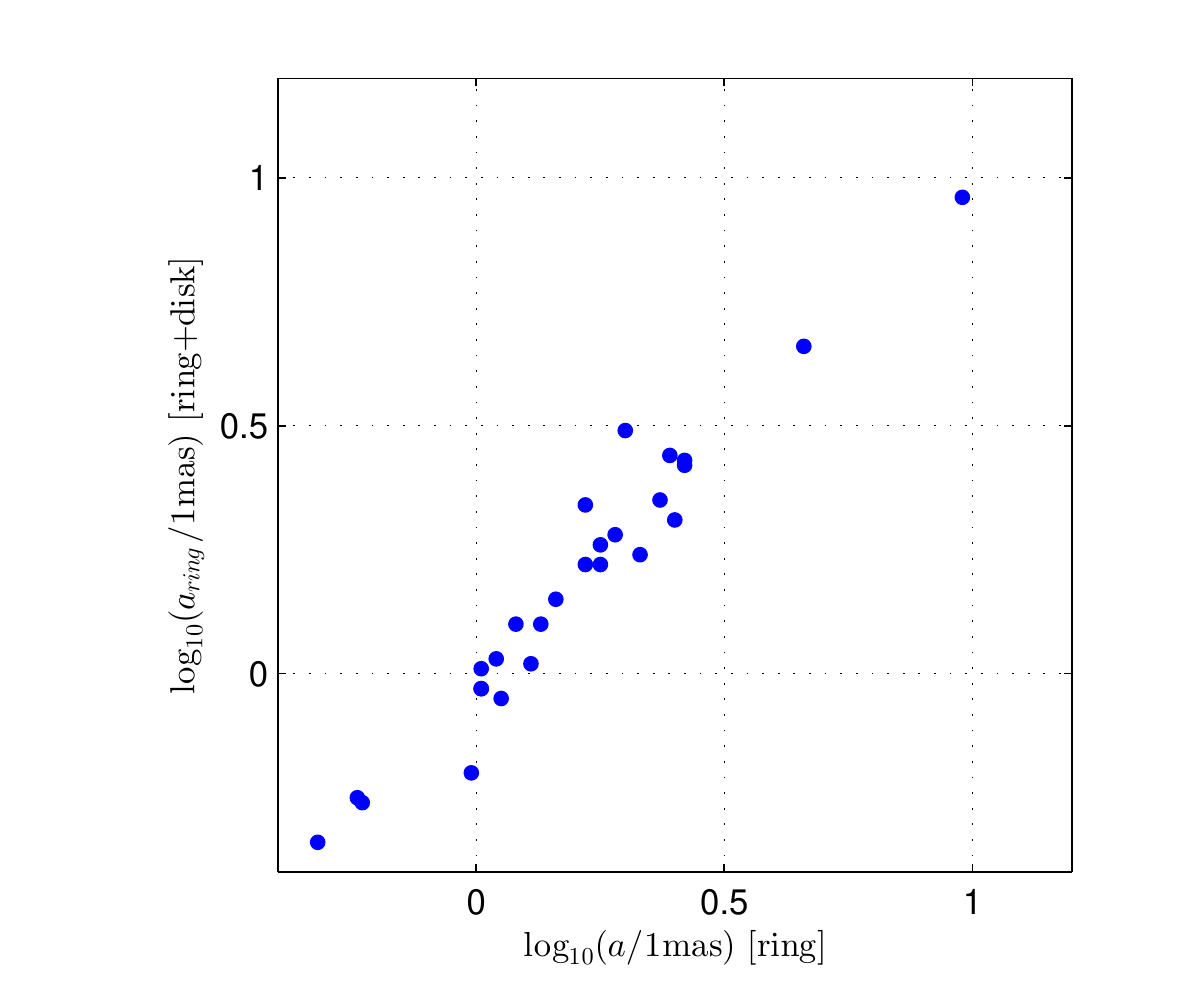} 
\caption{Relationship between ring radii derived in the ring+disk model and the previous Ring model.}
\label{radiusringdisk}
\end{figure}

We show just two results of this experiment. Figure~\ref{chi2ringdisk} shows the improvement in $\chi^2$ from an Ellipsoid fit to a ring+disk model. Comparing with Fig.\ref{chi2ringnoring}, one can see that the two are quite similar. Figure~\ref{radiusringdisk} compares the values obtained for the ring radius in the present versus the Ring model fit; again, the values are quite similar: the ratio has a mean of 0.975 and a standard deviation of 0.107. 

What can we gather from this limited test of the ring+disk model? The results are quite close to those of the Ring model. With our data, we have typically just enough resolution and precision to distinguish between a sharp rim and any other structure with a more damped visibility versus resolution curve, whether that is a wide rim, or a narrow one combined with an inner disk, central clump, or smaller ring. 

\section{Summary and conclusions}

We have presented interferometric H-band observations of 51 Herbig AeBe objects, of which 44 have a non-binary star, and 27 have been classified high quality (HQ) on the basis of the resolution achieved; a dedicated and homogeneous set of photometric measurements was also acquired. 

The spatial resolution achieved for a significant part of our sample allowed us to go beyond just size determinations, but, for most objects, was not sufficient to produce images. In order to make the fullest use of our data set, we examined the statistics of structural parameters obtained by fitting simple geometrical templates: ellipsoids and rings. We separately generated a small number of physical models of dust disks irradiated by the central star; they were observed with a $uv$ coverage similar to actual sample object, and these synthetic observations were in turn fitted exactly like the real observations. This allowed us to (i) eliminate possible biases of out fitting process; (ii) make statements on physical models, and not just on structural parameters. 

We have established, or confirmed, a number of properties of the dusty environment of HAeBe objects, that we summarize below under headings corresponding to the four goals spelt out in the Introduction. 

\subsection{Can we constrain the radial and vertical structure of the inner disk?}

\subsubsection*{Radial structure.} 
We have confirmed the presence of a ring-shaped structure (that tends to be taken for granted) in the Herbig objects of our sample, in several ways. First, for a substantial number of objects, fitting a ring structure results in a lower reduced $\chi^2$ than fitting an ellipsoid, this being more marked for the better resolved objects. Second, performing a parametric fit that allows both an ellipsoid or a ring shape, with a continuous transition, for a number of objects, a ring shape is preferred. Also, when a ring shape is preferred, it is a wide ring, with $w>0.4$, where $w$ is the ring width normalized by the global half-light radius. This excludes ``fat'' rim geometries with a substantial curvature radius and favors wedge-shaped rims, concurring with several recent physical models. 
\subsubsection*{Luminosity-radius relation.} 
We confirm the known luminosity-radius relation. The observed spread can be explained in part by variations in the rim geometry and in the maximum grain size, but a few objects have radii that are smaller than expected. The observed spread in sublimation temperatures can only be a minor contributor to the luminosity-radius scatter. 

\subsubsection*{Vertical structure.} 
The statistics of the reprocessed fraction (from the star's bolometric flux to the dust's NIR flux) allow us to infer a thickness $z/r\approx 0.2$ for the inner (meaning close and hot enough to radiate in the NIR) region of the dusty disk. The statistics of inferred inclination   show a histogram populated only at $\cos i >0.45$, that we interpret as resulting from self-obscuration by the outer parts of a flared ($z/r \approx 0.5$) disk.

\subsection{Can we constrain the rim dust temperature? }

We find that the dust responsible for the NIR emission of our sample objects has a sublimation temperature definitely larger than 1500~K, the favored value being 1800~K. This result is obtained by using radiative transfer models to calibrate out both the spread of temperatures in any real object, and possible biases in our fit procedure. Relating a sublimation temperature to dust properties is a difficult task, considering the range of  the physical conditions and possible time-dependent effects. Accordingly, we abstain from discussing dust composition.

\subsection{Can we constrain the presence of additional emitting structures?}

\subsubsection*{Extended component.} 
We find in a significant fraction of our sample a spatial component that is fully resolved at the shortest spacings 
($B/\lambda \approx 5\times10^6$), i.e. an angular scale larger than 40~mas. We also find that the presence and relative flux of that component are correlated with the Herbig group, either the original categorical form or based on infrared colors. Diffusion of starlight off a flared disk might be the cause of such extended components, but there may be other viable mechanisms. 

\subsubsection*{Inner disk.} 
Our conclusion (above) that rims are wide results essentially from the (near-)absence of bounce in the visibility curve beyond the first minimum. A number of authors have interpreted similar observations as resulting from  an extra component (hot gas, very refractory grains?) inside the dust sublimation radius. Such an alternate explanation, while  viable, is loosely constrained, lacking a predictive physical model.

\subsection{Can we reveal deviation from axisymmetry? }

\subsubsection*{Azimuthal modulation.} 
Fit models that include azimuthal modulation result in  lower values of $\chi^2_\mathrm{r}$; such azimuthal modulation is expected from self-shadowing of an inclined rim. To probe further that agreement between observations and models, we also tested the alignment of the modulation; we find a suggestive, but not overwhelming, agreement with the expected alignment. 

\subsubsection*{Phase closures.} 
This item is closely related to the previous one, except (a)~model-independent, and (b)~only odd azimuthal orders contribute. Phase closures versus resolution show, as expected, a maximum near dimensionless resolution $\rho'\approx 0.5$. The value of the maximum is of order $\phi_\mathrm{cp}\approx 10\degr$; one of our physical models (IN) comes close to this value, but the other one (THM) peaks much lower, at $\phi_\mathrm{cp}\approx 2\degr$; we leave this point open for future work.

\subsection{Outlook}

Our understanding of the structure and physics of Herbig AeBe objects would benefit, in our opinion, from: (a)~an increase in angular resolution of NIR interferometric observations by a modest factor (between two and three), that would provide a decisive improvement for all the objects that are marginally resolved with current data; (b)~the continuation and extension of theoretical or numerical models of the inner disk regions, notably with the following issues in mind: the shape of the rim; its vertical structure and hydrostatic equilibrium; a comprehensive treatment of the dusty rim and the inner gaseous disk; the deviations from axisymmetry. 

\begin{acknowledgements}
Based on observations made with ESO Telescopes at the La Silla Paranal Observatory under programme ID 190.C-0963. Generous time allocations by SAAO are gratefully acknowledged. We are grateful to Francois Van Wyk for diligently carrying out the infrared part of the photometry at Sutherland. JK acknowledges support from a Marie Sklodowska-Curie CIG grant (Ref. 618910, PI: Stefan Kraus). JDM and FB acknowledge support from NSF-AST 1210972. Part of the research was carried out at the Jet Propulsion Laboratory, California Institute of Technology, under a contract with the National Aeronautics and Space Administration. This research has made use of: the Jean-Marie Mariotti Center \texttt{Aspro2} and \texttt{SearchCal} services \footnote{Available at http://www.jmmc.fr/aspro}; the SIMBAD database, operated at CDS, Strasbourg, France; and data products from the Two Micron All Sky Survey, which is a joint project of the University of Massachusetts and the Infrared Processing and Analysis Center/California Institute of Technology. This work has made use of data from the European Space Agency (ESA) mission {\it Gaia} (\url{http://www.cosmos.esa.int/gaia})
\end{acknowledgements}

\bibliographystyle{aa}
\bibliography{HAeBeLP}

\begin{appendix}
\section{MCFOST models used for double-weighing the parametric fits}
\label{McfostModelDetails}

\begin{figure}
\centering
\includegraphics[width=85mm]{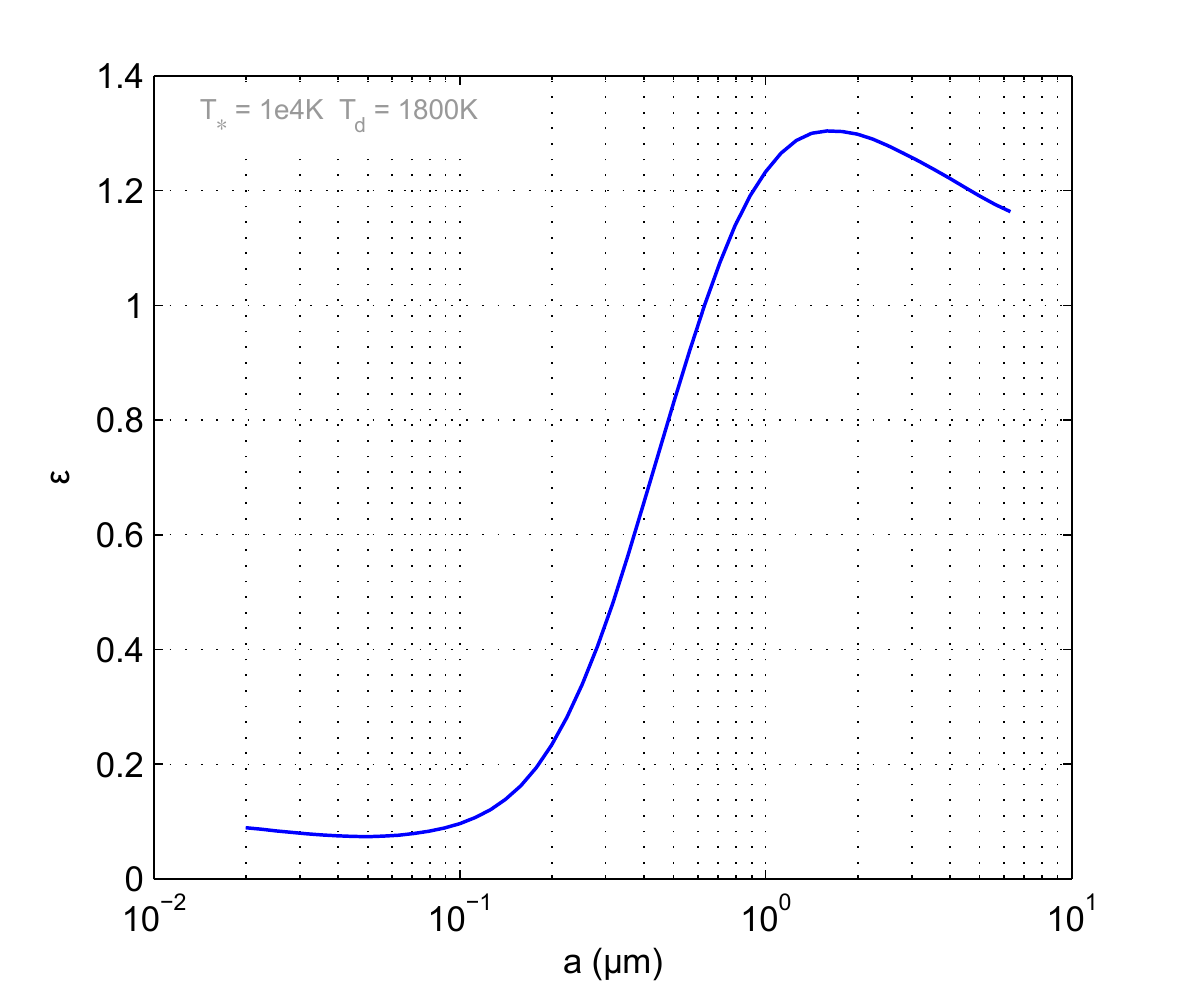}
\caption{
Cooling efficiency $\epsilon$ as a function of grain size, for the material adopted in our MCFOST simulations: amorphous carbon. The plot is made for $T_\mathrm{d}=1800K$ and $T_\ast = 10\,000K$. 
}
\label{cooleff}
\end{figure}

Here we describe in more details the MCFOST models used in conjunction with the parametric fits. We populate the inner rim region with grains having the highest sublimation temperature, meaning carbon grains (see Sect.~\ref{dust_temp_photom}), and the lowest equilibrium temperature, meaning large grains; altogether, the grains at the inner rim are the fittest for survival. We use in the MCFOST program amorphous carbon grains, with the optical constants from \citet{1991ApJ...377..526R}. Figure~\ref{cooleff} shows, versus grain radius $a$, the cooling efficiency $\epsilon$, the ratio of the Planck-weighted averages of the absorption cross-section at, respectively, the dust and star temperatures. 

\begin{figure*}
\centering
\includegraphics[width=170mm]{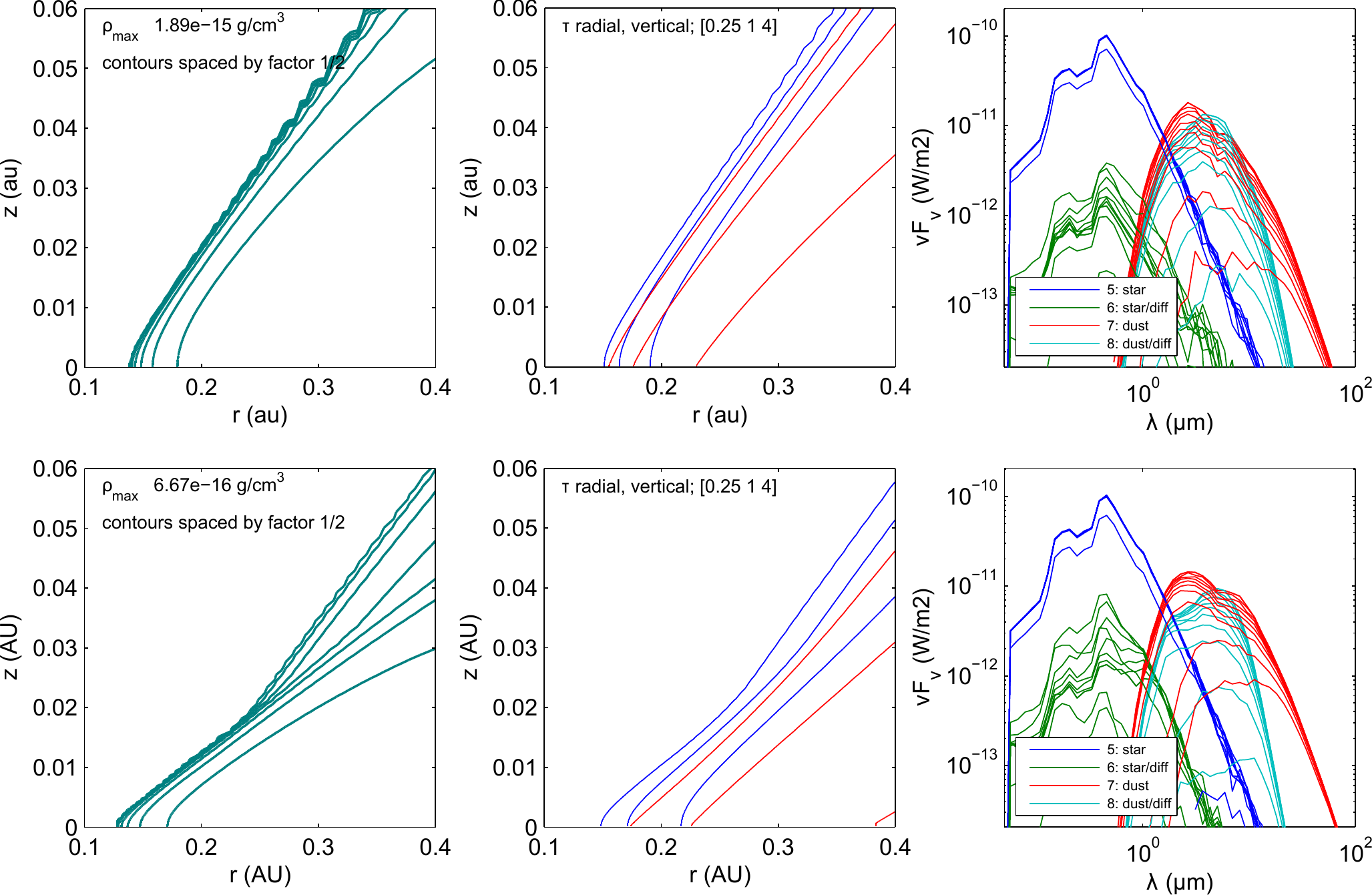}
\caption{
Properties of the IN and THM models, complementing those of Fig.~\ref{inthimages}. Top row: IN; bottom: THM. Left to right: dust mass density contours;  contours of radial (blue) and vertical (red) optical depth at values $\tau=0.25,\,1.0,\,4.0$;  several components of the SED at each of ten  inclinations; blue: stellar photosphere; green: ditto, diffused by dust; red: thermal dust emission; cyan: ditto, diffused by dust. 
}
\label{inthm_maps}
\end{figure*}

Our IN05-inspired model, labeled IN, has a curved inner rim and a single grain radius: 1.2$\mu$m. Our THM07-inspired model, labeled THM, has two grain radii: 1.2$\mu$m and 0.35$\mu$m; the smaller grains, because of their smaller emissivity in the NIR, cannot survive quite as close to the star as the larger grains, but, where they survive, their equilibrium temperature is higher than for larger grains, which helps to achieve a radially extended H-band emission. As explained in \citet{2007ApJ...661..374T}, the respective spatial distributions of the two grain species result naturally from differential sedimentation and differential sublimation. In the present work, however, the spatial distributions are ad hoc, subject to the condition that the maximum temperature for each species is equal to the assumed sublimation temperature $T_\mathrm{sub}=1800K$. Both the geometry and the range of grain sizes contribute to the THM model having a broader radial distribution of H-band emission than the IN model. The information provided in 
Fig.~\ref{inthm_maps} complements that already given in 
Fig.~\ref{inthimages}.

We simulate observations of each MCFOST model as follows:
(1)~we vary the inclination with values of $\cos(i)$ at the centers of 10 uniform bins;
(2)~we vary the distance in steps of $10^{0.1}$ subject (for each inclination) to the magnitude limits of our $HQ$ sub-sample: $4 \leq H \leq 8$; note that we claim neither that our sample is complete within these limits, nor that the distribution of distances in our simulated sample matches the observed sample, but we believe that this serves sufficiently well our purpose: to reveal trends while avoiding possible systematic biases caused by parametric fitting;
(3)~for each combination of inclination and distance, the $uv$ plane is sampled following the coverage of a typical Large Program object (HD~144668); noise is added; values of visibility squared $V^{2}$ and phase closure $\phi_\mathrm{cp}$ are derived; 
(4)~inclinations for which the $V$ band extinction of the central star is in excess of three magnitudes are ignored; while the limit is somewhat arbitrary, heavily obscured central stars will not appear in lists of HAeBe objects. 

Simulated noise is added to the model's complex visibilities; the resulting  noise for the (real) observables $V^{2}$ and $\phi_\mathrm{cp}$  has median values  slightly larger than the median values for the error estimates derived by the data reduction software for the real LP objects. We feel that matching the median error between real and simulated data is adequate for our purposes, and that a discussion of noise models for interferometric data is outside the scope of this work. We do not attempt to simulate the calibration errors or other systematic errors. 

Finally, the simulated observations of the MCFOST models are subject to the same parametric model fitting as the observational data. 

\begin{table}[ht]
\centering
\caption{Comparison of statistical error estimates for LP data and actual added noise for MCFOST models.}
\label{noisemedians}
\begin{tabular}{lcc}
\hline
                        & vis. squared  & ph. closure   \\
                        & $V^2$                 & $\phi_\mathrm{cp}$                      \\
\hline
LP                      & 0.016                         & 1.52$\degr$    \\
Models          & 0.019                         & 1.84$\degr$    \\
\hline
\end{tabular}
\tablefoot{
We use the median metric rather than root-mean-square to avoid giving an excessive weight to a few atypically bad observations.
}
\end{table}

\clearpage
\onecolumn
\section{Results of photometric and interferometric fits}

\begin{table*}[hbp]
\caption{Results of fits to the SED of program sources. The second value in a pair of columns is the 1-$\sigma$ estimate of the error derived from the scatter of the marginal distribution.}
\label{tablePhotomFit}
\centering
\begin{tabular}{lr@{\hskip 22pt}rr@{\hskip 22pt}rr@{\hskip 22pt}rr@{\hskip 22pt}rr@{\hskip 22pt}rr}
\hline \hline
    Object  &   $\chi^2_r $  &  
    \multicolumn{2}{c}{$\log_{10} \, F_{\nu \mathrm{s} V}$}  & 
    \multicolumn{2}{c}{$A_{\mathrm{mod},V}$} &  
    \multicolumn{2}{c}{$\log_{10} \, F_{\nu \mathrm{d} H}$}  &   
    \multicolumn{2}{c}{$\log_{10} \, T_\mathrm{d}$} &      
    \multicolumn{2}{c}{$f_\mathrm{d}$} \\
\hline
HD~17081      &  0.124  &  -6.386  &   0.011  & 0.036 & 0.028 & -10.841 & 0.848 & 3.134 & 0.280 & 0.001 & 0.002  \\ 
AB~AUR        &  1.002  &  -7.391  &   0.043  & 0.333 & 0.095 &  -7.864 & 0.017 & 3.260 & 0.017 & 0.780 & 0.021  \\ 
HD~31648      &  1.012  &  -7.721  &   0.044  & 0.136 & 0.092 &  -8.442 & 0.049 & 3.221 & 0.041 & 0.592 & 0.046  \\ 
HD~293782     &  0.589  &  -8.442  &   0.025  & 0.460 & 0.054 &  -9.082 & 0.023 & 3.192 & 0.018 & 0.666 & 0.021  \\ 
HD~34282      &  0.234  &  -8.490  &   0.027  & 0.445 & 0.059 &  -9.314 & 0.030 & 3.244 & 0.024 & 0.607 & 0.028  \\ 
CO~ORI        &  1.010  &  -8.747  &   0.044  & 1.728 & 0.090 &  -8.727 & 0.023 & 3.244 & 0.021 & 0.790 & 0.021  \\ 
HD~35929      &  0.967  &  -7.848  &   0.020  & 0.276 & 0.046 &  -9.330 & 0.159 & 3.145 & 0.088 & 0.145 & 0.049  \\ 
MWC~758       &  0.210  &  -7.898  &   0.026  & 0.324 & 0.056 &  -8.496 & 0.023 & 3.217 & 0.020 & 0.658 & 0.023  \\ 
HD~244604     &  0.084  &  -8.347  &   0.025  & 0.342 & 0.054 &  -9.110 & 0.027 & 3.224 & 0.023 & 0.623 & 0.025  \\ 
HD~36917      &  0.511  &  -7.595  &   0.029  & 0.778 & 0.062 &  -8.987 & 0.083 & 3.301 & 0.067 & 0.330 & 0.054  \\ 
T~ORI         &  1.005  &  -8.756  &   0.126  & 0.469 & 0.277 &  -8.705 & 0.067 & 3.196 & 0.048 & 0.915 & 0.026  \\ 
HD~37258      &  0.181  &  -8.410  &   0.026  & 0.330 & 0.056 &  -9.292 & 0.029 & 3.245 & 0.025 & 0.580 & 0.028  \\ 
HD~37357      &  0.372  &  -8.163  &   0.027  & 0.254 & 0.059 &  -9.194 & 0.039 & 3.251 & 0.031 & 0.490 & 0.035  \\ 
V~1247~ORI    &  0.355  &  -8.525  &   0.026  & 0.319 & 0.055 &  -9.208 & 0.031 & 3.216 & 0.025 & 0.595 & 0.028  \\ 
HD~37411      &  0.035  &  -8.452  &   0.027  & 0.468 & 0.060 &  -9.236 & 0.023 & 3.229 & 0.022 & 0.635 & 0.024  \\ 
HD~37806      &  1.136  &  -7.820  &   0.025  & 0.062 & 0.052 &  -8.499 & 0.037 & 3.180 & 0.030 & 0.676 & 0.026  \\ 
HD~38087      &  1.250  &  -7.626  &   0.039  & 0.996 & 0.084 &  -9.164 & 0.084 & 3.461 & 0.040 & 0.323 & 0.059  \\ 
HD~39014      &  0.344  &  -6.411  &   0.011  & 0.058 & 0.032 & -10.536 & 1.020 & 3.097 & 0.310 & 0.001 & 0.003  \\ 
HD~250550     &  0.616  &  -8.308  &   0.029  & 0.567 & 0.063 &  -8.791 & 0.016 & 3.245 & 0.015 & 0.802 & 0.014  \\ 
HD~45677      &  1.001  &  -7.285  &   0.024  & 0.880 & 0.055 &  -8.262 & 0.045 & 3.032 & 0.027 & 0.655 & 0.030  \\ 
HD~259431     &  1.002  &  -7.670  &   0.052  & 1.225 & 0.109 &  -8.429 & 0.031 & 3.245 & 0.029 & 0.720 & 0.035  \\ 
MWC~158       &  0.257  &  -7.196  &   0.024  & 0.386 & 0.056 &  -7.929 & 0.020 & 3.213 & 0.017 & 0.714 & 0.018  \\ 
HD~53367      &  1.308  &  -6.863  &   0.074  & 1.530 & 0.160 &  -8.271 & 0.157 & 3.426 & 0.080 & 0.382 & 0.117  \\ 
HD~56895      &  0.617  &  -7.972  &   0.027  & 0.204 & 0.061 &  -7.749 & 0.014 & 3.189 & 0.010 & 0.905 & 0.006  \\ 
HD~58647      &  0.091  &  -7.280  &   0.022  & 0.362 & 0.051 &  -8.605 & 0.064 & 3.148 & 0.039 & 0.364 & 0.043  \\ 
HD~85567      &  1.008  &  -7.816  &   0.061  & 0.668 & 0.134 &  -8.482 & 0.046 & 3.217 & 0.046 & 0.748 & 0.039  \\ 
HD~95881      &  0.094  &  -7.807  &   0.026  & 0.483 & 0.058 &  -8.403 & 0.020 & 3.211 & 0.018 & 0.734 & 0.018  \\ 
HD~97048      &  1.008  &  -7.641  &   0.032  & 1.259 & 0.070 &  -8.640 & 0.043 & 3.242 & 0.035 & 0.539 & 0.040  \\ 
HD~98922      &  0.130  &  -7.264  &   0.024  & 0.359 & 0.055 &  -8.047 & 0.026 & 3.173 & 0.020 & 0.657 & 0.023  \\ 
HD~100453     &  0.288  &  -7.804  &   0.014  & 0.036 & 0.031 &  -8.785 & 0.049 & 3.109 & 0.030 & 0.375 & 0.031  \\ 
HD~100546     &  0.904  &  -7.355  &   0.013  & 0.023 & 0.023 &  -8.710 & 0.059 & 3.189 & 0.042 & 0.318 & 0.034  \\ 
HD~104237     &  0.197  &  -7.296  &   0.027  & 0.183 & 0.059 &  -8.034 & 0.029 & 3.260 & 0.028 & 0.577 & 0.028  \\ 
HD~139614     &  0.370  &  -7.989  &   0.017  & 0.044 & 0.036 &  -9.081 & 0.053 & 3.161 & 0.034 & 0.350 & 0.033  \\ 
HD~141569     &  0.583  &  -7.428  &   0.028  & 0.296 & 0.058 &  -9.225 & 0.182 & 3.391 & 0.101 & 0.144 & 0.056  \\ 
HD~142666     &  0.632  &  -7.904  &   0.024  & 0.672 & 0.055 &  -8.657 & 0.035 & 3.177 & 0.025 & 0.524 & 0.030  \\ 
HD~142527     &  0.681  &  -7.714  &   0.025  & 0.798 & 0.057 &  -8.267 & 0.032 & 3.225 & 0.025 & 0.535 & 0.029  \\ 
HD~143006     &  0.304  &  -8.666  &   0.027  & 0.196 & 0.058 &  -9.039 & 0.038 & 3.215 & 0.031 & 0.488 & 0.035  \\ 
HD~144432     &  0.183  &  -7.857  &   0.027  & 0.287 & 0.058 &  -8.615 & 0.034 & 3.250 & 0.030 & 0.521 & 0.033  \\ 
HD~144668     &  0.281  &  -7.289  &   0.029  & 0.705 & 0.063 &  -7.874 & 0.021 & 3.283 & 0.021 & 0.693 & 0.022  \\ 
HD~145718     &  0.197  &  -7.921  &   0.022  & 0.757 & 0.052 &  -9.066 & 0.060 & 3.205 & 0.045 & 0.348 & 0.040  \\ 
HD~149914     &  1.600  &  -6.985  &   0.044  & 1.019 & 0.101 &  -8.973 & 0.862 & 3.400 & 0.172 & 0.105 & 0.098  \\ 
HD~150193     &  0.214  &  -7.719  &   0.033  & 1.288 & 0.068 &  -8.173 & 0.017 & 3.297 & 0.019 & 0.770 & 0.018  \\ 
AK~SCO        &  0.293  &  -8.185  &   0.063  & 0.470 & 0.138 &  -8.901 & 0.095 & 3.191 & 0.055 & 0.408 & 0.085  \\ 
HD~158643     &  0.236  &  -6.576  &   0.022  & 0.097 & 0.049 &  -8.507 & 0.341 & 3.161 & 0.184 & 0.109 & 0.067  \\ 
HD~163296     &  0.215  &  -7.386  &   0.024  & 0.117 & 0.053 &  -8.083 & 0.023 & 3.203 & 0.020 & 0.669 & 0.021  \\ 
HD~169142     &  1.371  &  -7.935  &   0.013  & 0.021 & 0.021 &  -9.416 & 0.145 & 3.217 & 0.100 & 0.147 & 0.042  \\ 
MWC~297       &  1.001  &  -6.756  &   0.093  & 7.189 & 0.205 &  -7.040 & 0.061 & 3.220 & 0.044 & 0.914 & 0.020  \\ 
VV~SER        &  1.006  &  -8.211  &   0.033  & 3.052 & 0.069 &  -8.513 & 0.015 & 3.236 & 0.015 & 0.880 & 0.009  \\ 
R~CRA         &  1.003  &  -8.480  &   0.098  & 3.308 & 0.201 &  -7.568 & 0.051 & 3.127 & 0.025 & 0.992 & 0.002  \\ 
HD~179218     &  0.256  &  -7.560  &   0.017  & 0.271 & 0.037 &  -8.749 & 0.048 & 3.173 & 0.034 & 0.406 & 0.032  \\ 
HD~190073     &  1.002  &  -7.795  &   0.019  & 0.068 & 0.040 &  -8.584 & 0.031 & 3.195 & 0.026 & 0.601 & 0.024  \\ 
\hline
\end{tabular}
\end{table*}

\begin{table*}[tbp]
\caption{Values for the eight parameters of the fit of an Ellipsoid model. }
\label{tableellFit}
\centering
\begin{tabular}{lrrrrrrrrr}
\hline \hline
Object  & $\chi^{2}r$ & $k_\mathrm{c}$ & $f_\mathrm{s}$ & $f_\mathrm{h}$ & 
$\cos i$ & $\theta$ & $f_\mathrm{Lor}$ & $l_\mathrm{a}$ & $f_\mathrm{c}$ \\
\hline
HD~17081     &  0.17 &  1.66 & 0.96 &  0.00 & 0.76 & -0.20 & 0.00 & -0.13 & 0.04  \\ 
AB~AUR       &  0.40 & -5.85 & 0.19 &  0.03 & 0.98 &  2.14 & 0.27 &  0.32 & 0.78  \\ 
HD~31648     &  0.45 & -0.48 & 0.42 &  0.01 & 0.78 &  2.23 & 0.16 &  0.28 & 0.57  \\ 
HD~293782    &  3.33 &   -   & 0.33 &  0.00 & 0.87 &  1.54 & 0.66 & -0.10 & 0.67  \\ 
HD~34282     &  2.13 &   -   & 0.30 &  0.08 & 0.41 &  2.29 & 0.33 & -0.08 & 0.62  \\ 
CO~ORI       &  1.06 & -5.22 & 0.22 &  0.00 & 0.66 &  0.67 & 0.42 & -0.23 & 0.78  \\ 
HD~35929     &  0.59 & -1.06 & 0.86 &  0.00 & 0.85 &  0.27 & 0.00 & -0.25 & 0.14  \\ 
MWC~758      &  0.11 & -3.32 & 0.33 &  0.02 & 0.66 &  1.74 & 0.05 &  0.13 & 0.65  \\ 
HD~244604    &  0.00 &   -   & 0.38 &  0.00 & 0.58 &  0.35 & 0.64 & -0.24 & 0.62  \\ 
HD~36917     &  1.19 & -5.82 & 0.69 &  0.03 & 0.01 &  1.67 & 0.50 & -0.58 & 0.29  \\ 
HD~37258     &  1.51 &   -   & 0.42 &  0.00 & 0.41 &  2.48 & 0.01 & -1.63 & 0.58  \\ 
V~1247~ORI   &  0.99 &   -   & 0.38 &  0.03 & 0.89 &  3.08 & 0.14 & -0.36 & 0.59  \\ 
HD~37411     &  0.29 &   -   & 0.36 &  0.00 & 0.45 &  2.00 & 0.46 & -0.18 & 0.64  \\ 
HD~37806     &  3.87 & -3.54 & 0.23 &  0.01 & 0.77 &  0.75 & 0.37 &  0.16 & 0.76  \\ 
HD~39014     &  1.85 &  1.97 & 0.91 &  0.01 & 0.59 &  0.65 & 0.03 &  0.07 & 0.08  \\ 
HD~250550    &  0.36 &   -   & 0.17 &  0.03 & 0.62 &  2.31 & 0.17 & -0.13 & 0.79  \\ 
HD~45677     & 13.77 & -4.11 & 0.40 &  0.00 & 0.73 &  1.22 & 0.00 &  0.99 & 0.60  \\ 
HD~259431    &  0.99 & -5.39 & 0.25 &  0.02 & 0.91 &  0.66 & 0.59 & -0.34 & 0.73  \\ 
MWC~158      &  7.77 & -5.75 & 0.14 &  0.02 & 0.56 &  1.20 & 0.98 &  0.31 & 0.84  \\ 
HD~58647     &  1.36 & -4.90 & 0.59 &  0.02 & 0.48 &  0.23 & 0.24 &  0.19 & 0.39  \\ 
HD~85567     &  1.37 & -3.36 & 0.23 &  0.00 & 0.76 &  2.02 & 0.30 & -0.27 & 0.77  \\ 
HD~95881     &  1.13 & -5.50 & 0.27 &  0.04 & 0.65 &  2.74 & 0.75 & -0.00 & 0.69  \\ 
HD~97048     &  0.84 & -4.18 & 0.47 &  0.04 & 0.53 &  2.90 & 0.71 &  0.03 & 0.49  \\ 
HD~98922     &  6.81 & -6.00 & 0.17 &  0.01 & 0.92 &  0.98 & 0.78 &  0.25 & 0.82  \\ 
HD~100453    &  2.00 & -5.97 & 0.58 &  0.11 & 0.67 &  1.42 & 0.09 &  0.39 & 0.31  \\ 
HD~100546    &  4.44 & -3.42 & 0.48 &  0.11 & 0.66 &  2.58 & 0.57 &  0.37 & 0.42  \\ 
HD~139614    &  1.06 & -5.74 & 0.54 &  0.06 & 0.83 &  0.18 & 0.89 &  0.11 & 0.41  \\ 
HD~141569    &  0.61 &  0.22 & 0.86 &  0.00 & 0.21 &  1.92 & 0.80 & -0.53 & 0.13  \\ 
HD~142666    &  1.53 & -3.64 & 0.44 &  0.02 & 0.50 &  2.82 & 0.41 &  0.03 & 0.54  \\ 
HD~142527    &  2.52 & -3.90 & 0.41 &  0.03 & 0.83 &  0.12 & 0.43 &  0.06 & 0.56  \\ 
HD~143006    &  0.85 & -4.64 & 0.48 &  0.03 & 0.89 &  0.01 & 0.19 & -0.30 & 0.49  \\ 
HD~144432    &  1.63 & -2.66 & 0.43 &  0.00 & 0.90 &  1.38 & 0.86 &  0.09 & 0.57  \\ 
HD~144668    &  2.84 & -2.64 & 0.38 &  0.00 & 0.59 &  2.18 & 0.25 &  0.27 & 0.62  \\ 
HD~145718    &  1.02 & -2.88 & 0.59 &  0.03 & 0.45 &  0.03 & 1.00 &  0.15 & 0.38  \\ 
HD~149914    &  0.51 &  1.76 & 0.88 &  0.00 & 0.87 &  0.85 & 0.05 & -0.46 & 0.12  \\ 
HD~150193    &  1.43 & -3.18 & 0.30 &  0.00 & 0.82 &  2.12 & 0.77 &  0.21 & 0.70  \\ 
HD~158643    &  0.57 & -2.25 & 0.83 &  0.01 & 0.65 &  2.13 & 0.88 &  0.03 & 0.16  \\ 
HD~163296    &  1.56 & -3.72 & 0.24 &  0.00 & 0.68 &  2.24 & 0.41 &  0.34 & 0.76  \\ 
HD~169142    &  1.00 & -5.99 & 0.76 &  0.08 & 0.93 &  1.81 & 0.88 & -0.01 & 0.17  \\ 
MWC~297      &  9.84 & -5.64 & 0.11 &  0.01 & 0.89 &  1.65 & 0.00 &  0.36 & 0.87  \\ 
VV~SER       &  2.35 & -5.97 & 0.08 &  0.02 & 0.52 &  0.02 & 0.50 &  0.03 & 0.90  \\ 
R~CRA        &  4.88 & -1.42 & 0.04 &  0.06 & 0.69 &  0.12 & 0.52 &  0.63 & 0.90  \\ 
HD~179218    &  2.55 & -2.46 & 0.58 &  0.02 & 0.66 &  0.50 & 0.03 &  1.38 & 0.40  \\ 
HD~190073    &  1.83 & -3.19 & 0.34 &  0.00 & 0.93 &  2.88 & 0.00 &  0.19 & 0.66  \\ 
\hline
\end{tabular}
\tablefoot{
 This fit was performed on the 44 non-binary program objects. See Table~\ref{parfitell} for symbol details. No value is given for the spectral index $k_\mathrm{c}$ of the circumstellar component for those objects where 50\% or more of the observations were made in non-dispersed mode. Error bars are not shown because of page width limitation; a complete version of this table can be found online at \textbf{(CDS TBD)}
}
\end{table*}

\begin{table*}[h]
\caption{Values for the eleven parameters of the fit of a Ring model with first order azimuthal modulation. }
\label{table_fitM1}
\centering
\begin{tabular}{lrrrrrrrrrrrr}
\hline
Object  & $\chi^{2}r$ & $k_\mathrm{c}$ & $f_\mathrm{s}$ & $f_\mathrm{h}$ & 
$\cos i$ & $\theta$ & $f_\mathrm{Lor}$ & $l_\mathrm{a}$ & $f_\mathrm{r}$ &
$l_\mathrm{kr}$ & $c_1$ & $s_1$ \\
\hline
 AB~AUR      &     0.79 & -5.75 & 0.24 & 0.03 & 0.91 & 1.40 & 0.31 &  0.37 & 0.73 &  0.08 &  0.97 &  0.18   \\
 HD~37806    &     2.61 & -1.73 & 0.34 & 0.00 & 0.75 & 0.79 & 0.94 &  0.22 & 0.66 & -0.18 &  0.08 & -0.13   \\
 HD~45677    &     5.59 & -4.12 & 0.42 & 0.03 & 0.63 & 1.20 & 1.00 &  0.98 & 0.55 & -0.26 & -0.18 &  0.98   \\
 HD~259431   &     1.43 & -5.18 & 0.27 & 0.02 & 0.98 & 0.82 & 0.99 & -0.35 & 0.71 & -0.13 &  0.05 & -1.00   \\
 MWC~158     &    21.53 & -3.25 & 0.24 & 0.12 & 0.53 & 1.26 & 1.00 &  0.33 & 0.63 & -0.08 &  1.00 & -0.02   \\
 HD~58647    &     1.41 & -3.98 & 0.65 & 0.02 & 0.47 & 0.23 & 0.66 &  0.25 & 0.33 & -0.30 & -0.07 & -0.26   \\
 HD~85567    &     1.15 & -2.04 & 0.29 & 0.00 & 0.79 & 2.10 & 0.38 & -0.22 & 0.70 &  0.09 &  0.28 & -0.96   \\
 HD~95881    &     1.41 & -4.44 & 0.32 & 0.05 & 0.62 & 2.73 & 0.85 &  0.05 & 0.64 &  0.15 & -0.99 & -0.15   \\
 HD~97048    &     0.81 & -3.84 & 0.48 & 0.04 & 0.56 & 2.89 & 0.85 &  0.05 & 0.48 &  0.07 &  0.60 &  0.78   \\
 HD~98922    &     5.83 & -5.85 & 0.21 & 0.03 & 0.94 & 1.20 & 0.90 &  0.25 & 0.76 &  0.25 &  0.42 & -0.66   \\
 HD~100453   &     1.29 & -5.31 & 0.61 & 0.09 & 0.66 & 1.41 & 0.83 &  0.42 & 0.30 & -0.21 &  0.01 &  0.03   \\
 HD~100546   &     3.96 & -2.43 & 0.55 & 0.11 & 0.69 & 2.66 & 1.00 &  0.42 & 0.34 & -0.22 &  0.17 & -0.17   \\
 HD~139614   &     1.02 & -5.56 & 0.57 & 0.06 & 0.85 & 0.25 & 1.00 &  0.14 & 0.37 &  0.16 &  0.16 & -0.59   \\
 HD~142666   &     1.36 & -4.18 & 0.45 & 0.02 & 0.50 & 2.84 & 0.58 &  0.03 & 0.53 &  0.13 & -0.04 &  0.00   \\
 HD~142527   &     2.27 & -3.67 & 0.42 & 0.03 & 0.84 & 0.09 & 0.42 &  0.08 & 0.55 &  0.54 & -0.60 & -0.57   \\
 HD~143006   &     0.86 & -4.26 & 0.50 & 0.03 & 0.88 & 2.94 & 0.66 & -0.25 & 0.47 & -0.51 &  0.61 &  0.20   \\
 HD~144432   &     1.75 & -2.41 & 0.45 & 0.00 & 0.91 & 1.35 & 0.99 &  0.09 & 0.55 &  0.32 & -0.18 &  0.81   \\
 HD~144668   &     3.27 & -2.48 & 0.40 & 0.00 & 0.61 & 2.22 & 0.21 &  0.30 & 0.60 &  0.20 & -0.39 &  0.92   \\
 HD~145718   &     1.11 & -2.82 & 0.60 & 0.03 & 0.47 & 0.03 & 1.00 &  0.16 & 0.37 &  0.77 &  0.18 & -0.86   \\
 HD~150193   &     1.84 & -2.69 & 0.33 & 0.00 & 0.85 & 2.13 & 0.82 &  0.22 & 0.67 &  0.39 &  0.24 & -0.97   \\
 HD~158643   &     0.64 & -2.38 & 0.81 & 0.01 & 0.66 & 2.15 & 0.98 & -0.06 & 0.18 &  0.47 & -0.82 & -0.51   \\
 HD~163296   &     2.18 & -2.64 & 0.32 & 0.00 & 0.67 & 2.20 & 0.63 &  0.39 & 0.68 & -0.04 & -0.54 & -0.35   \\
 HD~169142   &     0.96 & -6.00 & 0.76 & 0.08 & 0.92 & 2.18 & 0.86 &  0.00 & 0.16 &  0.54 &  0.45 & -0.60   \\
 MWC~297     &     9.34 & -3.39 & 0.15 & 0.00 & 0.88 & 1.82 & 0.40 &  0.39 & 0.85 & -0.11 &  0.08 & -0.11   \\
 VV~SER      &     1.98 & -4.36 & 0.12 & 0.00 & 0.60 & 0.08 & 0.99 &  0.01 & 0.88 & -0.10 &  0.09 & -0.99   \\
 R~CRA       &     4.79 &  1.17 & 0.08 & 0.00 & 0.72 & 0.26 & 0.91 &  0.67 & 0.91 &  0.07 & -0.69 &  0.72   \\ 
 HD~190073   &     1.21 & -2.15 & 0.42 & 0.00 & 0.86 & 2.78 & 0.00 &  0.27 & 0.58 & -0.26 & -0.16 &  0.10   \\
\hline
\end{tabular}
\tablefoot{
This fit was performed on the HQ subsample. See Tables~\ref{parfitring} and \ref{paramradiuswidth} for symbol details. Error bars are not shown because of page width limitation; a complete version of this table can be found online at \textbf{(CDS TBD)} 
}
\end{table*}

\clearpage

\section{Log of observations}
\label{logobs}
\nopagebreak[3]
\begin{longtab}
\begin{longtable}{llllll}
\caption{Log of observations for the program objects.} \\
\hline
Target & Obs. date & Configuration & MJD & nV$^{2}$ & nCP \\
\hline
\endhead
\multicolumn{6}{r}{{\em Continued on next page}} \\ 
\endfoot
\hline
\endlastfoot
AB AUR & 2012-12-22 & A1-B2-C1-D0 & 56283.1 & 18 & 12\\ 
AB AUR & 2012-12-20 & A1-B2-C1-D0 & 56281.2 & 18 & 12\\ 
\hline
AK SCO & 2013-06-17 & D0-G1-H0-I1 & 56460.2 & 15 & 12\\ 
AK SCO & 2013-07-03 & A1-B2-C1-D0 & 56476.1 & 18 & 12\\ 
AK SCO & 2013-06-06 & A1-G1-J3-K0 & 56449.3 & 18 & 12\\ 
AK SCO & 2013-06-10 & A1-G1-J3-K0 & 56453.3 & 18 & 12\\ 
\hline
CO ORI & 2013-01-28 & A1-G1-J3-K0 & 56320.1 & 6 & 4\\ 
CO ORI & 2012-12-22 & A1-C1-D0 & 56283.1 & 9 & 3\\ 
CO ORI & 2012-12-21 & A1-B2-C1-D0 & 56282.2 & 15 & 9\\ 
\hline
HD100453 & 2013-01-29 & A1-G1-J3-K0 & 56321.3 & 18 & 12\\ 
HD100453 & 2013-01-28 & A1-G1-J3-K0 & 56320.4 & 18 & 12\\ 
HD100453 & 2012-12-22 & A1-B2-C1-D0 & 56283.3 & 18 & 12\\ 
HD100453 & 2013-01-27 & A1-G1-J3-K0 & 56319.3 & 27 & 12\\ 
HD100453 & 2012-12-19 & A1-B2-C1-D0 & 56280.3 & 36 & 24\\ 
HD100453 & 2013-01-30 & A1-G1-J3-K0 & 56322.3 & 12 & 3\\ 
HD100453 & 2013-01-31 & A1-G1-J3-K0 & 56323.2 & 18 & 12\\ 
HD100453 & 2013-02-17 & D0-G1-H0-I1 & 56340.3 & 36 & 24\\ 
HD100453 & 2013-02-20 & D0-G1-H0-I1 & 56343.2 & 36 & 24\\ 
HD100453 & 2013-02-18 & D0-G1-H0-I1 & 56341.3 & 72 & 45\\ 
HD100453 & 2013-02-01 & A1-G1-J3-K0 & 56324.3 & 78 & 36\\ 
\hline
HD100546 & 2013-01-29 & A1-J3-K0 & 56321.3 & 18 & 6\\ 
HD100546 & 2012-12-22 & A1-B2-C1-D0 & 56283.3 & 18 & 12\\ 
HD100546 & 2012-12-20 & A1-B2-C1-D0 & 56281.4 & 18 & 12\\ 
HD100546 & 2013-02-20 & D0-G1-H0-I1 & 56343.3 & 18 & 12\\ 
HD100546 & 2013-01-30 & A1-J3-K0 & 56322.3 & 18 & 6\\ 
HD100546 & 2013-01-31 & A1-K0 & 56323.3 & 3 & 0\\ 
HD100546 & 2013-02-17 & D0-G1-H0-I1 & 56340.4 & 18 & 12\\ 
HD100546 & 2013-02-19 & D0-G1-H0-I1 & 56342.3 & 72 & 48\\ 
HD100546 & 2013-02-18 & D0-G1-H0-I1 & 56341.3 & 18 & 12\\ 
HD100546 & 2013-02-01 & A1-G1-J3-K0 & 56324.3 & 36 & 24\\ 
\hline
HD104237 & 2013-04-11 & D0-G1-H0-I1 & 56393.2 & 378 & 252\\ 
HD104237 & 2013-02-19 & D0-G1-H0-I1 & 56342.2 & 36 & 24\\ 
HD104237 & 2013-02-18 & D0-G1-H0-I1 & 56341.4 & 18 & 12\\ 
HD104237 & 2012-12-22 & A1-B2-C1-D0 & 56283.3 & 18 & 12\\ 
HD104237 & 2012-12-21 & A1-B2-C1-D0 & 56282.4 & 18 & 12\\ 
\hline
HD139614 & 2013-07-03 & A1-B2-C1-D0 & 56477.0 & 18 & 12\\ 
HD139614 & 2013-06-16 & D0-G1-H0-I1 & 56460.0 & 36 & 21\\ 
HD139614 & 2013-06-06 & A1-G1-J3-K0 & 56449.1 & 18 & 12\\ 
\hline
HD141569 & 2013-06-06 & A1-G1-J3-K0 & 56449.2 & 18 & 12\\ 
\hline
HD142527 & 2013-06-17 & D0-G1-H0-I1 & 56460.2 & 15 & 12\\ 
HD142527 & 2013-06-16 & D0-G1-H0-I1 & 56459.6 & 36 & 24\\ 
HD142527 & 2013-06-15 & D0-G1-H0-I1 & 56458.4 & 54 & 36\\ 
HD142527 & 2013-06-04 & A1-G1-J3-K0 & 56447.2 & 18 & 12\\ 
HD142527 & 2013-05-13 & A1-B2-C1-D0 & 56425.2 & 5 & 4\\ 
HD142527 & 2013-06-06 & A1-G1-J3-K0 & 56449.1 & 18 & 12\\ 
HD142527 & 2013-06-07 & A1-G1-J3-K0 & 56450.1 & 36 & 24\\ 
HD142527 & 2013-07-03 & A1-B2-C1-D0 & 56476.1 & 18 & 12\\ 
HD142527 & 2013-06-10 & A1-G1-J3-K0 & 56453.0 & 18 & 12\\ 
HD142527 & 2013-07-04 & A1-B2-C1-D0 & 56477.1 & 18 & 12\\ 
HD142527 & 2013-02-20 & D0-G1-H0-I1 & 56343.3 & 18 & 12\\ 
HD142527 & 2013-06-05 & A1-G1-J3-K0 & 56448.2 & 54 & 36\\ 
\hline
HD142666 & 2013-06-17 & D0-G1-H0-I1 & 56460.0 & 18 & 12\\ 
HD142666 & 2013-07-03 & A1-B2-C1-D0 & 56476.1 & 18 & 12\\ 
HD142666 & 2013-06-06 & A1-G1-J3-K0 & 56449.2 & 18 & 12\\ 
\hline
HD143006 & 2013-06-17 & D0-G1-H0-I1 & 56460.1 & 18 & 12\\ 
HD143006 & 2013-06-06 & A1-G1-J3-K0 & 56449.2 & 18 & 12\\ 
HD143006 & 2013-07-04 & A1-B2-C1-D0 & 56477.0 & 18 & 12\\ 
\hline
HD144432 & 2013-06-17 & D0-G1-H0-I1 & 56460.1 & 18 & 12\\ 
HD144432 & 2013-06-16 & D0-H0 & 56459.3 & 3 & 0\\ 
HD144432 & 2013-06-04 & A1-G1-J3-K0 & 56447.2 & 18 & 12\\ 
HD144432 & 2013-06-10 & A1-G1-J3-K0 & 56453.3 & 18 & 12\\ 
HD144432 & 2013-07-03 & A1-B2-C1-D0 & 56476.2 & 18 & 12\\ 
HD144432 & 2013-06-07 & A1-G1-J3-K0 & 56450.2 & 36 & 24\\ 
\hline
HD144668 & 2013-06-15 & D0-G1-H0-I1 & 56458.2 & 36 & 24\\ 
HD144668 & 2013-06-04 & A1-G1-J3-K0 & 56447.3 & 18 & 12\\ 
HD144668 & 2013-06-05 & A1-G1-J3-K0 & 56448.2 & 54 & 36\\ 
HD144668 & 2013-06-10 & A1-G1-J3-K0 & 56453.1 & 36 & 24\\ 
HD144668 & 2013-07-03 & A1-B2-C1-D0 & 56476.1 & 18 & 12\\ 
HD144668 & 2013-06-07 & A1-G1-J3-K0 & 56450.3 & 18 & 12\\ 
HD144668 & 2013-07-04 & A1-B2-C1-D0 & 56477.1 & 15 & 12\\ 
\hline
HD145718 & 2013-06-17 & D0-G1-H0-I1 & 56460.0 & 18 & 12\\ 
HD145718 & 2013-07-03 & A1-B2-C1-D0 & 56476.2 & 18 & 12\\ 
HD145718 & 2013-06-06 & A1-G1-J3-K0 & 56449.3 & 18 & 12\\ 
HD145718 & 2013-06-10 & A1-G1-J3-K0 & 56453.2 & 18 & 12\\ 
\hline
HD149914 & 2013-06-15 & D0-G1-H0-I1 & 56458.2 & 18 & 12\\ 
HD149914 & 2013-06-04 & A1-G1-J3-K0 & 56447.3 & 18 & 12\\ 
HD149914 & 2013-06-06 & A1-G1-J3-K0 & 56449.1 & 18 & 12\\ 
HD149914 & 2013-06-10 & A1-G1-J3-K0 & 56453.2 & 18 & 12\\ 
\hline
HD150193 & 2013-06-17 & D0-G1-H0-I1 & 56460.2 & 18 & 12\\ 
HD150193 & 2013-07-04 & A1-B2-C1-D0 & 56477.2 & 18 & 12\\ 
HD150193 & 2013-06-06 & A1-G1-J3-K0 & 56449.3 & 18 & 12\\ 
HD150193 & 2013-06-10 & A1-G1-J3-K0 & 56453.2 & 18 & 12\\ 
\hline
HD158643 & 2013-06-15 & D0-G1-H0-I1 & 56458.1 & 36 & 24\\ 
HD158643 & 2013-06-04 & A1-G1-J3-K0 & 56447.3 & 18 & 12\\ 
HD158643 & 2013-06-06 & A1-G1-J3-K0 & 56449.4 & 18 & 12\\ 
HD158643 & 2013-06-07 & A1-G1-J3-K0 & 56450.2 & 36 & 24\\ 
HD158643 & 2013-07-03 & A1-B2-C1-D0 & 56476.6 & 36 & 24\\ 
HD158643 & 2013-07-06 & A1-B2-C1-D0 & 56479.0 & 18 & 12\\ 
HD158643 & 2013-07-04 & A1-B2-C1-D0 & 56477.2 & 36 & 24\\ 
\hline
HD163296 & 2013-07-03 & A1-B2-C1-D0 & 56476.2 & 18 & 12\\ 
HD163296 & 2013-06-15 & D0-G1-H0-I1 & 56458.2 & 54 & 36\\ 
\hline
HD169142 & 2013-06-17 & D0-G1-H0-I1 & 56460.2 & 15 & 12\\ 
HD169142 & 2013-06-09 & A1-G1-J3-K0 & 56452.2 & 18 & 12\\ 
HD169142 & 2013-07-03 & A1-B2-C1-D0 & 56476.3 & 18 & 12\\ 
HD169142 & 2013-06-04 & A1-G1-J3-K0 & 56447.3 & 18 & 12\\ 
\hline
HD17081 & 2013-01-21 & A1-G1-J3-K0 & 56313.1 & 18 & 12\\ 
HD17081 & 2013-01-26 & A1-G1-J3-K0 & 56318.0 & 18 & 12\\ 
HD17081 & 2012-12-19 & A1-B2-C1-D0 & 56280.1 & 18 & 12\\ 
\hline
HD179218 & 2013-06-06 & A1-G1-J3-K0 & 56449.4 & 18 & 12\\ 
HD179218 & 2013-07-04 & A1-B2-C1-D0 & 56477.2 & 48 & 33\\ 
\hline
HD190073 & 2013-06-06 & A1-G1-J3-K0 & 56449.4 & 18 & 12\\ 
HD190073 & 2013-07-04 & A1-B2-C1-D0 & 56477.3 & 33 & 24\\ 
\hline
HD244604 & 2013-01-28 & A1-G1-J3-K0 & 56320.2 & 4 & 1\\ 
\hline
HD250550 & 2013-01-28 & A1-G1-J3-K0 & 56320.2 & 5 & 4\\ 
\hline
HD259431 & 2013-01-30 & A1-G1-J3-K0 & 56322.2 & 18 & 12\\ 
HD259431 & 2012-12-20 & A1-B2-C1-D0 & 56281.2 & 18 & 12\\ 
\hline
HD293782 & 2013-02-20 & D0-G1-H0-I1 & 56343.1 & 6 & 4\\ 
HD293782 & 2013-01-29 & G1-J3-K0 & 56321.1 & 2 & 0\\ 
HD293782 & 2012-12-22 & A1-B2-C1-D0 & 56283.2 & 5 & 4\\ 
HD293782 & 2013-02-19 & H0-I1 & 56342.1 & 1 & 0\\ 
HD293782 & 2012-12-21 & A1-B2-C1-D0 & 56282.3 & 6 & 4\\ 
\hline
HD31648 & 2012-12-21 & A1-B2-C1-D0 & 56282.1 & 18 & 12\\ 
\hline
HD34282 & 2013-02-20 & D0-G1-H0-I1 & 56343.0 & 6 & 4\\ 
HD34282 & 2013-02-19 & D0-H0 & 56342.1 & 1 & 0\\ 
HD34282 & 2013-02-01 & A1-G1-J3-K0 & 56324.1 & 6 & 4\\ 
\hline
HD35929 & 2013-01-26 & A1-G1-J3-K0 & 56318.1 & 18 & 12\\ 
HD35929 & 2012-12-19 & A1-B2-C1-D0 & 56280.1 & 18 & 12\\ 
HD35929 & 2012-12-21 & A1-B2-C1-D0 & 56282.2 & 18 & 12\\ 
\hline
HD36917 & 2013-01-27 & A1-G1-J3-K0 & 56319.1 & 15 & 12\\ 
HD36917 & 2012-12-19 & A1-B2-C1-D0 & 56280.1 & 18 & 12\\ 
\hline
HD37258 & 2012-12-21 & A1-B2-C1-D0 & 56282.3 & 4 & 4\\ 
\hline
HD37357 & 2012-12-22 & A1-B2-C1-D0 & 56283.2 & 6 & 4\\ 
HD37357 & 2013-01-31 & A1-G1-J3-K0 & 56323.2 & 6 & 2\\ 
HD37357 & 2012-12-21 & A1-B2-C1-D0 & 56282.3 & 11 & 8\\ 
\hline
HD37411 & 2013-02-01 & A1-G1-J3-K0 & 56324.1 & 5 & 3\\ 
HD37411 & 2012-12-21 & A1-B2-C1-D0 & 56282.3 & 5 & 4\\ 
\hline
HD37806 & 2012-12-22 & A1-B2-C1-D0 & 56283.0 & 18 & 12\\ 
HD37806 & 2013-01-27 & G1-K0 & 56319.1 & 3 & 0\\ 
HD37806 & 2013-01-26 & A1-G1-J3-K0 & 56318.2 & 54 & 36\\ 
HD37806 & 2012-12-21 & A1-B2-C1-D0 & 56282.0 & 18 & 12\\ 
HD37806 & 2013-01-31 & A1-G1-J3-K0 & 56323.1 & 18 & 9\\ 
HD37806 & 2013-02-18 & D0-G1-H0-I1 & 56341.1 & 18 & 12\\ 
\hline
HD38087 & 2013-01-26 & A1-G1-J3-K0 & 56318.2 & 90 & 60\\ 
HD38087 & 2012-12-19 & A1-B2-C1-D0 & 56280.2 & 18 & 12\\ 
\hline
HD39014 & 2013-01-29 & A1-G1-J3-K0 & 56321.1 & 33 & 24\\ 
HD39014 & 2013-01-28 & A1-G1-J3-K0 & 56320.0 & 18 & 12\\ 
HD39014 & 2013-01-22 & A1-G1-J3-K0 & 56314.0 & 36 & 24\\ 
HD39014 & 2013-01-27 & A1-G1-J3-K0 & 56319.0 & 36 & 24\\ 
HD39014 & 2013-01-26 & A1-G1-J3-K0 & 56318.0 & 18 & 12\\ 
HD39014 & 2013-01-25 & A1-J3-K0 & 56317.1 & 9 & 3\\ 
HD39014 & 2012-12-21 & A1-B2-C1-D0 & 56282.1 & 18 & 12\\ 
HD39014 & 2013-01-30 & A1-G1-J3-K0 & 56322.0 & 18 & 12\\ 
HD39014 & 2013-01-31 & A1-G1-J3-K0 & 56323.0 & 18 & 12\\ 
\hline
HD45677 & 2013-01-29 & A1-G1-J3-K0 & 56321.2 & 27 & 15\\ 
HD45677 & 2012-12-22 & A1-B2-C1-D0 & 56283.2 & 18 & 12\\ 
HD45677 & 2013-01-27 & A1-G1-J3-K0 & 56319.2 & 30 & 15\\ 
HD45677 & 2013-01-26 & A1-G1-J3-K0 & 56318.3 & 18 & 12\\ 
HD45677 & 2012-12-20 & A1-B2-C1-D0 & 56281.2 & 54 & 36\\ 
HD45677 & 2013-02-20 & D0-G1-H0-I1 & 56343.1 & 18 & 12\\ 
HD45677 & 2013-01-30 & A1-G1-J3-K0 & 56322.1 & 36 & 21\\ 
HD45677 & 2013-01-31 & A1-G1-J3-K0 & 56323.1 & 18 & 12\\ 
HD45677 & 2013-02-22 & D0-G1-H0-I1 & 56345.2 & 18 & 12\\ 
HD45677 & 2013-02-17 & D0-G1-H0-I1 & 56340.0 & 15 & 6\\ 
HD45677 & 2013-02-19 & D0-G1-H0-I1 & 56342.1 & 18 & 12\\ 
HD45677 & 2013-02-18 & D0-G1-H0-I1 & 56341.1 & 36 & 24\\ 
HD45677 & 2013-02-01 & A1-G1-J3-K0 & 56324.1 & 18 & 9\\ 
\hline
HD53367 & 2013-01-27 & A1-G1-J3-K0 & 56319.1 & 18 & 12\\ 
HD53367 & 2013-02-20 & D0-G1-H0-I1 & 56343.2 & 18 & 12\\ 
HD53367 & 2012-12-22 & A1-B2-C1-D0 & 56283.2 & 18 & 12\\ 
HD53367 & 2012-12-19 & A1-B2-C1-D0 & 56280.3 & 18 & 12\\ 
\hline
%HD56895 & 2013-01-29 & G1-J3-K0 & 56321.3 & 6 & 3\\ 
%HD56895 & 2013-01-27 & A1-G1-J3-K0 & 56319.2 & 12 & 12\\ 
%HD56895 & 2012-12-19 & A1-B2-C1-D0 & 56280.2 & 18 & 12\\ 
%\hline
HD58647 & 2012-12-22 & A1-B2-C1-D0 & 56283.1 & 18 & 12\\ 
HD58647 & 2013-01-27 & A1-G1-J3-K0 & 56319.2 & 18 & 12\\ 
HD58647 & 2012-12-19 & A1-B2-C1-D0 & 56280.2 & 18 & 12\\ 
HD58647 & 2013-02-17 & D0-G1-H0-I1 & 56340.2 & 18 & 12\\ 
HD58647 & 2013-02-19 & D0-G1-H0-I1 & 56342.2 & 18 & 12\\ 
HD58647 & 2013-02-18 & D0-G1-H0-I1 & 56341.1 & 18 & 12\\ 
\hline
HD85567 & 2013-06-04 & A1-G1-J3-K0 & 56448.0 & 18 & 12\\ 
HD85567 & 2013-06-05 & A1-G1-J3-K0 & 56448.0 & 18 & 12\\ 
HD85567 & 2013-06-06 & A1-G1-J3-K0 & 56449.5 & 36 & 24\\ 
HD85567 & 2013-06-09 & A1-G1-J3-K0 & 56453.0 & 18 & 12\\ 
HD85567 & 2013-01-26 & A1-G1-J3-K0 & 56318.3 & 15 & 12\\ 
HD85567 & 2012-12-19 & A1-B2-C1-D0 & 56280.3 & 18 & 12\\ 
HD85567 & 2013-02-17 & D0-G1-H0-I1 & 56340.2 & 18 & 12\\ 
HD85567 & 2013-02-20 & D0-G1-H0-I1 & 56343.3 & 18 & 12\\ 
HD85567 & 2013-02-18 & D0-G1-H0-I1 & 56341.4 & 18 & 12\\ 
HD85567 & 2013-02-19 & D0-G1-H0-I1 & 56342.0 & 18 & 12\\ 
\hline
HD95881 & 2012-12-22 & A1-B2-C1-D0 & 56283.3 & 18 & 12\\ 
HD95881 & 2012-12-19 & A1-B2-C1-D0 & 56280.4 & 18 & 12\\ 
HD95881 & 2013-01-30 & A1-J3-K0 & 56322.3 & 9 & 3\\ 
HD95881 & 2013-02-17 & D0-G1-H0-I1 & 56340.3 & 36 & 24\\ 
HD95881 & 2013-02-19 & D0-G1-H0-I1 & 56342.3 & 18 & 12\\ 
HD95881 & 2013-02-20 & D0-G1-H0-I1 & 56343.2 & 18 & 12\\ 
\hline
HD97048 & 2013-02-17 & D0-G1-H0-I1 & 56340.3 & 18 & 12\\ 
HD97048 & 2012-12-22 & A1-B2-C1-D0 & 56283.4 & 18 & 12\\ 
HD97048 & 2012-12-20 & A1-B2-C1-D0 & 56281.3 & 18 & 12\\ 
\hline
HD98922 & 2013-01-28 & A1-G1-J3-K0 & 56320.4 & 54 & 36\\ 
HD98922 & 2012-12-22 & A1-B2-C1-D0 & 56283.4 & 18 & 12\\ 
HD98922 & 2013-01-27 & A1-G1-J3-K0 & 56319.3 & 51 & 30\\ 
HD98922 & 2013-01-26 & A1-G1-J3-K0 & 56318.4 & 18 & 12\\ 
HD98922 & 2012-12-20 & A1-B2-C1-D0 & 56281.3 & 36 & 24\\ 
HD98922 & 2013-02-20 & D0-G1-H0-I1 & 56343.2 & 36 & 24\\ 
HD98922 & 2013-01-30 & A1-G1-J3-K0 & 56322.3 & 24 & 12\\ 
HD98922 & 2013-01-31 & A1-G1-J3-K0 & 56323.2 & 36 & 24\\ 
HD98922 & 2013-02-17 & D0-G1-H0-I1 & 56340.2 & 54 & 36\\ 
HD98922 & 2013-02-19 & D0-G1-H0-I1 & 56342.3 & 18 & 12\\ 
HD98922 & 2013-02-18 & D0-G1-H0-I1 & 56341.3 & 72 & 48\\ 
HD98922 & 2013-02-01 & A1-G1-J3-K0 & 56324.2 & 54 & 36\\ 
\hline
MWC158 & 2013-01-28 & A1-G1-J3-K0 & 56320.3 & 18 & 12\\ 
MWC158 & 2013-01-27 & A1-G1-J3-K0 & 56319.2 & 54 & 36\\ 
MWC158 & 2013-01-26 & A1-G1-J3-K0 & 56318.3 & 18 & 12\\ 
MWC158 & 2012-12-20 & A1-B2-C1-D0 & 56281.1 & 18 & 12\\ 
MWC158 & 2013-01-30 & A1-G1-J3-K0 & 56322.1 & 18 & 12\\ 
MWC158 & 2012-12-19 & A1-B2-C1-D0 & 56280.2 & 36 & 24\\ 
MWC158 & 2013-02-17 & D0-G1-H0-I1 & 56340.1 & 18 & 12\\ 
MWC158 & 2013-02-20 & D0-G1-H0-I1 & 56343.2 & 18 & 12\\ 
MWC158 & 2013-02-18 & D0-G1-H0-I1 & 56341.2 & 18 & 12\\ 
MWC158 & 2013-02-01 & A1-G1-J3-K0 & 56324.1 & 36 & 24\\ 
MWC158 & 2013-01-31 & A1-G1-J3-K0 & 56323.1 & 18 & 12\\ 
\hline
MWC297 & 2013-06-17 & D0-G1-H0-I1 & 56460.3 & 18 & 12\\ 
MWC297 & 2013-06-15 & D0-G1-H0-I1 & 56458.3 & 18 & 12\\ 
MWC297 & 2013-06-04 & A1-G1-J3-K0 & 56447.4 & 18 & 12\\ 
MWC297 & 2013-06-10 & A1-G1-J3-K0 & 56453.4 & 18 & 12\\ 
MWC297 & 2013-07-03 & A1-B2-C1-D0 & 56476.3 & 18 & 12\\ 
MWC297 & 2013-06-09 & A1-G1-J3-K0 & 56452.3 & 18 & 12\\ 
MWC297 & 2013-06-07 & A1-G1-J3-K0 & 56450.4 & 90 & 60\\ 
MWC297 & 2013-07-05 & A1-B2-C1-D0 & 56479.0 & 18 & 12\\ 
MWC297 & 2013-07-04 & A1-B2-C1-D0 & 56477.3 & 18 & 12\\ 
\hline
MWC758 & 2012-12-20 & A1-B2-C1-D0 & 56281.2 & 18 & 12\\ 
\hline
R CRA & 2013-06-07 & G1-K0 & 56450.4 & 3 & 0\\ 
R CRA & 2013-06-15 & D0-G1-H0 & 56458.2 & 6 & 0\\ 
R CRA & 2013-06-17 & D0-G1-H0-I1 & 56460.4 & 9 & 6\\ 
R CRA & 2013-07-03 & A1-B2-C1-D0 & 56476.4 & 54 & 36\\ 
R CRA & 2013-07-04 & A1-B2-C1-D0 & 56477.4 & 36 & 24\\ 
R CRA & 2013-07-06 & A1-B2-C1-D0 & 56479.1 & 36 & 24\\ 
\hline
T ORI & 2013-01-31 & A1-K0 & 56323.1 & 3 & 0\\ 
T ORI & 2013-02-18 & D0-G1-H0-I1 & 56341.1 & 18 & 12\\ 
T ORI & 2013-01-26 & A1-G1-J3-K0 & 56318.1 & 18 & 12\\ 
T ORI & 2012-12-19 & A1-B2-C1-D0 & 56280.2 & 18 & 12\\ 
T ORI & 2012-12-21 & A1-B2-C1-D0 & 56282.2 & 18 & 12\\ 
\hline
V 1247 ORI & 2013-01-31 & A1-G1-J3-K0 & 56323.2 & 5 & 3\\ 
V 1247 ORI & 2012-12-21 & A1-B2-C1-D0 & 56282.3 & 12 & 8\\ 
\hline
VV SER & 2013-07-03 & A1-B2-C1-D0 & 56476.3 & 18 & 12\\ 
VV SER & 2013-06-04 & A1-G1-J3-K0 & 56447.4 & 24 & 15\\ 
VV SER & 2013-06-20 & U1-U2-U3-U4 & 56463.1 & 18 & 12\\ 
VV SER & 2013-06-07 & A1-G1-J3-K0 & 56450.3 & 18 & 12\\ 
\hline
\end{longtable}
\end{longtab}

\clearpage

\section{Summary plots for  program objects}

\begin{figure}[h]
\centering
\includegraphics[width=85mm]{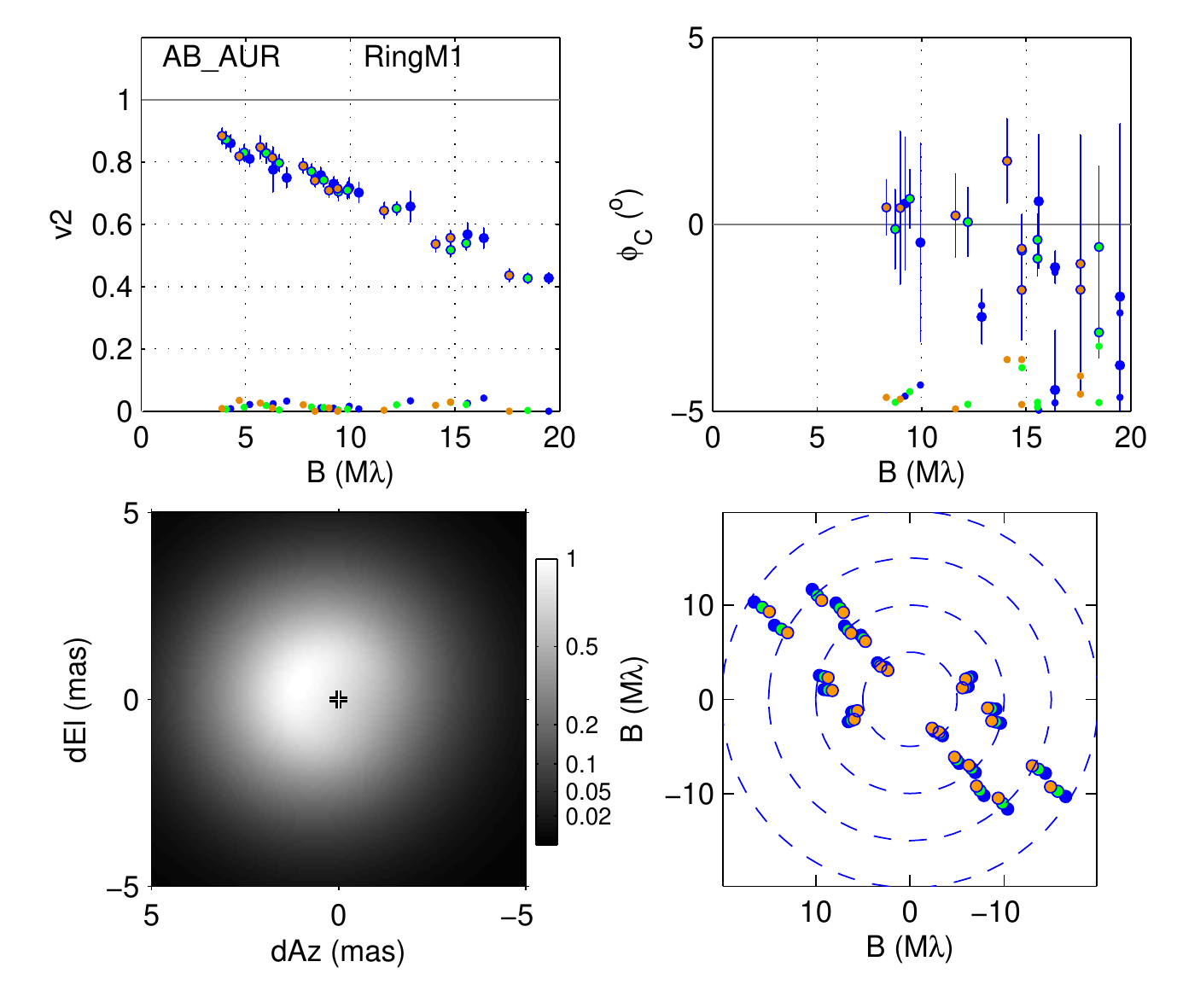}
\includegraphics[width=85mm]{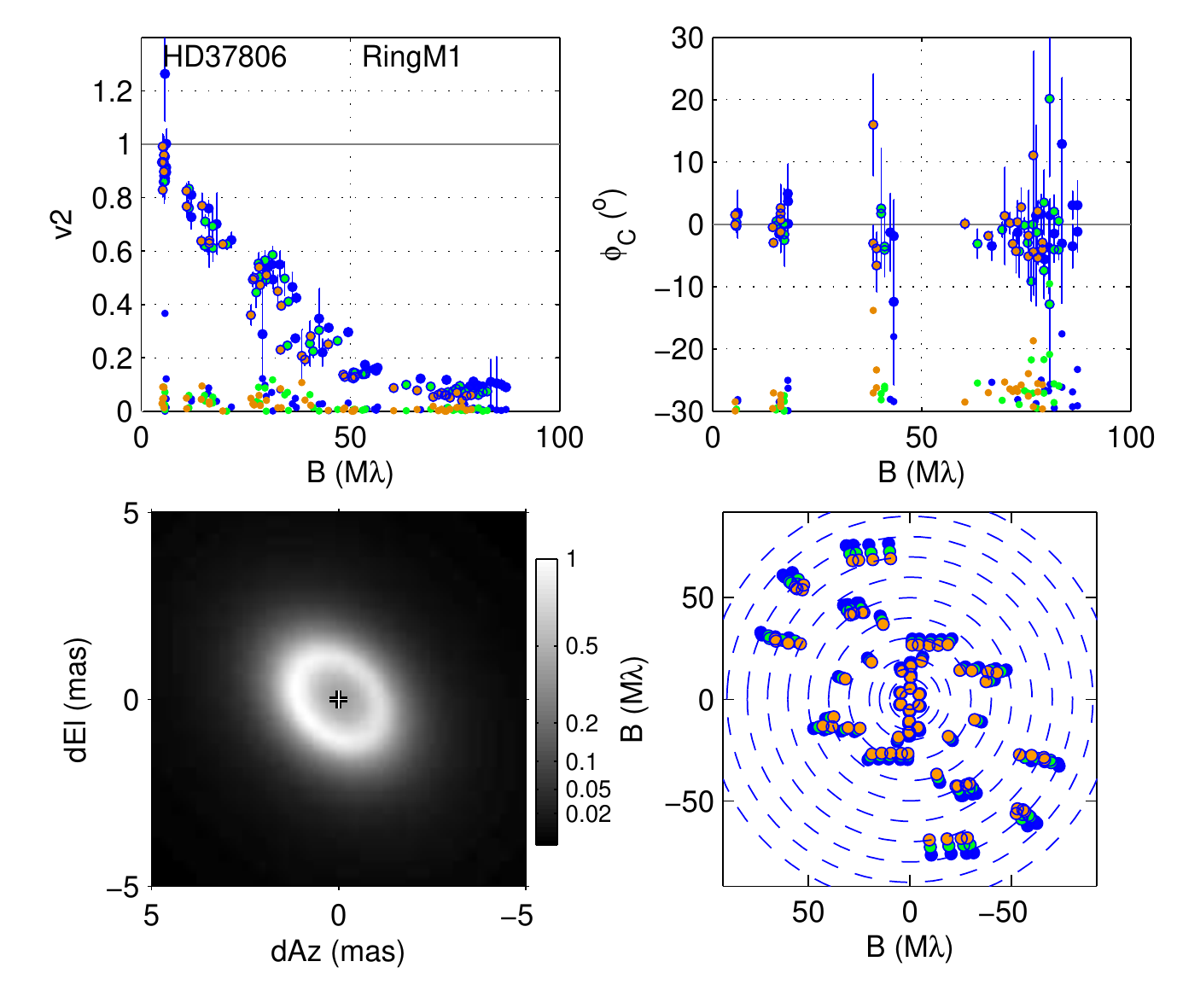}

\includegraphics[width=85mm]{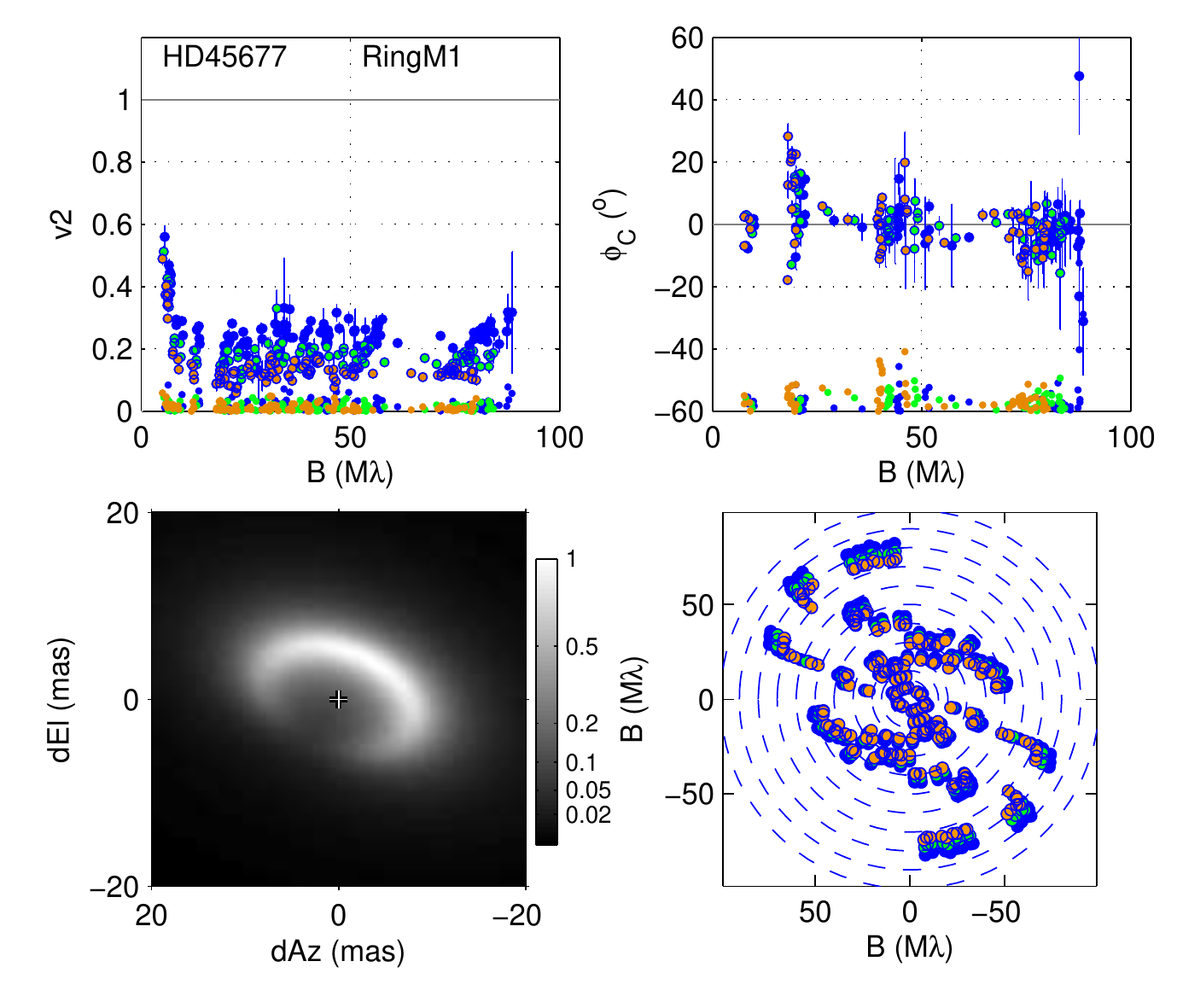}
\includegraphics[width=85mm]{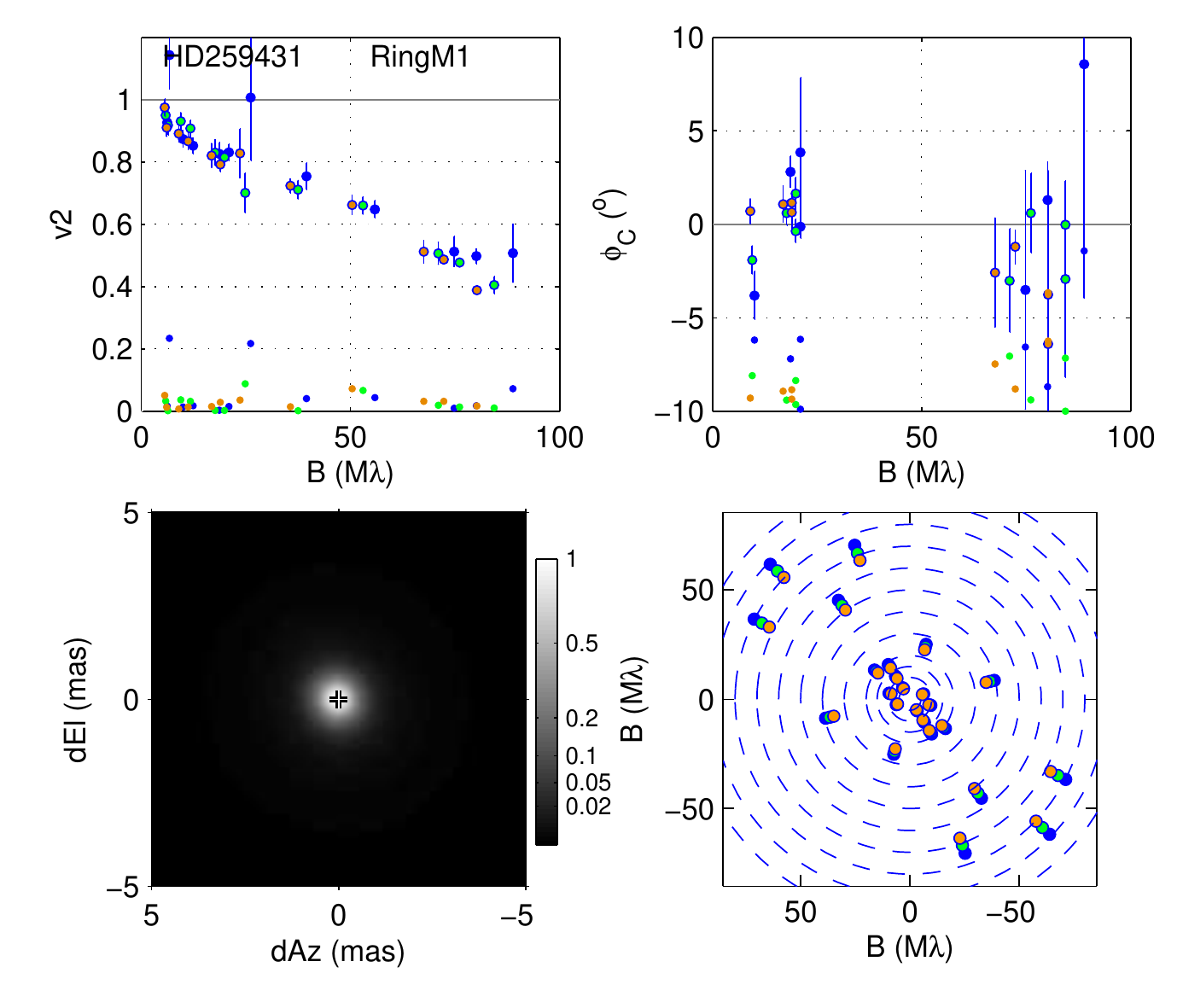}
\caption{
Summary plots for the \emph{HQ} objects. 
Four panels are shown for each object: (1)~visibility squared $V^2$ versus projected baseline in units of $10^6 \lambda$; (2)~phase closure $\phi_c$ versus the largest projected baseline of each triangle; (3)~a halftone image of the circumstellar component resulting from a \emph{Ring} fit with $m=1$ azimuthal modulation, the star's position is shown by a cross symbol; (4)~the $(uv)$ plane coverage. For spectrally dispersed observations, the red, green, blue colors represent wavelength from longest to shortest.  For the $V^2$ and $\phi_c$ plots, the absolute value of the fit residuals is shown by the dots at the bottom of the plot. 
\label{summplotHQ}
}
\end{figure}

\begin{figure}[h]
\centering
\includegraphics[width=85mm]{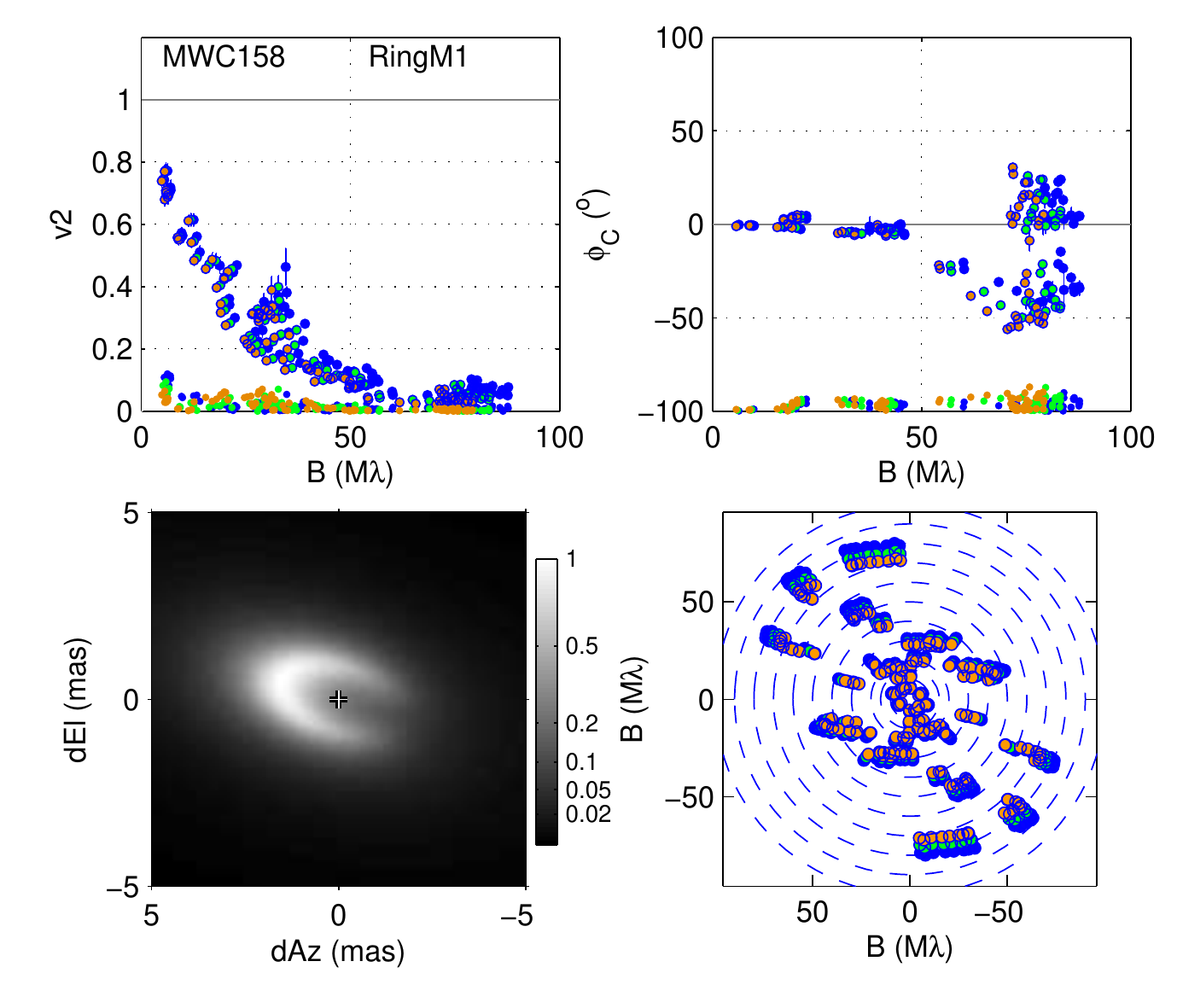}
\includegraphics[width=85mm]{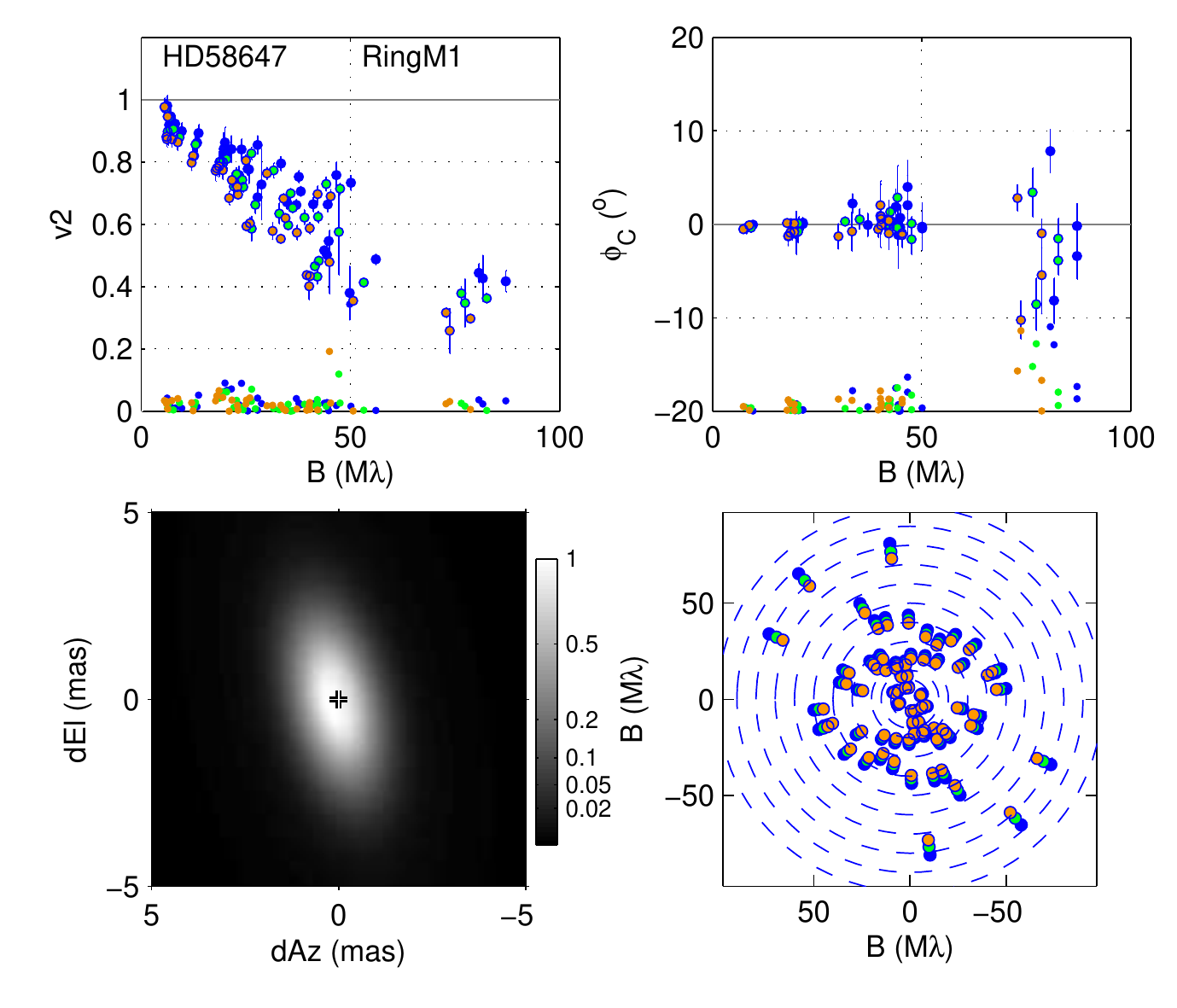}

\includegraphics[width=85mm]{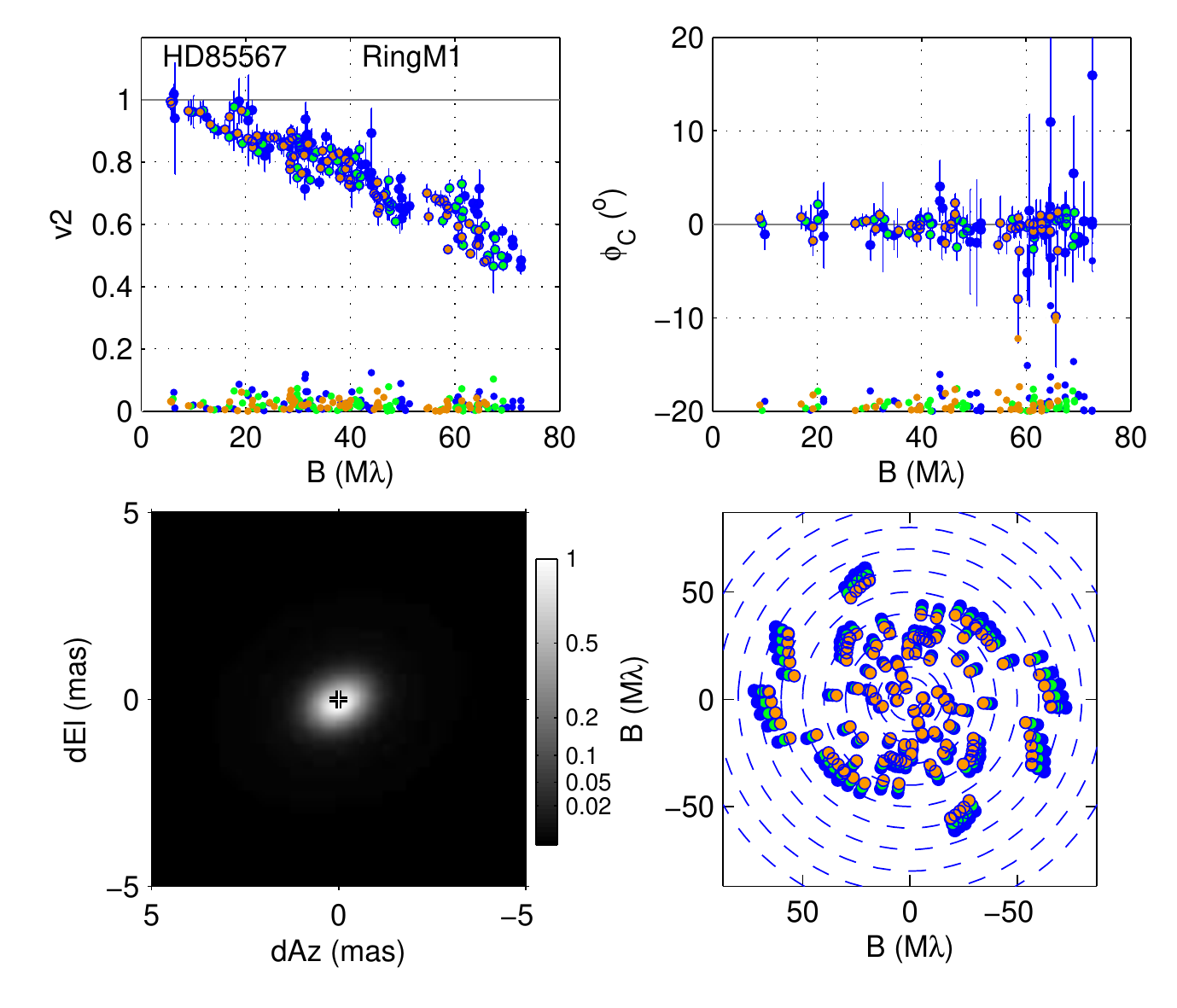}
\includegraphics[width=85mm]{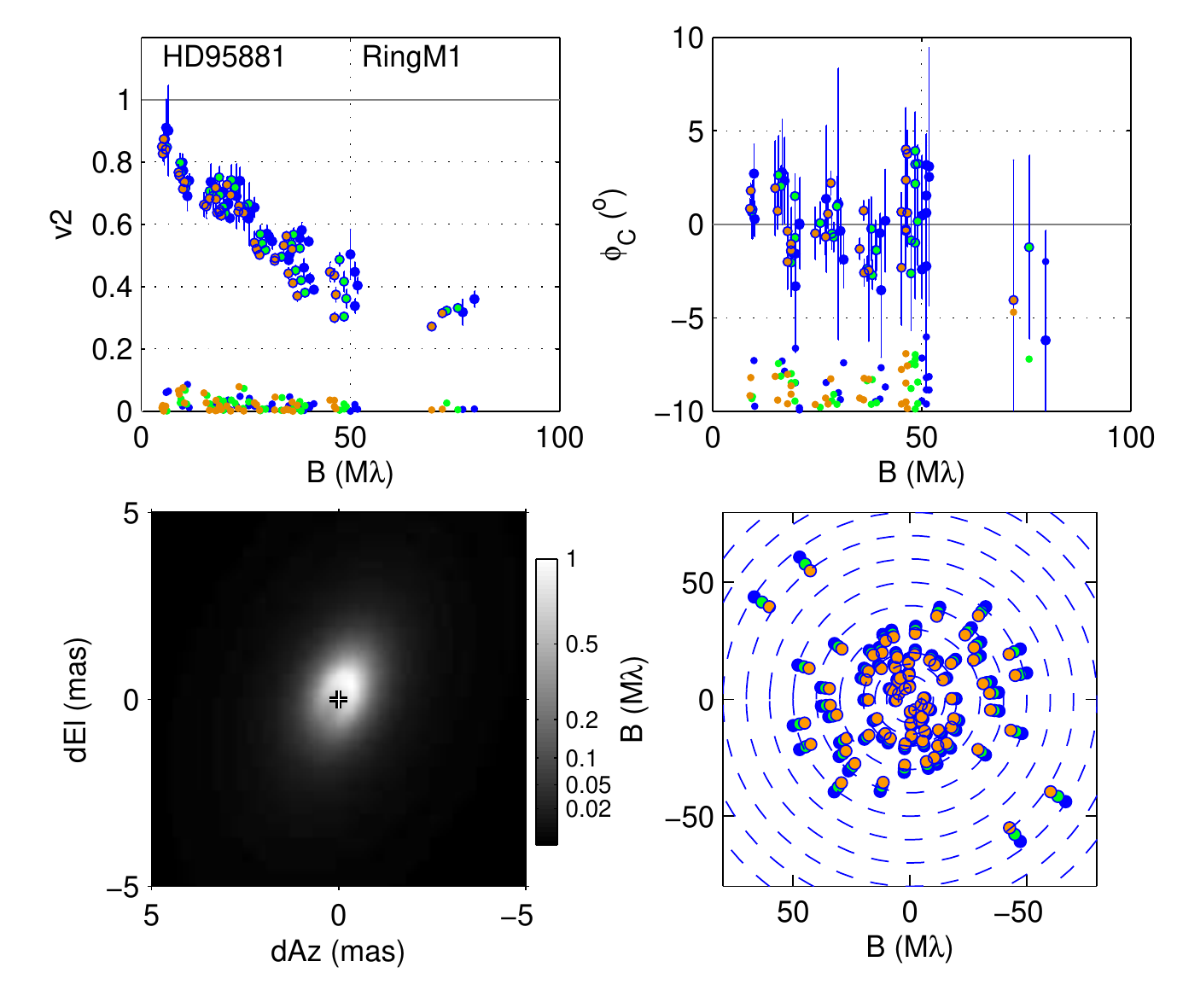}

\includegraphics[width=85mm]{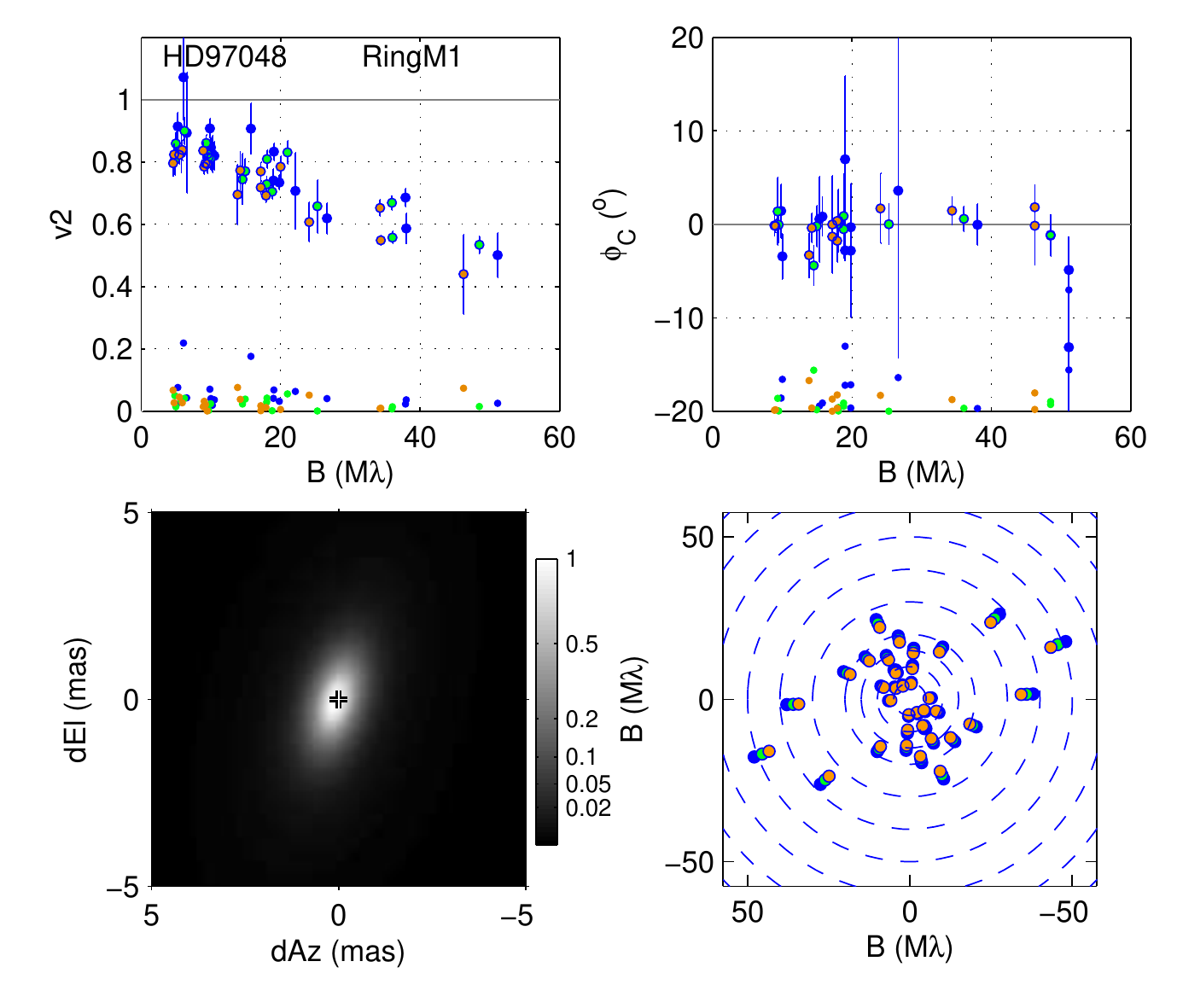}
\includegraphics[width=85mm]{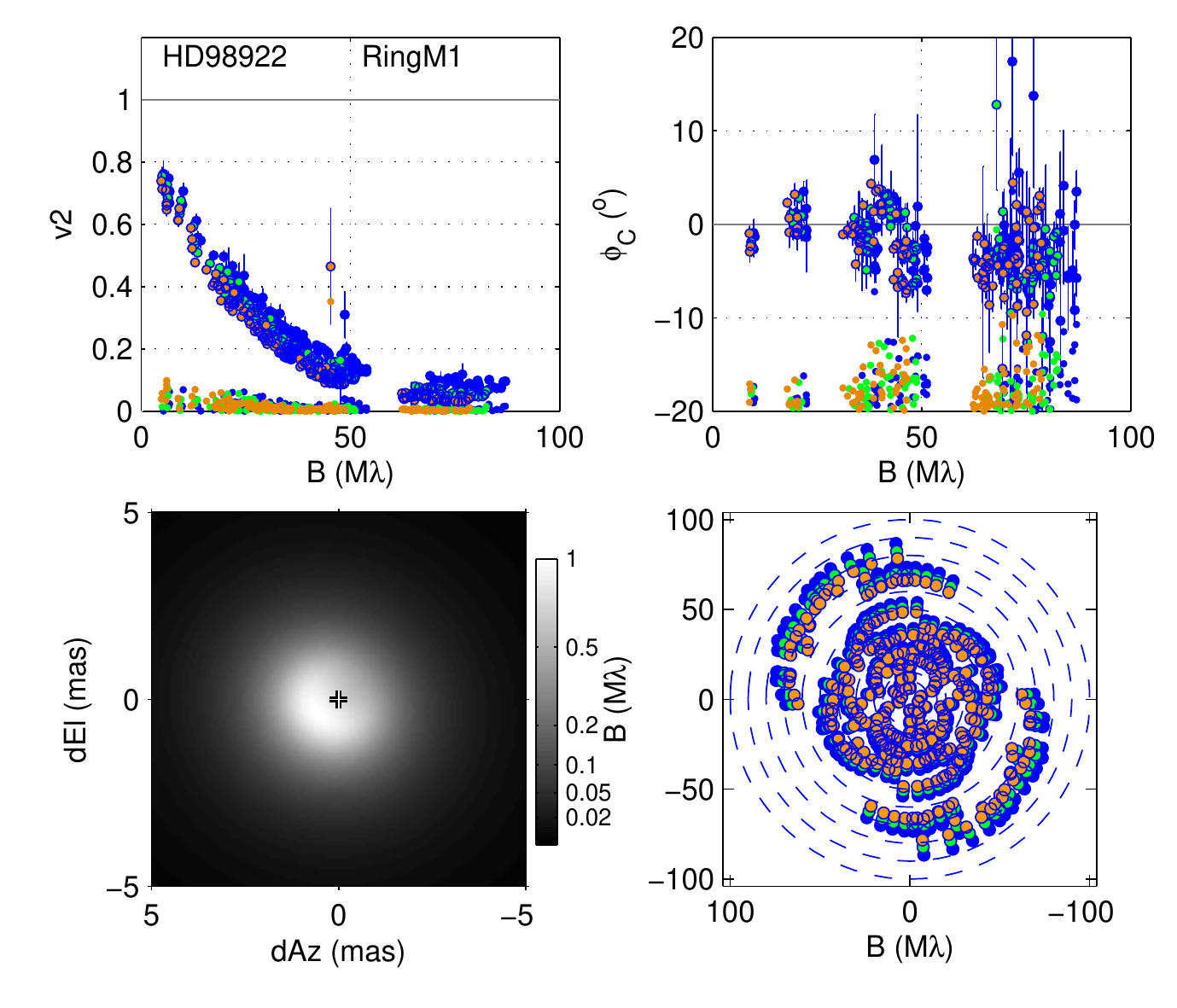}
\caption{
Summary plots for the \emph{HQ} objects (continued). 
}
\end{figure}

\begin{figure}[h]
\centering
\includegraphics[width=85mm]{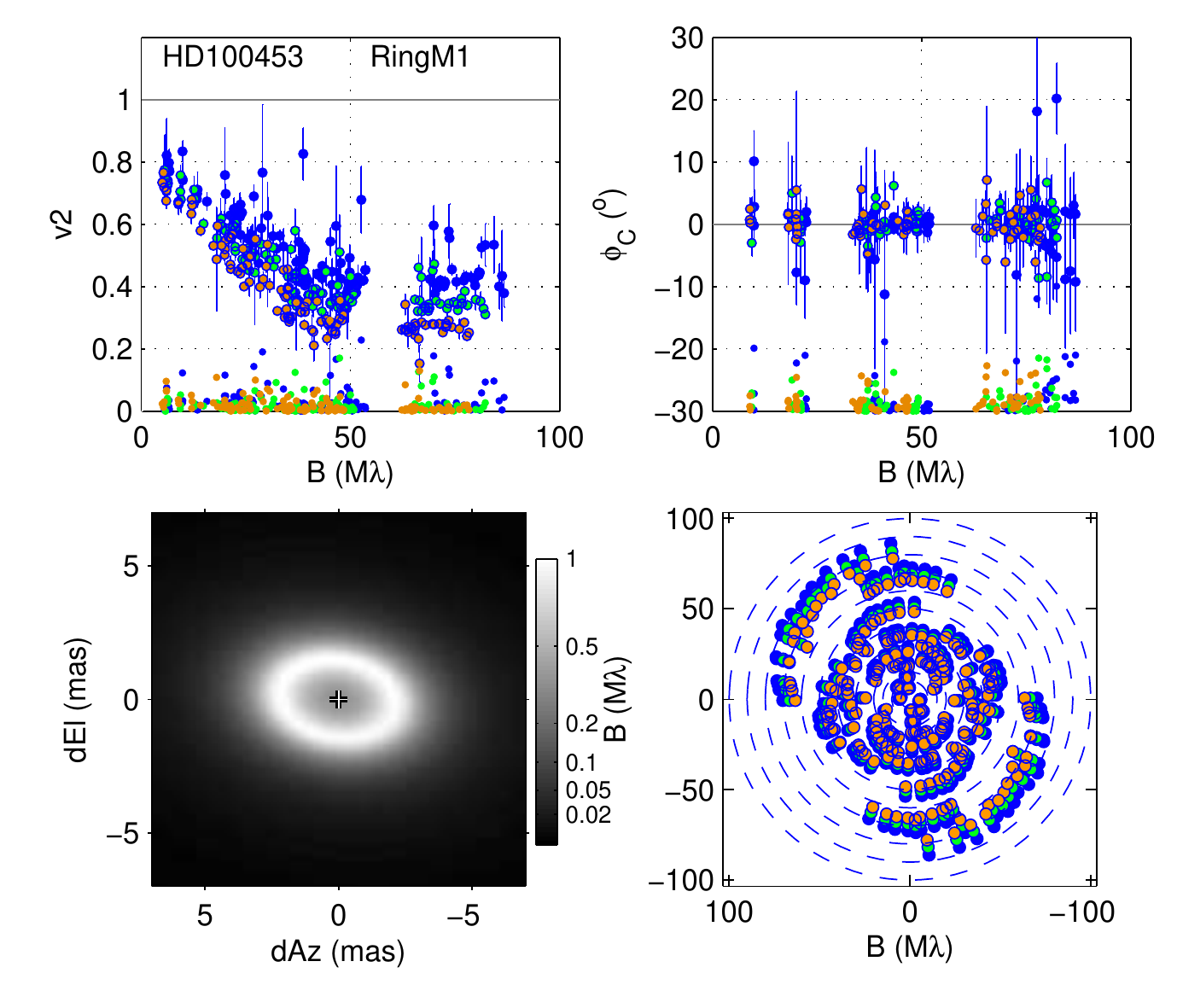}
\includegraphics[width=85mm]{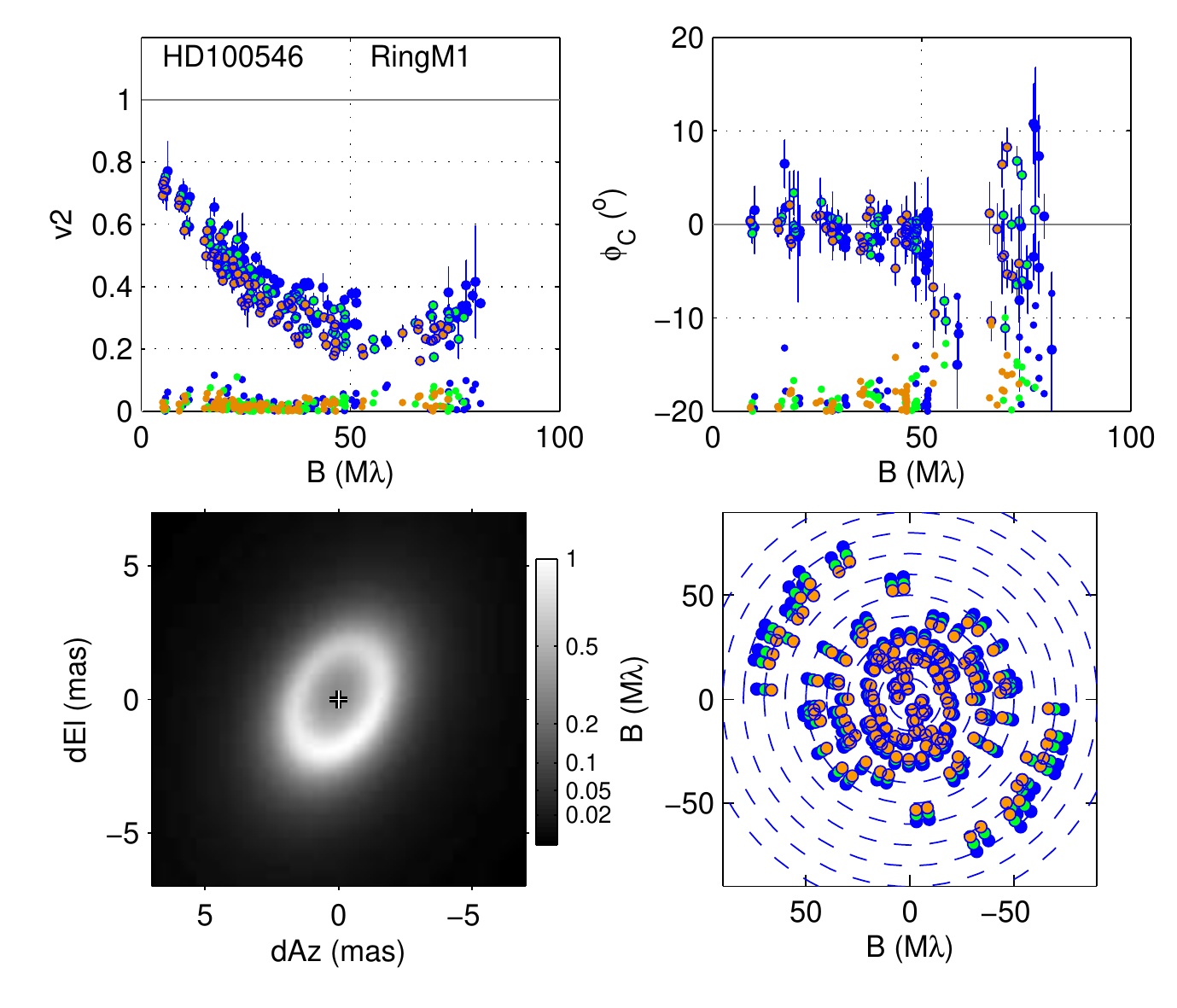}

\includegraphics[width=85mm]{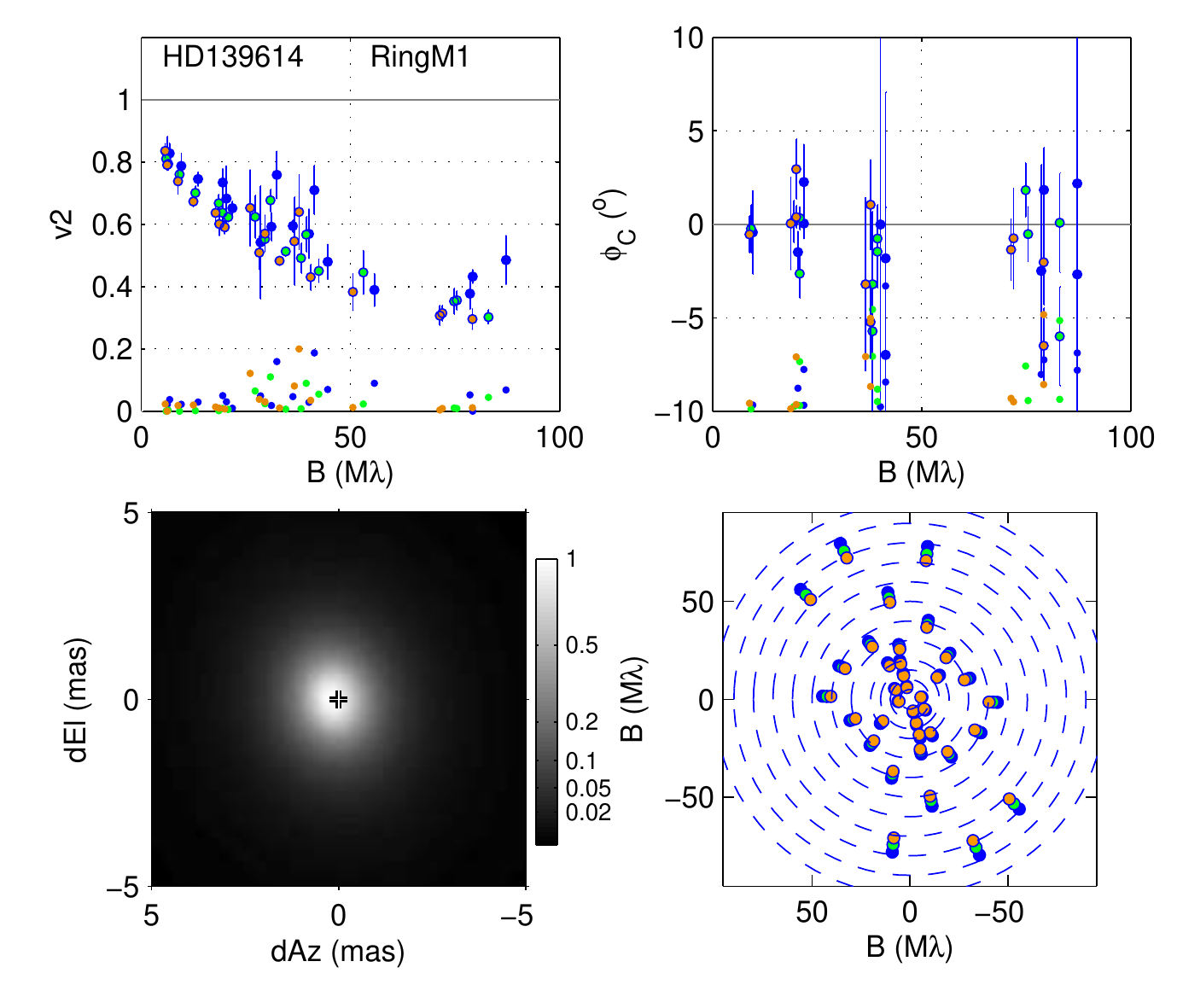}
\includegraphics[width=85mm]{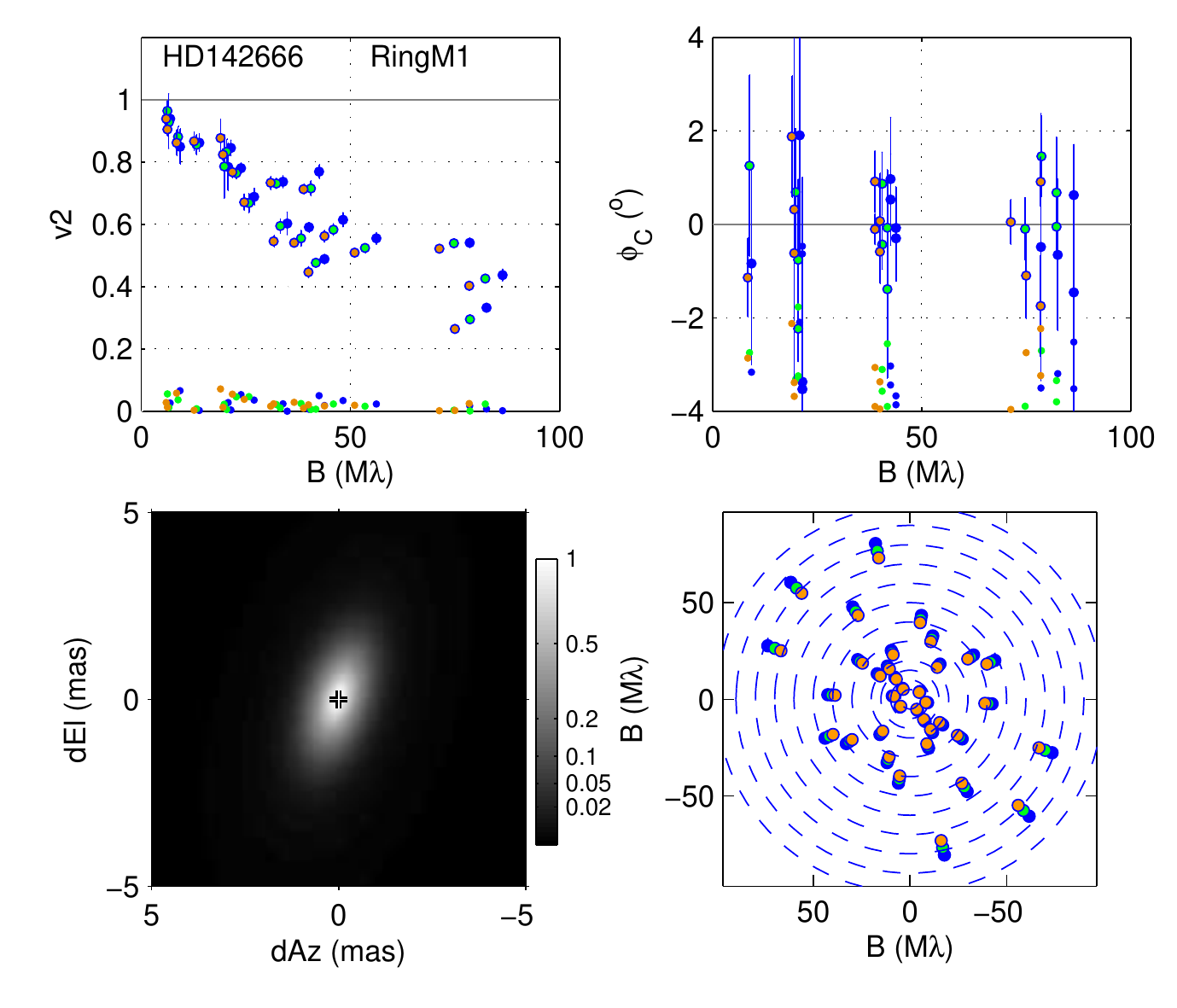}

\includegraphics[width=85mm]{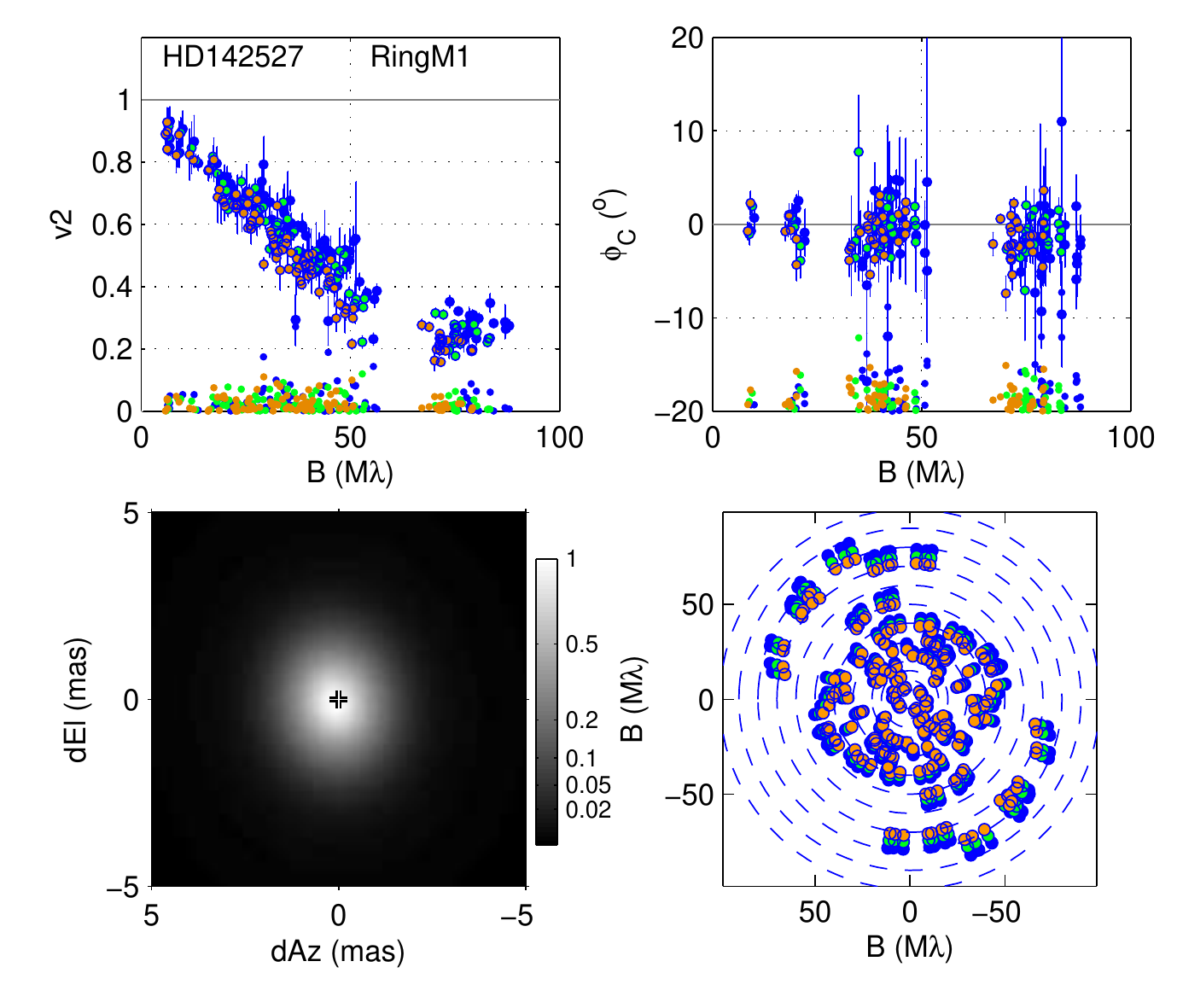}
\includegraphics[width=85mm]{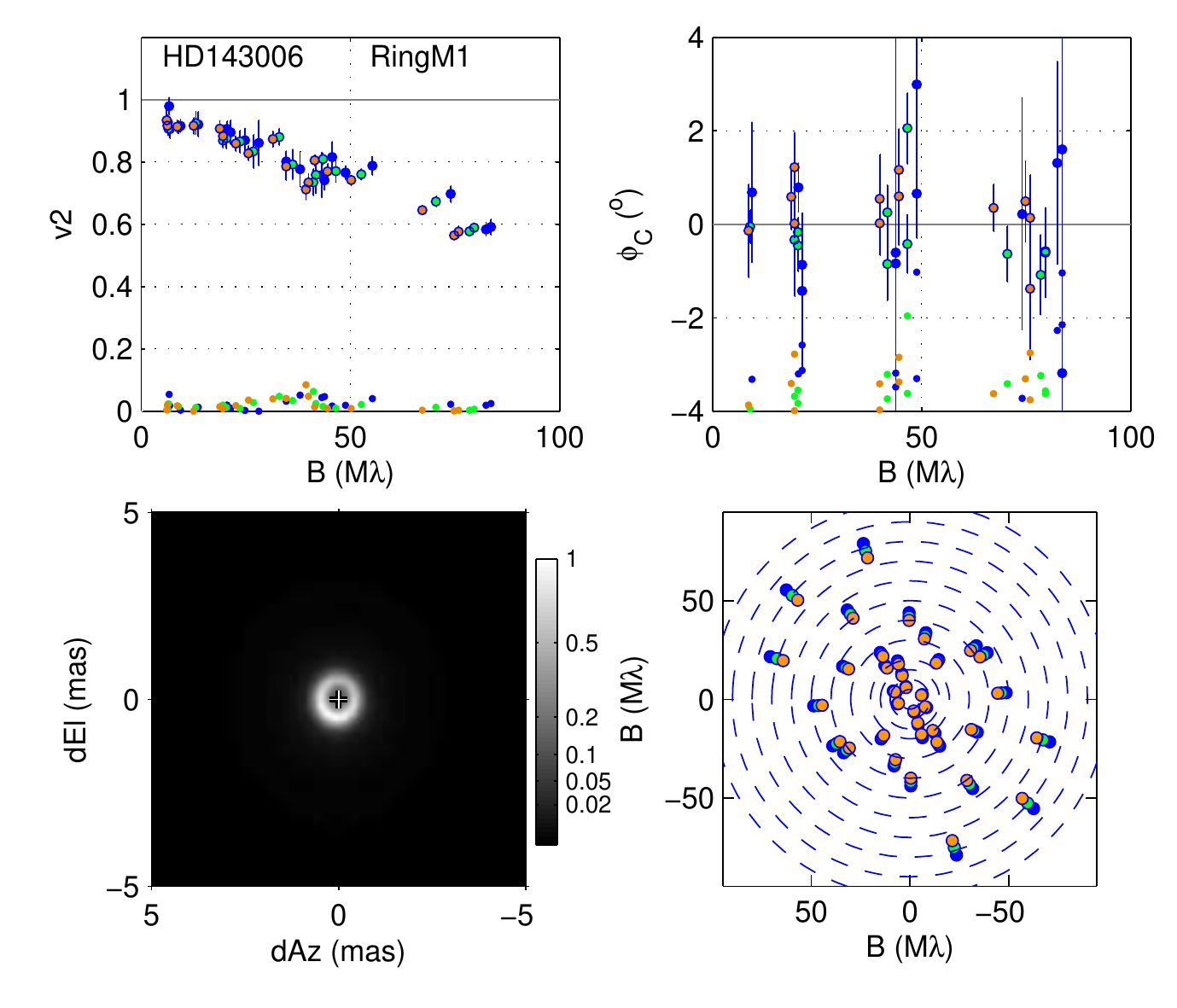}
\caption{
Summary plots for the \emph{HQ} objects (continued). 
}
\end{figure}

\begin{figure}[h]
\centering
\includegraphics[width=85mm]{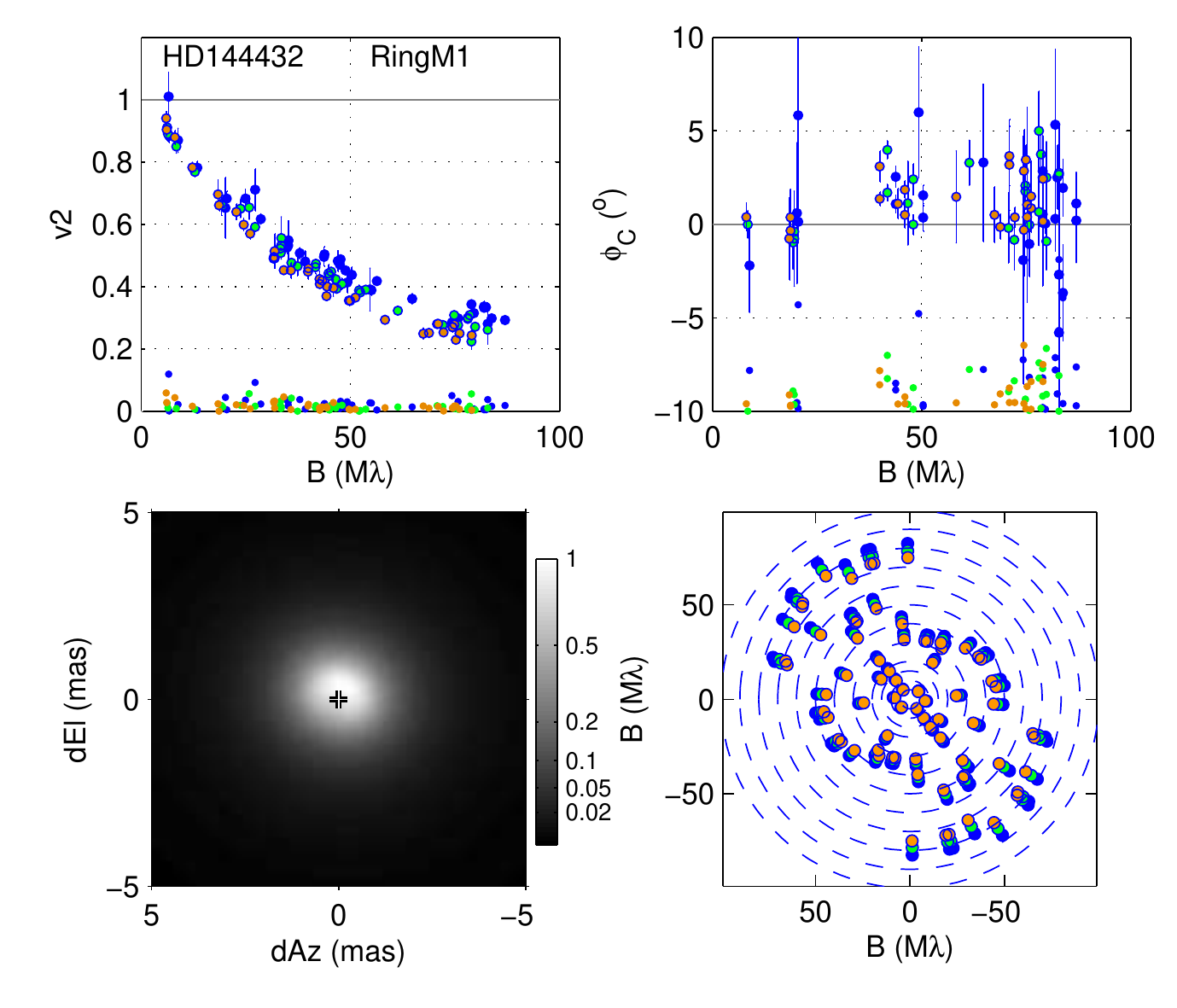}
\includegraphics[width=85mm]{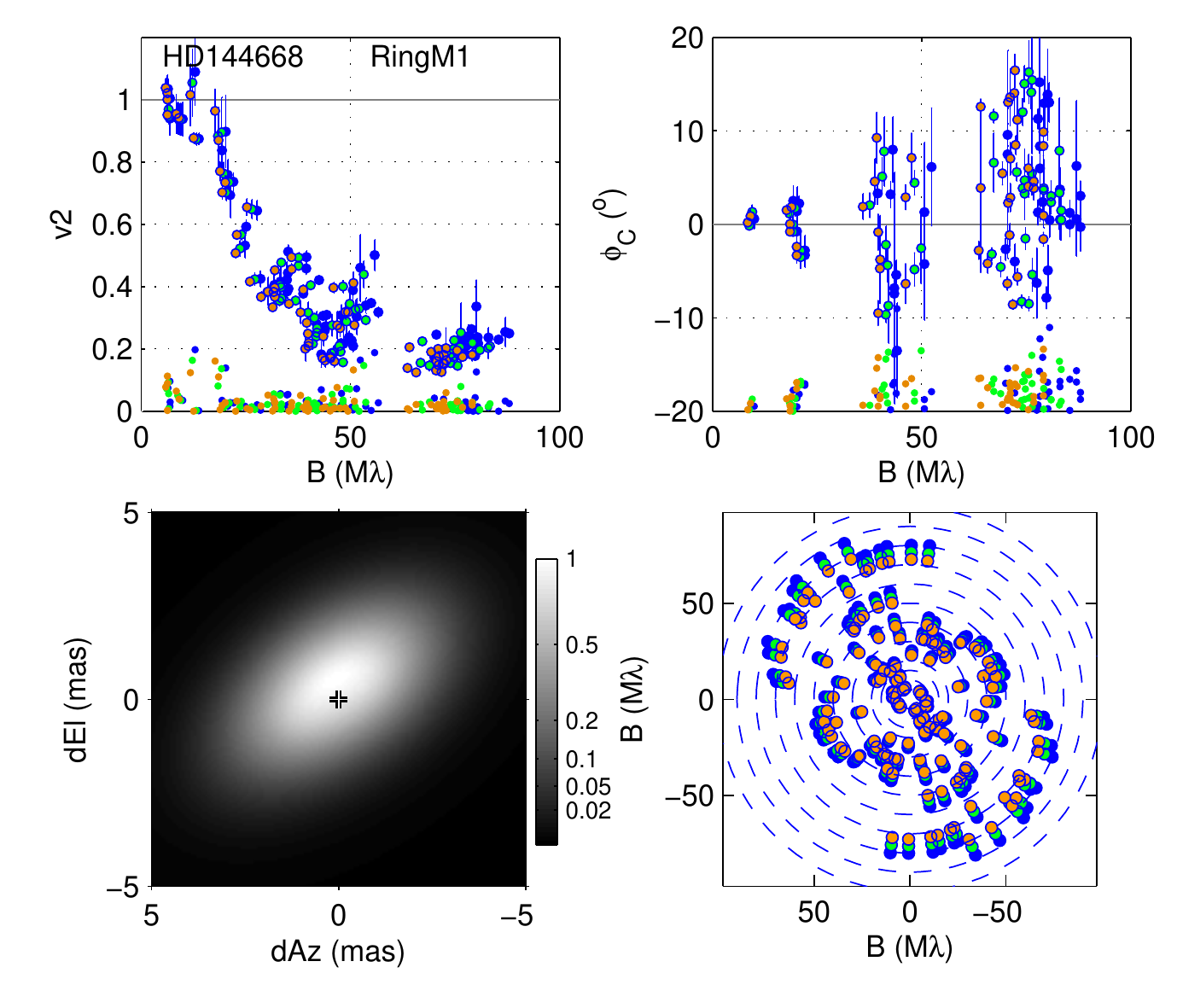}

\includegraphics[width=85mm]{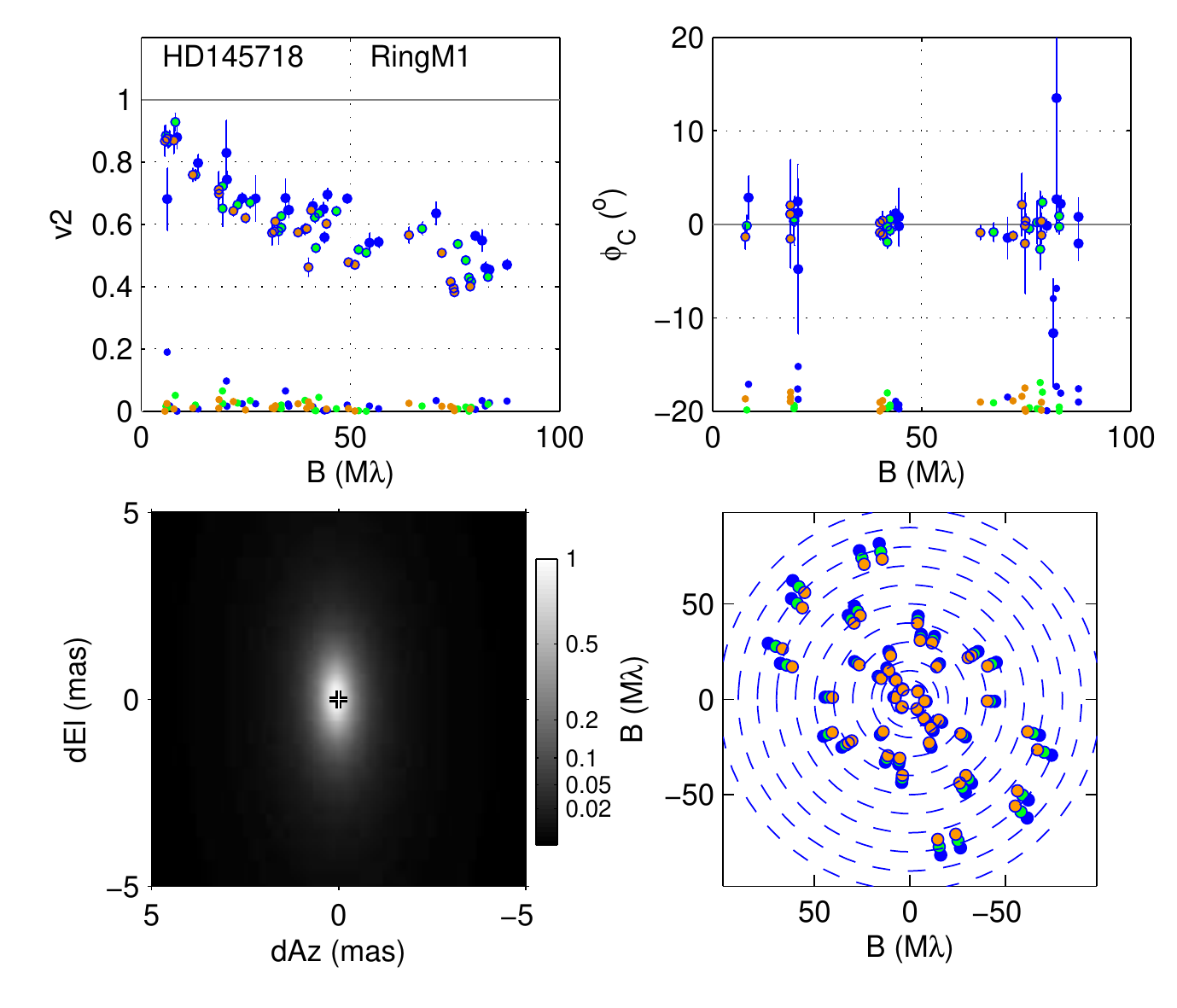}
\includegraphics[width=85mm]{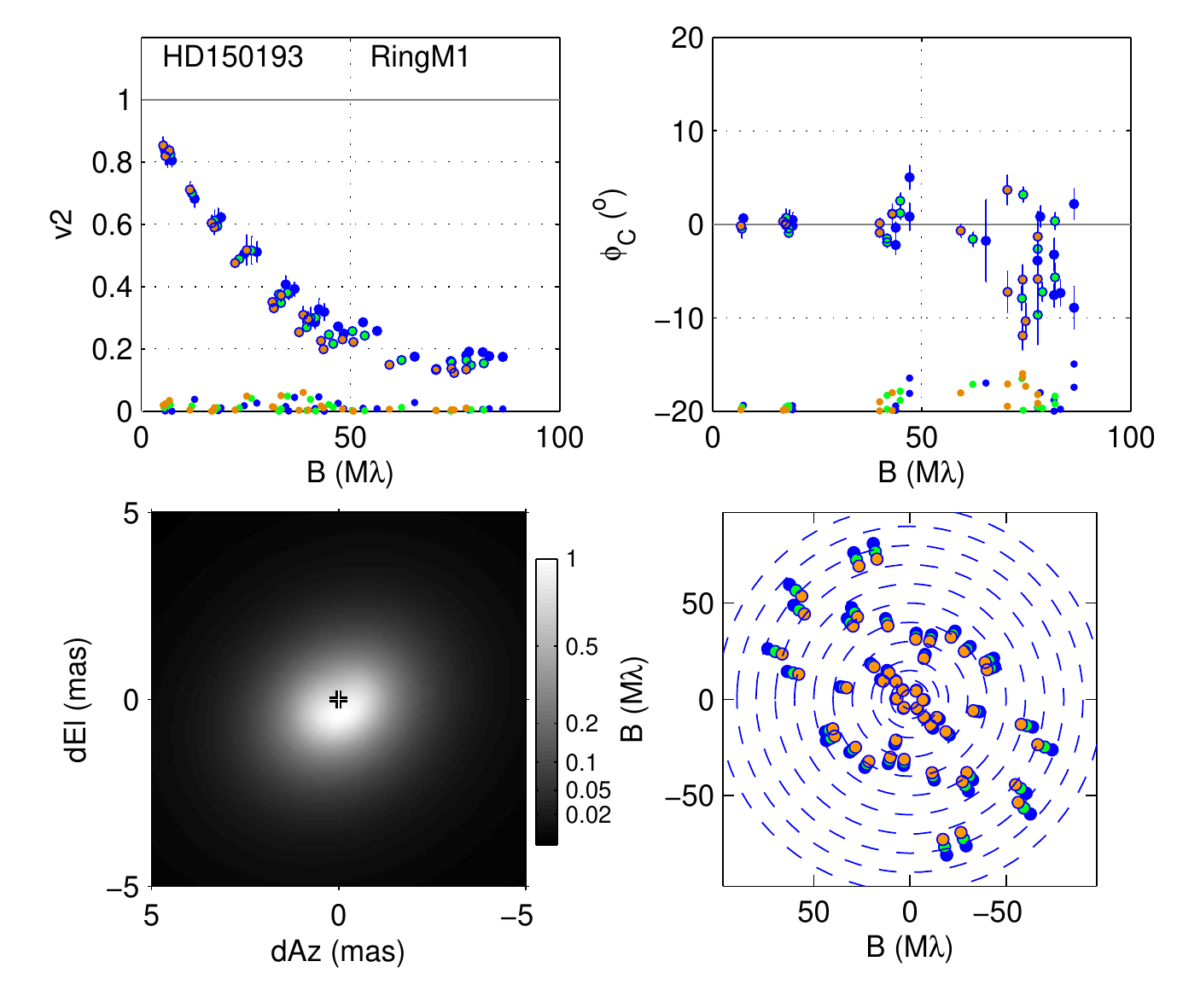}

\includegraphics[width=85mm]{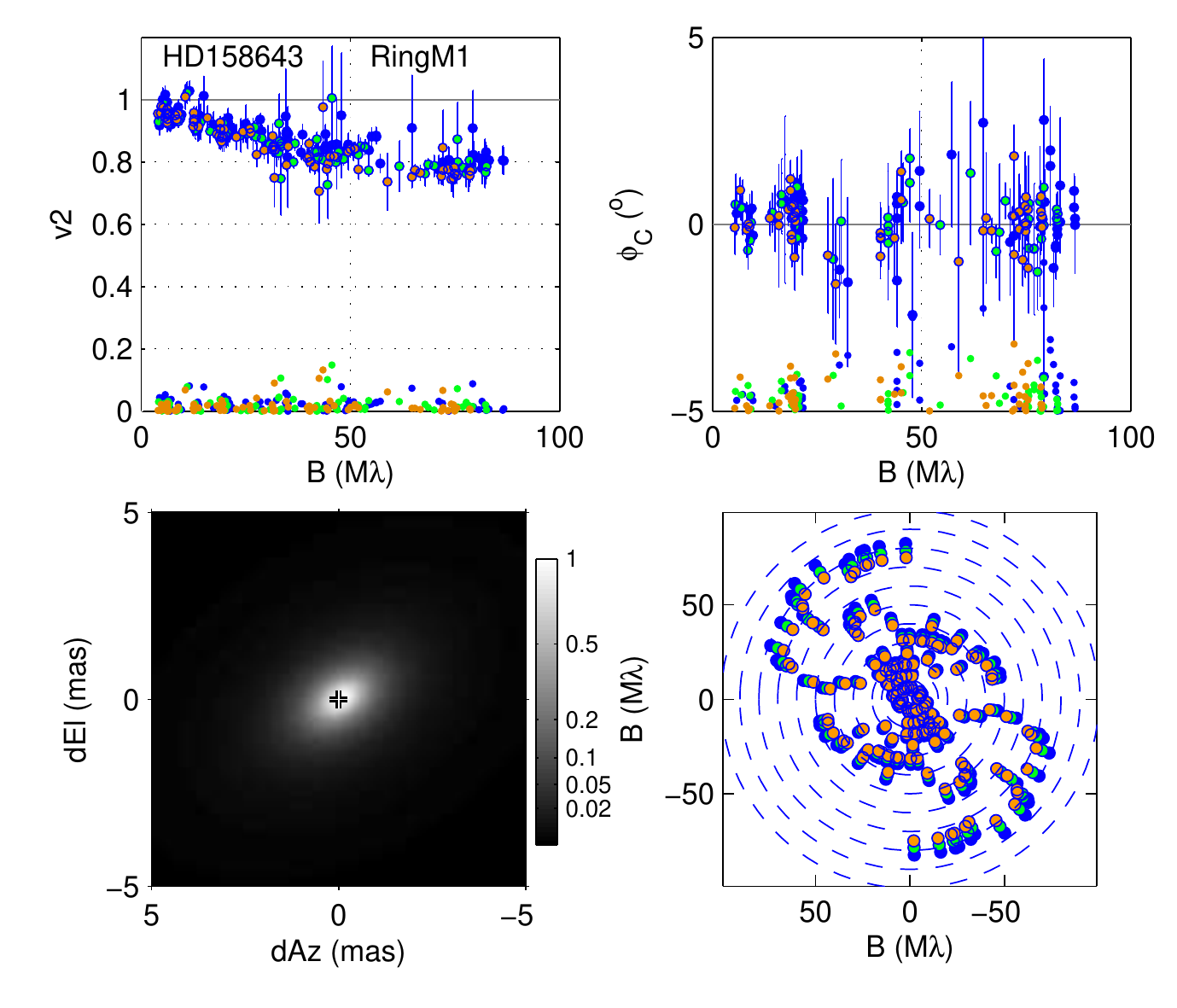}
\includegraphics[width=85mm]{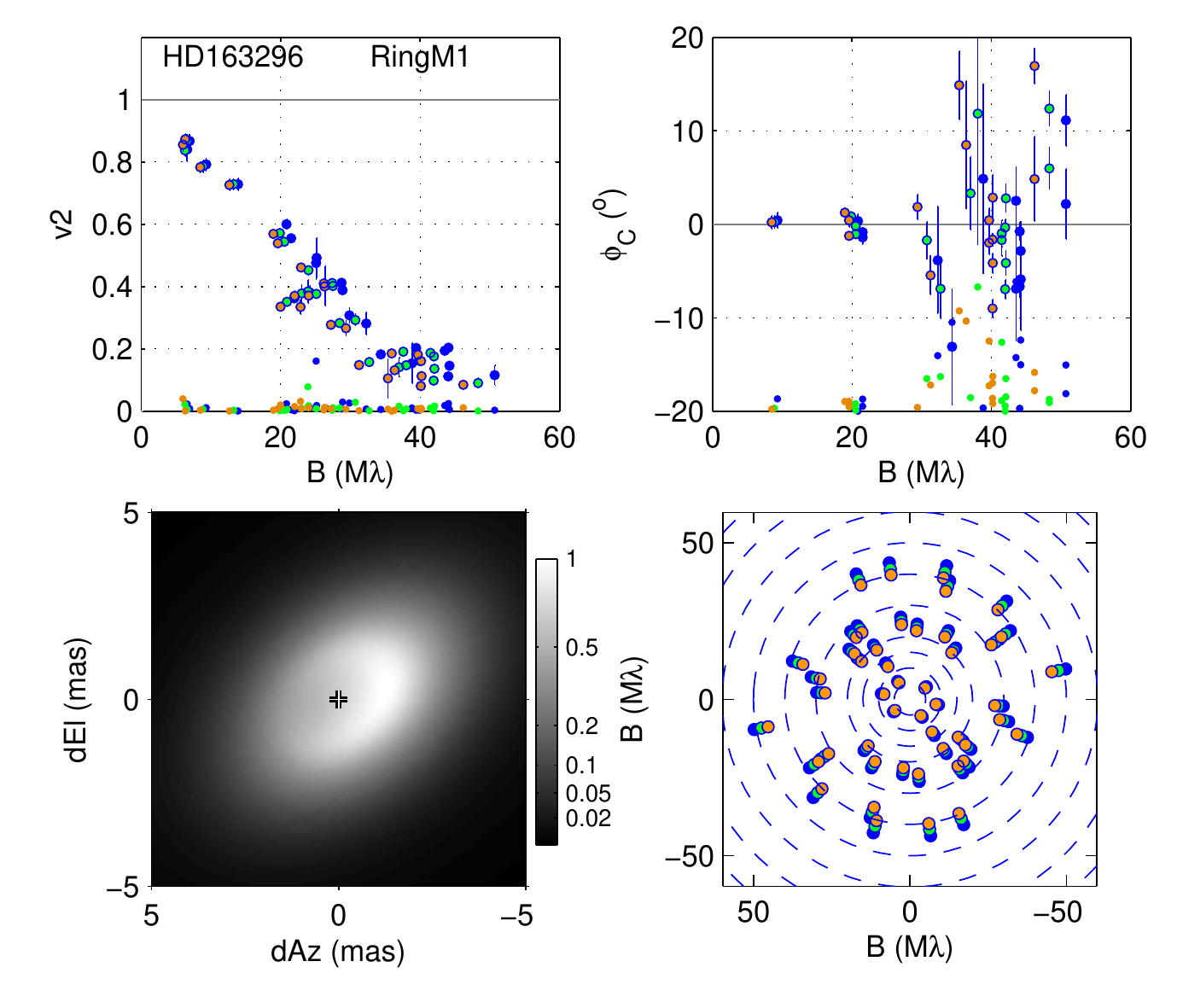}
\caption{
Summary plots for the \emph{HQ} objects (continued). 
}
\end{figure}

\begin{figure}[h]
\centering
\includegraphics[width=85mm]{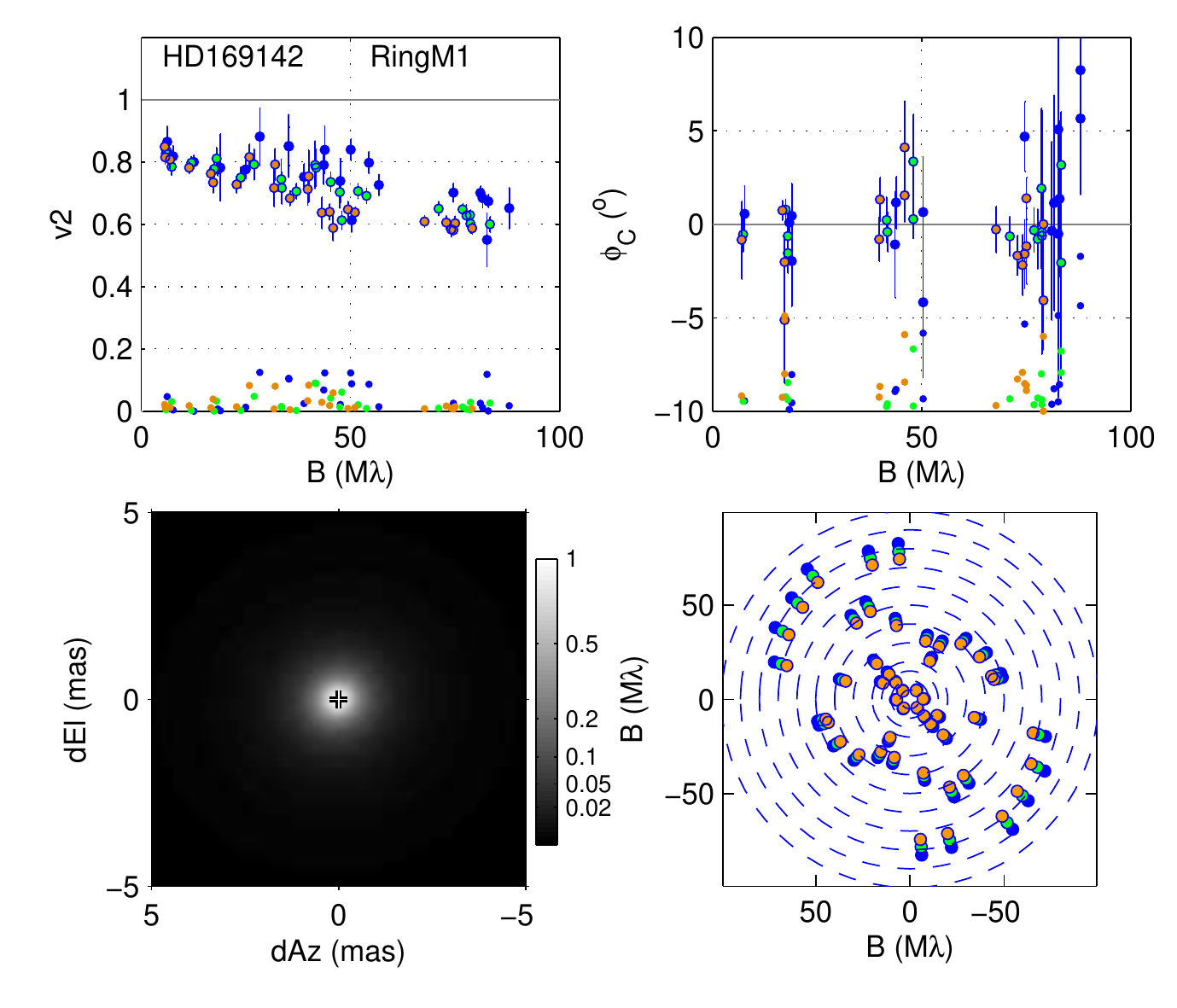}
\includegraphics[width=85mm]{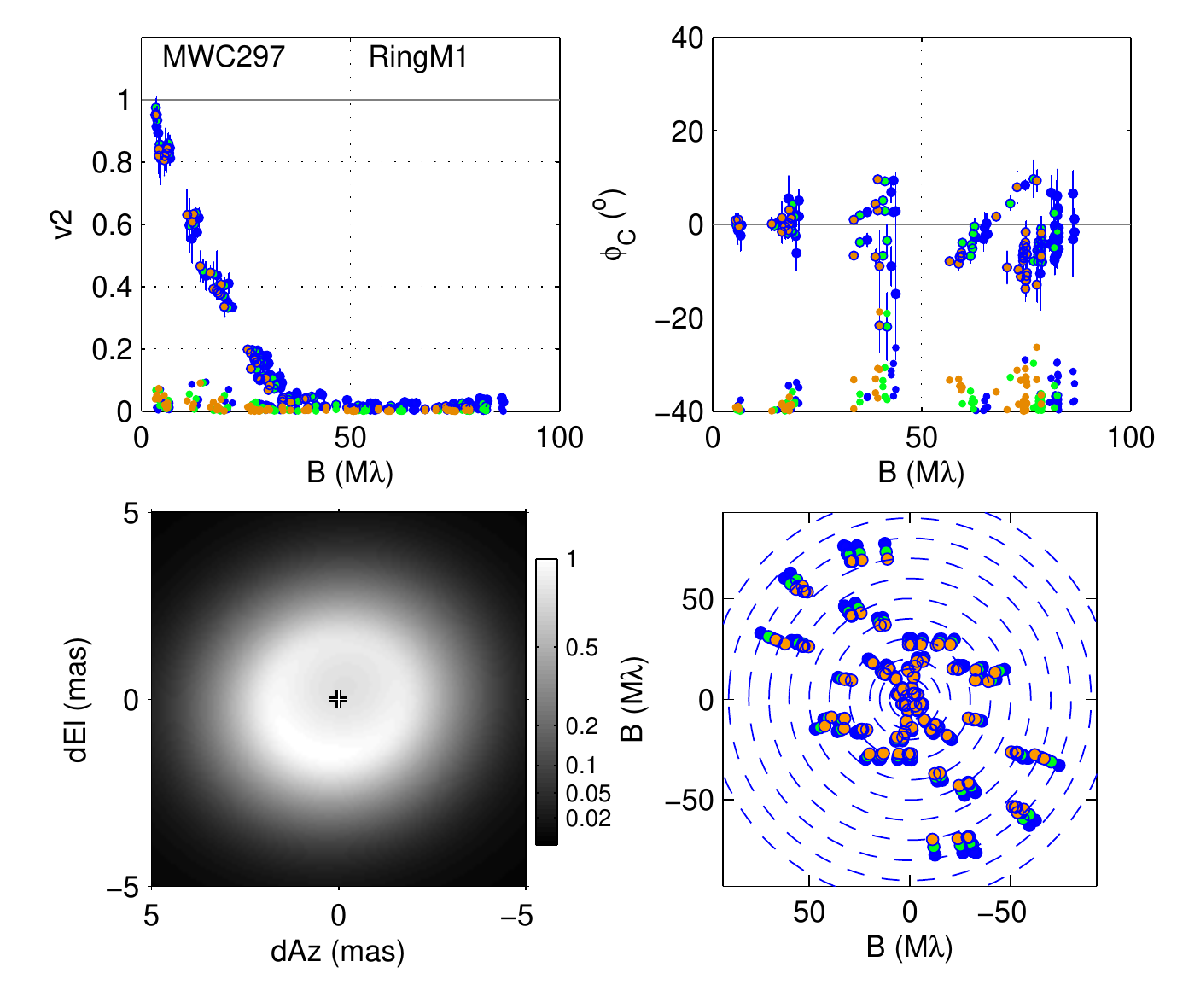}

\includegraphics[width=85mm]{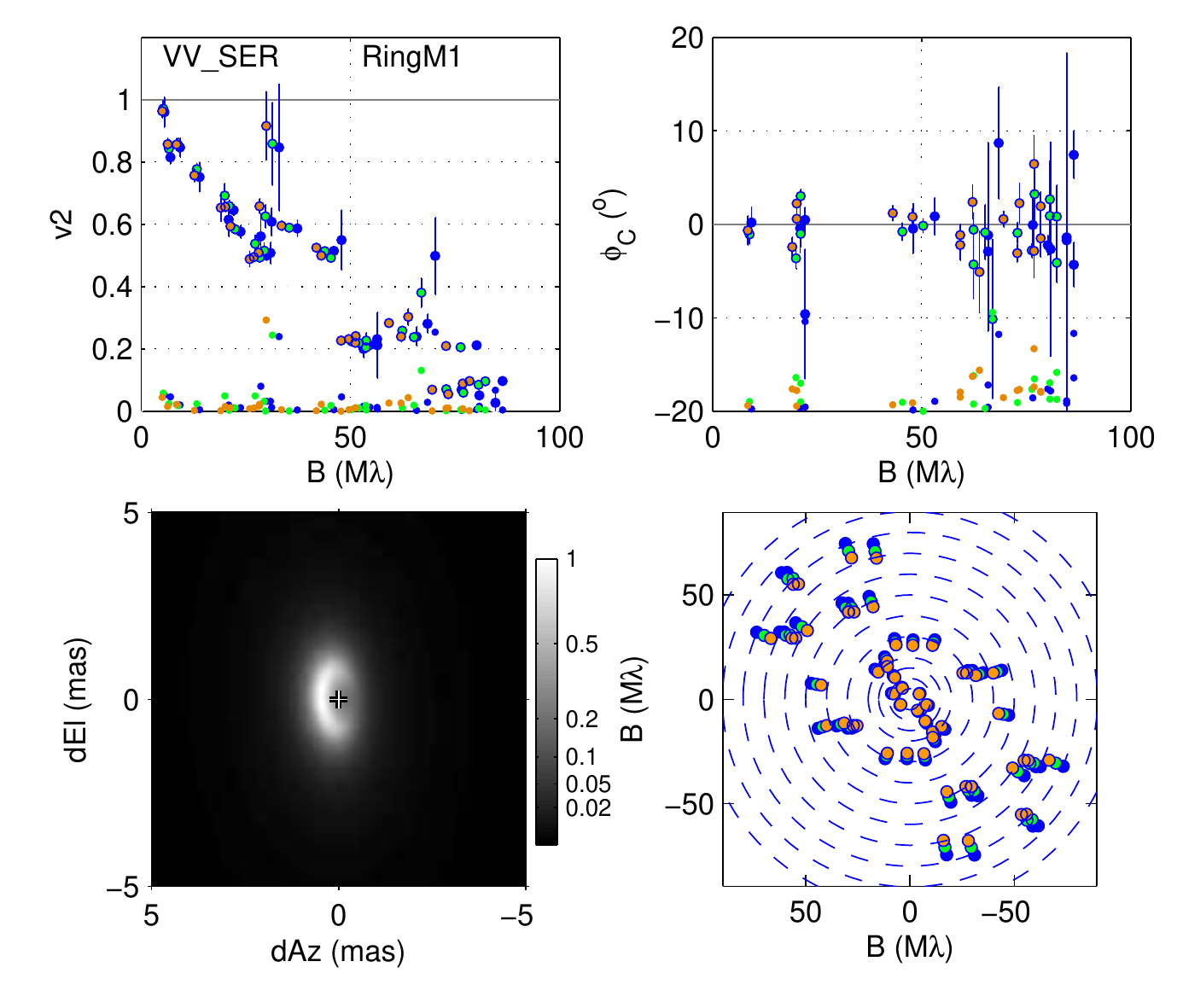}
\includegraphics[width=85mm]{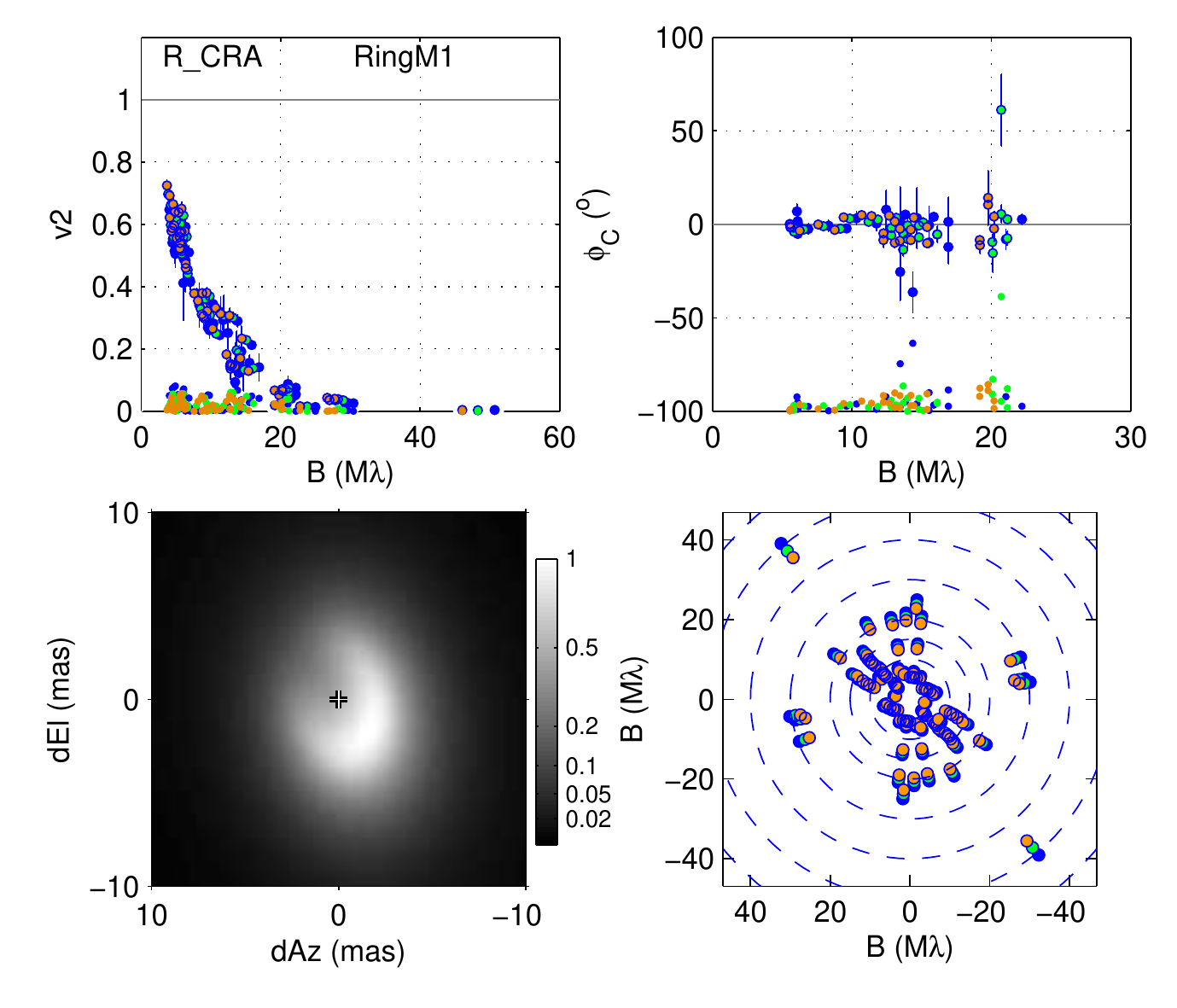}
\\
\includegraphics[width=85mm]{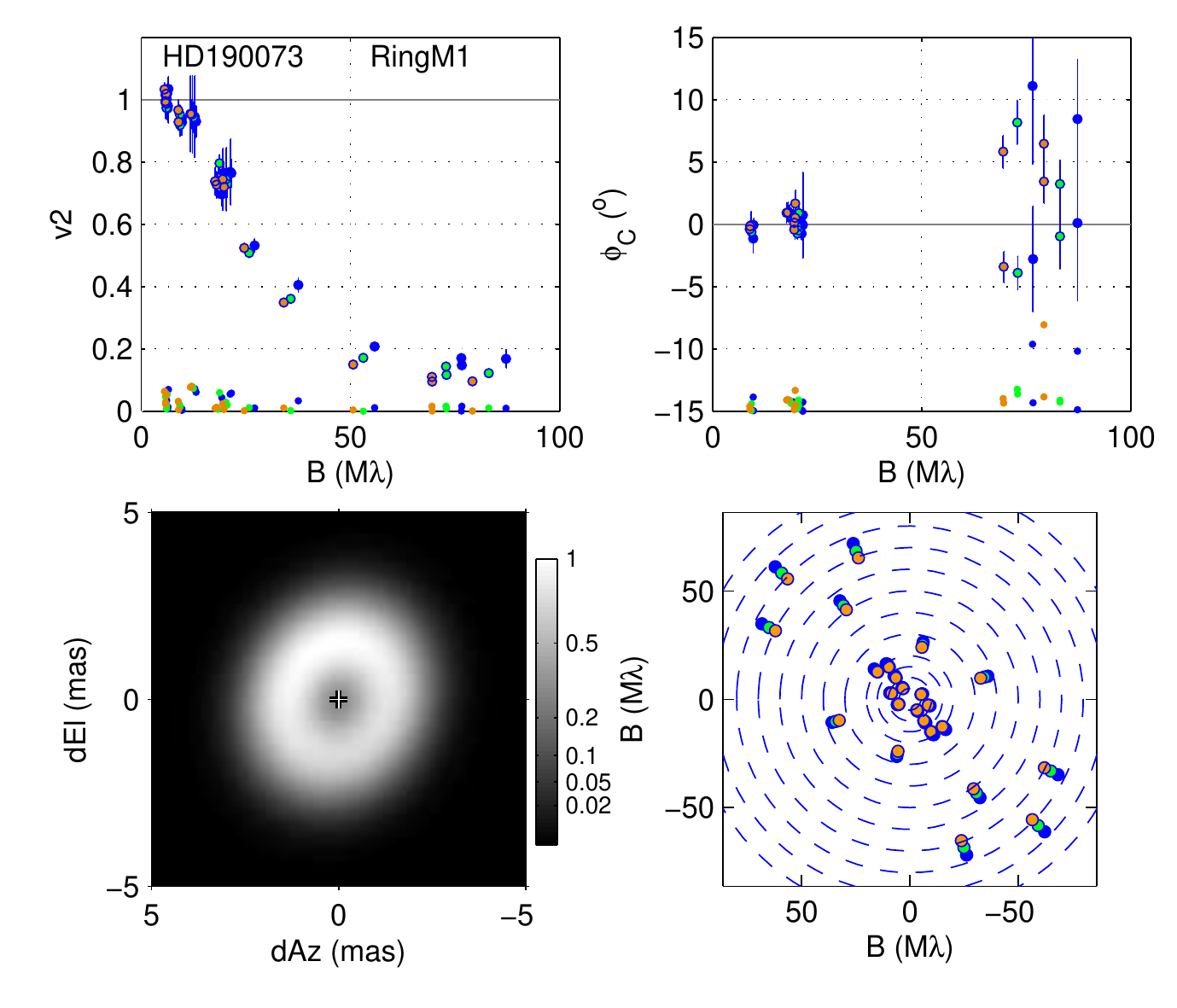}
\hfill
\caption{
Summary plots for the \emph{HQ} objects (continued). 
}
\end{figure}

\clearpage

\begin{figure}[h]
\centering
\includegraphics[width=85mm]{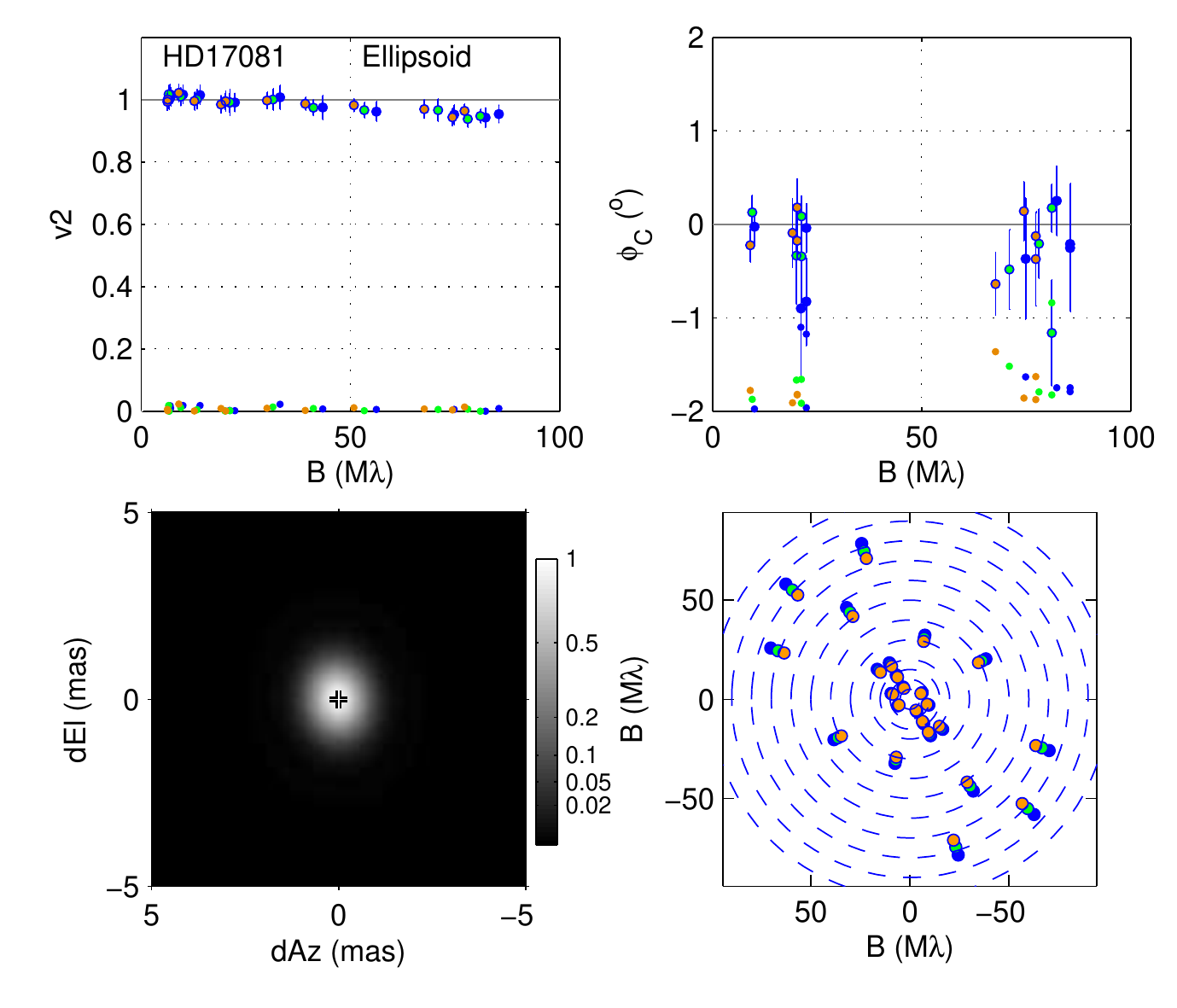}
\includegraphics[width=85mm]{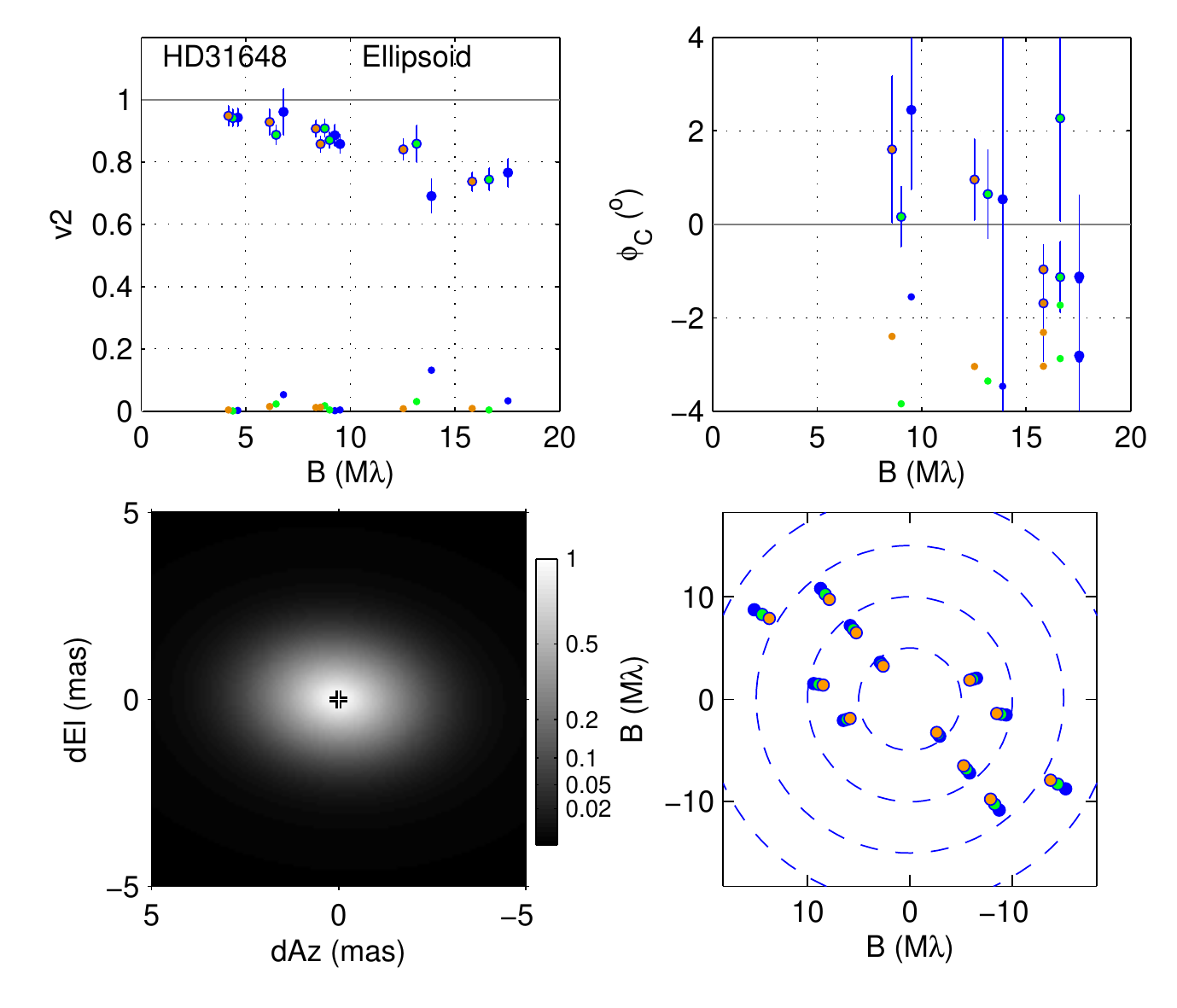}

\includegraphics[width=85mm]{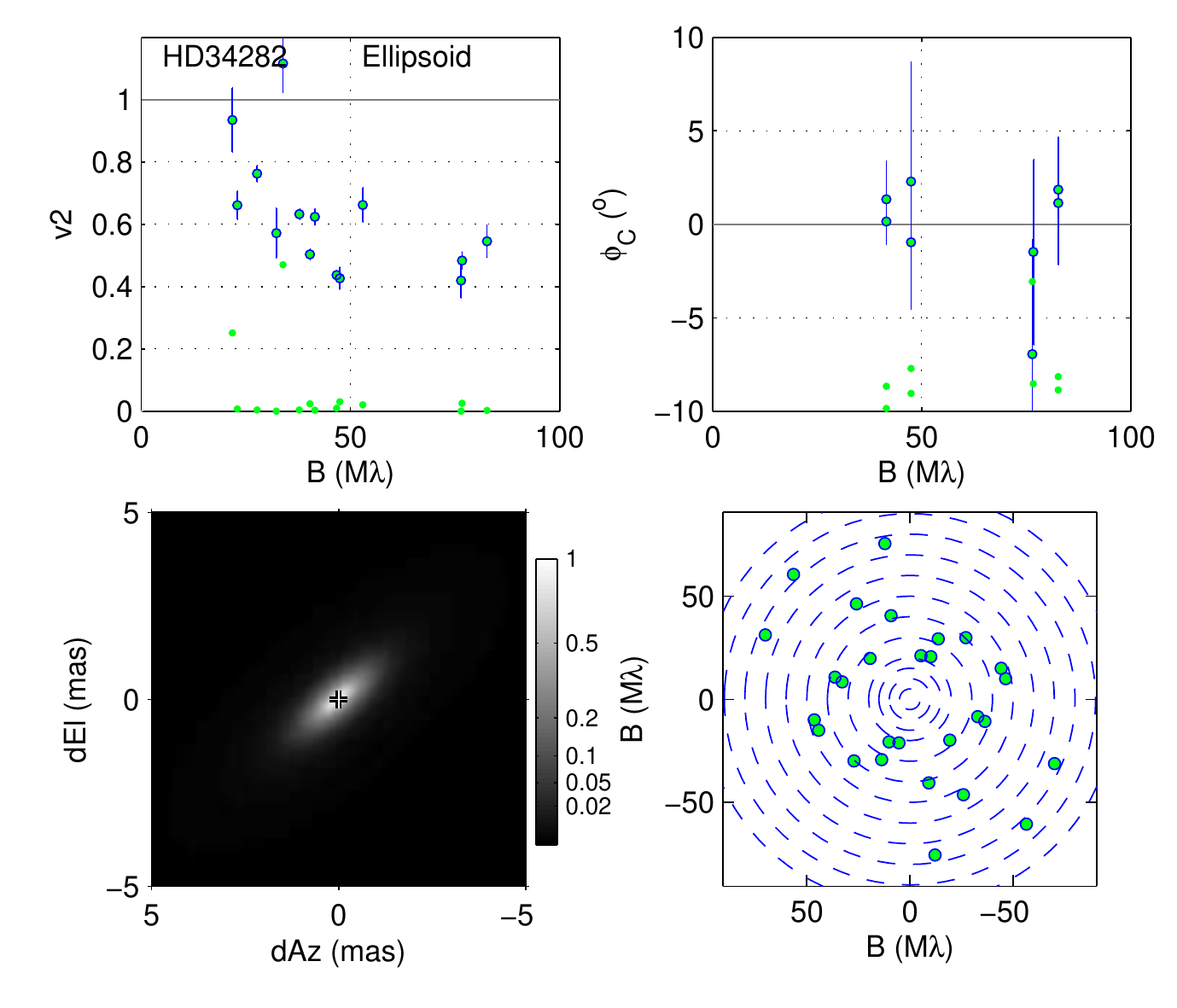}
\includegraphics[width=85mm]{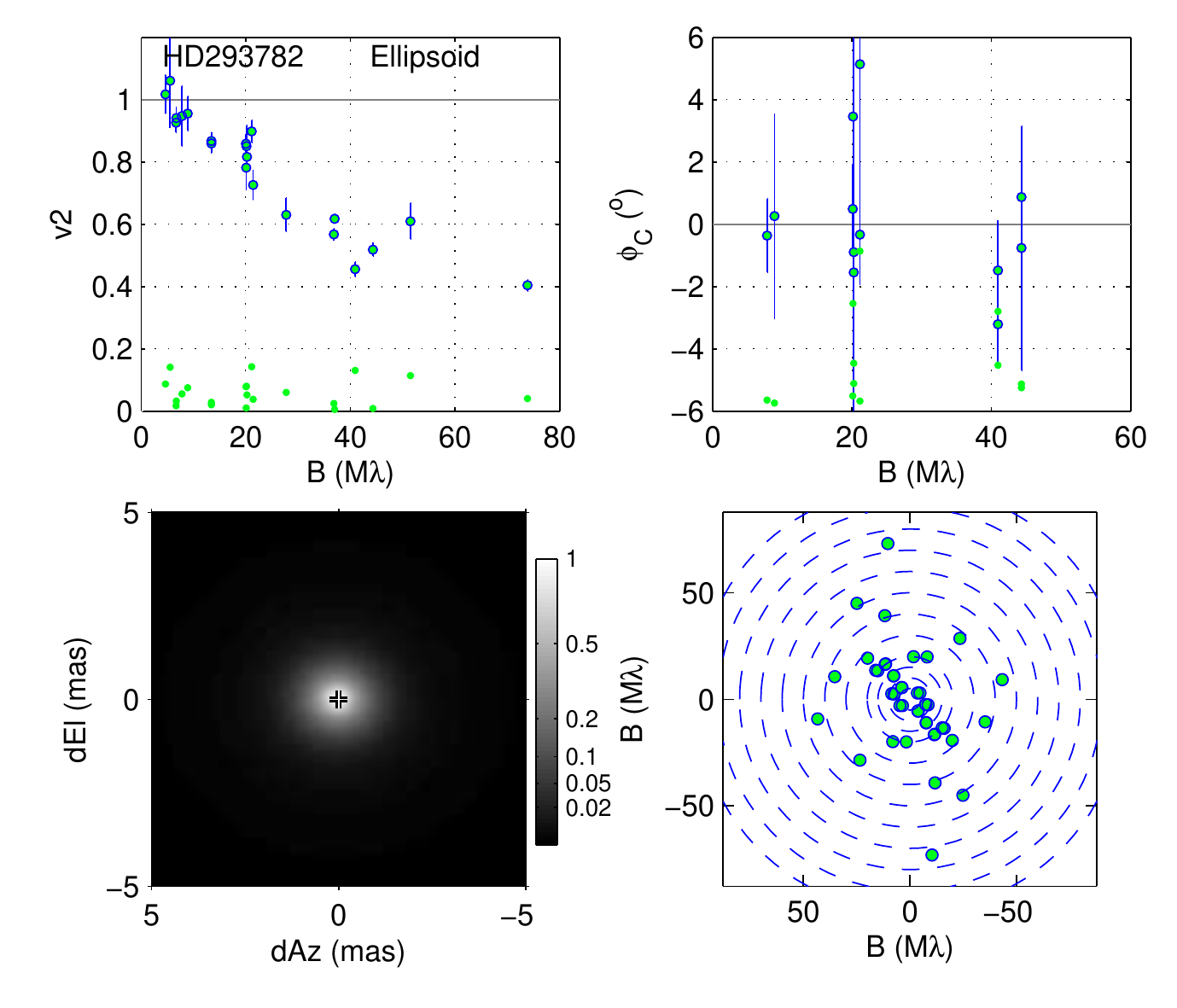}
\caption{
Summary plots for the non-\emph{HQ} objects. 
Similar plots as in Fig.~\ref{summplotHQ} above, for the rest of the sample. For these objects, because of the limited resolution achieved or other limitations, a simpler \emph{Ellipsoid} fit was made. For objects tagged as binary, no fit was attempted, and the image sub-panel is absent. 
\label{summplotNonHQ}
}
\end{figure}

\begin{figure}[h]
\centering
\includegraphics[width=85mm]{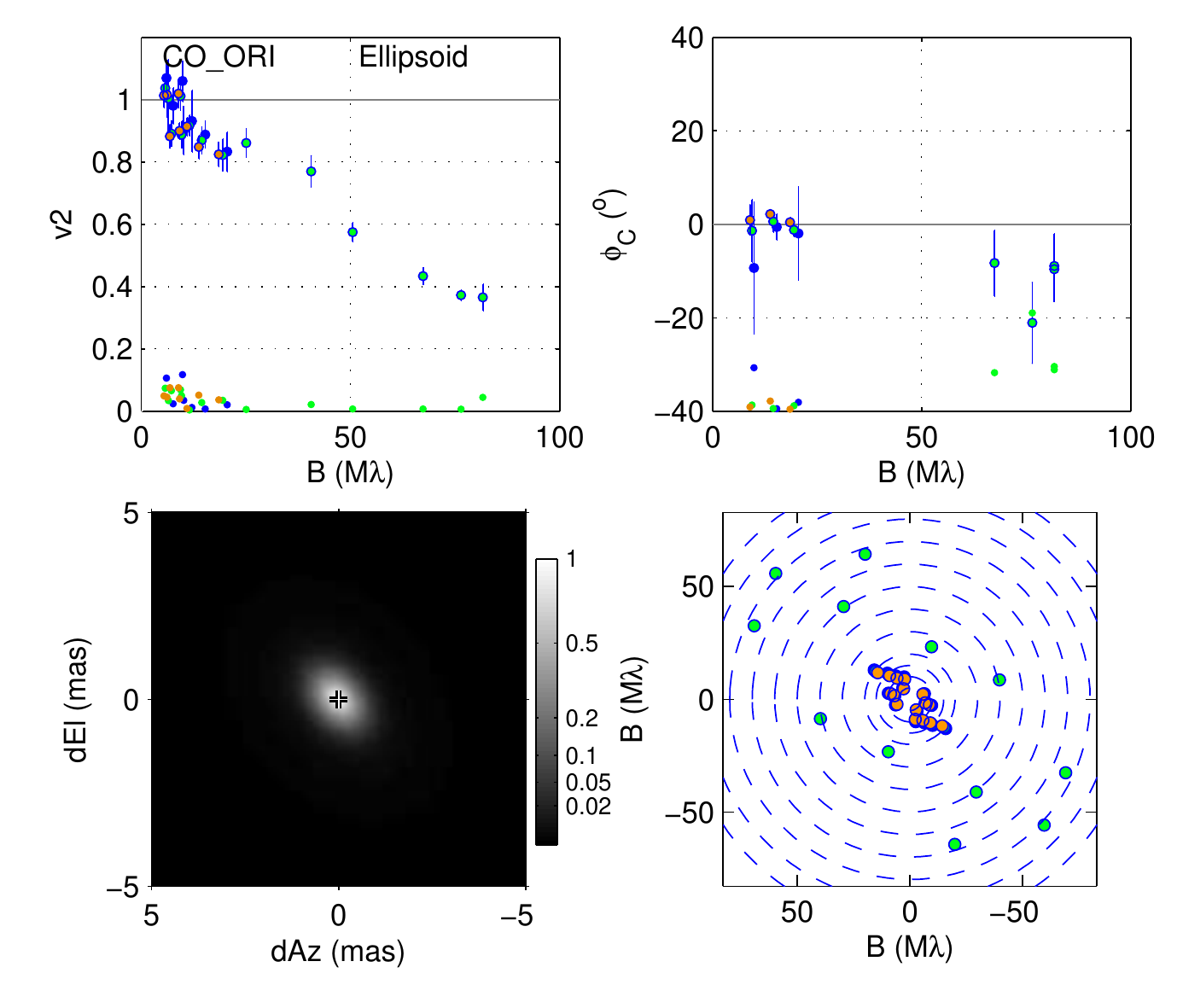}
\includegraphics[width=85mm]{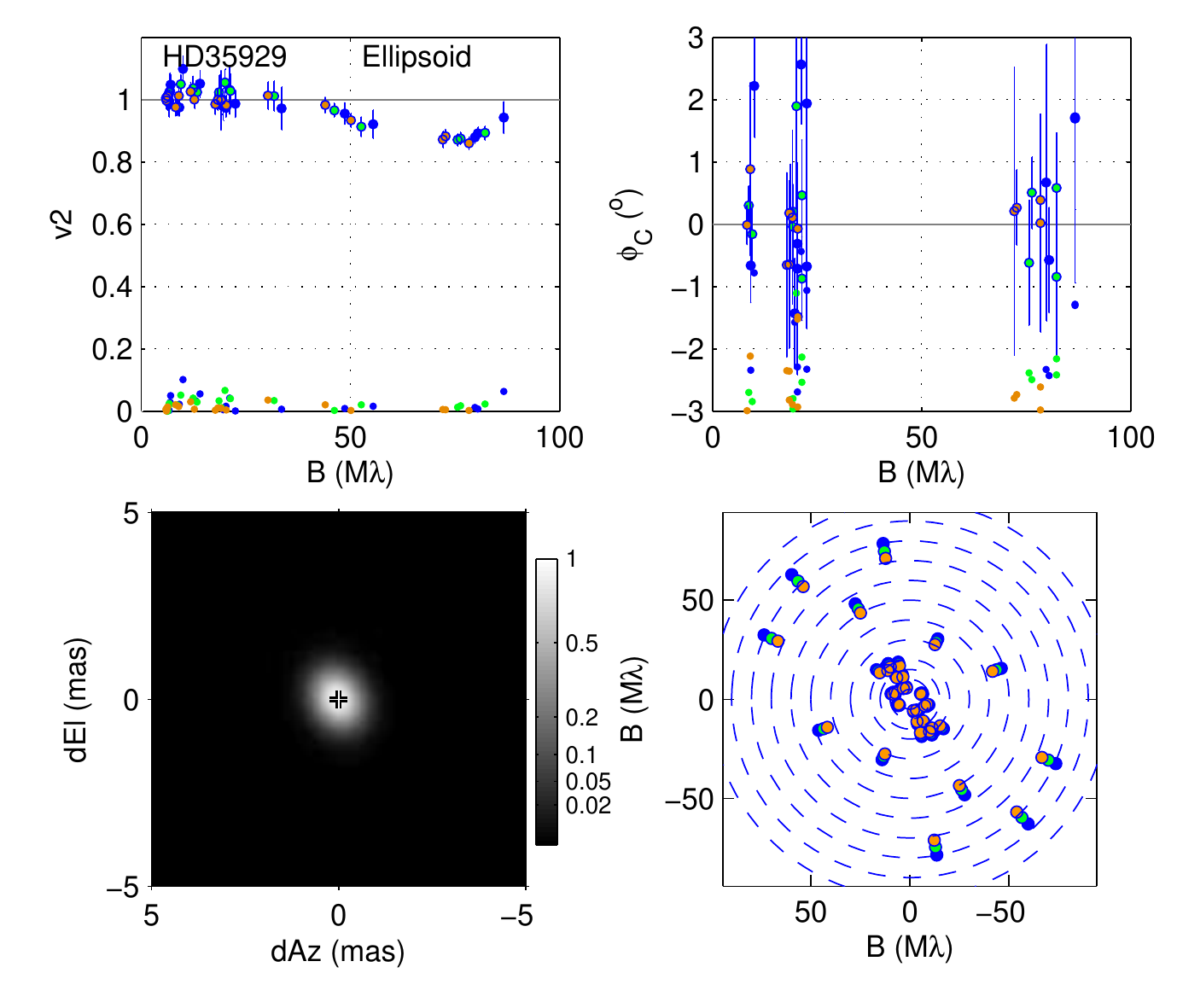}

\includegraphics[width=85mm]{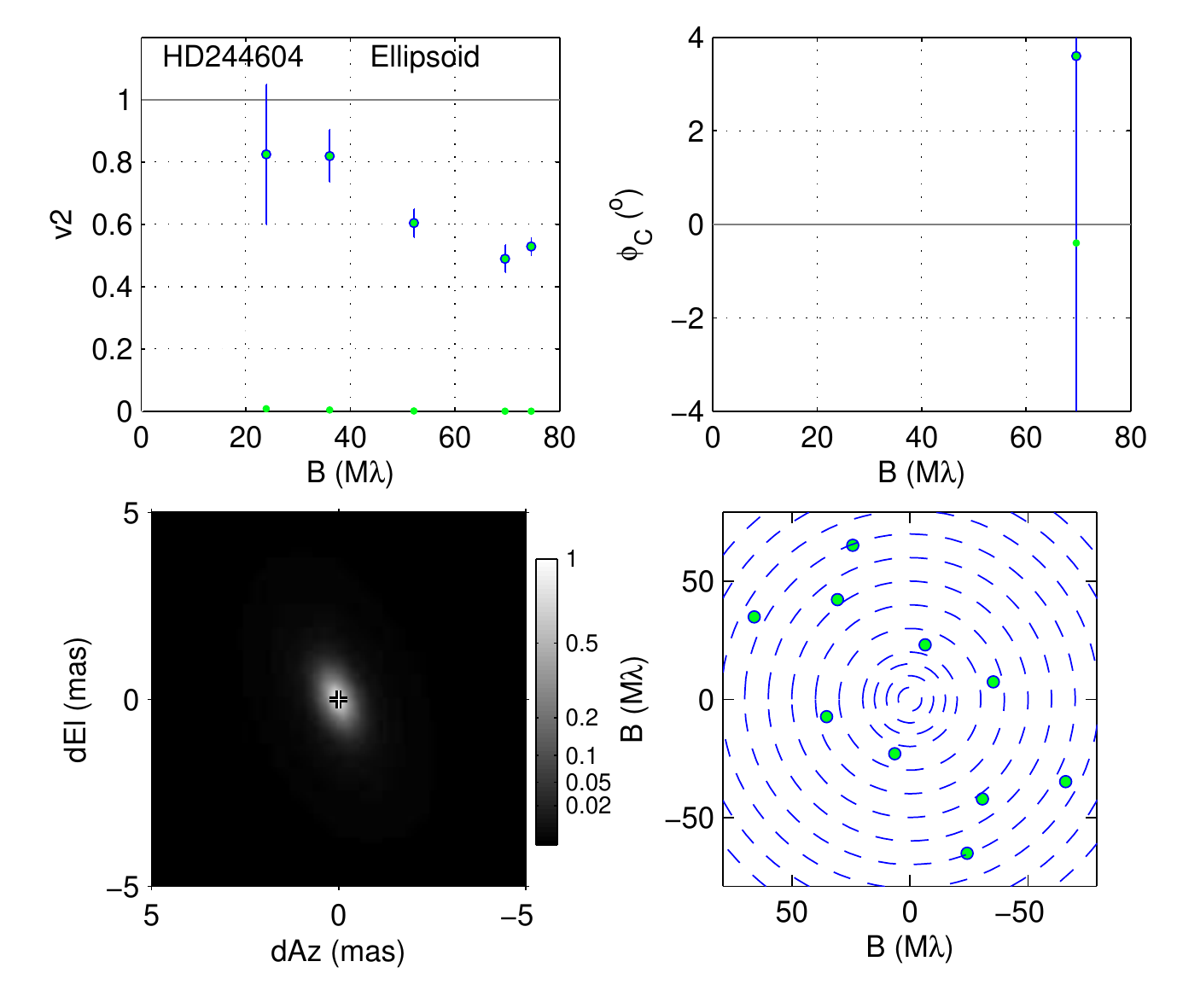}
\includegraphics[width=85mm]{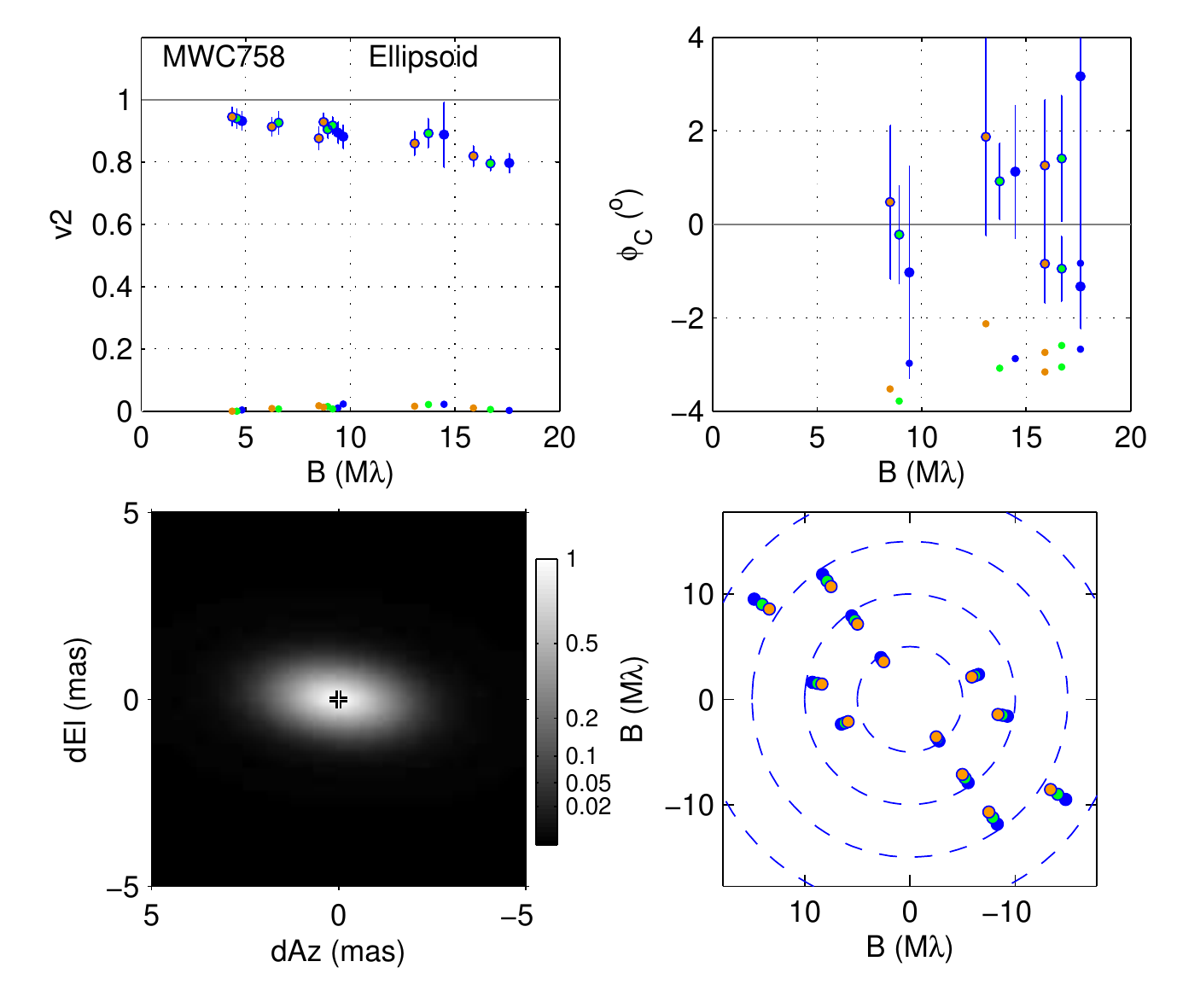}

\includegraphics[width=85mm]{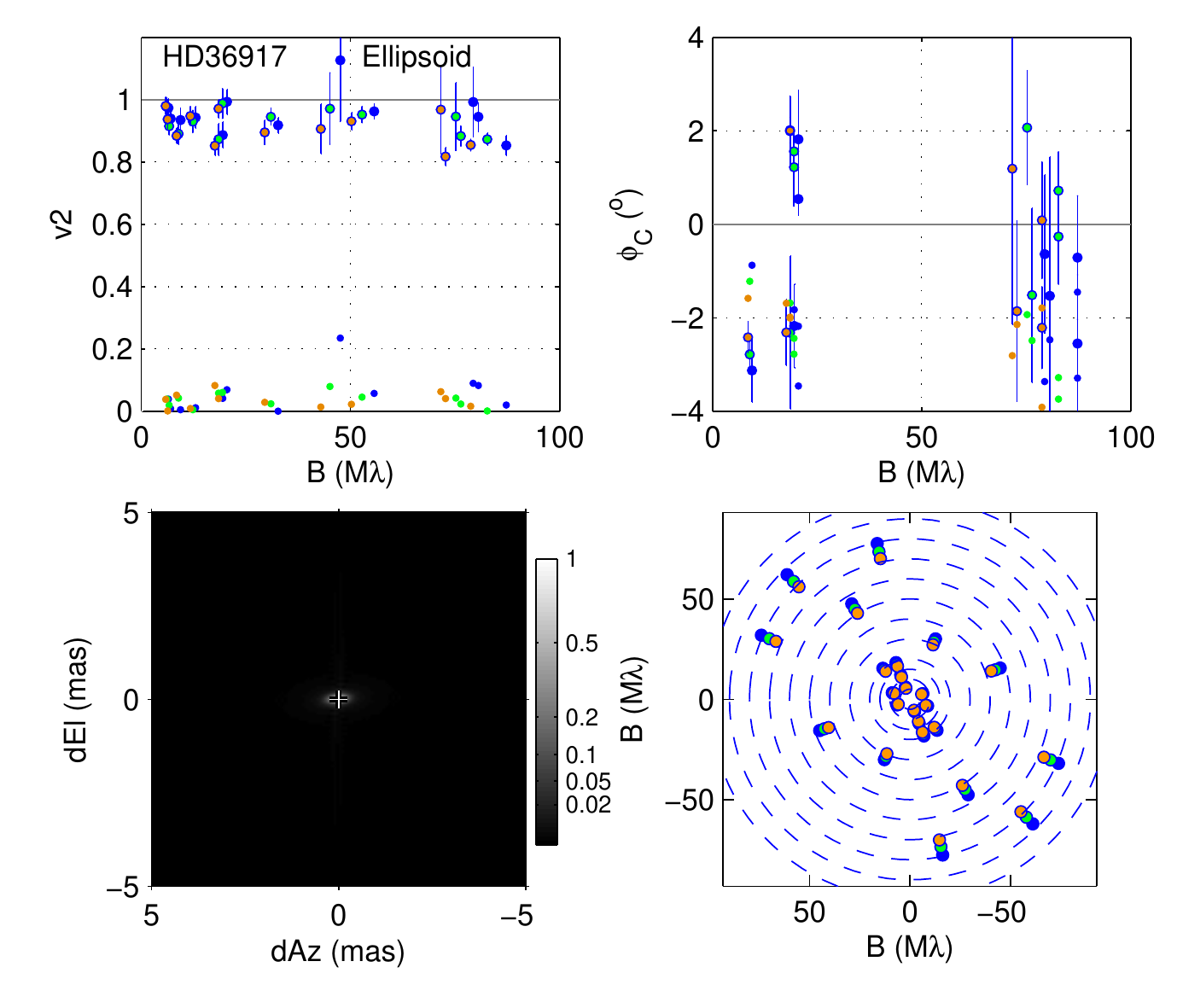}
\includegraphics[width=85mm]{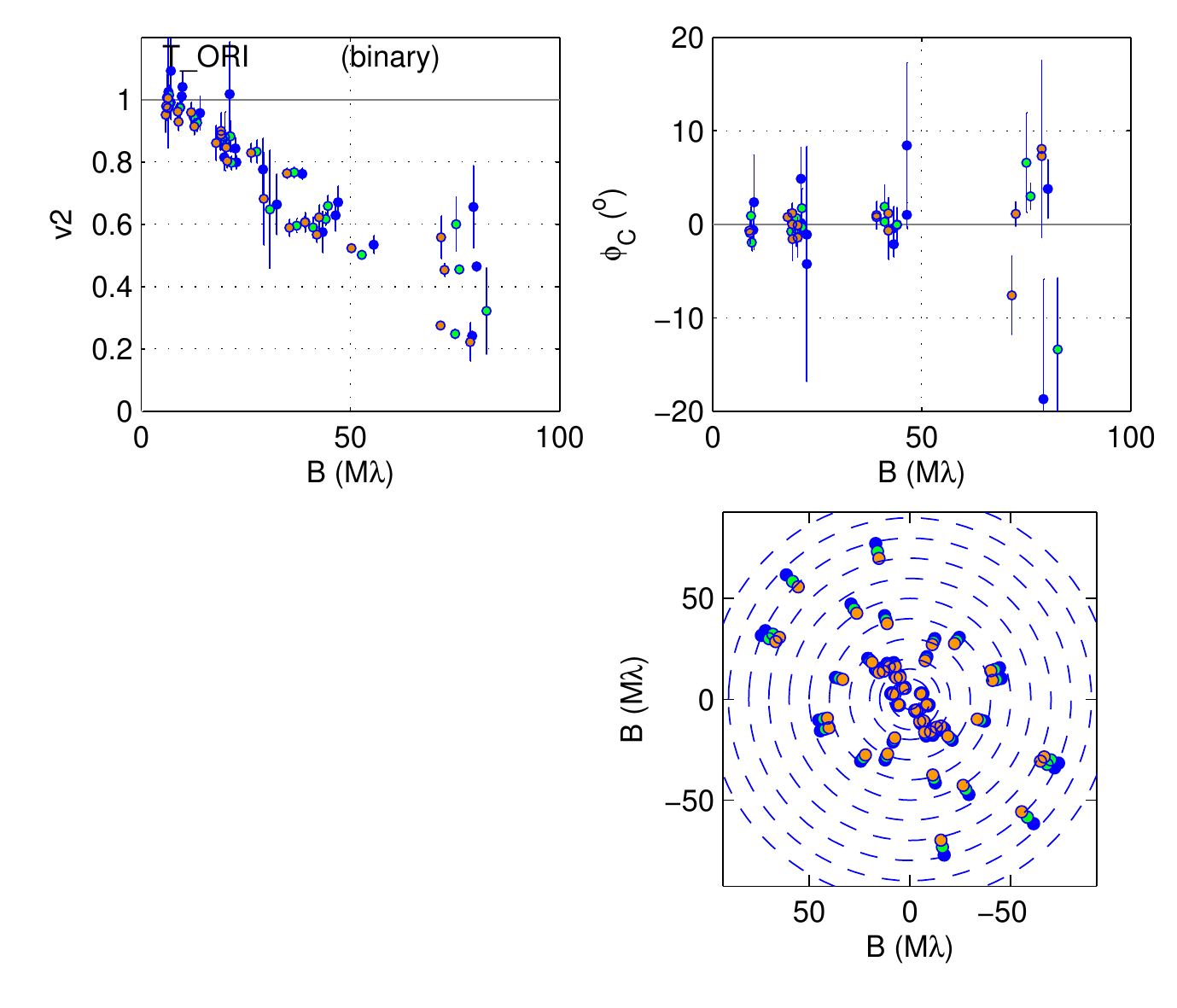}
\caption{Summary plots for the non-\emph{HQ} objects (continued).}
\end{figure}

\begin{figure}[h]
\centering
\includegraphics[width=85mm]{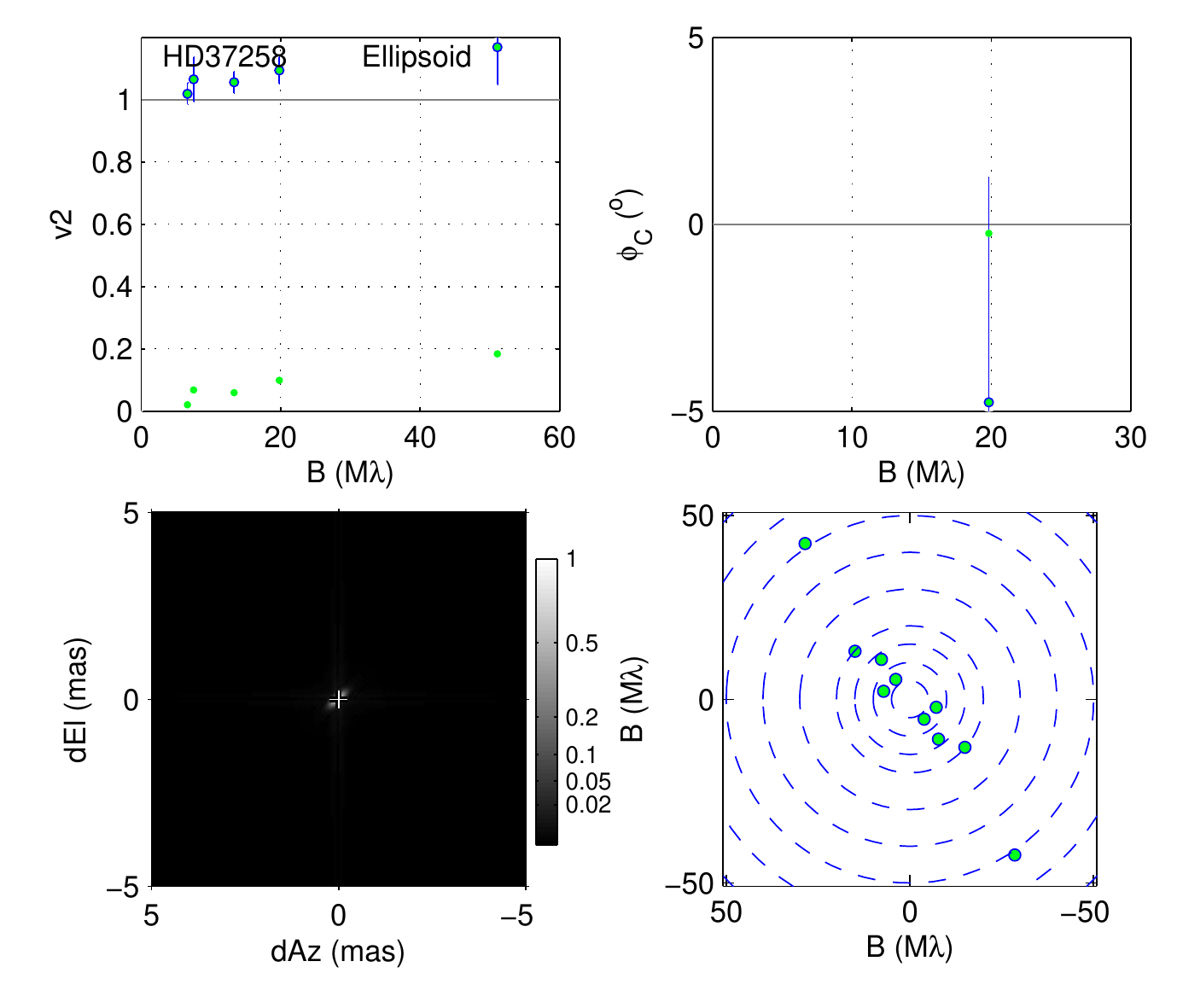}
\includegraphics[width=85mm]{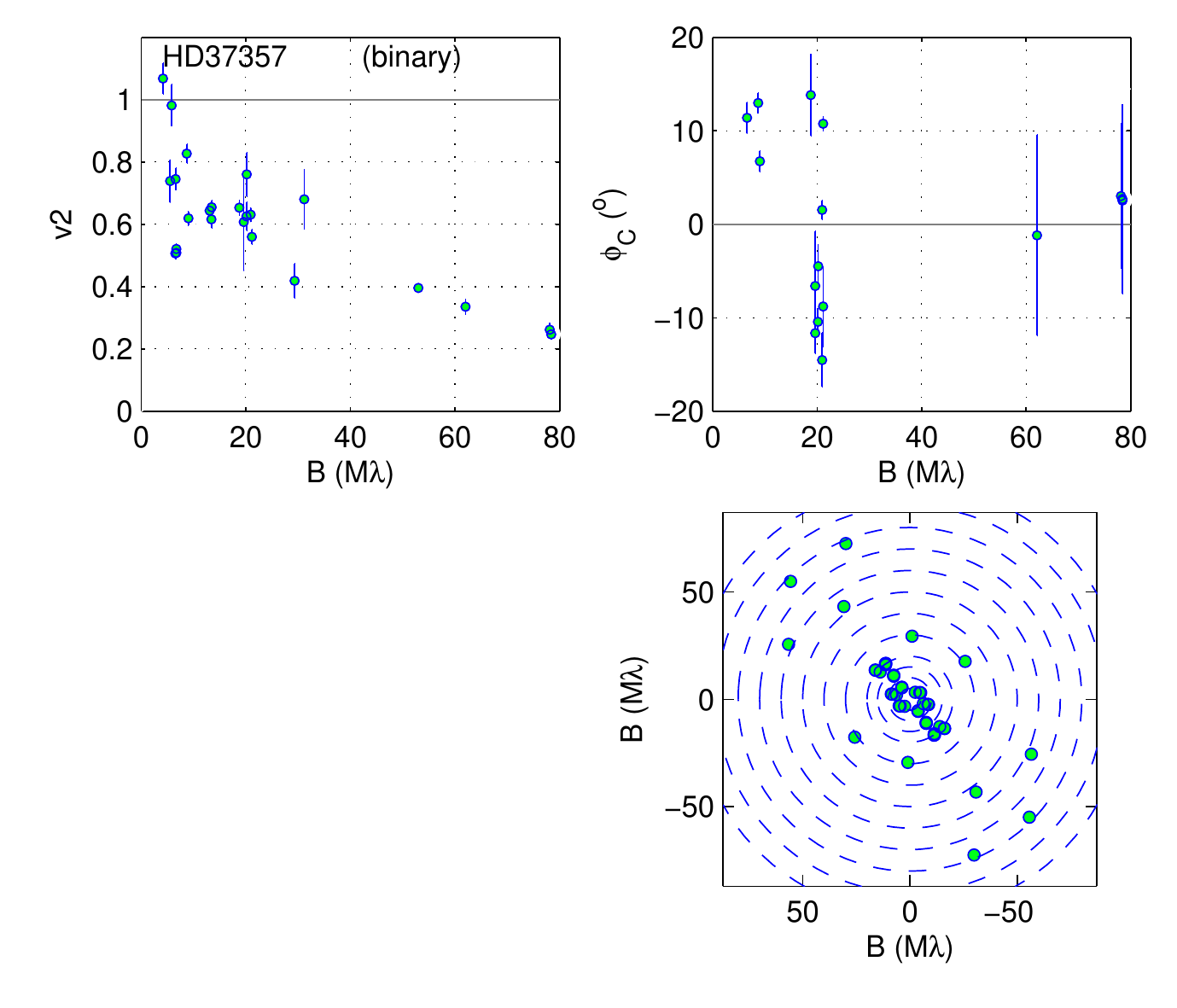}

\includegraphics[width=85mm]{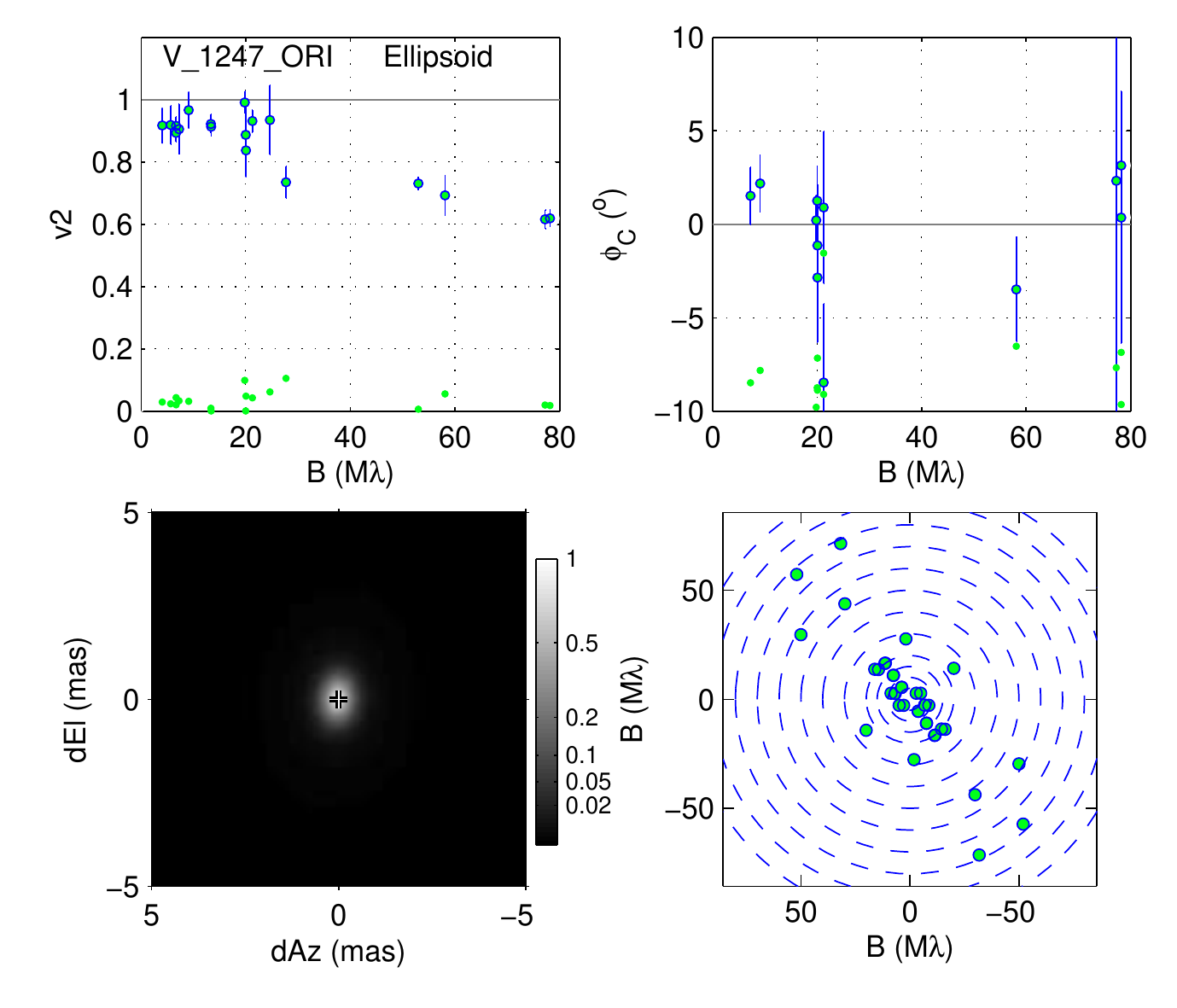}
\includegraphics[width=85mm]{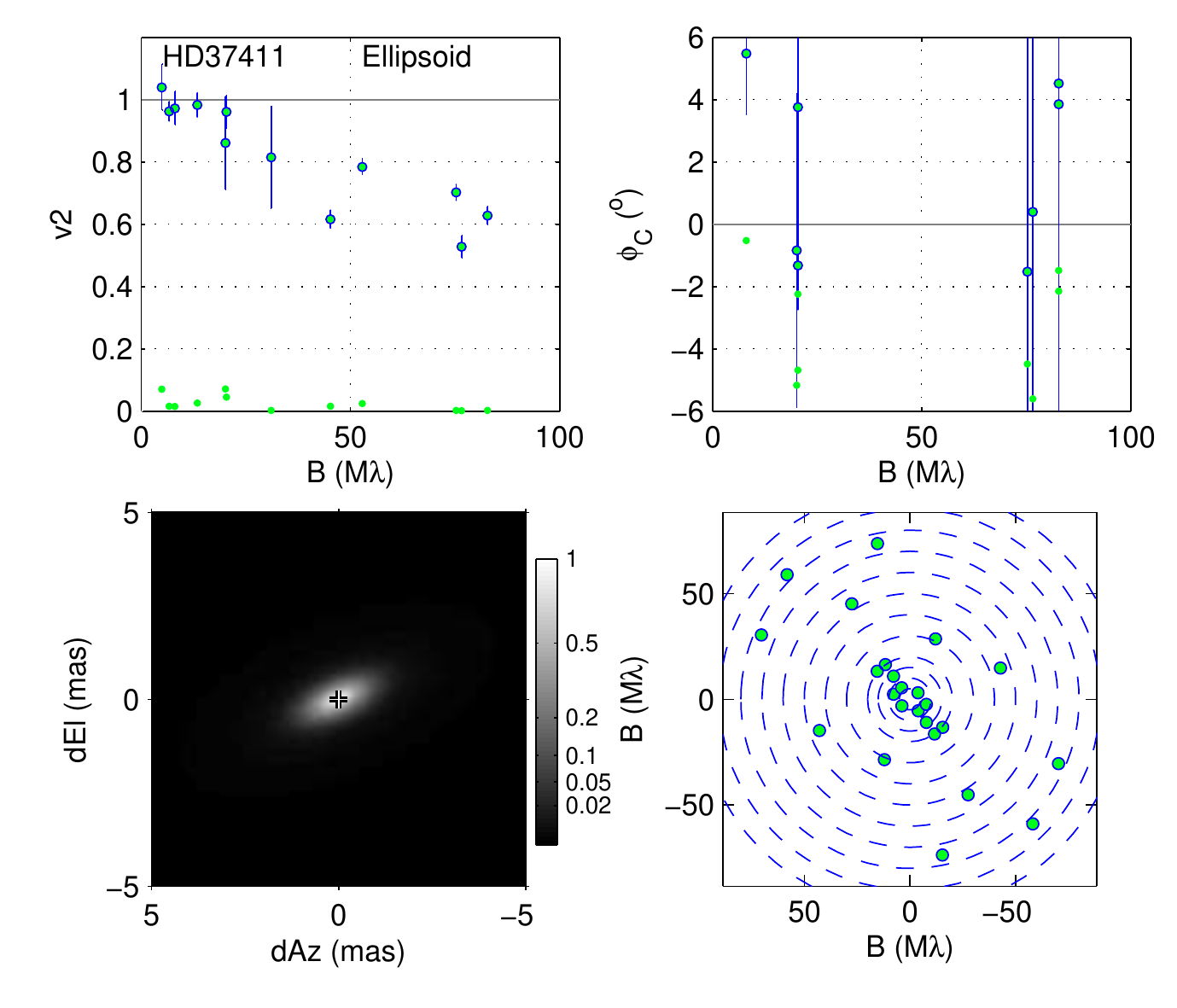}

\includegraphics[width=85mm]{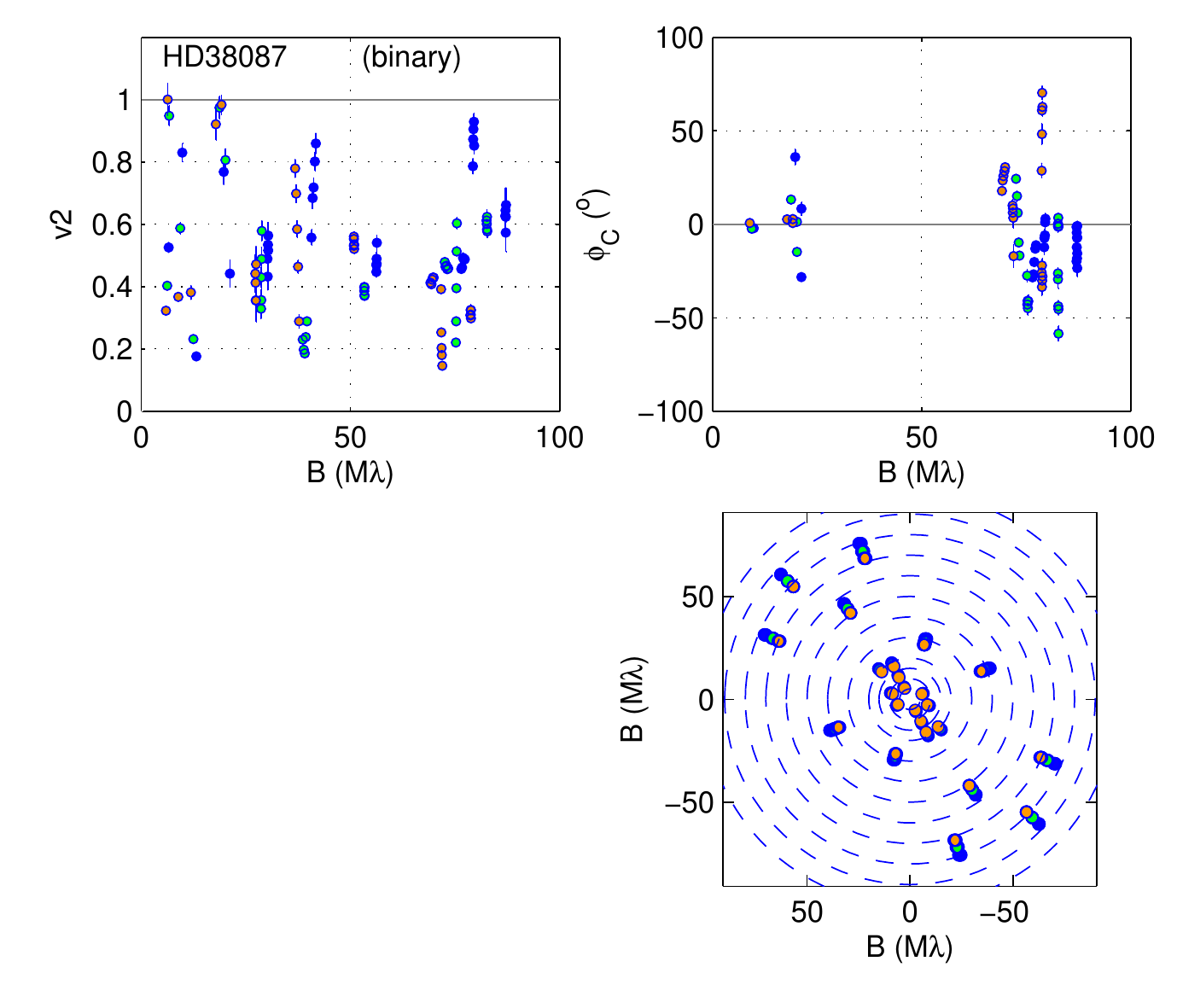}
\includegraphics[width=85mm]{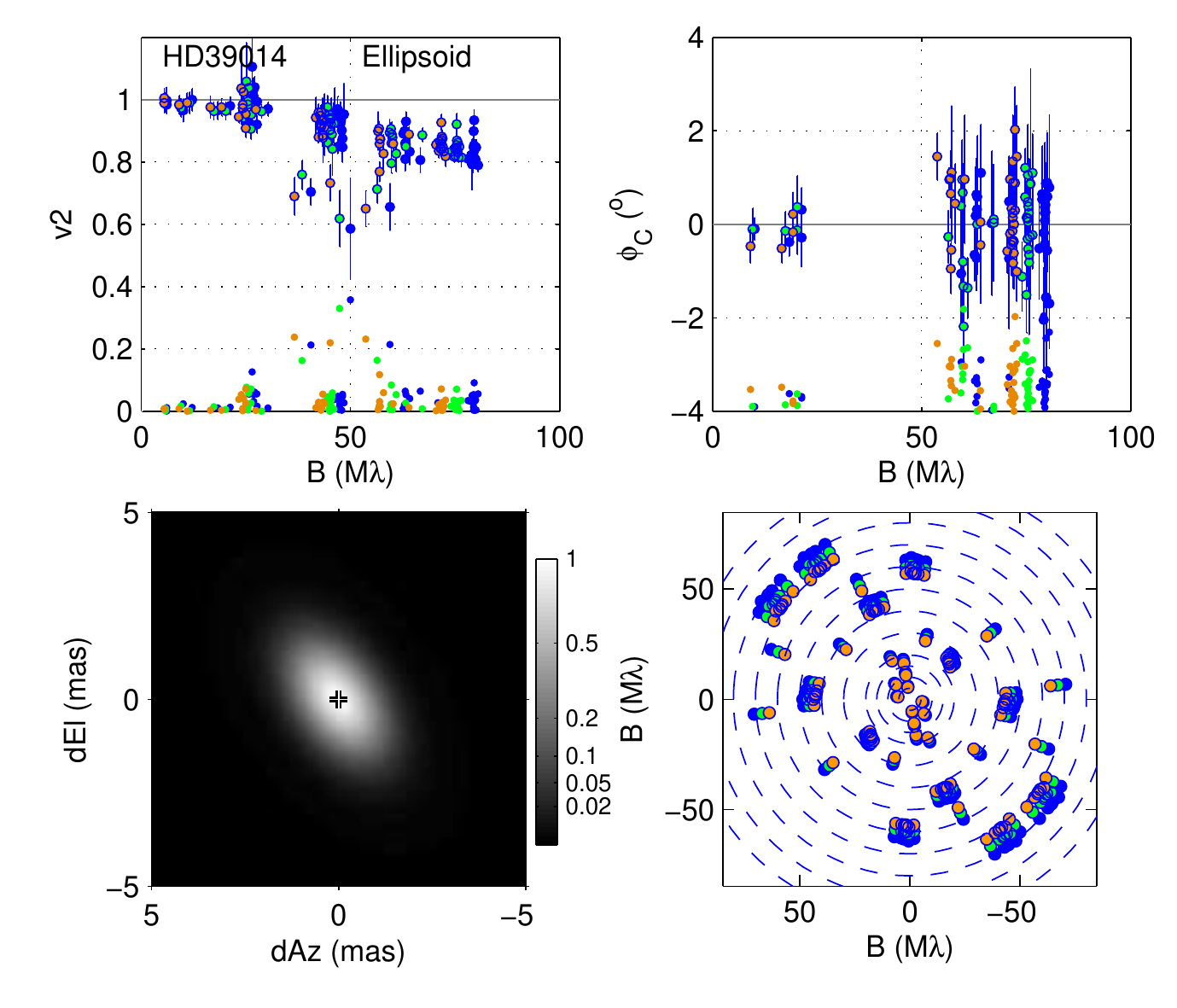}
\caption{Summary plots for the non-\emph{HQ} objects (continued).}
\end{figure}

\begin{figure}[h]
\centering
\includegraphics[width=85mm]{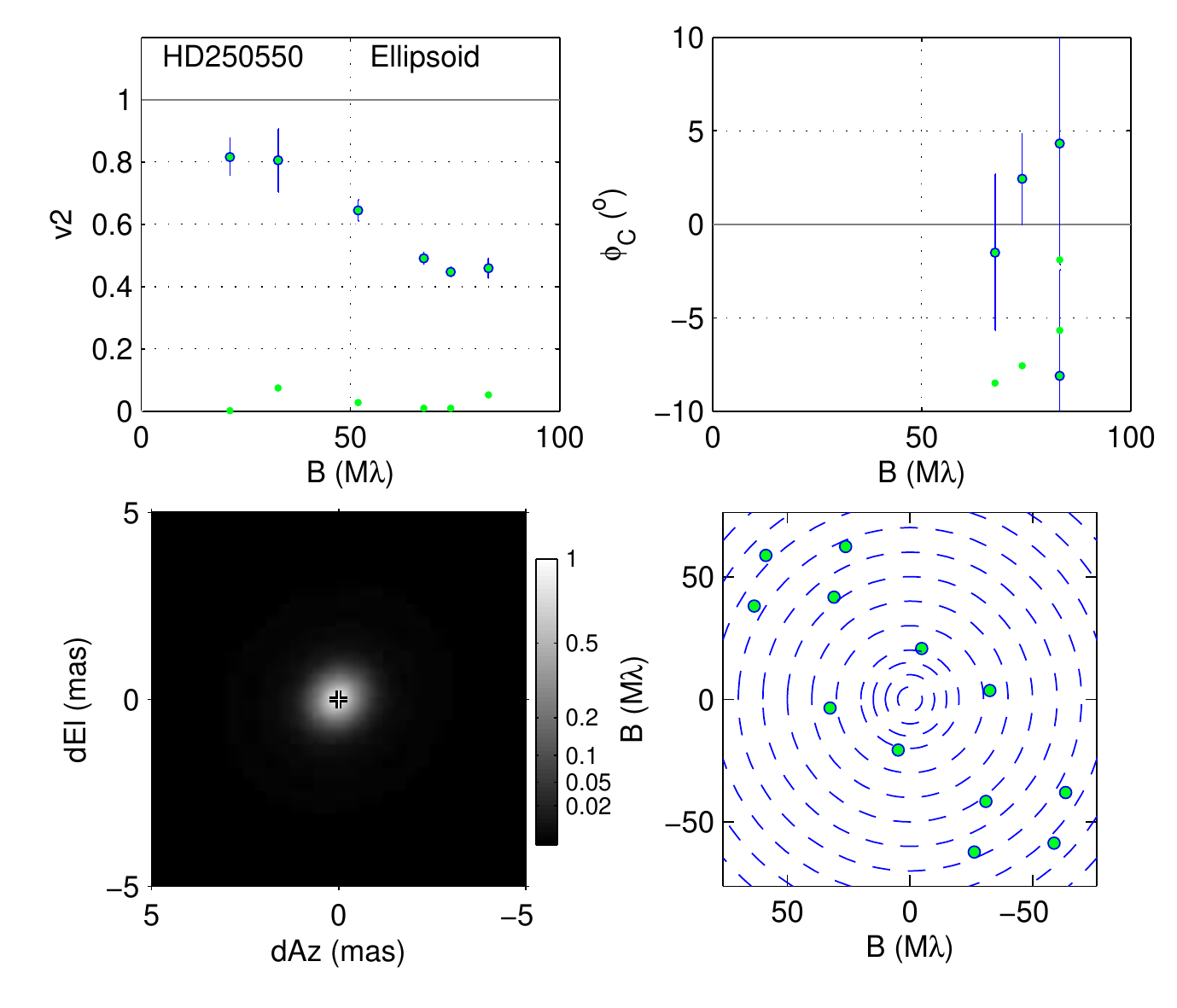}
\includegraphics[width=85mm]{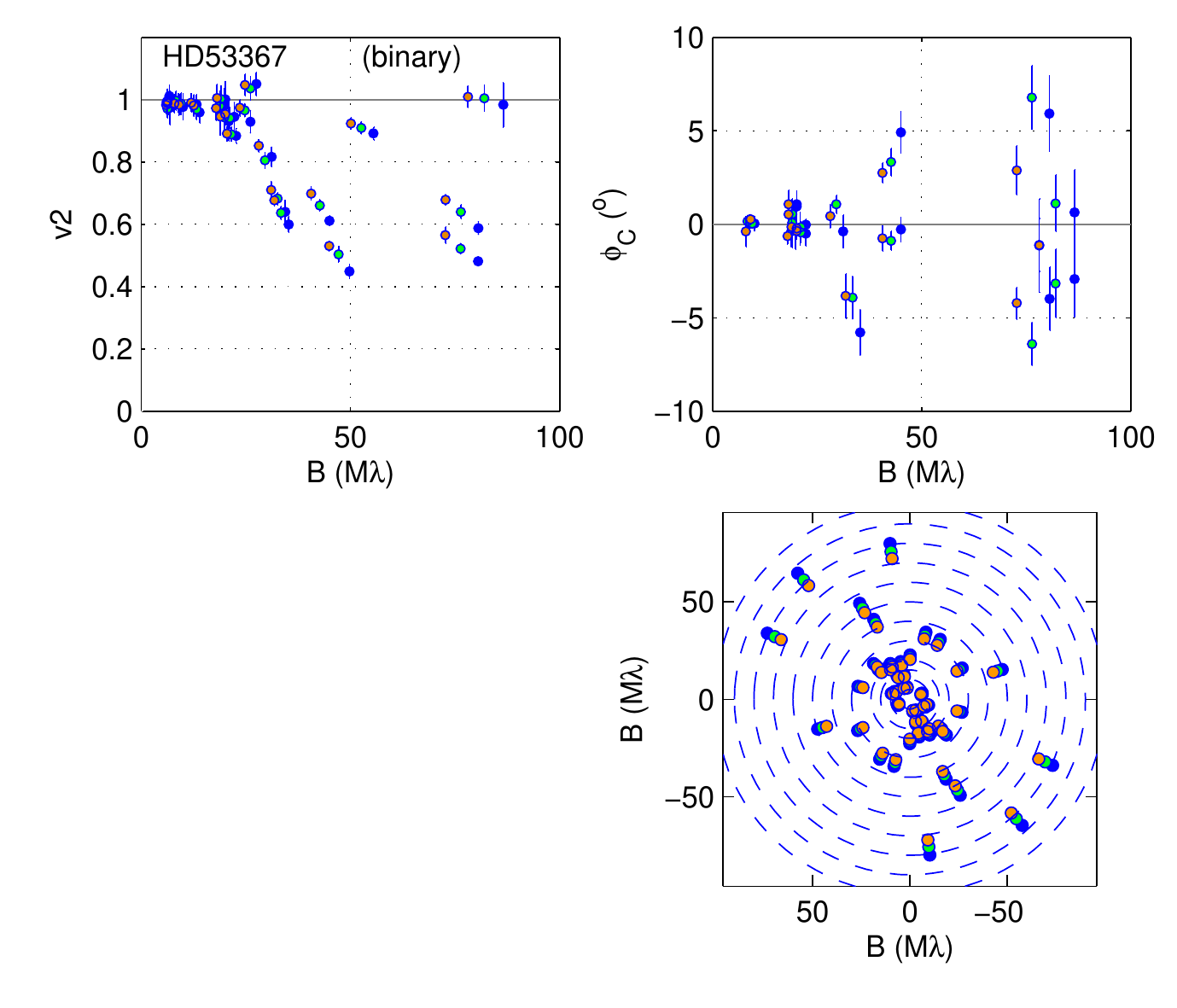}

\includegraphics[width=85mm]{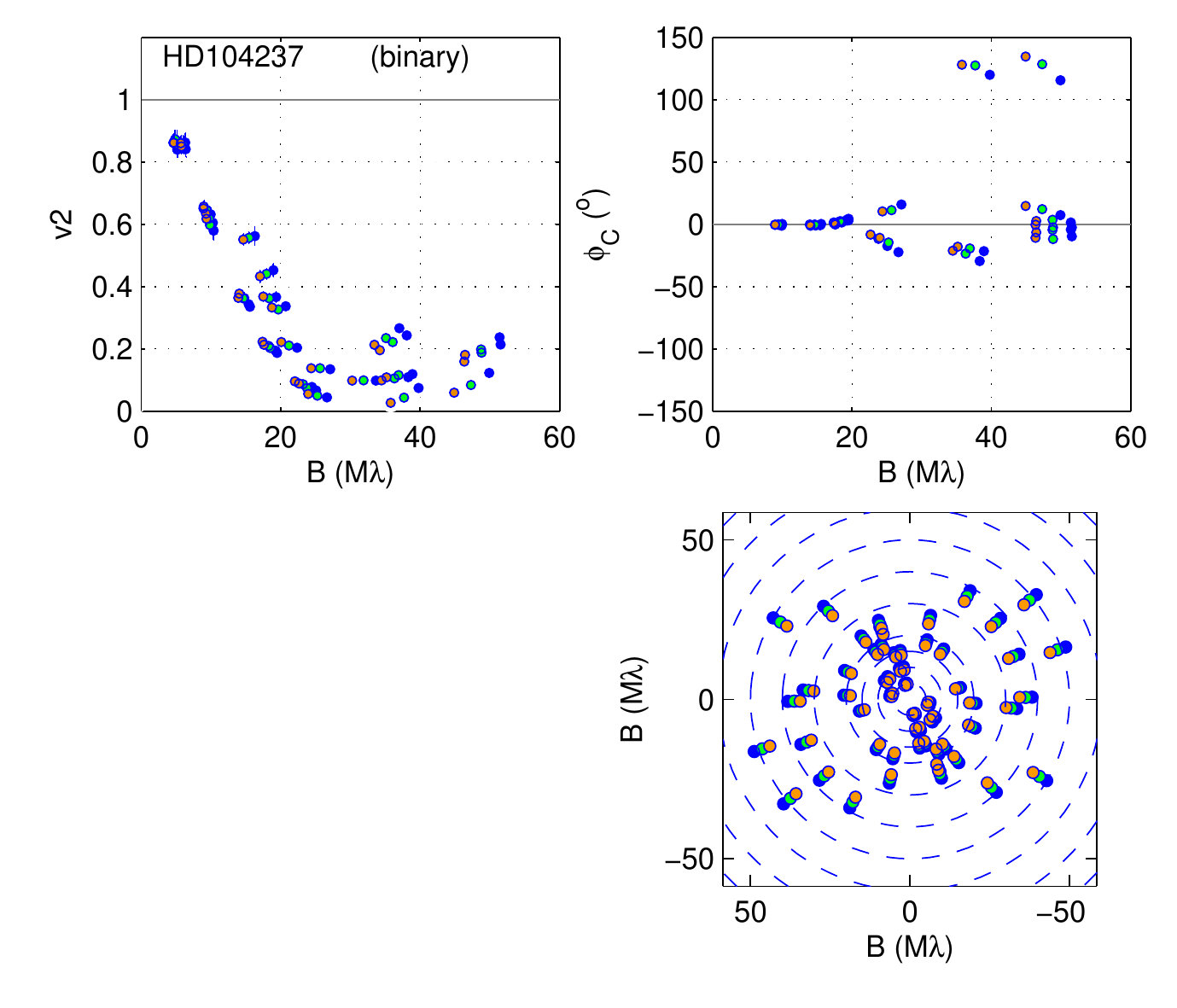}
\includegraphics[width=85mm]{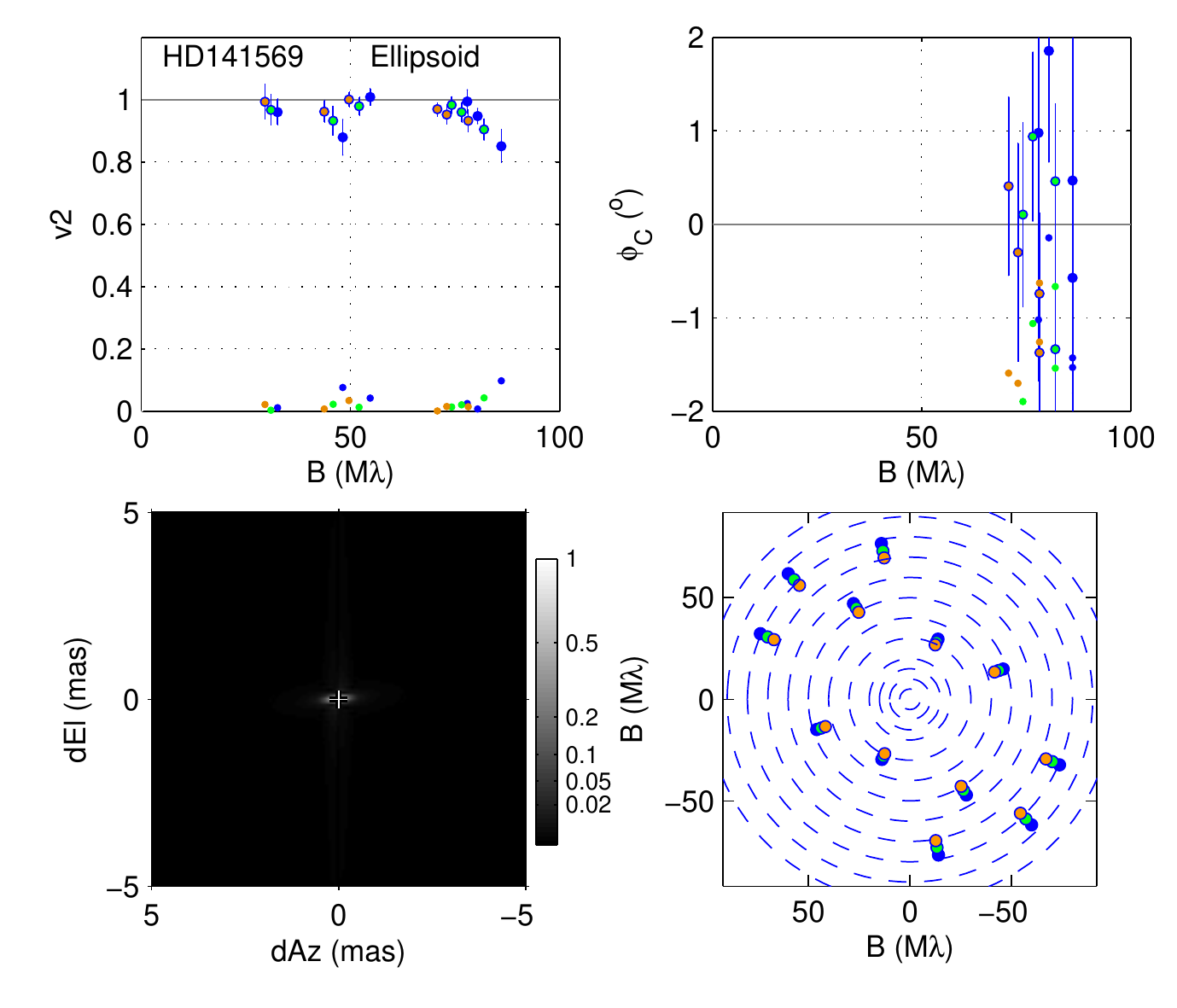}

\includegraphics[width=85mm]{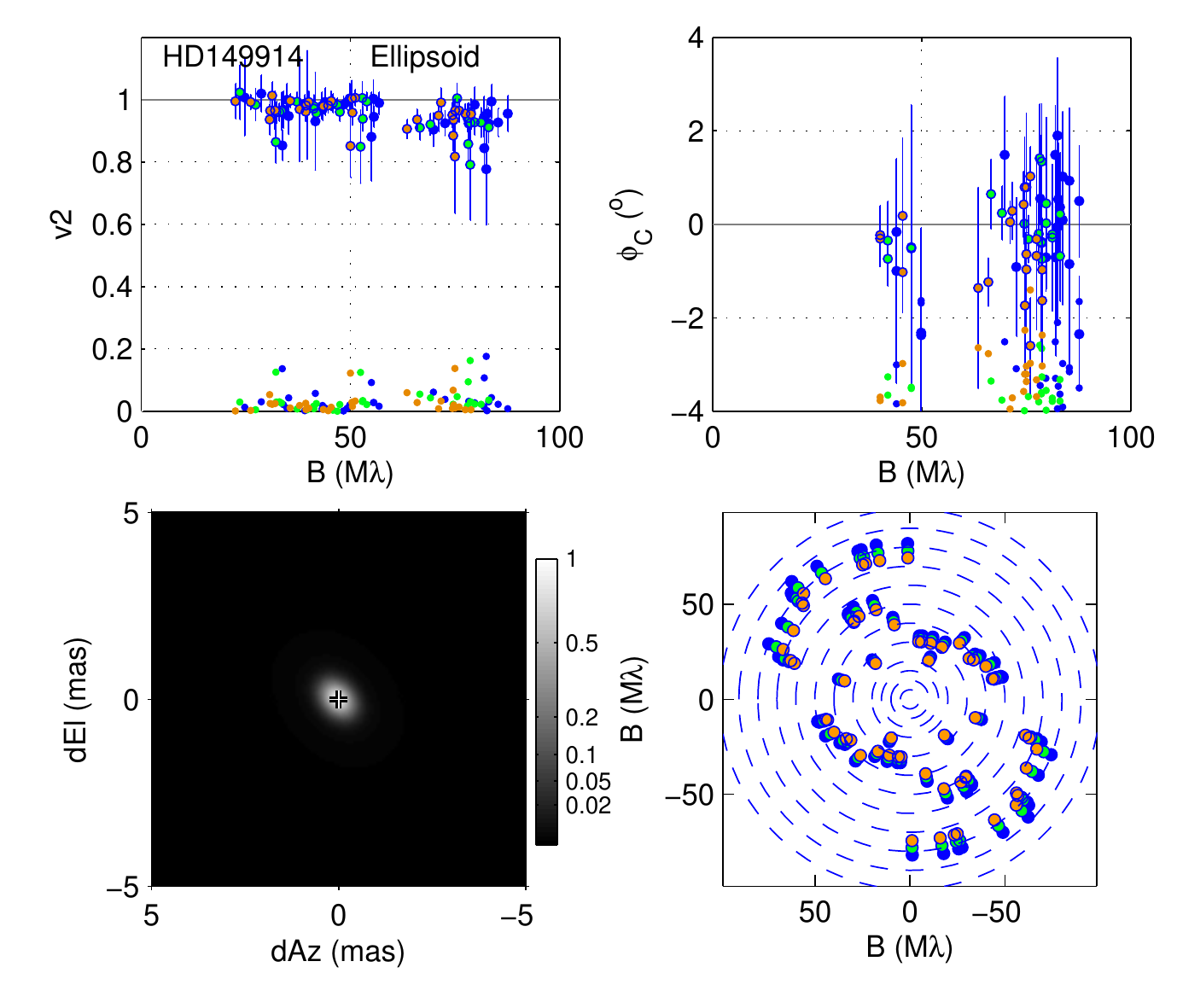}
\includegraphics[width=85mm]{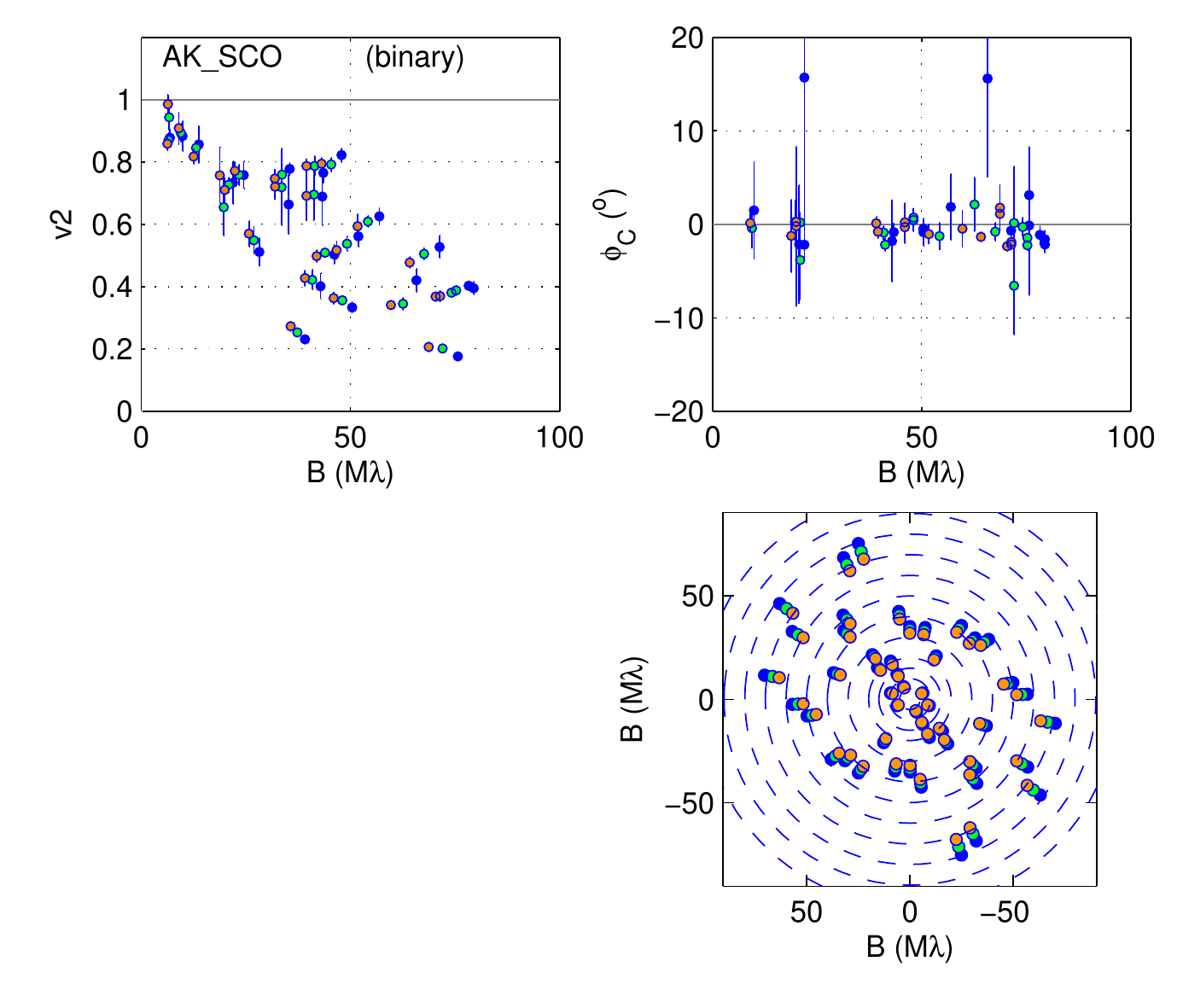}
\caption{Summary plots for the non-\emph{HQ} objects (continued).}
\end{figure}

\begin{figure}[h]
\centering
\includegraphics[width=85mm]{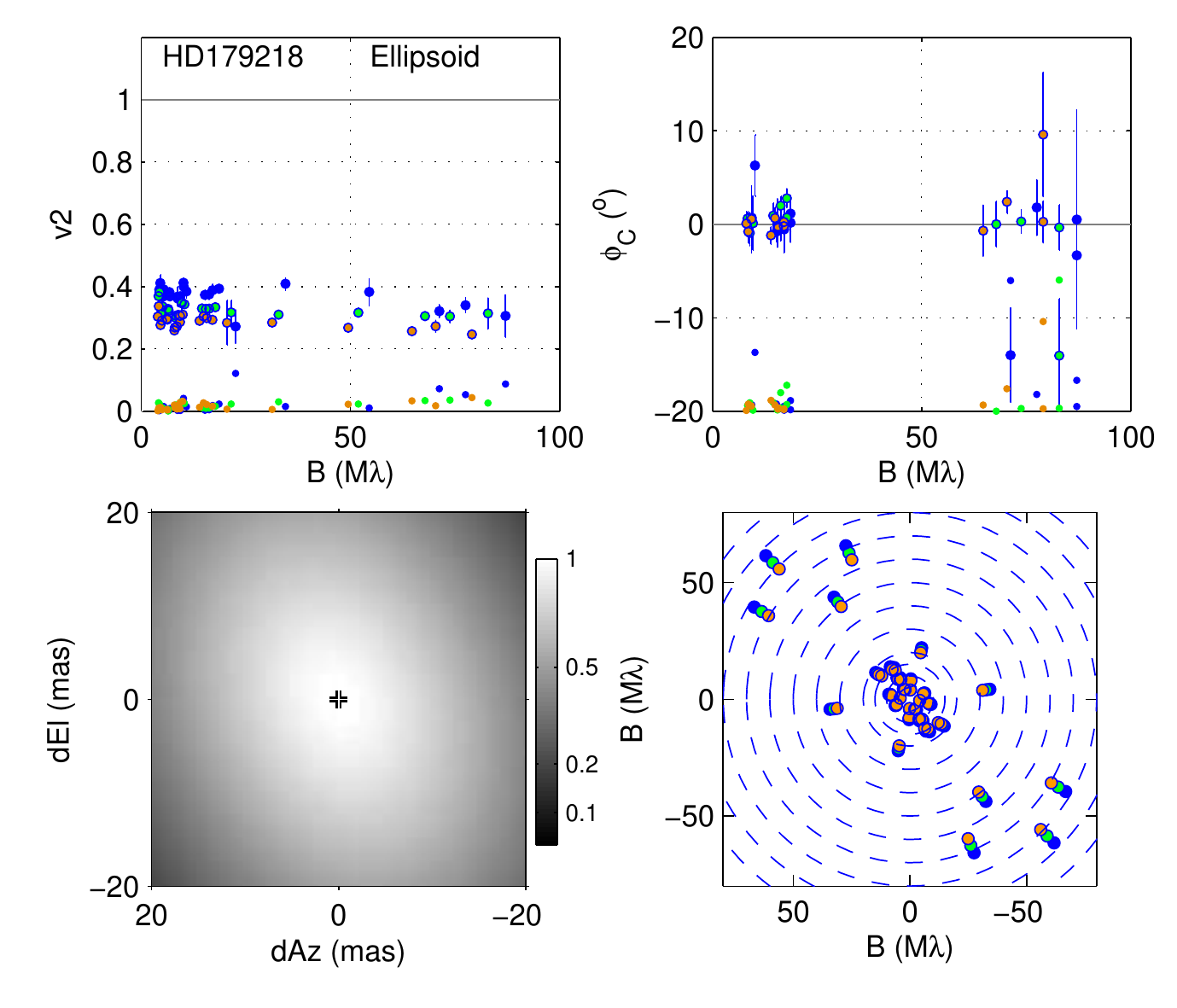}
\caption{Summary plots for the non-\emph{HQ} objects (continued).}
\end{figure}

\end{appendix}

\end{document}